\documentclass[11pt,english]{article}
\usepackage{amsmath}
\usepackage{amsthm}
\usepackage{amssymb}
\usepackage{fontspec}
\usepackage{unicode-math}
\usepackage{color}
\usepackage{verbatim}
\usepackage{zref-clever}
\usepackage{mathtools}
\usepackage{enumitem}
\usepackage{makeidx}
\makeindex
\usepackage{graphicx}
\usepackage[letterpaper]{geometry}
\usepackage{microtype}
\usepackage[pdfusetitle,
 bookmarks=true,bookmarksnumbered=true,bookmarksopen=true,bookmarksopenlevel=3,
 breaklinks=true,pdfborder={0 0 0},pdfborderstyle={},backref=false,colorlinks=true]
 {hyperref}
\hypersetup{
 linktoc=all}

\makeatletter

\providecommand{\tabularnewline}{\\}

\numberwithin{equation}{section}
\numberwithin{figure}{section}
\numberwithin{table}{section}
\newlist{problemparts}{enumerate}{1}
\setlist[problemparts,1]{label=\Alph*.,ref=\Alph*,font=\bfseries}

\@ifundefined{date}{}{\date{}}
\usepackage[margin=1cm]{caption}
\usepackage{tocloft}
\cftsetindents{subsubsection}{4em}{4em}
\DeclareMathOperator{\e}{e}

\DeclareMathOperator{\ii}{i}
\DeclareMathOperator{\im}{Im}

\DeclareMathOperator{\Log}{Log}

\DeclareMathOperator{\re}{Re}

\DeclareMathOperator{\SU}{SU}
\DeclareMathOperator{\tr}{tr}

\usepackage{fontawesome5}
\AtBeginDocument{\setcounter{biburlnumpenalty}{9000}\setcounter{biburlucpenalty}{9000}\setcounter{biburllcpenalty}{9000}}

\makeatother

\providecommand\theoremname{Theorem}
\theoremstyle{plain}
\newtheorem{thm}{\protect\theoremname}[section]
\providecommand\exercisename{Exercise}
\theoremstyle{definition}
\newtheorem{xca}[thm]{\protect\exercisename}
\providecommand\problemname{Problem}
\newtheorem{problem}[thm]{\protect\problemname}
\usepackage{polyglossia}
\setdefaultlanguage[variant=american]{english}
\usepackage[style=numeric-comp,sorting=none,giveninits=false,doi=true,eprint=true,backref=true]{biblatex}
\zcsetup{countertype={thm=theorem}}
\AddToHook{env/xca/begin}{\zcsetup{countertype={thm=xca}}}
\AddToHook{env/problem/begin}{\zcsetup{countertype={thm=problem}}}
\zcRefTypeSetup{problem}{
Name-sg = Problem,
name-sg = problem,
Name-pl = Problems,
name-pl = problems}
\zcRefTypeSetup{xca}{
Name-sg = Exercise,
name-sg = exercise,
Name-pl = Exercises,
name-pl = exercises}

\begin{document}
\begin{comment}
Macros 2026-08-29
\end{comment}

\begin{comment}
Bold Uppercase Latin Letters: \textbackslash (letter)
\end{comment}
\global\long\def\A{\mathbf{A}}%
\global\long\def\B{\mathbf{B}}%
\global\long\def\C{\mathbf{C}}%
\global\long\def\D{\mathbf{D}}%
\global\long\def\E{\mathbf{E}}%
\global\long\def\F{\mathbf{F}}%
\global\long\def\G{\mathbf{G}}%
\global\long\def\H{\mathbf{H}}%
\global\long\def\I{\mathbf{I}}%
\global\long\def\J{\mathbf{J}}%
\global\long\def\K{\mathbf{K}}%
\global\long\def\LL{\mathbf{L}}%
\global\long\def\M{\mathbf{M}}%
\global\long\def\N{\mathbf{N}}%
\global\long\def\OO{\mathbf{O}}%
\global\long\def\P{\mathbf{P}}%
\global\long\def\Q{\mathbf{Q}}%
\global\long\def\RR{\mathbf{R}}%
\global\long\def\SS{\mathbf{S}}%
\global\long\def\T{\mathbf{T}}%
\global\long\def\U{\mathbf{U}}%
\global\long\def\V{\mathbf{V}}%
\global\long\def\W{\mathbf{W}}%
\global\long\def\X{\mathbf{X}}%
\global\long\def\Y{\mathbf{Y}}%
\global\long\def\Z{\mathbf{Z}}%

\begin{comment}
Bold Lowercase Latin Letters: \textbackslash (letter)
\end{comment}
\global\long\def\a{\mathbf{a}}%
\global\long\def\b{\mathbf{b}}%
\global\long\def\c{\mathbf{c}}%
\global\long\def\dd{\mathbf{d}}%
\global\long\def\ee{\mathbf{e}}%
\global\long\def\f{\mathbf{f}}%
\global\long\def\g{\mathbf{g}}%
\global\long\def\h{\mathbf{h}}%
\global\long\def\iii{\mathbf{i}}%
\global\long\def\j{\mathbf{j}}%
\global\long\def\k{\mathbf{k}}%
\global\long\def\l{\boldsymbol{l}}%
\global\long\def\el{\boldsymbol{\ell}}%
\global\long\def\m{\mathbf{m}}%
\global\long\def\n{\mathbf{n}}%
\global\long\def\o{\mathbf{o}}%
\global\long\def\p{\mathbf{p}}%
\global\long\def\q{\mathbf{q}}%
\global\long\def\r{\mathbf{r}}%
\global\long\def\s{\mathbf{s}}%
\global\long\def\t{\mathbf{t}}%
\global\long\def\u{\mathbf{u}}%
\global\long\def\v{\mathbf{v}}%
\global\long\def\w{\mathbf{w}}%
\global\long\def\x{\mathbf{x}}%
\global\long\def\y{\mathbf{y}}%
\global\long\def\z{\mathbf{z}}%

\begin{comment}
Bold Uppercase Greek Letters: \textbackslash (first two characters)
\end{comment}
\global\long\def\Ga{\boldsymbol{\Gamma}}%
\global\long\def\De{\boldsymbol{\Delta}}%
\global\long\def\Th{\boldsymbol{\Theta}}%
\global\long\def\La{\boldsymbol{\Lambda}}%
\global\long\def\Xii{\boldsymbol{\Xi}}%
\global\long\def\Pii{\boldsymbol{\Pi}}%
\global\long\def\Si{\boldsymbol{\Sigma}}%
\global\long\def\Ph{\boldsymbol{\Phi}}%
\global\long\def\Ps{\boldsymbol{\Psi}}%
\global\long\def\Om{\boldsymbol{\Omega}}%

\begin{comment}
Bold Lowercase Greek Letters: \textbackslash (first two characters)
\end{comment}
\global\long\def\al{\boldsymbol{\alpha}}%
\global\long\def\be{\boldsymbol{\beta}}%
\global\long\def\ga{\boldsymbol{\gamma}}%
\global\long\def\de{\boldsymbol{\delta}}%
\global\long\def\ep{\boldsymbol{\epsilon}}%
\global\long\def\vep{\boldsymbol{\varepsilon}}%
\global\long\def\ze{\boldsymbol{\zeta}}%
\global\long\def\et{\boldsymbol{\eta}}%
\global\long\def\th{\boldsymbol{\theta}}%
\global\long\def\io{\boldsymbol{\iota}}%
\global\long\def\ka{\boldsymbol{\kappa}}%
\global\long\def\la{\boldsymbol{\lambda}}%
\global\long\def\muu{\boldsymbol{\mu}}%
\global\long\def\nuu{\boldsymbol{\nu}}%
\global\long\def\xii{\boldsymbol{\xi}}%
\global\long\def\pii{\boldsymbol{\pi}}%
\global\long\def\rhh{\boldsymbol{\rho}}%
\global\long\def\si{\boldsymbol{\sigma}}%
\global\long\def\ta{\boldsymbol{\tau}}%
\global\long\def\ups{\boldsymbol{\upsilon}}%
\global\long\def\ph{\boldsymbol{\phi}}%
\global\long\def\vph{\boldsymbol{\varphi}}%
\global\long\def\ch{\boldsymbol{\chi}}%
\global\long\def\ps{\boldsymbol{\psi}}%
\global\long\def\om{\boldsymbol{\omega}}%

\begin{comment}
Bold Calligraphic: \textbackslash (letter)(letter)b
\end{comment}
\global\long\def\AAb{\boldsymbol{\mathcal{A}}}%
\global\long\def\BBb{\boldsymbol{\mathcal{B}}}%
\global\long\def\CCb{\boldsymbol{\mathcal{C}}}%
\global\long\def\DDb{\boldsymbol{\mathcal{D}}}%
\global\long\def\EEb{\boldsymbol{\mathcal{E}}}%
\global\long\def\FFb{\boldsymbol{\mathcal{F}}}%
\global\long\def\GGb{\boldsymbol{\mathcal{G}}}%
\global\long\def\HHb{\boldsymbol{\mathcal{H}}}%
\global\long\def\IIb{\boldsymbol{\mathcal{I}}}%
\global\long\def\JJb{\boldsymbol{\mathcal{J}}}%
\global\long\def\KKb{\boldsymbol{\mathcal{K}}}%
\global\long\def\LLb{\boldsymbol{\mathcal{L}}}%
\global\long\def\MMb{\boldsymbol{\mathcal{M}}}%
\global\long\def\NNb{\boldsymbol{\mathcal{N}}}%
\global\long\def\OOb{\boldsymbol{\mathcal{O}}}%
\global\long\def\PPb{\boldsymbol{\mathcal{P}}}%
\global\long\def\QQb{\boldsymbol{\mathcal{Q}}}%
\global\long\def\RRb{\boldsymbol{\mathcal{R}}}%
\global\long\def\SSb{\boldsymbol{\mathcal{S}}}%
\global\long\def\TTb{\boldsymbol{\mathcal{T}}}%
\global\long\def\UUb{\boldsymbol{\mathcal{U}}}%
\global\long\def\VVb{\boldsymbol{\mathcal{V}}}%
\global\long\def\WWb{\boldsymbol{\mathcal{W}}}%
\global\long\def\XXb{\boldsymbol{\mathcal{X}}}%
\global\long\def\YYb{\boldsymbol{\mathcal{Y}}}%
\global\long\def\ZZb{\boldsymbol{\mathcal{Z}}}%

\begin{comment}
Bar Uppercase Latin Letters: \textbackslash (letter)b
\end{comment}
\global\long\def\Ab{\bar{A}}%
\global\long\def\Bb{\bar{B}}%
\global\long\def\Cb{\bar{C}}%
\global\long\def\Db{\bar{D}}%
\global\long\def\Eb{\bar{E}}%
\global\long\def\Fb{\bar{F}}%
\global\long\def\Gb{\bar{G}}%
\global\long\def\Hb{\bar{H}}%
\global\long\def\Ib{\bar{I}}%
\global\long\def\Jb{\bar{J}}%
\global\long\def\Kb{\bar{K}}%
\global\long\def\Lb{\bar{L}}%
\global\long\def\Mb{\bar{M}}%
\global\long\def\Nb{\bar{N}}%
\global\long\def\Ob{\bar{O}}%
\global\long\def\Pb{\bar{P}}%
\global\long\def\Qb{\bar{Q}}%
\global\long\def\Rb{\bar{R}}%
\global\long\def\Sb{\bar{S}}%
\global\long\def\Tb{\bar{T}}%
\global\long\def\Ub{\bar{U}}%
\global\long\def\Vb{\bar{V}}%
\global\long\def\Wb{\bar{W}}%
\global\long\def\Xb{\bar{X}}%
\global\long\def\Yb{\bar{Y}}%
\global\long\def\Zb{\bar{Z}}%

\begin{comment}
Bar Lowercase Latin Letters: \textbackslash (letter)b
\end{comment}
\global\long\def\ab{\bar{a}}%
\global\long\def\bb{\bar{b}}%
\global\long\def\cb{\bar{c}}%
\global\long\def\db{\bar{d}}%
\global\long\def\eb{\bar{e}}%
\global\long\def\fb{\bar{f}}%
\global\long\def\gb{\bar{g}}%
\global\long\def\hb{\bar{h}}%
\global\long\def\ib{\bar{i}}%
\global\long\def\jb{\bar{j}}%
\global\long\def\kb{\bar{k}}%
\global\long\def\lb{\bar{l}}%
\global\long\def\elb{\bar{\ell}}%
\global\long\def\mb{\bar{m}}%
\global\long\def\nb{\bar{n}}%
\global\long\def\ob{\bar{o}}%
\global\long\def\pb{\bar{p}}%
\global\long\def\qb{\bar{q}}%
\global\long\def\rb{\bar{r}}%
\global\long\def\ssb{\bar{s}}%
\global\long\def\tb{\bar{t}}%
\global\long\def\ub{\bar{u}}%
\global\long\def\vb{\bar{v}}%
\global\long\def\wb{\bar{w}}%
\global\long\def\xb{\bar{x}}%
\global\long\def\yb{\bar{y}}%
\global\long\def\zb{\bar{z}}%

\begin{comment}
Bar Uppercase Greek Letters: \textbackslash (first two characters)b
\end{comment}
\global\long\def\Gab{\bar{\Gamma}}%
\global\long\def\Deb{\bar{\Delta}}%
\global\long\def\Thb{\bar{\Theta}}%
\global\long\def\Lab{\bar{\Lambda}}%
\global\long\def\Xib{\bar{\Xi}}%
\global\long\def\Pib{\bar{\Pi}}%
\global\long\def\Sib{\bar{\Sigma}}%
\global\long\def\Phb{\bar{\Phi}}%
\global\long\def\Psb{\bar{\Psi}}%
\global\long\def\Thb{\bar{\Theta}}%

\begin{comment}
Bar Lowercase Greek Letters: \textbackslash (first two characters)b
\end{comment}
\global\long\def\alb{\bar{\alpha}}%
\global\long\def\beb{\bar{\beta}}%
\global\long\def\gab{\bar{\gamma}}%
\global\long\def\deb{\bar{\delta}}%
\global\long\def\epb{\bar{\epsilon}}%
\global\long\def\vepb{\bar{\varepsilon}}%
\global\long\def\zeb{\bar{\zeta}}%
\global\long\def\etb{\bar{\eta}}%
\global\long\def\thb{\bar{\theta}}%
\global\long\def\iob{\bar{\iota}}%
\global\long\def\kab{\bar{\kappa}}%
\global\long\def\lab{\bar{\lambda}}%
\global\long\def\mub{\bar{\mu}}%
\global\long\def\nub{\bar{\nu}}%
\global\long\def\xib{\bar{\xi}}%
\global\long\def\pib{\bar{\pi}}%
\global\long\def\rhb{\bar{\rho}}%
\global\long\def\sib{\bar{\sigma}}%
\global\long\def\tab{\bar{\tau}}%
\global\long\def\upb{\bar{\upsilon}}%
\global\long\def\phb{\bar{\phi}}%
\global\long\def\vphb{\bar{\varphi}}%
\global\long\def\chb{\bar{\chi}}%
\global\long\def\psb{\bar{\psi}}%
\global\long\def\omb{\bar{\omega}}%

\begin{comment}
Dot \& Double Dot Lowercase Latin Letters: \textbackslash (letter)d or dd
\end{comment}
\global\long\def\adt{\dot{a}}%
\global\long\def\add{\ddot{a}}%
\global\long\def\bd{\dot{b}}%
\global\long\def\bdd{\ddot{b}}%
\global\long\def\cd{\dot{c}}%
\global\long\def\cdd{\ddot{c}}%
\global\long\def\ddd{\dot{d}}%
\global\long\def\dddd{\ddot{d}}%
\global\long\def\ed{\dot{e}}%
\global\long\def\edd{\ddot{e}}%
\global\long\def\fd{\dot{f}}%
\global\long\def\fdd{\ddot{f}}%
\global\long\def\gd{\dot{g}}%
\global\long\def\gdd{\ddot{g}}%
\global\long\def\hd{\dot{h}}%
\global\long\def\hdd{\ddot{h}}%
\global\long\def\kd{\dot{k}}%
\global\long\def\kdd{\ddot{k}}%
\global\long\def\ld{\dot{l}}%
\global\long\def\ldd{\ddot{l}}%
\global\long\def\eld{\dot{\ell}}%
\global\long\def\eldd{\ddot{\ell}}%
\global\long\def\md{\dot{m}}%
\global\long\def\mdd{\ddot{m}}%
\global\long\def\nd{\dot{n}}%
\global\long\def\ndd{\ddot{n}}%
\global\long\def\od{\dot{o}}%
\global\long\def\odd{\ddot{o}}%
\global\long\def\pd{\dot{p}}%
\global\long\def\pdd{\ddot{p}}%
\global\long\def\qd{\dot{q}}%
\global\long\def\qdd{\ddot{q}}%
\global\long\def\rd{\dot{r}}%
\global\long\def\rdd{\ddot{r}}%
\global\long\def\sd{\dot{s}}%
\global\long\def\sdd{\ddot{s}}%
\global\long\def\td{\dot{t}}%
\global\long\def\tdd{\ddot{t}}%
\global\long\def\ud{\dot{u}}%
\global\long\def\udd{\ddot{u}}%
\global\long\def\vd{\dot{v}}%
\global\long\def\vdd{\ddot{v}}%
\global\long\def\wdt{\dot{w}}%
\global\long\def\wdd{\ddot{w}}%
\global\long\def\xd{\dot{x}}%
\global\long\def\xdd{\ddot{x}}%
\global\long\def\yd{\dot{y}}%
\global\long\def\ydd{\ddot{y}}%
\global\long\def\zd{\dot{z}}%
\global\long\def\zdd{\ddot{z}}%

\begin{comment}
Dot \& Double Dot Uppercase Latin Letters: \textbackslash (letter)d or dd
\end{comment}
\global\long\def\Adt{\dot{A}}%
\global\long\def\Add{\ddot{A}}%
\global\long\def\Bd{\dot{B}}%
\global\long\def\Bdd{\ddot{B}}%
\global\long\def\Cd{\dot{C}}%
\global\long\def\Cdd{\ddot{C}}%
\global\long\def\Dd{\dot{D}}%
\global\long\def\Ddd{\ddot{D}}%
\global\long\def\Ed{\dot{E}}%
\global\long\def\Edd{\ddot{E}}%
\global\long\def\Fd{\dot{F}}%
\global\long\def\Fdd{\ddot{F}}%
\global\long\def\Gd{\dot{G}}%
\global\long\def\Gdd{\ddot{G}}%
\global\long\def\Hd{\dot{H}}%
\global\long\def\Hdd{\ddot{H}}%
\global\long\def\Id{\dot{I}}%
\global\long\def\Idd{\ddot{I}}%
\global\long\def\Jd{\dot{J}}%
\global\long\def\Jdd{\ddot{J}}%
\global\long\def\Kd{\dot{K}}%
\global\long\def\Kdd{\ddot{K}}%
\global\long\def\Ld{\dot{L}}%
\global\long\def\Ldd{\ddot{L}}%
\global\long\def\Md{\dot{M}}%
\global\long\def\Mdd{\ddot{M}}%
\global\long\def\Nd{\dot{N}}%
\global\long\def\Ndd{\ddot{N}}%
\global\long\def\Od{\dot{O}}%
\global\long\def\Odd{\ddot{O}}%
\global\long\def\Pd{\dot{P}}%
\global\long\def\Pdd{\ddot{P}}%
\global\long\def\Qd{\dot{Q}}%
\global\long\def\Qdd{\ddot{Q}}%
\global\long\def\Rd{\dot{R}}%
\global\long\def\Rdd{\ddot{R}}%
\global\long\def\Sd{\dot{S}}%
\global\long\def\Sdd{\ddot{S}}%
\global\long\def\Td{\dot{T}}%
\global\long\def\Tdd{\ddot{T}}%
\global\long\def\Ud{\dot{U}}%
\global\long\def\Udd{\ddot{U}}%
\global\long\def\Vd{\dot{R}}%
\global\long\def\Vdd{\ddot{R}}%
\global\long\def\Wd{\dot{W}}%
\global\long\def\Wdd{\ddot{W}}%
\global\long\def\Xd{\dot{X}}%
\global\long\def\Xdd{\ddot{X}}%
\global\long\def\Yd{\dot{Y}}%
\global\long\def\Ydd{\ddot{Y}}%
\global\long\def\Zd{\dot{Z}}%
\global\long\def\Zdd{\ddot{Z}}%

\begin{comment}
Dot \& Double Dot Uppercase Greek Letters: \textbackslash (first two characters)d or dd
\end{comment}
\global\long\def\Gad{\dot{\Gamma}}%
\global\long\def\Gadd{\ddot{\Gamma}}%
\global\long\def\Ded{\dot{\Delta}}%
\global\long\def\Dedd{\ddot{\Delta}}%
\global\long\def\Thd{\dot{\Theta}}%
\global\long\def\Thdd{\ddot{\Theta}}%
\global\long\def\Lad{\dot{\Lambda}}%
\global\long\def\Ladd{\ddot{\Lambda}}%
\global\long\def\Xid{\dot{\Xi}}%
\global\long\def\Xidd{\ddot{\Xi}}%
\global\long\def\Pid{\dot{\Pi}}%
\global\long\def\Pidd{\ddot{\Pi}}%
\global\long\def\Sid{\dot{\Sigma}}%
\global\long\def\Sidd{\ddot{\Sigma}}%
\global\long\def\Phd{\dot{\Phi}}%
\global\long\def\Phdd{\ddot{\Phi}}%
\global\long\def\Psd{\dot{\Psi}}%
\global\long\def\Psdd{\ddot{\Psi}}%
\global\long\def\Thd{\dot{\Theta}}%
\global\long\def\Thdd{\ddot{\Theta}}%

\begin{comment}
Dot \& Double Dot Lowercase Greek Letters: \textbackslash (first two characters)d or dd
\end{comment}
\global\long\def\ald{\dot{\alpha}}%
\global\long\def\aldd{\ddot{\alpha}}%
\global\long\def\bed{\dot{\beta}}%
\global\long\def\bedd{\ddot{\beta}}%
\global\long\def\gad{\dot{\gamma}}%
\global\long\def\gadd{\ddot{\gamma}}%
\global\long\def\ded{\dot{\delta}}%
\global\long\def\dedd{\ddot{\delta}}%
\global\long\def\epd{\dot{\epsilon}}%
\global\long\def\epdd{\ddot{\epsilon}}%
\global\long\def\vepd{\dot{\varepsilon}}%
\global\long\def\vepdd{\ddot{\varepsilon}}%
\global\long\def\zed{\dot{\zeta}}%
\global\long\def\zedd{\ddot{\zeta}}%
\global\long\def\etd{\dot{\eta}}%
\global\long\def\etdd{\ddot{\eta}}%
\global\long\def\thd{\dot{\theta}}%
\global\long\def\thdd{\ddot{\theta}}%
\global\long\def\iod{\dot{\iota}}%
\global\long\def\iodd{\ddot{\iota}}%
\global\long\def\kad{\dot{\kappa}}%
\global\long\def\kadd{\ddot{\kappa}}%
\global\long\def\lad{\dot{\lambda}}%
\global\long\def\ladd{\ddot{\lambda}}%
\global\long\def\mud{\dot{\mu}}%
\global\long\def\mudd{\ddot{\mu}}%
\global\long\def\nud{\dot{\nu}}%
\global\long\def\nudd{\ddot{\nu}}%
\global\long\def\xid{\dot{\xi}}%
\global\long\def\xidd{\ddot{\xi}}%
\global\long\def\pid{\dot{\pi}}%
\global\long\def\pidd{\ddot{\pi}}%
\global\long\def\rhod{\dot{\rho}}%
\global\long\def\rhodd{\ddot{\rho}}%
\global\long\def\sid{\dot{\sigma}}%
\global\long\def\sidd{\ddot{\sigma}}%
\global\long\def\tad{\dot{\tau}}%
\global\long\def\tadd{\ddot{\tau}}%
\global\long\def\upd{\dot{\upsilon}}%
\global\long\def\updd{\ddot{\upsilon}}%
\global\long\def\phd{\dot{\phi}}%
\global\long\def\phdd{\ddot{\phi}}%
\global\long\def\vpd{\dot{\varphi}}%
\global\long\def\vpdd{\ddot{\varphi}}%
\global\long\def\chd{\dot{\chi}}%
\global\long\def\chdd{\ddot{\chi}}%
\global\long\def\psd{\dot{\psi}}%
\global\long\def\psdd{\ddot{\psi}}%
\global\long\def\omd{\dot{\omega}}%
\global\long\def\omdd{\ddot{\omega}}%

\begin{comment}
Dot \& Double Dot Bold Letters
\end{comment}
\global\long\def\xxd{\dot{\mathbf{x}}}%
\global\long\def\xxdd{\ddot{\mathbf{x}}}%
\global\long\def\vvd{\dot{\mathbf{v}}}%

\begin{comment}
Blackboard: \textbackslash BB(letter)
\end{comment}
\global\long\def\BBA{\mathbb{A}}%
\global\long\def\BBB{\mathbb{B}}%
\global\long\def\BBC{\mathbb{C}}%
\global\long\def\BBD{\mathbb{D}}%
\global\long\def\BBE{\mathbb{E}}%
\global\long\def\BBF{\mathbb{F}}%
\global\long\def\BBG{\mathbb{G}}%
\global\long\def\BBH{\mathbb{H}}%
\global\long\def\BBI{\mathbb{I}}%
\global\long\def\BBJ{\mathbb{J}}%
\global\long\def\BBK{\mathbb{K}}%
\global\long\def\BBL{\mathbb{L}}%
\global\long\def\BBM{\mathbb{M}}%
\global\long\def\BBN{\mathbb{N}}%
\global\long\def\BBO{\mathbb{O}}%
\global\long\def\BBP{\mathbb{P}}%
\global\long\def\BBQ{\mathbb{Q}}%
\global\long\def\BBR{\mathbb{R}}%
\global\long\def\BBS{\mathbb{S}}%
\global\long\def\BBT{\mathbb{T}}%
\global\long\def\BBU{\mathbb{U}}%
\global\long\def\BBV{\mathbb{V}}%
\global\long\def\BBW{\mathbb{W}}%
\global\long\def\BBX{\mathbb{X}}%
\global\long\def\BBY{\mathbb{Y}}%
\global\long\def\BBZ{\mathbb{Z}}%

\begin{comment}
Calligraphic: \textbackslash (letter)(letter)
\end{comment}
\global\long\def\AA{\mathcal{A}}%
\global\long\def\BB{\mathcal{B}}%
\global\long\def\CC{\mathcal{C}}%
\global\long\def\DD{\mathcal{D}}%
\global\long\def\EE{\mathcal{E}}%
\global\long\def\FF{\mathcal{F}}%
\global\long\def\GG{\mathcal{G}}%
\global\long\def\HH{\mathcal{H}}%
\global\long\def\II{\mathcal{I}}%
\global\long\def\JJ{\mathcal{J}}%
\global\long\def\KK{\mathcal{K}}%
\global\long\def\LLL{\mathcal{L}}%
\global\long\def\MM{\mathcal{M}}%
\global\long\def\NN{\mathcal{N}}%
\global\long\def\OOO{\mathcal{O}}%
\global\long\def\PP{\mathcal{P}}%
\global\long\def\QQ{\mathcal{Q}}%
\global\long\def\RRR{\mathcal{R}}%
\global\long\def\SSS{\mathcal{S}}%
\global\long\def\TT{\mathcal{T}}%
\global\long\def\UU{\mathcal{U}}%
\global\long\def\VV{\mathcal{V}}%
\global\long\def\WW{\mathcal{W}}%
\global\long\def\XX{\mathcal{X}}%
\global\long\def\YY{\mathcal{Y}}%
\global\long\def\ZZ{\mathcal{Z}}%

\begin{comment}
Tilde Uppercase Latin Letters: \textbackslash (letter)t
\end{comment}
\global\long\def\At{\tilde{A}}%
\global\long\def\Bt{\tilde{B}}%
\global\long\def\Ct{\tilde{C}}%
\global\long\def\Dt{\tilde{D}}%
\global\long\def\Et{\tilde{E}}%
\global\long\def\Ft{\tilde{F}}%
\global\long\def\Gt{\tilde{G}}%
\global\long\def\Ht{\tilde{H}}%
\global\long\def\It{\tilde{I}}%
\global\long\def\Jt{\tilde{J}}%
\global\long\def\Kt{\tilde{K}}%
\global\long\def\Lt{\tilde{L}}%
\global\long\def\Mt{\tilde{M}}%
\global\long\def\Nt{\tilde{N}}%
\global\long\def\Ot{\tilde{O}}%
\global\long\def\Pt{\tilde{P}}%
\global\long\def\Qt{\tilde{Q}}%
\global\long\def\Rt{\tilde{R}}%
\global\long\def\St{\tilde{S}}%
\global\long\def\Tt{\tilde{T}}%
\global\long\def\Ut{\tilde{U}}%
\global\long\def\Vt{\tilde{V}}%
\global\long\def\Wt{\tilde{W}}%
\global\long\def\Xt{\tilde{X}}%
\global\long\def\Yt{\tilde{Y}}%
\global\long\def\Zt{\tilde{Z}}%

\begin{comment}
Tilde Lowercase Latin Letters: \textbackslash (letter)t
\end{comment}
\global\long\def\at{\tilde{a}}%
\global\long\def\bt{\tilde{b}}%
\global\long\def\ct{\tilde{c}}%
\global\long\def\dt{\tilde{d}}%
\global\long\def\eet{\tilde{e}}%
\global\long\def\ft{\tilde{f}}%
\global\long\def\gt{\tilde{g}}%
\global\long\def\hht{\tilde{h}}%
\global\long\def\it{\tilde{i}}%
\global\long\def\jt{\tilde{j}}%
\global\long\def\kt{\tilde{k}}%
\global\long\def\lt{\tilde{l}}%
\global\long\def\elt{\tilde{\ell}}%
\global\long\def\mt{\tilde{m}}%
\global\long\def\nt{\tilde{n}}%
\global\long\def\ot{\tilde{o}}%
\global\long\def\pt{\tilde{p}}%
\global\long\def\qt{\tilde{q}}%
\global\long\def\rt{\tilde{r}}%
\global\long\def\st{\tilde{s}}%
\global\long\def\tt{\tilde{t}}%
\global\long\def\ut{\tilde{u}}%
\global\long\def\vt{\tilde{v}}%
\global\long\def\wt{\tilde{w}}%
\global\long\def\xt{\tilde{x}}%
\global\long\def\yt{\tilde{y}}%
\global\long\def\zt{\tilde{z}}%

\begin{comment}
Fraktur: \textbackslash mf(letter)
\end{comment}
\global\long\def\mfA{\mathfrak{A}}%
\global\long\def\mfB{\mathfrak{B}}%
\global\long\def\mfC{\mathfrak{C}}%
\global\long\def\mfD{\mathfrak{D}}%
\global\long\def\mfE{\mathfrak{E}}%
\global\long\def\mfF{\mathfrak{F}}%
\global\long\def\mfG{\mathfrak{G}}%
\global\long\def\mfH{\mathfrak{H}}%
\global\long\def\mfI{\mathfrak{I}}%
\global\long\def\mfJ{\mathfrak{J}}%
\global\long\def\mfK{\mathfrak{K}}%
\global\long\def\mfL{\mathfrak{L}}%
\global\long\def\mfM{\mathfrak{M}}%
\global\long\def\mfN{\mathfrak{N}}%
\global\long\def\mfO{\mathfrak{O}}%
\global\long\def\mfP{\mathfrak{P}}%
\global\long\def\mfQ{\mathfrak{Q}}%
\global\long\def\mfR{\mathfrak{R}}%
\global\long\def\mfS{\mathfrak{S}}%
\global\long\def\mfT{\mathfrak{T}}%
\global\long\def\mfU{\mathfrak{U}}%
\global\long\def\mfV{\mathfrak{V}}%
\global\long\def\mfW{\mathfrak{W}}%
\global\long\def\mfX{\mathfrak{X}}%
\global\long\def\mfY{\mathfrak{Y}}%
\global\long\def\mfZ{\mathfrak{Z}}%
\global\long\def\mfa{\mathfrak{a}}%
\global\long\def\mfb{\mathfrak{b}}%
\global\long\def\mfc{\mathfrak{c}}%
\global\long\def\mfd{\mathfrak{d}}%
\global\long\def\mfe{\mathfrak{e}}%
\global\long\def\mff{\mathfrak{f}}%
\global\long\def\mfg{\mathfrak{g}}%
\global\long\def\mfh{\mathfrak{h}}%
\global\long\def\mfi{\mathfrak{i}}%
\global\long\def\mfj{\mathfrak{j}}%
\global\long\def\mfk{\mathfrak{k}}%
\global\long\def\mfl{\mathfrak{l}}%
\global\long\def\mfm{\mathfrak{m}}%
\global\long\def\mfn{\mathfrak{n}}%
\global\long\def\mfo{\mathfrak{o}}%
\global\long\def\mfp{\mathfrak{p}}%
\global\long\def\mfq{\mathfrak{q}}%
\global\long\def\mfr{\mathfrak{r}}%
\global\long\def\mfs{\mathfrak{s}}%
\global\long\def\mft{\mathfrak{t}}%
\global\long\def\mfu{\mathfrak{u}}%
\global\long\def\mfv{\mathfrak{v}}%
\global\long\def\mfw{\mathfrak{w}}%
\global\long\def\mfx{\mathfrak{x}}%
\global\long\def\mfy{\mathfrak{y}}%
\global\long\def\mfz{\mathfrak{z}}%

\begin{comment}
Roman
\end{comment}
\global\long\def\d{\mathrm{d}}%
\global\long\def\DDD{\mathrm{D}}%
\global\long\def\EEE{\mathrm{E}}%
\global\long\def\i{\ii}%
\global\long\def\MMM{\mathrm{M}}%
\global\long\def\OOOO{\mathrm{O}}%
\global\long\def\RRRR{\mathrm{R}}%
\global\long\def\TTT{\mathrm{T}}%
\global\long\def\UUU{\mathrm{U}}%

\begin{comment}
Hat
\end{comment}
\global\long\def\hx{\hat{x}}%
\global\long\def\hp{\hat{p}}%
\global\long\def\hxx{\hat{\mathbf{x}}}%
\global\long\def\hvv{\hat{\mathbf{v}}}%

\begin{comment}
Lie Groups \& Algebras
\end{comment}
\global\long\def\GL{\mathrm{GL}}%
\global\long\def\ISU{\mathrm{ISU}}%
\global\long\def\ISUT{\mathrm{ISU}\left(2\right)}%
\global\long\def\SL{\mathrm{SL}}%
\global\long\def\SO{\mathrm{SO}}%
\global\long\def\SOH{\mathrm{SO}\left(3\right)}%
\global\long\def\SOT{\mathrm{SO}\left(2\right)}%
\global\long\def\Sp{\mathrm{Sp}}%
\global\long\def\SU{\mathrm{SU}}%
\global\long\def\SUT{\mathrm{SU}\left(2\right)}%
\global\long\def\UO{\mathrm{U}\left(1\right)}%
\global\long\def\gl{\mathfrak{gl}}%
\global\long\def\sl{\mathfrak{sl}}%
\global\long\def\sso{\mathfrak{so}}%
\global\long\def\soh{\mathfrak{so}\left(3\right)}%
\global\long\def\su{\mathfrak{su}}%
\global\long\def\sut{\mathfrak{su}\left(2\right)}%
\global\long\def\isut{\mathfrak{isu}\left(2\right)}%

\begin{comment}
Arrows
\end{comment}
\global\long\def\so{\Rightarrow}%
\global\long\def\os{\Leftarrow}%
\global\long\def\to{\rightarrow}%
\global\long\def\ot{\leftarrow}%
\global\long\def\soo{\Longrightarrow}%
\global\long\def\oos{\Longleftarrow}%
\global\long\def\too{\longrightarrow}%
\global\long\def\oot{\longleftarrow}%
\global\long\def\sos{\Leftrightarrow}%
\global\long\def\tot{\leftrightarrow}%
\global\long\def\soos{\Longleftrightarrow}%
\global\long\def\toot{\longleftrightarrow}%
\global\long\def\mt{\mapsto}%
\global\long\def\mtt{\longmapsto}%
\global\long\def\dn{\downarrow}%
\global\long\def\up{\uparrow}%
\global\long\def\updn{\updownarrow}%
\global\long\def\sea{\searrow}%
\global\long\def\nea{\nearrow}%
\global\long\def\nwa{\nwarrow}%
\global\long\def\swa{\swarrow}%
\global\long\def\soosp{\quad\Longrightarrow\quad}%
\global\long\def\oossp{\quad\Longleftarrow\quad}%
\global\long\def\soossp{\quad\Longleftrightarrow\quad}%

\begin{comment}
Multiline Brackets
\end{comment}
\global\long\def\multibrl#1{\left(#1\right.}%
\global\long\def\multibrr#1{\left.#1\right)}%
\global\long\def\multisql#1{\left[#1\right.}%
\global\long\def\multisqr#1{\left.#1\right]}%
\global\long\def\multicul#1{\left\{  #1\right.}%
\global\long\def\multicur#1{\left.#1\right\}  }%

\begin{comment}
Fractions
\end{comment}
\global\long\def\hf{\frac{1}{2}}%
\global\long\def\trd{\frac{1}{3}}%
\global\long\def\fr{\frac{1}{4}}%
\global\long\def\ff{\frac{1}{5}}%
\global\long\def\sxt{\frac{1}{6}}%
\global\long\def\sv{\frac{1}{7}}%
\global\long\def\ei{\frac{1}{8}}%
\global\long\def\nt{\frac{1}{9}}%
\global\long\def\hfp{\frac{\pi}{2}}%
\global\long\def\fp{\frac{\pi}{4}}%

\begin{comment}
Vertical Lines
\end{comment}
\global\long\def\bl{\big|}%
\global\long\def\bll{\Big|}%
\global\long\def\blll{\bigg|}%
\global\long\def\bllll{\Bigg|}%

\begin{comment}
Middle Bar
\end{comment}
\global\long\def\ma#1#2{\left\langle #1\thinspace\middle|\thinspace#2\right\rangle }%
\global\long\def\mma#1#2#3{\left\langle #1\thinspace\middle|\thinspace#2\thinspace\middle|\thinspace#3\right\rangle }%
\global\long\def\mc#1#2{\left\{  #1\thinspace\middle|\thinspace#2\right\}  }%
\global\long\def\mmc#1#2#3{\left\{  #1\thinspace\middle|\thinspace#2\thinspace\middle|\thinspace#3\right\}  }%
\global\long\def\mr#1#2{\left(#1\thinspace\middle|\thinspace#2\right) }%
\global\long\def\mmr#1#2#3{\left(#1\thinspace\middle|\thinspace#2\thinspace\middle|\thinspace#3\right)}%

\begin{comment}
Misc
\end{comment}
\global\long\def\pr{\parallel}%
\global\long\def\xx{\times}%
\global\long\def\dg{\lyxmathsym{\textdegree}}%
\global\long\def\sp{,\qquad}%
\global\long\def\sq{\square}%
\global\long\def\pt{\propto}%
\global\long\def\lrc{\lrcorner\thinspace}%
\global\long\def\pexp{\overrightarrow{\exp}}%
\global\long\def\dui#1#2#3{#1_{#2}{}^{#3}}%
\global\long\def\udi#1#2#3{#1^{#2}{}_{#3}}%
\global\long\def\pab{\bar{\partial}}%
\global\long\def\zr{\mathbf{0}}%
\global\long\def\on{\mathbf{1}}%
\global\long\def\na{\boldsymbol{\nabla}}%
\global\long\def\hf{\frac{1}{2}}%
\global\long\def\trd{\frac{1}{3}}%
\global\long\def\fr{\frac{1}{4}}%
\global\long\def\ei{\frac{1}{8}}%
\global\long\def\ap{\approx}%
\global\long\def\eqm{\overset{?}{=}}%
\global\long\def\fa{\forall}%
\global\long\def\ex{\exists}%
\global\long\def\ept{\tilde{\epsilon}}%
\global\long\def\sci#1#2#3{\unit[#1\xx10^{#2}]{#3}}%

\title{“Thinking Quantum”: Lectures on Quantum Theory}
\author{\textbf{Barak Shoshany}{\small}\\
{\small\faIcon{envelope} \href{mailto:bshoshany@brocku.ca}{bshoshany@brocku.ca}$\qquad$\faIcon{orcid} \href{https://orcid.org/0000-0003-2222-127X}{0000-0003-2222-127X}$\qquad$\faIcon{globe} \href{https://baraksh.com/}{https://baraksh.com/}}\\
{\small\faIcon{university} \href{https://brocku.ca/}{Department of Physics, Brock University}}\\
{\small\faIcon{map-marker-alt} \href{https://goo.gl/maps/qscBMigohESxxczM7}{1812 Sir Isaac Brock Way, St. Catharines, Ontario, L2S 3A1, Canada}}}
\maketitle
\begin{abstract}
We present a conceptually clear introduction to quantum theory, deriving the theory from scratch from the point of view of quantum information, with focus on developing intuition and dispelling misconceptions. Different subsets of these lectures were taught to a wide variety of audiences, including exceptional high-school students in the International Summer School for Young Physicists (ISSYP) at Perimeter Institute, 2nd-year physics undergraduates at the University of Toronto, and 4th-year physics and math undergraduate and graduate students at Brock University. The lectures are completely self-contained, including all the necessary mathematical background: complex numbers, linear algebra, and probability theory. They cover topics such as the axioms of quantum theory, qubits, superposition, entanglement, the uncertainty principle, unitary transformations and evolution, quantum gates, projective measurements, Bell's theorem, interpretations of quantum mechanics, the no-cloning theorem, quantum teleportation, quantum algorithms, density operators, mixed states, von Neumann entropy, the Bloch sphere, quantum decoherence, Hamiltonians, the Schrödinger equation, canonical quantization, quantum harmonic oscillators, wavefunctions, and much more. The lectures also contain 111 proof-based problems and 51 computational exercises.
\end{abstract}
\textbf{Author's note:} These lecture notes will soon be published as a textbook by World Scientific, titled “Thinking Quantum”! The book will contain plenty of additional content, as well as supplementary materials such as solutions to problems and exercises. This is the last version to be uploaded to arXiv; any further corrections or additions will be made only as errata or additional supplements for the book. If you enjoyed these notes, please consider purchasing the book, or ask your library to make it available to you.

\tableofcontents{}

\listoffigures

\section{Introduction}

\subsection{Outline}

These lecture notes were developed between 2018 and 2026 at Perimeter Institute for Theoretical Physics, the University of Toronto, and Brock University. Different subsets of these lectures were taught to a wide variety of audiences, from exceptional high-school students to graduate students, as detailed in the abstract. However, the current version is generally best suited for upper-year physics undergraduates, and aims to serve as good preparation for graduate-level study and research.

My goal is that by the end of these notes, you will gain a deep and intuitive understanding of the foundations of quantum theory, from the modern point of view of 21st-century theoretical physics – as it is currently understood by researchers in cutting-edge fields such as quantum foundations, quantum information, quantum computation, quantum field theory, and quantum gravity. Such an understanding will be absolutely crucial if you want to be a theorist, and will also be extremely useful if you want to be an experimentalist.

Before you begin reading these notes, you should forget everything you learned about quantum mechanics previously! We will re-learn quantum theory from scratch, developing it in an axiomatic and mathematically rigorous way from first principles. We will see that there is nothing particularly complex or mysterious about quantum mechanics (with the crucial exception of its philosophical interpretations), and obtain insight that will allow us to understand how it makes the universe work at the most fundamental level.

These notes were explicitly designed to be complete and self-contained; any required concepts in mathematics or physics will be defined within the notes. This means you do not need any specific background, except for basic college-level math and physics, in order to understand them, and you should not need to consult any external sources.

\subsection{Exercises and Problems}

Throughout these notes, you will find many \textbf{exercises} and \textbf{problems}.
\begin{itemize}
\item \textbf{Exercises} are usually just calculations. They are meant to verify that you understand how to calculate things, and they are usually simple and straightforward.
\item \textbf{Problems} are usually proof-based. They are meant to verify that you understand the more abstract relations between the concepts we will introduce, and they often require some thought.
\end{itemize}

\subsection{Overview: Quantum vs. Classical Mechanics}

\begin{figure}[tbph]
\centering
\begin{centering}
\includegraphics[width=0.9\textwidth]{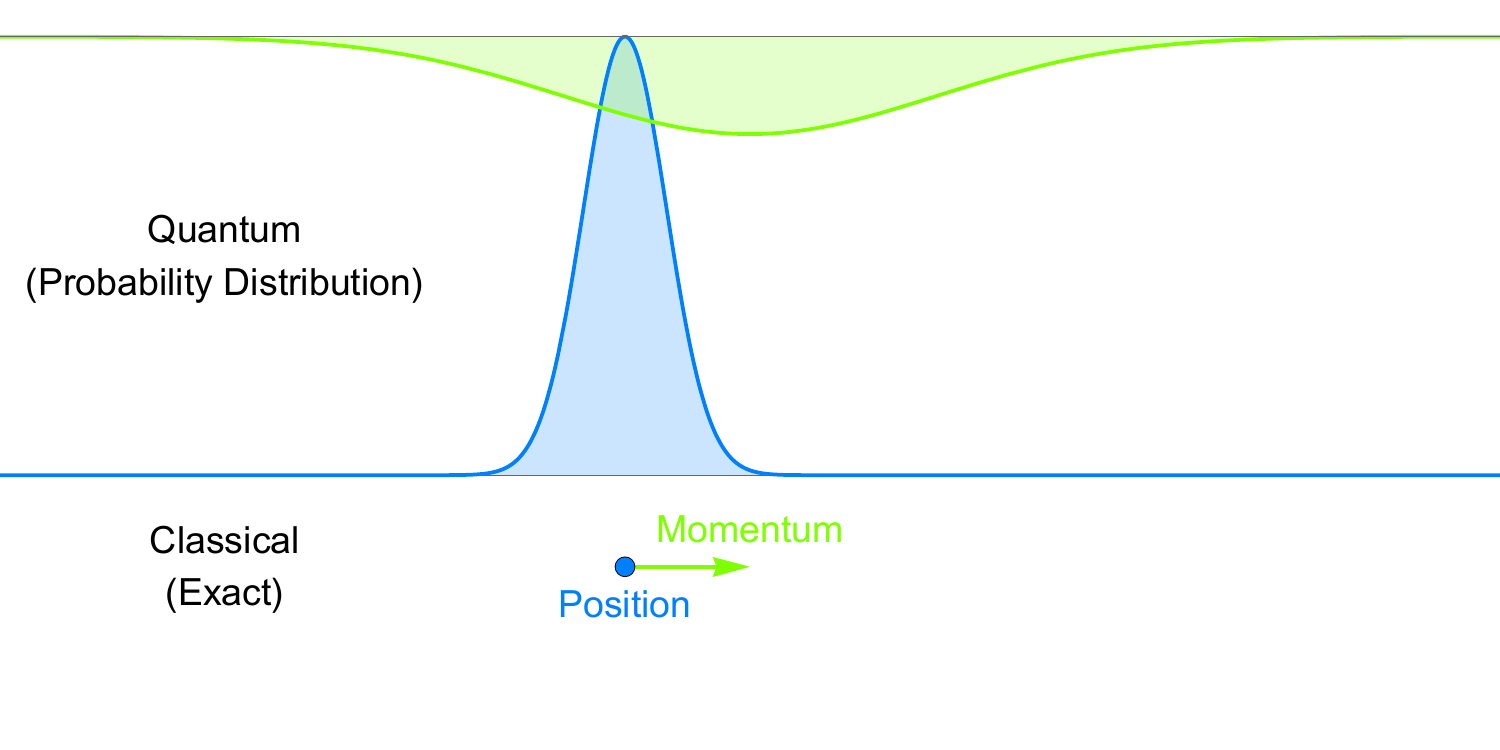}
\par\end{centering}
\caption[Quantum mechanics is probabilistic]{Quantum mechanics is probabilistic; position and momentum are probability distributions, not exact values as in classical mechanics.}\label{fig:Quantum-probability}
\end{figure}

To set the stage, let us first give a non-technical overview of the most important features of quantum mechanics and how they differ from their classical-mechanical counterparts. We will go into much more detail about these features later, once we are done with the mathematical preliminaries, but you should keep them in the back of your mind until then.
\begin{enumerate}
\item \textbf{Quantum mechanics is, as far as we know, the exact and fundamental theory of reality.} Classical mechanics turns out to be just an approximation to this theory. This means that, in general, all modern theories of physics must be quantum theories if they intend to be fundamental. One important exception to that rule is general relativity, which we do not yet know how to describe as a complete and consistent quantum theory; if we did, we would call that theory \emph{quantum gravity}\index{Quantum gravity}. However, this is usually not a problem, since gravity can be treated classically in the vast majority of cases of interest; quantum gravity is expected to become important only in extreme cases, like near the center of a black hole.
\item \textbf{Quantum mechanics is the theory of the smallest things.} This includes elementary particles, atoms, and molecules. Since all big things are made of small things, quantum mechanics also describes humans, planets, galaxies, and the whole universe. However, this is exactly where the \emph{classical limit\index{Classical limit}} comes in; when many small quantum systems make up one big system, classical mechanics generally turns out to be a good enough description for all practical purposes. This is similar to how relativity is always the correct way to describe physics, but at low velocities, much smaller than the speed of light, Newtonian physics is a good enough approximation.
\item \textbf{Quantum mechanics usually involves discrete things.} This is in contrast with classical mechanics, which usually involves continuous things. In fact, continuous classical things generally turn out to be made of discrete quantum things. One famous example of this is that light – a continuous electromagnetic field – is actually made of discrete photons. Similarly, angular momentum, which is continuous in the classical theory, has discrete values in the quantum theory. Intrinsic angular momentum, called \emph{spin}\index{Spin}, is a unique feature of quantum theory, and it is also discrete.
\item \textbf{Quantum mechanics is a probabilistic theory.} Classical mechanics, on the other hand, is a deterministic theory. For example, in classical mechanics, given a particle's exact position and momentum at any one time, we can (in principle) predict its position and momentum at any other time – with absolute certainty. However, in quantum mechanics, the most we can ever hope to know is the probability distribution for finding the particle at a certain position or with a certain momentum. This is illustrated in \zcref[S]{fig:Quantum-probability}.
\item \textbf{Quantum mechanics allows for superposition of states.} In classical mechanics, the state of a particle is simply given by the exact values of its position and momentum. In contrast, in quantum mechanics the particle can – in fact, usually \textbf{must} – be in a \emph{superposition\index{Superposition}} of possible positions and momenta. Each one of the possibilities in the superposition has a probability assigned to it, and this is where the probability distribution in \zcref[S]{fig:Quantum-probability} comes from.
\item \textbf{Quantum mechanics features uncertainty in measurements.} This is called the \emph{uncertainty principle}\index{Uncertainty principle}. In classical mechanics, at least theoretically, we can precisely know both the position and momentum of the particle. However, in quantum mechanics, the more we know about the position, the less we know about the momentum – and vice versa. If the position probability distribution is narrow and concentrated in a certain region, meaning that there is low uncertainty in the position, then one can prove that the momentum probability distribution must be wide, meaning that there is high uncertainty in the momentum. The opposite is also true. This is again illustrated in \zcref[S]{fig:Quantum-probability}.
\item \textbf{Quantum mechanics has a stronger type of correlation called entanglement.} Classical mechanics also allows for correlation. For example, let's say I have two sealed envelopes with notes inside them, one with the number 0 and the other with the number 1. I give one to Alice and one to Bob. If Alice opens her envelope and sees the number 0, she can be sure that Bob has the envelope with the number 1, and vice versa. The results are clearly correlated. However, if I instead place the pair of notes in a quantum superposition of “Alice has 0 and Bob has 1” and “Alice has 1 and Bob has 0”, the envelopes are now correlated more strongly via \emph{quantum entanglement}\index{Quantum entanglement}. We will learn in exactly what way quantum entanglement is stronger than classical correlation, and that this fact is part of what gives quantum computers\index{Quantum computer} their power.
\end{enumerate}

\section{Mathematical Background}

Quantum theory is the theoretical framework believed to describe all aspects of our universe at the most fundamental level. Mathematically, as we will see, it is relatively simple, although much more abstract than classical physics. However, conceptually, it is very hard to understand using the classical intuition we have from our daily lives. In these lectures we will learn to develop quantum intuition.

In this chapter we shall learn some basic mathematical concepts, focusing on complex numbers, linear algebra, and probability theory, which will be used extensively throughout the notes. Even if you are already familiar with these concepts, it is still a good idea to go over this chapter, since the unique notation commonly used in quantum mechanics is different than the notation used elsewhere in mathematics and physics.

\subsection{Complex Numbers}

Complex numbers are at the very core of the mathematical formulation of quantum theory. In this section we will give a review of complex numbers and present some definitions and results that will be used throughout the notes.

\subsubsection{Motivation}

In real life, we only encounter \emph{real numbers}\index{Real number}. These numbers form a \emph{field}\index{Field (algebra)}, denoted $\BBR$. Intuitively, a field is a set of elements with operations of addition, subtraction, multiplication, and division that work the way we defined them in elementary school. Geometrically, we can imagine $\BBR$ as a 1-dimensional line, stretching from $-\infty$ to $+\infty$.

Unfortunately, it turns out that the field of real numbers has a serious flaw. One can write down completely reasonable-looking quadratic equations, with only real coefficients, which nonetheless have no solutions in $\BBR$. Consider the most general quadratic equation:
\[
ax^{2}+bx+c=0\sp a,b,c\in\BBR\sp a\ne0.
\]
One can easily prove (by completing the square) that there are two potential solutions, given by
\[
x_{\pm}\equiv\frac{-b\pm\sqrt{b^{2}-4ac}}{2a}.
\]
Here, one solution corresponds to the choice $+$ and the other one to $-$. However, the square root $\sqrt{b^{2}-4ac}$ poses a problem, because the square of a real number is always non-negative\footnote{Here, $\fa$ means “for all”.}:
\[
x^{2}\ge0\sp\fa x\in\BBR.
\]
The number (and existence) of real solutions is thus determined by the sign of the expression inside the square root, called the \emph{discriminant}\index{Discriminant} $\Delta\equiv b^{2}-4ac$:
\[
\begin{cases}
\Delta>0:\vphantom{\blll} & \textrm{two real roots }{\displaystyle x_{\pm}=\frac{-b\pm\sqrt{\Delta}}{2a}},\\
\Delta=0:\vphantom{\blll} & \textrm{one real root }{\displaystyle x=-\frac{b}{2a}},\\
\Delta<0:\vphantom{\blll} & \textrm{no real roots}.
\end{cases}
\]
It would be very convenient (not to mention more elegant) to have a field of numbers that is \emph{algebraically closed}\index{Algebraically closed field}, meaning that every non-constant polynomial (and in particular, a quadratic polynomial) with coefficients in the field has a root in the field.

Since the problem stems from the fact that no real number can square to a negative number, let us simply extend our field with just one number, the \emph{imaginary unit\index{Imaginary unit}}, denoted\footnote{We use non-italic font exclusively for $\i$ in order to distinguish it from $i$, which will be used for labels and variables. Of course, it is usually a wise idea not to have both $\i$ and $i$ in the same equation in the first place, but sometimes that is unavoidable.} $\i$, whose sole purpose is to square to a negative number. The most natural choice is for $\i$ to square to $-1$:
\[
\i^{2}\equiv-1.
\]
The new field created by extending $\BBR$ with $\i$ is the field of \emph{complex numbers}\index{Complex number}, denoted $\BBC$. A general complex number is written
\[
z=a+\i b\sp z\in\BBC\sp a,b\in\BBR,
\]
where $a$ is called the \emph{real part} and $b$ is called the \emph{imaginary part}\index{Real and imaginary parts of a complex number}, both real numbers.

Now, in the quadratic equation, having $\sqrt{\Delta}$ with a negative $\Delta$ is no longer a problem, since the number $\i\sqrt{-\Delta}$ squares to $\Delta$:
\[
\left(\i\sqrt{-\Delta}\right)^{2}=\i^{2}\left(-\Delta\right)=\left(-1\right)\left(-\Delta\right)=\Delta.
\]
Therefore, we conclude that \textbf{every} quadratic equation has a solution in the field of complex numbers\footnote{Note that real numbers are a special case of complex numbers, so the two real roots are also two complex roots.}:
\begin{equation}
\begin{cases}
\Delta>0:\vphantom{\bllll} & \textrm{two real roots }{\displaystyle x_{\pm}=\frac{-b\pm\sqrt{\Delta}}{2a}},\\
\Delta=0:\vphantom{\bllll} & \textrm{one real root }{\displaystyle x=-\frac{b}{2a}},\\
\Delta<0:\vphantom{\bllll} & \textrm{two complex roots }{\displaystyle x_{\pm}=-\frac{b}{2a}\pm\i\frac{\sqrt{-\Delta}}{2a}}.
\end{cases}\label{eq:quadratic-roots}
\end{equation}
As a matter of fact, this is a special case of the \emph{fundamental theorem of algebra}\index{Fundamental theorem of algebra}: any polynomial of degree $n\ge1$ with complex coefficients\footnote{Again, real numbers are a special case of complex numbers, so the coefficients can be all real.} has at least one, and at most $n$, distinct complex roots\footnote{Or equivalently, it has exactly $n$ not necessarily distinct complex roots, accounting for possible degeneracy/multiplicity. For example, for $\Delta=0$ the quadratic equation has two degenerate roots, or one root of multiplicity 2.}. The quadratic equation corresponds to the case $n=2$.
\begin{xca}\leavevmode
\begin{problemparts}
\item Solve the quadratic equation
\[
x^{2}-6x+25=0.
\]
\item Find the quadratic equation whose solutions are $z=7\pm2\i$.
\end{problemparts}
\end{xca}

\subsubsection{Operations on Complex Numbers}

Complex numbers can be added to and multiplied by other complex numbers. There is really nothing special about these operations, except that it is customary to group the imaginary terms (i.e. anything that is a multiple of $\i$) together and turn $\i^{2}$ into $-1$ in the final result:
\[
\left(a+\i b\right)+\left(c+\i d\right)=\left(a+c\right)+\i\left(b+d\right),
\]
\[
\left(a+\i b\right)\left(c+\i d\right)=\left(ac-bd\right)+\i\left(ad+bc\right).
\]
Next, note that the two solutions to a quadratic equation with $\Delta<0$ are the same, up to the sign of $\i$. That is, if we replace $\i$ with $-\i$ in one of the solutions, we get the other solution. Such numbers are called \emph{complex conjugates}\index{Complex conjugate}, and the process of replacing $\i$ with $-\i$ is called \emph{complex conjugation\index{Complex conjugation}}. The complex conjugate of $z$ is denoted $z^{*}$:
\[
z=a+\i b\soosp z^{*}=a-\i b.
\]
Of course, the conjugate of the conjugate is the original number:
\[
\left(z^{*}\right)^{*}=z.
\]
This means that the complex conjugation operation is an \emph{involution}\index{Involution}, that is, its own inverse.

Complex conjugation allows us to write a general formula for the real and imaginary parts of a complex number\index{Real and imaginary parts of a complex number}, denoted $\re z$ and $\im z$ respectively:
\begin{equation}
\re z\equiv\frac{z+z^{*}}{2}\sp\im z\equiv\frac{z-z^{*}}{2\i}.\label{eq:re-im}
\end{equation}
You can check that if $z=a+\i b$ then we get $\re z=a$ and $\im z=b$, as expected.

Given two complex numbers $z=a+\i b$ and $w=c+\i d$, we have
\begin{align*}
\left(z+w\right)^{*} & =\left(a+\i b+c+\i d\right)^{*}\\
 & =\left(\left(a+c\right)+\i\left(b+d\right)\right)^{*}\\
 & =\left(a+c\right)-\i\left(b+d\right)\\
 & =\left(a-\i b\right)+\left(c-\i d\right)\\
 & =z^{*}+w^{*}.
\end{align*}
Therefore, complex conjugation \emph{commutes}\index{Commutative operation} with addition (that is, the two operations can be taken in any order). Similarly, we have
\begin{align*}
\left(zw\right)^{*} & =\left(\left(a+\i b\right)\left(c+\i d\right)\right)^{*}\\
 & =\left(\left(ac-bd\right)+\i\left(ad+bc\right)\right)^{*}\\
 & =\left(ac-bd\right)-\i\left(ad+bc\right)\\
 & =ac-bd-\i ad-\i bc\\
 & =\left(a-\i b\right)\left(c-\i d\right)\\
 & =z^{*}w^{*},
\end{align*}
so complex conjugation also commutes with multiplication.
\begin{xca}\leavevmode
What are the real and imaginary parts of $4-7\i$? What is its complex conjugate?
\end{xca}

\begin{problem}\leavevmode
\label{prob:quadratic}In \zcref[S]{eq:quadratic-roots} we saw that the equation $ax^{2}+bx+c=0$ with $a,b,c\in\BBR$ can either have two real solutions, one real solution, or two complex solutions that are conjugates of each other.
\begin{problemparts}
\item \emph{Imaginary numbers\index{Imaginary number}}\footnote{Sometimes also called \emph{purely imaginary numbers}\index{Purely imaginary number}.} are numbers of the form $\i b$ for $b\in\BBR$. What kind of equation has two imaginary solutions that are complex conjugates of each other?
\item What kind of equation has two imaginary solutions that are in general \textbf{not }complex conjugates of each other?
\item What kind of equation has two arbitrary complex solutions that are in general not complex conjugates of each other?
\end{problemparts}
\textbf{Note:} In all of the above, don't just find a specific equation that has this property – find a \textbf{family} of equations with arbitrary parameters of certain types.
\end{problem}

\begin{problem}\leavevmode
If a number is the complex conjugate of itself, can you say anything interesting about that number? What about if a number is minus the complex conjugate of itself?
\end{problem}

\subsubsection{The Complex Plane and Real 2-Vectors}\label{subsec:The-Complex-Plane}

\begin{figure}[tbph]
\centering
\begin{centering}
\includegraphics[width=0.9\textwidth]{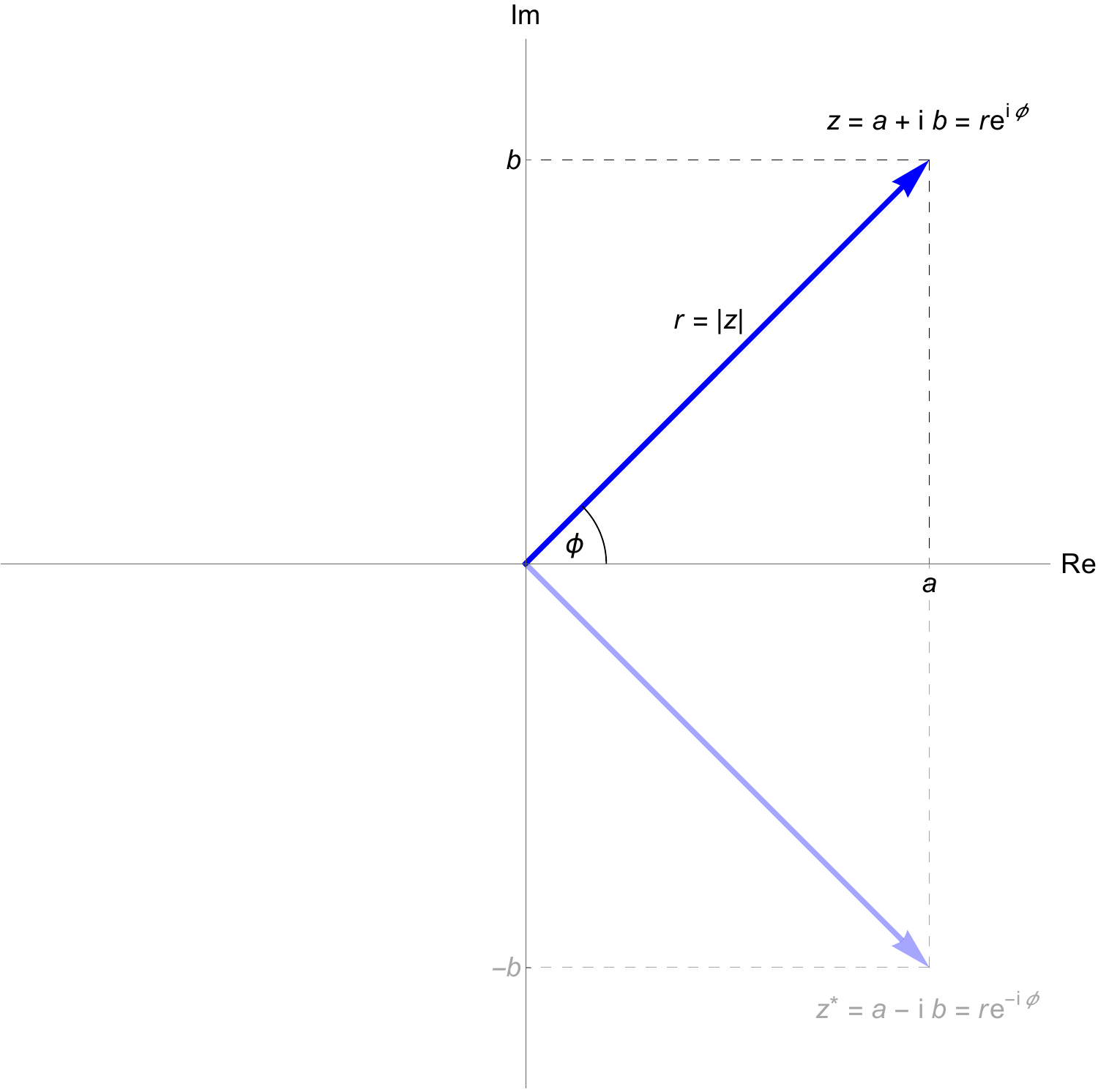}
\par\end{centering}
\caption[The complex plane]{The complex plane, with a complex number $z=a+\protect\i b$ and its conjugate $z^{*}=a-\protect\i b$. Also shown is the polar representation of both numbers (see \zcref[S]{subsec:Polar-Coordinates-and}).}\label{fig:The-complex-plane}
\end{figure}

Recall that the field of real numbers $\BBR$ is geometrically a line. The space $\BBR^{n}$ is an $n$-dimensional space which is home to \emph{real $n$-vectors}\index{Real $n$-vector}, that is, ordered lists of $n$ real numbers of the form $\left(v_{1},\ldots,v_{n}\right)$. In particular, $\BBR^{2}$ is geometrically a plane, with vectors of the form $\left(x,y\right)$.

The \emph{complex plane\index{Complex plane} $\BBC$} is similar to $\BBR^{2}$, except that instead of the $x$ and $y$ axes we have the real and imaginary axes respectively. The real unit $1$, which squares to $+1$, defines the positive direction of the real axis, while the imaginary unit $\i$, which squares to $-1$, defines the positive direction of the imaginary axis. This is illustrated in \zcref[S]{fig:The-complex-plane}.

Since $\BBC$ is a plane, we can define vectors on it, just like on $\BBR^{2}$. A real 2-vector $\left(a,b\right)$ is an arrow in $\BBR^{2}$ which points from the origin $\left(0,0\right)$ to the point that is $a$ steps in the direction of the $x$ axis and $b$ steps in the direction of the $y$ axis. A complex number $z=a+\i b$ is similarly an arrow in $\BBC$ which points from the origin $0$ to the point that is $a$ steps along the real axis and $b$ steps along the imaginary axis.

The complex conjugate $z^{*}=a-\i b$ is obtained by replacing $\i$ with $-\i$. Since $\i$ defines the direction of the imaginary axis, this is equivalent to flipping the imaginary axis. In other words, $z^{*}$ is the reflection of $z$ across the real axis, as shown in \zcref[S]{fig:The-complex-plane}.

From the Pythagorean theorem, we know that the magnitude (or length) of the real 2-vector $\left(a,b\right)$ is $\sqrt{a^{2}+b^{2}}$. The \emph{magnitude\index{Magnitude of a complex number}} or \emph{absolute value}\index{Absolute value} $\left|z\right|$ of the complex number $z=a+\i b$ is also $\sqrt{a^{2}+b^{2}}$. (Inspect \zcref[S]{fig:The-complex-plane} to see how the Pythagorean theorem fits in.) Furthermore, since $z^{*}$ is just a reflection of $z$, they both have the same magnitude. A convenient way to calculate the magnitude of either $z$ or $z^{*}$ is to multiply them with each other:
\[
\left|z\right|^{2}=\left|z^{*}\right|^{2}\equiv z^{*}z=\left(a+\i b\right)\left(a-\i b\right)=a^{2}-\i^{2}b^{2}=a^{2}+b^{2},
\]
so
\[
\left|z\right|=\left|z^{*}\right|=\sqrt{a^{2}+b^{2}}.
\]
For an abstract complex number (where we don't necessarily know the explicit values of the real and imaginary parts) one can also write
\begin{equation}
\left|z\right|=\left|z^{*}\right|=\sqrt{\left(\re z\right)^{2}+\left(\im z\right)^{2}}.\label{eq:mag-re-im}
\end{equation}
Given two numbers $z,w\in\BBC$, the magnitude-squared of their product is:
\[
\left|zw\right|^{2}=\left(zw\right)^{*}\left(zw\right)=z^{*}w^{*}zw=z^{*}zw^{*}w=\left|z\right|^{2}\left|w\right|^{2},
\]
so
\[
\left|zw\right|=\left|z\right|\left|w\right|,
\]
and we see that the magnitude of a product is the product of magnitudes. For a sum, things are more complicated. Let's start with the magnitude-squared of the sum:
\begin{align*}
\left|z+w\right|^{2} & =\left(z+w\right)^{*}\left(z+w\right)\\
 & =\left(z^{*}+w^{*}\right)\left(z+w\right)\\
 & =z^{*}z+z^{*}w+w^{*}z+w^{*}w.
\end{align*}
The first term is $\left|z\right|^{2}$ and the last term is $\left|w\right|^{2}$. For the middle two terms, from \zcref[S]{eq:re-im} we see that
\[
z^{*}w+w^{*}z=z^{*}w+\left(z^{*}w\right)^{*}=2\re\left(z^{*}w\right).
\]
In conclusion,
\begin{equation}
\left|z+w\right|^{2}=\left|z\right|^{2}+\left|w\right|^{2}+2\re\left(z^{*}w\right).\label{eq:z-plus-w}
\end{equation}
Now, from \zcref[S]{eq:mag-re-im} we can see that, for any $u\in\BBC$,
\[
\left|u\right|^{2}=\left(\re u\right)^{2}+\left(\im u\right)^{2}\ge\left(\re u\right)^{2},
\]
since $\left(\im u\right)^{2}\ge0$. Taking the square root, and noting that $\sqrt{\left(\re u\right)^{2}}=\left|\re u\right|$, we get
\[
\left|u\right|\ge\left|\re u\right|.
\]
Taking $u=z^{*}w$, we see that
\[
2\re\left(z^{*}w\right)\le2\left|z^{*}w\right|=2\left|z\right|\left|w\right|,
\]
and if we plug that back into \zcref[S]{eq:z-plus-w} we get
\[
\left|z+w\right|^{2}\le\left|z\right|^{2}+\left|w\right|^{2}+2\left|z\right|\left|w\right|=\left(\left|z\right|+\left|w\right|\right)^{2}.
\]
Taking the square root, we find
\[
\left|z+w\right|\le\left|z\right|+\left|w\right|.
\]
This is known as the \emph{triangle inequality}\index{Triangle inequality}. To see why, consider $z$ and $w$ as arrows on the complex plane. The sum $z+w$ is then an arrow obtained by placing the tail of $w$ at the head of $z$. The three arrows $z$ (starting at the origin), $w$ (starting at $z$), and $z+w$ (starting at the origin) form a triangle, as you can check by drawing them. The triangle inequality therefore states that the length of one side of a triangle can never be larger than the sum of the lengths of the other two sides. Equality is only reached when one of $z$ and $w$ is zero, or when they both point in the same direction, in which case all three sides overlap and we get a degenerate triangle.

The triangle inequality can be extended to a set of complex numbers $z_{1},z_{2},\ldots$ as follows:
\begin{equation}
\left|\sum_{i}z_{i}\right|\le\sum_{i}\left|z_{i}\right|.\label{eq:triangle-inequality-general}
\end{equation}
This is correct even for an infinite number of complex numbers, as long as the sums converge.

Finally, we note that there is an \emph{isomorphism\index{Isomorphism}} between complex numbers and real 2-vectors. An isomorphism between two spaces is a mapping between the spaces that can be taken in either direction (i.e. is invertible), and preserves the structure of each space. The isomorphism between $\BBC$ and $\BBR^{2}$ is given by:
\[
a+\i b\toot\left(a,b\right).
\]
We have already seen that the magnitude is preserved. Similarly, addition of complex numbers
\[
\left(a+\i b\right)+\left(c+\i d\right)=\left(a+c\right)+\i\left(b+d\right)
\]
maps into addition of 2-vectors
\[
\left(a,b\right)+\left(c,d\right)=\left(a+c,b+d\right).
\]

\begin{xca}\leavevmode
Let $z=5+6\i$ and $w=7+8\i$.
\begin{problemparts}
\item Calculate $z^{*}$, $w^{*}$, $\left|z\right|$, $\left|w\right|$, $z+w$, $z-w$, $\left|z+w\right|$, $\left|z-w\right|$, and $zw$.
\item Find the 2-vectors isomorphic to $z$ and $w$.
\end{problemparts}
\end{xca}

\begin{problem}\leavevmode
Let the vector $\v\equiv\left(a,b\right)\in\BBR^{2}$ be isomorphic to the complex number $z\equiv a+\i b\in\BBC$. Show explicitly that the following operations on $\v$ map to equivalent operations on $z$:
\begin{enumerate}
\item Multiplication of $\v$ by a real number $\lambda\in\BBR$.
\item Reflection of $\v$ with respect to the $x$ axis.
\item Reflection of $\v$ with respect to the $y$ axis.
\item Reflection of $\v$ with respect to both axes.
\end{enumerate}
Formulate the equivalent operations on $z$ in terms of $z$ itself, \textbf{without }using $a$, $b$, $\re z$, or $\im z$.
\end{problem}

\begin{problem}\leavevmode
Prove \zcref[S]{eq:triangle-inequality-general} for a finite number $n$ of summands by induction on $n$.
\end{problem}

\subsubsection{Polar Coordinates and Phases}\label{subsec:Polar-Coordinates-and}

A vector in $\BBR^{2}$ can be converted from Cartesian coordinates $\left(x,y\right)$ to \emph{polar coordinates}\index{Polar coordinates} $\left(r,\phi\right)$. The $r$ coordinate is the magnitude of the vector, and the $\phi$ coordinate is the angle that the vector makes with respect to the $x$ axis. The relation between the coordinate systems is given by
\begin{equation}
x=r\cos\phi\sp y=r\sin\phi,\label{eq:xy-rphi}
\end{equation}
\[
r=\sqrt{x^{2}+y^{2}}\sp\phi=\arctan\frac{y}{x}.
\]
This simply follows from the definitions of $\cos\phi$ and $\sin\phi$, since the vector creates a right triangle with the $x$ axis (see \zcref[S]{fig:The-complex-plane}). For example, the vector $\left(x,y\right)=\left(1,\sqrt{3}\right)$ in Cartesian coordinates corresponds to $r=2$ and $\phi=\pi/3$.

$x$ and $y$ can be any real numbers, but $r$ must be non-negative and $\phi$ must be in the range $\left(-\pi,\pi\right]$ (in radians) where $\phi=0$ corresponds to the $x$ axis. However, there is a subtlety here: the range of the $\arctan$ function is $\left(-\pi/2,\pi/2\right)$, so $\phi$ needs to be further adjusted according to the quadrant. One can instead use a more complicated definition that automatically takes the quadrant into account:
\[
\phi=\begin{cases}
\arctan(\frac{y}{x}) & \text{if }x>0,\\
\arctan(\frac{y}{x})+\pi & \text{if }x<0\text{ and }y\ge0,\\
\arctan(\frac{y}{x})-\pi & \text{if }x<0\text{ and }y<0,\\
+\frac{\pi}{2} & \text{if }x=0\text{ and }y>0,\\
-\frac{\pi}{2} & \text{if }x=0\text{ and }y<0,\\
\text{undefined} & \text{if }x=0\text{ and }y=0.
\end{cases}
\]
This function is sometimes called $\textrm{atan2}\left(y,x\right)$, and it is implemented in most programming languages. Note that $\phi$ is undefined at the origin since a vector of length zero does not point in any direction.

Given that complex numbers are isomorphic to real 2-vectors, we should be able to write complex numbers in polar coordinates as well. Looking at \zcref[S]{eq:xy-rphi}, and replacing $x$ and $y$ with $a$ and $b$, we see that
\[
z=a+\i b=r\left(\cos\phi+\i\sin\phi\right).
\]
We can write this more compactly using \emph{Euler's formula}\index{Euler's formula} (see \zcref[S]{prob:e-1}):
\[
\e^{\i\phi}=\cos\phi+\i\sin\phi\soosp z=r\e^{\i\phi}.
\]
This is illustrated in \zcref[S]{fig:The-complex-plane}. In this context, the angle $\phi$ is called the \emph{phase}\index{Phase of a complex number} (or \emph{argument}\index{Argument of a complex number}) of $z$, and the factor $\e^{\i\phi}$ is called a \emph{phase factor}\index{Phase factor}. Both are of extreme importance in quantum mechanics, as we shall see.
\begin{xca}\leavevmode
Write $2\i-3$ in polar coordinates.
\end{xca}

\begin{problem}\leavevmode
Prove, using Euler's formula, that $\left|\e^{\i\phi}\right|=1$, that is, the magnitude of the complex number $\e^{\i\phi}$ is $1$. If $z=r\e^{\i\phi}$, what is $\left|z\right|$?
\end{problem}

\subsubsection{Exponentials and Logarithms}\label{subsec:Exponentials-and-Logarithms}

The \emph{exponential function}\index{Exponential function} is defined on arbitrary complex numbers $z\in\BBC$ using a \emph{power series}\index{Power series} as follows:
\begin{equation}
\e^{z}\equiv\sum^{\infty}_{n=0}\frac{z^{n}}{n!}=1+z+\hf z^{2}+\frac{1}{3!}z^{3}+\cdots.\label{eq:exp-power}
\end{equation}
The complex number $z$ is called the \emph{exponent}\index{Exponent}. If the exponent is zero, then all the terms in the series vanish except the first one, and we get $\e^{0}=1$. If the exponent is a natural number $n\in\BBN$, then \zcref[S]{eq:exp-power} turns out to be the same as taking the real number\footnote{But of course, in order to know the value of the number $\e$ in the first place, we need to calculate the power series in \zcref[S]{eq:exp-power} for $z=1$!} $\e\ap2.718$ to the power of $n$, that is, multiplying it by itself $n$ times. This can then be extended to negative integers using the formula
\[
\e^{-n}\equiv\frac{1}{\e^{n}},
\]
and to rational numbers using
\[
\frac{a}{b}\in\BBQ\soosp\e^{a/b}\equiv\sqrt[b]{\e^{a}}.
\]
However, for arbitrary real or complex numbers, we generally use the power series definition in \zcref[S]{eq:exp-power} directly, or an equivalent definition such as the ones you will prove in \zcref[S]{prob:By-assuming-a,prob:e-2} below.

By taking the complex conjugate of the series in \zcref[S]{eq:exp-power}, we get:
\begin{equation}
\left(\e^{z}\right)^{*}=\e^{z^{*}},\label{eq:exp-zconj}
\end{equation}
so the conjugate operation commutes\index{Commutative operation} with taking the exponential. In particular, given a complex number in the polar representation (see \zcref[S]{subsec:Polar-Coordinates-and}), we have
\[
z=r\e^{\i\phi}\soosp z^{*}=r\e^{-\i\phi}\sp r,\phi\in\BBR.
\]
This indeed makes sense, as taking the complex conjugate means reflecting $z$ across the real line, and thus turns the angle $\phi$, which is the angle with respect to the real line, into its negative – see \zcref[S]{fig:The-complex-plane}.

One can also prove from the definition in \zcref[S]{eq:exp-power} that $\e^{z+w}=\e^{z}\e^{w}$, so
\[
\e^{\i\phi}\e^{-\i\phi}=\e^{\i\phi-\i\phi}=\e^{0}=1.
\]
Therefore the magnitude of $z$ is
\[
\left|z\right|=\sqrt{zz^{*}}=\sqrt{r\e^{\i\phi}\cdot r\e^{-\i\phi}}=\sqrt{r^{2}}=r,
\]
as expected.

The exponential function is its own derivative:
\[
\frac{\d}{\d z}\e^{z}=\e^{z}.
\]
In fact, it can be \textbf{defined} using this property, as you will prove in \zcref[S]{prob:By-assuming-a}. Using the chain rule, we get the more general result
\begin{equation}
\frac{\d}{\d z}\e^{\lambda z}=\frac{\d}{\d z}\left(\lambda z\right)\e^{\lambda z}=\lambda\e^{\lambda z},\label{eq:exp-lambda-t}
\end{equation}
where $\lambda$ is any \textbf{constant} complex number (i.e. independent of $z$).

For $x\in\BBR$, it is easy to see that the function $\e^{x}$ is continuous. It is also strictly increasing, because its derivative is $\e^{x}>0$, and has the limits
\[
\lim_{x\to-\infty}\e^{x}=0\sp\lim_{x\to+\infty}\e^{x}=+\infty.
\]
So $\e^{x}$ is a map from $\BBR$ to the interval $\left(0,\infty\right)$, and it is:
\begin{itemize}
\item \emph{Injective (one-to-one)}\index{Injective function}\index{One-to-one function}: If $\e^{x}=\e^{y}$ then we must have $x=y$, because the exponential is strictly increasing, so it never produces the same value twice.
\item \emph{Surjective (onto)}\index{Surjective function}\index{Onto function}: For every $y\in\left(0,\infty\right)$ there is some $x\in\BBR$ such that $\e^{x}=y$, because the exponential is continuous and tends to $0$ as $x\to-\infty$ and to $+\infty$ as $x\to+\infty$, so it must cover the entire interval $\left(0,\infty\right)$.
\end{itemize}
Therefore, $\e^{x}$ is a \emph{bijection}\index{Bijection}, meaning it maps each element of $\BBR$ to exactly one element of $\left(0,\infty\right)$ \textbf{and vice versa}. This means it is invertible, and we can define an inverse function $\log y$ from $\left(0,\infty\right)$ to $\BBR$, called the \emph{logarithm}\index{Logarithm}, such that
\begin{equation}
\log\e^{x}=x\sp\e^{\log y}=y\label{eq:log}
\end{equation}
for all $x\in\BBR$ and $y\in\left(0,\infty\right)$. This is also called the \emph{natural logarithm\index{Natural logarithm}}, since it is taken with respect to the “natural” \emph{base}\index{Base of a logarithm} $\e\ap2.718$. More generally, a logarithm with respect to the base $b$, where $b>0$ and $b\ne1$, satisfies
\[
\log_{b}b^{x}=x\sp b^{\log_{b}y}=y.
\]
For a general base $b$ we have
\[
\frac{\d}{\d x}b^{x}=b^{x}\log_{\e}b,
\]
and the extra factor becomes 1 when $b=\e$, since $\log_{\e}\e=1$. This explains why the base $\e\ap2.718$ is “natural”; it is the unique base for which the function $b^{x}$ is its own derivative, without the extra factor. Sometimes the notation $\ln$ is also used for the natural logarithm: $\ln\equiv\log_{\e}$.

For complex numbers, the definition of the logarithm becomes more complicated, because $\e^{\i\phi}$ for $\phi\in\BBR$ is \textbf{periodic}; if we increase or decrease the phase angle $\phi$ by $2\pi$, we get back to where we started:
\[
\e^{\i\left(\phi+2\pi k\right)}=\e^{\i\phi}\sp\fa k\in\BBZ.
\]
This means that for imaginary exponents, the exponential function is no longer injective: for $x=0$ and $y=2\pi\i$ we have $\e^{x}=\e^{y}$ even though $x\ne y$. Therefore, it is also not invertible. In fact, it turns out that every non-zero complex number
\[
z=r\e^{\i\phi}\sp r\in\left(0,\infty\right)\sp\phi\in\BBR,
\]
has an \textbf{infinite} number of logarithms, given by
\[
\log z=\log r+\i\left(\phi+2\pi k\right)\sp k\in\BBZ.
\]
One way to solve this problem is to treat each $k$ as a separate \emph{branch}\index{Branch (logarithm)}. We can then choose the \emph{principal branch}\index{Principal branch (logarithm)} to be $k=0$, and define the \emph{principal logarithm}\index{Principal logarithm} $\Log z$ as the function that takes any non-zero complex number $z=r\e^{\i\phi}$ and returns the unique value
\[
\textrm{Log}\thinspace z=\log r+\i\phi,
\]
where $\phi$ is shifted so that $\phi\in\left(-\pi,\pi\right]$.

The power series definition in \zcref[S]{eq:exp-power} and the principal logarithm can be used to define the power $z^{w}$ for any non-zero complex number $z$ and any complex number $w$ using the formula
\[
z^{w}=\left(\e^{\Log z}\right)^{w}=\e^{w\Log z}.
\]

\begin{problem}\leavevmode
\label{prob:By-assuming-a}By assuming a generic power series expansion
\[
f\left(z\right)=\sum^{\infty}_{n=0}a_{n}z^{n},
\]
prove that if $f\left(z\right)$ is its own derivative and satisfies $f\left(0\right)=1$, then it must be the exponential function, i.e. $a_{n}=1/n!$.
\end{problem}

\begin{problem}\leavevmode
\label{prob:e-1}The power series expansions of the trigonometric functions $\cos x$ and $\sin x$ are
\[
\cos x=\sum^{\infty}_{n=0}\frac{\left(-1\right)^{n}}{\left(2n\right)!}x^{2n}=1-\frac{1}{2}x^{2}+\frac{1}{4!}x^{4}+\cdots,
\]
\[
\sin x=\sum^{\infty}_{n=0}\frac{\left(-1\right)^{n}}{\left(2n+1\right)!}x^{2n+1}=x-\frac{1}{3!}x^{3}+\frac{1}{5!}x^{5}+\cdots.
\]
Use them to prove \emph{Euler's formula}\index{Euler's formula}
\[
\e^{\i x}=\cos x+\i\sin x.
\]
As a corollary, show that
\[
\cos x=\re\left(\e^{\i x}\right)=\frac{\e^{\i x}+\e^{-\i x}}{2},
\]
\[
\sin x=\im\left(\e^{\i x}\right)=\frac{\e^{\i x}-\e^{-\i x}}{2\i}.
\]
\end{problem}

\begin{problem}\leavevmode
\label{prob:e-2}The \emph{binomial theorem\index{Binomial theorem}} states that for $x,y\in\BBC$ and $n\in\BBN$:
\[
\left(x+y\right)^{n}=\sum^{n}_{k=0}\binom{n}{k}x^{n-k}y^{k},
\]
where the \emph{binomial coefficients}\index{Binomial coefficients} are defined as
\[
\binom{n}{k}\equiv\frac{n!}{k!\left(n-k\right)!}.
\]
So, explicitly, we have
\[
\left(x+y\right)^{n}=x^{n}+nx^{n-1}y+\frac{1}{2}n\left(n-1\right)x^{n-2}y^{2}+\cdots.
\]
Using the binomial theorem and the power series definition of the exponential in \zcref[S]{eq:exp-power}, prove the equivalent definition
\[
\e^{z}=\lim_{n\to\infty}\left(1+\frac{z}{n}\right)^{n}.
\]
\end{problem}

\subsection{Linear Algebra}

The most important and fundamental mathematical structure in quantum theory is the \emph{Hilbert space}\index{Hilbert space}, a type of complex vector space. In this section we will define Hilbert spaces and learn about many important concepts and results from linear algebra that apply to them.

\subsubsection{Complex Vector Spaces}\label{subsec:Complex-Vector-Spaces}

A real $n$-vector is an ordered list of $n$ real numbers. Analogously, a \emph{complex $n$-vector}\index{Complex $n$-vector} is an ordered list of $n$ complex numbers. For example, a complex 2-vector with two complex components $\Psi_{1}$ and $\Psi_{2}$ is written as:
\[
\left|\Psi\right\rangle \equiv\left(\begin{array}{c}
\Psi_{1}\\
\Psi_{2}
\end{array}\right).
\]
The notation $\left|\Psi\right\rangle $ is \textbf{unique to quantum mechanics}, and it is called \emph{bra-ket notation}\index{Bra-ket notation} or sometimes \emph{Dirac notation}\index{Dirac notation}. In this notation, we write a straight line $|$ and an angle bracket $\rangle$, and between them, a label. We will usually denote a general vector with the label $\Psi$; this label, and its lowercase counterpart $\psi$, are very commonly used in quantum mechanics. However, we can use whatever label we want to describe our vector – including letters, numbers, symbols, or even whole words and sentences, for example:
\[
\left|A\right\rangle ,\left|\beta\right\rangle ,\left|3\right\rangle ,\left|\clubsuit\right\rangle ,\left|\textrm{Bob}\right\rangle ,\left|\textrm{Schrödinger's Cat Is Alive}\right\rangle ,\ldots.
\]
This is a great advantage of the bra-ket notation, as it allows us to be very descriptive in the labels we choose for our vectors – which we can't do with the notation $\v$ or $\vec{v}$ commonly used for vectors in mathematics and physics.

A \emph{vector space}\index{Vector space} $\VV$ over a field\footnote{The field is usually taken to be $\BBR$ or $\BBC$. Naturally, for a complex vector space, it will be $\BBC$.} $\BBF$ is a set of vectors equipped with two operations: \emph{addition of vectors}\index{Addition of vectors} and \emph{multiplication of a vector by a scalar}\index{Multiplication of vector by scalar}, where a \emph{scalar}\index{Scalar} is any number from the field $\BBF$. Vector addition must satisfy the following conditions:
\begin{enumerate}
\item \emph{Closed\index{Closed operation}} – the sum of two vectors is another vector in the same space:
\[
\fa\left|\Psi\right\rangle ,\left|\Phi\right\rangle \in\VV:\qquad\left|\Psi\right\rangle +\left|\Phi\right\rangle \in\VV.
\]
\item \emph{Commutative\index{Commutative operation}} – the order of vectors doesn't matter:
\[
\fa\left|\Psi\right\rangle ,\left|\Phi\right\rangle \in\VV:\qquad\left|\Psi\right\rangle +\left|\Phi\right\rangle =\left|\Phi\right\rangle +\left|\Psi\right\rangle .
\]
\item \emph{Associative\index{Associative operation}} – if three vectors are added, it doesn't matter which two are added first:
\[
\fa\left|\Psi\right\rangle ,\left|\Phi\right\rangle ,\left|\Theta\right\rangle \in\VV:\qquad\left(\vphantom{\bll}\left|\Psi\right\rangle +\left|\Phi\right\rangle \right)+\left|\Theta\right\rangle =\left|\Psi\right\rangle +\left(\vphantom{\bll}\left|\Phi\right\rangle +\left|\Theta\right\rangle \right).
\]
\item \emph{Identity vector\index{Identity vector}} or \emph{zero vector}\index{Zero vector} – there is a (unique) vector\footnote{Note that here we are using a slight abuse of notation by denoting the zero vector as the \textbf{number} $0$, instead of using bra-ket notation. The reason is that $\left|0\right\rangle $ already has a special common meaning in quantum mechanics, as we will see later; in the context of that special meaning, $\left|0\right\rangle $ is \textbf{not }the zero vector.} $0$ which, when added to any vector, does not change it\footnote{Here, $\ex$ means “exists”, so $\ex0\in\VV$ means “there exists an element $0$ in $\VV$”.}:
\[
\ex0\in\VV:\qquad\fa\left|\Psi\right\rangle \in\VV:\qquad\left|\Psi\right\rangle +0=\left|\Psi\right\rangle .
\]
\item \emph{Inverse vector}\index{Inverse vector} – for every vector there exists another (unique) vector such that the two vectors sum to the zero vector:
\[
\fa\left|\Psi\right\rangle \in\VV:\qquad\ex\left(\vphantom{\bll}-\left|\Psi\right\rangle \right)\in\VV:\qquad\left|\Psi\right\rangle +\left(\vphantom{\bll}-\left|\Psi\right\rangle \right)=0.
\]
\end{enumerate}
Furthermore, multiplication by a scalar must satisfy the following conditions:
\begin{enumerate}
\item \emph{Closed}\index{Closed operation} – the product of a vector and a scalar is a vector in the same space:
\[
\fa\alpha\in\BBF\sp\fa\left|\Psi\right\rangle \in\VV:\qquad\alpha\left|\Psi\right\rangle \in\VV.
\]
\item \emph{Associative\index{Associative operation}} – if two scalars are multiplied by a vector, it doesn't matter whether we first multiply the two scalars or we first multiply one of the scalars with the vector:
\[
\fa\alpha,\beta\in\BBF\sp\fa\left|\Psi\right\rangle \in\VV:\qquad\left(\alpha\beta\right)\left|\Psi\right\rangle =\alpha\left(\beta\left|\Psi\right\rangle \right).
\]
\item \emph{Distributive}\index{Distributive operation} over addition of scalars:
\[
\fa\alpha,\beta\in\BBF\sp\fa\left|\Psi\right\rangle \in\VV:\qquad\left(\alpha+\beta\right)\left|\Psi\right\rangle =\alpha\left|\Psi\right\rangle +\beta\left|\Psi\right\rangle .
\]
\item \emph{Distributive}\index{Distributive operation} over addition of vectors:
\[
\fa\alpha\in\BBF\sp\fa\left|\Psi\right\rangle ,\left|\Phi\right\rangle \in\VV:\qquad\alpha\left(\left|\Psi\right\rangle +\left|\Phi\right\rangle \right)=\alpha\left|\Psi\right\rangle +\alpha\left|\Phi\right\rangle .
\]
\item \emph{Identity scalar}\index{Identity scalar} or \emph{unit scalar}\index{Unit scalar} – there is a (unique) scalar $1$ which, when multiplied by any vector, does not change it:
\[
\ex1\in\BBF:\qquad\fa\left|\Psi\right\rangle \in\VV:\qquad1\left|\Psi\right\rangle =\left|\Psi\right\rangle .
\]
\end{enumerate}
We now define a $2$-dimensional \emph{complex vector space}\index{Complex vector space}, which we denote $\BBC^{2}$, as the space of complex $2$-vectors over $\BBC$, with addition of vectors given by
\[
\left|\Psi\right\rangle \equiv\left(\begin{array}{c}
\Psi_{1}\\
\Psi_{2}
\end{array}\right)\in\BBC^{2}\sp\left|\Phi\right\rangle \equiv\left(\begin{array}{c}
\Phi_{1}\\
\Phi_{2}
\end{array}\right)\in\BBC^{2}\soosp\left|\Psi\right\rangle +\left|\Phi\right\rangle =\left(\begin{array}{c}
\Psi_{1}+\Phi_{1}\\
\Psi_{2}+\Phi_{2}
\end{array}\right),
\]
and multiplication of a vector by a scalar given by
\[
\left|\Psi\right\rangle \equiv\left(\begin{array}{c}
\Psi_{1}\\
\Psi_{2}
\end{array}\right)\in\BBC^{2}\sp\lambda\in\BBC\soosp\lambda\left|\Psi\right\rangle =\left(\begin{array}{c}
\lambda\Psi_{1}\\
\lambda\Psi_{2}
\end{array}\right).
\]
The $n$-dimensional complex vector space $\BBC^{n}$ is defined analogously. In these notes, we will mostly focus on $\BBC^{2}$ for simplicity, in particular when giving explicit examples.
\begin{xca}\leavevmode
Let
\[
\left|\Psi\right\rangle \equiv\left(\begin{array}{c}
3+\i\\
-9
\end{array}\right)\sp\left|\Phi\right\rangle \equiv\left(\begin{array}{c}
\i-1\\
-10\i
\end{array}\right)\sp\alpha=7\i-2\sp\beta=-4-8\i.
\]
Calculate $\alpha\left|\Psi\right\rangle +\beta\left|\Phi\right\rangle $.
\end{xca}

\begin{problem}\leavevmode
Check that the addition and multiplication as defined above indeed satisfy all of the required conditions for a vector space. You can do this just for $\BBC^{2}$, for simplicity.
\end{problem}

\subsubsection{Dual Vectors, Inner Products, Norms, and Hilbert Spaces}\label{subsec:Dual-Vectors}

A \emph{dual vector\index{Dual vector} }is defined by writing the vector as a row instead of a column, and replacing each component with its complex conjugate. We denote the dual vector of $\left|\Psi\right\rangle $ as follows:
\[
\left\langle \Psi\right|=\left(\begin{array}{cc}
\Psi^{*}_{1} & \Psi^{*}_{2}\end{array}\right).
\]
In terms of notation, there is now an opposite angle bracket $\langle$ on the left of the label, and the straight line $|$ is on the right. Addition and multiplication by a scalar are defined as for vectors, simply replacing columns with rows. However, you may \textbf{not }add vectors and dual vectors together – adding a row to a column is undefined!

If we are given a dual vector, we can take \textbf{its }dual to get a “normal” (column) vector. In this case, the operation of taking the dual involves writing the vector as a column instead of a row and taking the complex conjugates of the components. This means that the operation of taking the dual is an involution – taking the dual of a vector twice gives back the same vector, since $\left(z^{*}\right)^{*}=z$.

Using dual vectors, we may define the \emph{inner product}\index{Inner product}. This product allows us to take a vector and a dual vector and produce a (complex) number out of them, similarly to the dot product of real vectors\footnote{The \emph{dot product}\index{Dot product} of the real vectors $\v\equiv\left(v_{1},v_{2}\right)$ and $\w\equiv\left(w_{1},w_{2}\right)$ in $\BBR^{2}$ is defined as $\v\cdot\w\equiv v_{1}w_{1}+v_{2}w_{2}$. In principle, this definition does secretly involve a dual (row) vector and a (column) vector, but since we don't need to take the complex conjugate in this case, the distinction between the two vector types is not important.}. Importantly, the inner product requires matching a vector with a dual vector; that is, if we have two vectors and we want to calculate their inner product, we must turn one into a dual vector first. Then we multiply the components of the vector and the dual vector one by one, and add them up:
\[
\langle\Psi|\Phi\rangle=\left(\begin{array}{cc}
\Psi^{*}_{1} & \Psi^{*}_{2}\end{array}\right)\left(\begin{array}{c}
\Phi_{1}\\
\Phi_{2}
\end{array}\right)=\Psi^{*}_{1}\Phi_{1}+\Psi^{*}_{2}\Phi_{2}.
\]
In bra-ket notation, vectors $\left|\Psi\right\rangle $ are called \emph{“kets”\index{Ket}} and dual vectors $\left\langle \Psi\right|$ are called \emph{“bras”}\index{Bra}. Then the notation for $\langle\Psi|\Phi\rangle$ is called a \emph{“bra(c)ket”}\index{Bra-ket notation}.

We define the \emph{norm-squared}\index{Norm-squared} of a vector by taking its inner product with its dual (“squaring” it):
\[
\left\Vert \Psi\right\Vert ^{2}\equiv\langle\Psi|\Psi\rangle=\left(\begin{array}{cc}
\Psi^{*}_{1} & \Psi^{*}_{2}\end{array}\right)\left(\begin{array}{c}
\Psi_{1}\\
\Psi_{2}
\end{array}\right)=\left|\Psi_{1}\right|^{2}+\left|\Psi_{2}\right|^{2},
\]
where the magnitude-squared of a complex number $z$ was defined in \zcref[S]{subsec:The-Complex-Plane} as $\left|z\right|^{2}=z^{*}z$. Then we can define the \emph{norm}\index{Norm} as the square root of the norm-squared:
\[
\left\Vert \Psi\right\Vert \equiv\sqrt{\left\Vert \Psi\right\Vert ^{2}}=\sqrt{\langle\Psi|\Psi\rangle}.
\]
Observe how taking the dual of a vector generalizes taking the complex conjugate of a number, and taking the norm of a vector generalizes taking the magnitude of a number; indeed, for 1-dimensional vectors, these operations are the same!

A vector space with an inner product is called a \emph{Hilbert space}\index{Hilbert space}, provided that it is also a \emph{complete metric space}\footnote{A vector space is a \emph{complete metric space}\index{Complete metric space} if whenever an infinite series of vectors $\left|\Psi_{i}\right\rangle $ \emph{converges absolutely}\index{Absolute convergence}, that is, the series of the norms of the vectors converges:
\[
\sum^{\infty}_{i=0}\left\Vert \Psi_{i}\right\Vert <\infty,
\]
then the series of the vectors themselves converges as well, to some vector $\left|\Psi\right\rangle $ in the same vector space:
\[
\sum^{\infty}_{i=0}\left|\Psi_{i}\right\rangle =\left|\Psi\right\rangle .
\]
} and that the inner product satisfies the same properties (which you will derive in \zcref[S]{prob:inner1,prob:inner2,prob:inner3}) as the standard inner product on $\BBC^{n}$. In particular, $\BBC^{n}$ itself is a Hilbert space, but there are many other Hilbert spaces, some of them much more abstract. The usual notation for a general Hilbert space is $\HH$, and the notation for its dual is $\HH^{*}$; that is, vectors (kets) belong in $\HH$ and their dual vectors (bras) belong in $\HH^{*}$.
\begin{xca}\leavevmode
Let
\[
\left|\Psi\right\rangle \equiv\left(\begin{array}{c}
7+7\i\\
-7-2\i
\end{array}\right)\sp\left|\Phi\right\rangle \equiv\left(\begin{array}{c}
-2-7\i\\
\i
\end{array}\right).
\]
Calculate $\left\langle \Psi\right|$, $\left\langle \Phi\right|$, $\left\Vert \Psi\right\Vert $, $\left\Vert \Phi\right\Vert $, $\langle\Psi|\Phi\rangle$, and $\langle\Phi|\Psi\rangle$.
\end{xca}

\begin{problem}\leavevmode
\label{prob:inner1}Prove that the norm-squared $\left\Vert \Psi\right\Vert ^{2}$ is always non-negative, and it is zero if and only if $\left|\Psi\right\rangle $ is the zero vector, that is, the vector whose components are all zero. In other words, the inner product is \emph{positive-definite}\index{Positive-definite inner product}. As a corollary, explain why we must take the complex conjugate of the components when we convert a vector to a dual vector. (What would have happened if we didn't?)
\end{problem}

\begin{problem}\leavevmode
\label{prob:inner2}Prove that $\langle\Phi|\Psi\rangle=\langle\Psi|\Phi\rangle^{*}$, that is, if we swap the order of vectors in the inner product we get the complex conjugate of the original product. Thus, unlike the dot product, the inner product on $\BBC^{n}$ is not symmetric. However, it is \emph{conjugate-symmetric}\index{Conjugate-symmetric inner product}, and in particular, the magnitude of the inner product remains the same, since $\left|z\right|=\left|z^{*}\right|$.
\end{problem}

\begin{problem}\leavevmode
\label{prob:inner3}Prove that if $\alpha,\beta\in\BBC$ and $\left|\Psi\right\rangle ,\left|\Phi\right\rangle ,\left|\Theta\right\rangle \in\BBC^{n}$ then
\[
\left\langle \Psi\right|\left(\alpha\left|\Phi\right\rangle +\beta\left|\Theta\right\rangle \right)=\alpha\langle\Psi|\Phi\rangle+\beta\langle\Psi|\Theta\rangle,
\]
that is, the inner product is \emph{linear}\index{Linear inner product} in its second argument.
\end{problem}

\subsubsection{Orthonormal Bases}

An \emph{orthonormal basis\index{Orthonormal basis} }of $\BBC^{n}$ is a set of $n$ non-zero vectors $\left\{ \left|B_{1}\right\rangle ,\ldots,\left|B_{n}\right\rangle \right\} $ – which we will usually denote $\left|B_{i}\right\rangle $ for short, with the implication that $i\in\left\{ 1,\ldots,n\right\} $ – such that:
\begin{enumerate}
\item They \emph{span\index{Span}} $\BBC^{n}$, which means that any vector $\left|\Psi\right\rangle \in\BBC^{n}$ can be written as a \emph{linear combination\index{Linear combination}} of the basis vectors, that is, a sum of the vectors $\left|B_{i}\right\rangle $ multiplied by some complex numbers $\lambda_{i}\in\BBC$:
\[
\left|\Psi\right\rangle =\sum^{n}_{i=1}\lambda_{i}\left|B_{i}\right\rangle .
\]
This property ensures that the basis can be used to define any single vector in the space $\BBC^{n}$, not just part of that space.\\
As a simple example, in $\BBR^{3}$ the vector $\hat{\x}\equiv\left(1,0,0\right)$ pointing along the $x$ axis and the vector $\hat{\y}\equiv\left(0,1,0\right)$ pointing along the $y$ axis span the $xy$ plane, but \textbf{not }all of $\BBR^{3}$. To get a basis for all of $\BBR^{3}$, we must add an appropriate third vector, such as the vector $\hat{\z}\equiv\left(0,0,1\right)$ pointing along the $z$ axis. (But other vectors, such as $\left(1,2,3\right)$, would work as well.)
\item They are \emph{linearly independent}\index{Linearly independent}, in that if the zero vector is a linear combination of the basis vectors, then the coefficients in the linear combination must all be zero:
\[
\sum^{n}_{i=1}\lambda_{i}\left|B_{i}\right\rangle =0\soosp\lambda_{i}=0\sp\fa i.
\]
Linear independence means (as you will show in \zcref[S]{prob:Show-that-linear}) that no vector in the set can be written as a linear combination of the other vectors in the set. If we could have done so, then that vector would have been redundant, and we would have needed to remove it in order to obtain a basis.\\
As a simple example, the set composed of $\hat{\x}$, $\hat{\y}$, and $\left(1,2,0\right)$ is linearly dependent, since $\left(1,2,0\right)=\hat{\x}+2\hat{\y}$, but the set $\left\{ \hat{\x},\hat{\y},\hat{\z}\right\} $ is linearly independent.
\item They are all \emph{orthogonal\index{Orthogonal}} to each other, that is, the inner product of any two \textbf{different }vectors evaluates to zero:
\[
\langle B_{i}|B_{j}\rangle=0\sp\fa i\ne j.
\]
\item They are all \emph{unit vectors}\index{Unit vector}, that is, they have a norm (and norm-squared) of 1:
\[
\left\Vert B_{i}\right\Vert ^{2}=\langle B_{i}|B_{i}\rangle=1\sp\fa i.
\]
\end{enumerate}
In fact, properties 3 and 4 may be expressed more compactly as:
\begin{equation}
\langle B_{i}|B_{j}\rangle=\delta_{ij}=\begin{cases}
0 & \textrm{if }i\ne j,\\
1 & \textrm{if }i=j,
\end{cases}\label{eq:Kron}
\end{equation}
where $\delta_{ij}$ is called the \emph{Kronecker delta}\index{Kronecker delta}. If this combined property is satisfied, we say that the vectors are \emph{orthonormal}\index{Orthonormal vectors}\footnote{Actually, bases don't have to be orthonormal in general, but in quantum mechanics it is very convenient to use orthonormal bases, for reasons that will become clear later.}.

These requirements become much simpler in $n=2$ dimensions. An orthonormal basis for $\BBC^{2}$ is a set of 2 non-zero vectors $\left|B_{1}\right\rangle ,\left|B_{2}\right\rangle $ such that:
\begin{enumerate}
\item They span $\BBC^{2}$, which means that any vector $\left|\Psi\right\rangle \in\BBC^{2}$ can be written as a linear combination of the basis vectors:
\[
\left|\Psi\right\rangle =\lambda_{1}\left|B_{1}\right\rangle +\lambda_{2}\left|B_{2}\right\rangle ,
\]
for some $\lambda_{1},\lambda_{2}\in\BBC$.
\item They are linearly independent, which means that we cannot write one in terms of a scalar times the other, i.e.:
\[
\left|B_{1}\right\rangle \ne\lambda\left|B_{2}\right\rangle \sp\fa\lambda\in\BBC.
\]
\item They are orthonormal to each other, that is, the inner product between them evaluates to zero and both of them have unit norm:
\[
\langle B_{1}|B_{2}\rangle=0,
\]
\[
\left\Vert B_{1}\right\Vert ^{2}=\langle B_{1}|B_{1}\rangle=1\sp\left\Vert B_{2}\right\Vert ^{2}=\langle B_{2}|B_{2}\rangle=1.
\]
\end{enumerate}
A very important basis, the \emph{standard basis\index{Standard basis}} of $\BBC^{2}$, is defined as:
\[
\left|1_{1}\right\rangle \equiv\left(\begin{array}{c}
1\\
0
\end{array}\right)\sp\left|1_{2}\right\rangle \equiv\left(\begin{array}{c}
0\\
1
\end{array}\right).
\]
We similarly define the standard basis of $\BBC^{n}$ for any $n$ in the obvious way.
\begin{problem}\leavevmode
Show that the standard basis vectors satisfy the properties above.
\end{problem}

\begin{problem}\leavevmode
\label{prob:Show-that-linear}Show that linear independence means that no vector in the basis can be written as a linear combination of the other vectors in the basis.
\end{problem}

\begin{problem}\leavevmode
Any basis which is orthogonal but not orthonormal, that is, does \textbf{not }satisfy property 4, can be made orthonormal by \emph{normalizing\index{Normalizing a vector}} each basis vector, that is, dividing it by its norm:
\[
\left|B_{i}\right\rangle \mt\frac{\left|B_{i}\right\rangle }{\left\Vert B_{i}\right\Vert }.
\]
Show that if an orthogonal but not orthonormal basis satisfies properties 1–3, then it still satisfies them after normalizing it in this way.
\end{problem}

\begin{xca}\leavevmode
Consider the complex vector
\[
\left|\Psi\right\rangle \equiv\left(\begin{array}{c}
1+\i\\
2+2\i
\end{array}\right).
\]
Normalize $\left|\Psi\right\rangle $ and find another complex vector $\left|\Phi\right\rangle $ such that the set $\left\{ \left|\Psi\right\rangle ,\left|\Phi\right\rangle \right\} $ is an orthonormal basis of $\BBC^{2}$ (i.e. satisfies all of the properties above).
\end{xca}

\begin{problem}\leavevmode
Find an orthonormal basis of $\BBC^{3}$ which is \textbf{not} the standard basis or a scalar multiple of the standard basis. Show that it is indeed an orthonormal basis.
\end{problem}

\subsubsection{Matrices and the Adjoint}

A \emph{matrix\index{Matrix}} in $n$ dimensions is an $n\xx n$ array\footnote{In fact, matrices don't have to be square: they can have a different number of rows and columns, that is, $n\xx m$ where $n\ne m$; but non-square matrices are generally not of much interest in quantum mechanics.} of (complex) numbers. In $n=2$ dimensions we have
\[
A=\left(\begin{array}{cc}
A_{11} & A_{12}\\
A_{21} & A_{22}
\end{array}\right)\sp A_{11},A_{12},A_{21},A_{22}\in\BBC.
\]
A matrix can act on a vector to produce another vector. If it's a ket (a vertical/column vector), the result is another ket. If it's a bra (a horizontal/row dual vector), the result is another bra.

If the matrix acts on a ket, then it must act \textbf{from the left}, and the element at \textbf{row} $i$ of the resulting ket is obtained by taking the inner product of \textbf{row} $i$ of the matrix with the ket:
\begin{equation}
A\left|\Psi\right\rangle =\left(\begin{array}{cc}
A_{11} & A_{12}\\
A_{21} & A_{22}
\end{array}\right)\left(\begin{array}{c}
\Psi_{1}\\
\Psi_{2}
\end{array}\right)=\left(\begin{array}{c}
A_{11}\Psi_{1}+A_{12}\Psi_{2}\\
A_{21}\Psi_{1}+A_{22}\Psi_{2}
\end{array}\right).\label{eq:APhi}
\end{equation}
If the matrix acts on a bra, then it must act \textbf{from the right}, and the element at \textbf{column} $i$ of the resulting bra is obtained by taking the inner product of \textbf{column} $i$ of the matrix with the bra:
\[
\left\langle \Psi\right|A=\left(\begin{array}{cc}
\Psi^{*}_{1} & \Psi^{*}_{2}\end{array}\right)\left(\begin{array}{cc}
A_{11} & A_{12}\\
A_{21} & A_{22}
\end{array}\right)=\left(\begin{array}{cc}
\Psi^{*}_{1}A_{11}+\Psi^{*}_{2}A_{21}\thinspace\thinspace & \Psi^{*}_{1}A_{12}+\Psi^{*}_{2}A_{22}\end{array}\right).
\]
Note that the dual vector $\left\langle \Psi\right|A$ is, in general, \textbf{not} the dual of the vector $A\left|\Psi\right\rangle $, as you can see by taking the dual of \zcref[S]{eq:APhi}. However, we can define the \emph{adjoint\index{Adjoint} }of a matrix by transposing rows into columns and then taking the complex conjugate of all the components:
\[
A^{\dagger}=\left(\begin{array}{cc}
A^{*}_{11} & A^{*}_{21}\\
A^{*}_{12} & A^{*}_{22}
\end{array}\right),
\]
where the notation $\dagger$ for the adjoint is called \emph{dagger}\index{Dagger}. Then the vector dual to $A\left|\Psi\right\rangle $ is $\left\langle \Psi\right|A^{\dagger}$, as you will check in \zcref[S]{prob:Show-that-the}. Actually, taking the adjoint of a matrix is exactly the same operation as taking the dual of a vector! The only difference is that for a matrix we have $n$ columns to transpose into rows, while for a vector we only have one. Therefore, we have
\[
\left|\Psi\right\rangle ^{\dagger}=\left\langle \Psi\right|\sp\left\langle \Psi\right|^{\dagger}=\left|\Psi\right\rangle ,
\]
and we get the following nice relation:
\[
\left(A\left|\Psi\right\rangle \right)^{\dagger}=\left\langle \Psi\right|A^{\dagger}.
\]

The \emph{identity matrix}\index{Identity matrix}, which we will write simply as 1, is:
\[
1=\left(\begin{array}{cc}
1 & 0\\
0 & 1
\end{array}\right).
\]
Acting with it on any vector or dual vector does not change it: $1\left|\Psi\right\rangle =\left|\Psi\right\rangle $ and $\left\langle \Psi\right|1=\left\langle \Psi\right|$.
\begin{problem}\leavevmode
To rotate (real) vectors in $\BBR^{2}$ by an angle $\theta$, we take their product with the (real) \emph{rotation matrix}\index{Rotation matrix}:
\[
R\left(\theta\right)\equiv\left(\begin{array}{cc}
\cos\theta & -\sin\theta\\
\sin\theta & \cos\theta
\end{array}\right).
\]
\begin{problemparts}
\item Calculate the matrix $R\left(\pi/3\right)$.
\item Write down the vector resulting from rotating $\left(-5,9\right)$ by $\pi/3$ radians, in both Cartesian and polar coordinates.
\item Repeat (B) for rotating a general 2-vector $\left(x,y\right)$ by a general angle $\theta$.
\item Find the mapping between rotations of 2-vectors in $\BBR^{2}$ and rotations of complex numbers in $\BBC$, and explain what the analogue of the rotation matrix is in terms of complex numbers.
\end{problemparts}
\end{problem}

\begin{problem}\leavevmode
\label{prob:Show-that-the}Show that the vector dual to $A\left|\Psi\right\rangle $ is indeed $\left\langle \Psi\right|A^{\dagger}$.
\end{problem}

\begin{xca}\leavevmode
Let
\[
A\equiv\left(\begin{array}{cc}
1+5\i & 2\\
3-7\i & 4+8\i
\end{array}\right)\sp\left\langle \Psi\right|\equiv\left(\begin{array}{cc}
\i-2 & \i-3\end{array}\right).
\]
Calculate $A\left|\Psi\right\rangle $ and $\left\langle \Psi\right|A^{\dagger}$ separately, and then check that they are the dual of each other.
\end{xca}

\begin{problem}\leavevmode
Show that $\left(A^{\dagger}\right)^{\dagger}=A$. This means that the adjoint operation is an involution, exactly like complex conjugation and taking the dual of a vector. In fact, all three are the exact same operation. By choosing an appropriate matrix, explain how taking the complex conjugate of a number is a special case of taking the adjoint of a matrix.
\end{problem}

\begin{problem}\leavevmode
Show that the action of a matrix on a vector is \emph{linear}\index{Linear transformation}, that is,
\[
A\left(\alpha\left|\Psi\right\rangle +\beta\left|\Phi\right\rangle \right)=\alpha A\left|\Psi\right\rangle +\beta A\left|\Phi\right\rangle .
\]
\end{problem}

\subsubsection{The Outer Product}

We have seen that vectors and dual vectors may be combined to generate a complex number using the inner product. We can similarly combine a vector and a dual vector to generate a matrix, using the \emph{outer product}\index{Outer product}. Given
\[
\left\langle \Psi\right|\equiv\left(\begin{array}{cc}
\Psi^{*}_{1} & \Psi^{*}_{2}\end{array}\right)\sp\left|\Phi\right\rangle \equiv\left(\begin{array}{c}
\Phi_{1}\\
\Phi_{2}
\end{array}\right),
\]
we define the outer product as the matrix whose component at row $i$, column $j$ is given by multiplying the component at row $i$ of $\left|\Phi\right\rangle $ with the component at column $j$ of $\left\langle \Psi\right|$:
\[
|\Phi\rangle\langle\Psi|=\left(\begin{array}{c}
\Phi_{1}\\
\Phi_{2}
\end{array}\right)\left(\begin{array}{cc}
\Psi^{*}_{1} & \Psi^{*}_{2}\end{array}\right)=\left(\begin{array}{cc}
\Psi^{*}_{1}\Phi_{1} & \Psi^{*}_{2}\Phi_{1}\\
\Psi^{*}_{1}\Phi_{2} & \Psi^{*}_{2}\Phi_{2}
\end{array}\right).
\]
Note how when taking an inner product the straight lines $|$ face each other: $\langle\Psi|\Phi\rangle$, while when taking an outer product the angle brackets $\rangle\langle$ face each other. This shows some of the elegance of the Dirac notation! A bra-ket is an inner product, while a ket-bra is an outer product.

We can assign a\emph{ tensor rank}\index{Tensor rank} to scalars, vectors, and matrices:
\begin{itemize}
\item Scalars have tensor rank 0 since they have $n^{0}=1$ component and require 0 component indices: $\phi$.
\item Vectors have tensor rank 1 since they have $n^{1}=n$ components and require 1 component index: $v_{i}$ for $i\in\left\{ 1,\ldots,n\right\} $.
\item Matrices have tensor rank 2 since they have $n^{2}$ components and require 2 component indices: $M_{ij}$ for $i,j\in\left\{ 1,\ldots,n\right\} $.
\end{itemize}
Then the inner product reduces the rank of the vectors from 1 to 0, while the outer product increases the rank from 1 to 2.
\begin{xca}\leavevmode
Calculate the outer product $|\Psi\rangle\langle\Phi|$ for
\[
\left|\Psi\right\rangle =\left(\begin{array}{c}
1\\
2+\i
\end{array}\right)\sp\left|\Phi\right\rangle =\left(\begin{array}{c}
3-\i\\
4\i
\end{array}\right).
\]
Remember that when writing the dual vector, the components are complex conjugated!
\end{xca}

\begin{problem}\leavevmode
\label{prob:adjoint-outer}Prove that for any two vectors $\left|\Psi\right\rangle $ and $\left|\Phi\right\rangle $, the adjoint of the outer product simply exchanges the bra and the ket:
\[
\left(|\Psi\rangle\langle\Phi|\right)^{\dagger}=|\Phi\rangle\langle\Psi|.
\]
\end{problem}

\subsubsection{The Completeness Relation}

Let us write the vector $\left|\Psi\right\rangle $ as a linear combination of basis vectors:
\begin{equation}
\left|\Psi\right\rangle =\sum^{n}_{i=1}\lambda_{i}\left|B_{i}\right\rangle .\label{eq:linear-comb}
\end{equation}
Taking the inner product of \zcref[S]{eq:linear-comb} with $\langle B_{j}|$ and using the fact that the basis vectors are orthonormal,
\[
\langle B_{i}|B_{j}\rangle=\delta_{ij}=\begin{cases}
0 & \textrm{if }i\ne j,\\
1 & \textrm{if }i=j,
\end{cases}
\]
we get:
\[
\langle B_{j}|\Psi\rangle=\sum^{n}_{i=1}\lambda_{i}\langle B_{j}|B_{i}\rangle=\sum^{n}_{i=1}\lambda_{i}\delta_{ij}=\lambda_{j},
\]
since all of the terms in the sum vanish except the one with $i=j$. Therefore, the coefficients $\lambda_{i}$ in \zcref[S]{eq:linear-comb} are given, for any vector $\left|\Psi\right\rangle $ and for any orthonormal basis $\left|B_{i}\right\rangle $, by
\begin{equation}
\lambda_{i}=\langle B_{i}|\Psi\rangle.\label{eq:lambda_i}
\end{equation}
Now, since $\lambda_{i}$ is a scalar, and multiplication by a scalar is commutative (unlike the inner and outer products!), we can move it to the right in \zcref[S]{eq:linear-comb}:
\[
\left|\Psi\right\rangle =\sum^{n}_{i=1}\left|B_{i}\right\rangle \lambda_{i}.
\]
We haven't actually \textbf{done }anything here; where to write the scalar, on the left or right of the vector, is completely arbitrary – it's just conventional to write it on the left. Then, replacing $\lambda_{i}$ with $\langle B_{i}|\Psi\rangle$ as per \zcref[S]{eq:lambda_i}, we get
\[
\left|\Psi\right\rangle =\sum^{n}_{i=1}|B_{i}\rangle\langle B_{i}|\Psi\rangle.
\]
To make this even more suggestive, let us add parentheses:
\begin{equation}
\left|\Psi\right\rangle =\left(\sum^{n}_{i=1}|B_{i}\rangle\langle B_{i}|\right)|\Psi\rangle.\label{eq:Phi-completeness}
\end{equation}
Note that what we did here is go from a \textbf{vector} $\left|B_{i}\right\rangle $ times a\textbf{ complex number} $\langle B_{i}|\Psi\rangle$ to a \textbf{matrix} $|B_{i}\rangle\langle B_{i}|$ times a \textbf{vector} $\left|\Psi\right\rangle $, for each $i$. The fact that these two different products are actually equal to one another (as you will prove in \zcref[S]{prob:Provide-a-rigorous}) is not at all trivial, but it is one of the main reasons we like to use bra-ket notation! The notation now suggests (see \zcref[S]{prob:divide}) that
\begin{equation}
\sum^{n}_{i=1}|B_{i}\rangle\langle B_{i}|=1,\label{eq:completeness}
\end{equation}
where $|B_{i}\rangle\langle B_{i}|$ is the outer product defined above, and the 1 on the right-hand side is the identity matrix. This extremely useful result is called the \emph{completeness relation}\index{Completeness relation}.

In $\BBC^{2}$, we simply have
\begin{equation}
|B_{1}\rangle\langle B_{1}|+|B_{2}\rangle\langle B_{2}|=1.\label{eq:C2-completeness}
\end{equation}

\begin{xca}\leavevmode
Given the basis
\[
\left|B_{1}\right\rangle =\frac{1}{\sqrt{2}}\left(\begin{array}{c}
1\\
1
\end{array}\right)\sp\left|B_{2}\right\rangle =\frac{1}{\sqrt{2}}\left(\begin{array}{c}
1\\
-1
\end{array}\right),
\]
first show that it is indeed an orthonormal basis, and then show that it satisfies the completeness relation given by \zcref[S]{eq:C2-completeness}.
\end{xca}

\begin{problem}\leavevmode
\label{prob:Provide-a-rigorous}For any three vectors $\left|A\right\rangle ,\left|B\right\rangle ,\left|C\right\rangle \in\BBC^{n}$ and for any dimension $n$, the following equality is satisfied:
\[
|A\rangle\langle B|C\rangle=\left(\vphantom{\bll}|A\rangle\langle B|\right)|C\rangle.
\]
In other words, the product ket-bra-ket is \emph{associative\index{Associative operation}}. On the left-hand side we have a vector $\left|A\right\rangle $ times a scalar $\langle B|C\rangle$ (which is the result of an inner product), while on the right-hand side we have a matrix $|A\rangle\langle B|$ (which is the result of an outer product) times a vector $\left|C\right\rangle $, but the two sides are nonetheless equal.
\begin{problemparts}
\item Provide a rigorous proof of this associative property.
\item Use this associative property to prove that
\[
\sum^{n}_{i=1}|B_{i}\rangle\langle B_{i}|\Psi\rangle=\left(\sum^{n}_{i=1}|B_{i}\rangle\langle B_{i}|\right)|\Psi\rangle.
\]
\end{problemparts}
\end{problem}

\begin{problem}\leavevmode
\label{prob:divide}Importantly, we didn't “divide \zcref[S]{eq:Phi-completeness} by $\left|\Psi\right\rangle $” to get \zcref[S]{eq:completeness}! You can't do that with matrices and vectors. Instead, \zcref[S]{eq:completeness} follows from the fact that any matrix $A$ which satisfies $\left|\Psi\right\rangle =A\left|\Psi\right\rangle $ for \textbf{every} vector $\left|\Psi\right\rangle $ must necessarily be the identity matrix. Prove this.
\end{problem}

\subsubsection{Representing Vectors in Different Bases}\label{subsec:Representing-Vectors-in}

Let us consider a complex $n$-vector defined as follows:
\[
\left|\Psi\right\rangle \equiv\left(\begin{array}{c}
\Psi_{1}\\
\vdots\\
\Psi_{n}
\end{array}\right)\sp\Psi_{i}\in\BBC.
\]
Given an orthonormal basis $\left|B_{i}\right\rangle $, we have seen that we can write $\left|\Psi\right\rangle $ as a linear combination of the basis vectors:
\[
\left|\Psi\right\rangle =\sum^{n}_{i=1}\lambda_{i}\left|B_{i}\right\rangle .
\]
The coefficients $\lambda_{i}\in\BBC$ depend on $\left|\Psi\right\rangle $ and on the basis vectors, as we showed in \zcref[S]{eq:lambda_i}:
\[
\lambda_{i}=\langle B_{i}|\Psi\rangle\soosp\left|\Psi\right\rangle =\sum^{n}_{i=1}|B_{i}\rangle\langle B_{i}|\Psi\rangle.
\]
With these coefficients, we can \emph{represent} the vector $\left|\Psi\right\rangle $ in the basis $\left|B_{i}\right\rangle $\index{Representing a vector in a basis}. This representation will be a vector of the same dimension $n$, with the components being the coefficients $\lambda_{i}=\langle B_{i}|\Psi\rangle$, and will be denoted as follows:
\[
\left|\Psi\right\rangle \bllll_{B}\equiv\left(\begin{array}{c}
\langle B_{1}|\Psi\rangle\\
\vdots\\
\langle B_{n}|\Psi\rangle
\end{array}\right)=\left(\begin{array}{c}
\lambda_{1}\\
\vdots\\
\lambda_{n}
\end{array}\right).
\]
We say that $\lambda_{i}$ are the \emph{coordinates}\index{Coordinates of a vector in a basis} of $\left|\Psi\right\rangle $ with respect to the basis $\left|B_{i}\right\rangle $.

The correct way to understand the meaning of a vector is as an \textbf{abstract }entity, like an arrow in space, which does not depend on any particular basis – it is just there. However, if we want to do concrete calculations with a vector, we must somehow represent it numerically. This is done by choosing a basis and writing down the coordinates of the vector in that basis.

Therefore, whenever we define a vector using its components – as we have been doing throughout this chapter – there is always a specific basis in which the vector is represented, with the components being the coordinates in this basis. If no particular basis is explicitly specified, it is implied that it is the standard basis. But no representation is better than any other; we usually choose whatever basis is most convenient to work with. In quantum mechanics, we often choose a basis defined by some physical observable, as we will see below.
\begin{xca}\leavevmode
Let a vector $\left|\Psi\right\rangle $ be represented in the standard basis as
\[
\left|\Psi\right\rangle \equiv\left(\begin{array}{c}
1-9\i\\
7\i-2
\end{array}\right).
\]
Find its representation $\left|\Psi\right\rangle \bl_{B}$ in terms of the orthonormal basis
\[
\left|B_{1}\right\rangle =\frac{1}{\sqrt{2}}\left(\begin{array}{c}
1\\
1
\end{array}\right)\sp\left|B_{2}\right\rangle =\frac{1}{\sqrt{2}}\left(\begin{array}{c}
1\\
-1
\end{array}\right).
\]
\end{xca}

\begin{problem}\leavevmode
Prove that the inner product (and thus also the norm) is independent of the choice of basis. That is, for any two vectors $\left|\Psi\right\rangle $ and $\left|\Phi\right\rangle $ and any two orthonormal bases $\left|B_{i}\right\rangle $ and $\left|C_{i}\right\rangle $,
\[
\langle\Psi|\Phi\rangle\bllll_{B}=\langle\Psi|\Phi\rangle\bllll_{C}.
\]
\end{problem}

\subsubsection{Change of Basis}

Let the representation of a vector $\left|\Psi\right\rangle $ in the basis $\left|B_{i}\right\rangle $ be
\[
\left|\Psi\right\rangle \bllll_{B}=\left(\begin{array}{c}
\langle B_{1}|\Psi\rangle\\
\vdots\\
\langle B_{n}|\Psi\rangle
\end{array}\right).
\]
Given a different orthonormal basis $\left|C_{i}\right\rangle $, we have a different representation
\[
\left|\Psi\right\rangle \bllll_{C}=\left(\begin{array}{c}
\langle C_{1}|\Psi\rangle\\
\vdots\\
\langle C_{n}|\Psi\rangle
\end{array}\right).
\]
To find a relation between the two representations, we use the completeness relation, \zcref[S]{eq:completeness}:
\[
\sum^{n}_{j=1}|B_{j}\rangle\langle B_{j}|=1.
\]
Inserting it in the middle of the inner product representing the coordinates $\langle C_{i}|\Psi\rangle$, we get that for all $i$
\[
\langle C_{i}|\Psi\rangle=\langle C_{i}|\left(\sum^{n}_{j=1}|B_{j}\rangle\langle B_{j}|\right)|\Psi\rangle=\sum^{n}_{j=1}\langle C_{i}|B_{j}\rangle\langle B_{j}|\Psi\rangle.
\]
Again, the Dirac notation proves to be pretty convenient! This relation can be expressed in matrix form as follows:
\[
\left(\begin{array}{c}
\langle C_{1}|\Psi\rangle\\
\vdots\\
\langle C_{n}|\Psi\rangle
\end{array}\right)=\left(\begin{array}{ccc}
\langle C_{1}|B_{1}\rangle & \cdots & \langle C_{1}|B_{n}\rangle\\
\vdots & \ddots & \vdots\\
\langle C_{n}|B_{1}\rangle & \cdots & \langle C_{n}|B_{n}\rangle
\end{array}\right)\left(\begin{array}{c}
\langle B_{1}|\Psi\rangle\\
\vdots\\
\langle B_{n}|\Psi\rangle
\end{array}\right),
\]
or in other words,
\[
\left|\Psi\right\rangle \bllll_{C}=P_{C\ot B}\left|\Psi\right\rangle \bllll_{B},
\]
where the \emph{change-of-basis matrix\index{Change-of-basis matrix}} from $\left|B_{i}\right\rangle $ to $\left|C_{i}\right\rangle $, denoted $P_{C\ot B}$, is defined as
\begin{equation}
P_{C\ot B}\equiv\left(\begin{array}{ccc}
\langle C_{1}|B_{1}\rangle & \cdots & \langle C_{1}|B_{n}\rangle\\
\vdots & \ddots & \vdots\\
\langle C_{n}|B_{1}\rangle & \cdots & \langle C_{n}|B_{n}\rangle
\end{array}\right).\label{eq:c-o-b}
\end{equation}

\begin{xca}\leavevmode
Consider the two bases
\[
\left|B_{1}\right\rangle =\frac{1}{\sqrt{2}}\left(\begin{array}{c}
1\\
1
\end{array}\right)\sp\left|B_{2}\right\rangle =\frac{1}{\sqrt{2}}\left(\begin{array}{c}
1\\
-1
\end{array}\right),
\]
\[
\left|C_{1}\right\rangle =\frac{1}{\sqrt{2}}\left(\begin{array}{c}
1\\
\i
\end{array}\right)\sp\left|C_{2}\right\rangle =\frac{1}{\sqrt{2}}\left(\begin{array}{c}
-\i\\
-1
\end{array}\right).
\]
\begin{problemparts}
\item The vector $\left|\Psi\right\rangle $ is represented in the standard basis as
\[
\left|\Psi\right\rangle =\left(\begin{array}{c}
-3\\
2+\i
\end{array}\right).
\]
Find its representations in the bases $\left|B_{i}\right\rangle $ and $\left|C_{i}\right\rangle $.
\item Find the change-of-basis matrix $P_{C\ot B}$. Calculate $P_{C\ot B}\left|\Psi\right\rangle \bl_{B}$ and verify that the result is equal to the expression you obtained in (A) for $\left|\Psi\right\rangle \bl_{C}$.
\end{problemparts}
\end{xca}

\subsubsection{Multiplication and Inverse of Matrices}\label{subsec:Multiplication-and-Inverse}

The \emph{matrix product}\index{Matrix product} of two matrices is another matrix. The element of that matrix at row $i$, column $j$ is calculated by taking the dot product of row $i$ of the left matrix with column $j$ of the right matrix:
\[
AB=\left(\begin{array}{cc}
A_{11} & A_{12}\\
A_{21} & A_{22}
\end{array}\right)\left(\begin{array}{cc}
B_{11} & B_{12}\\
B_{21} & B_{22}
\end{array}\right)=\left(\begin{array}{cc}
A_{11}B_{11}+A_{12}B_{21}\thinspace\thinspace & A_{11}B_{12}+A_{12}B_{22}\\
A_{21}B_{11}+A_{22}B_{21}\thinspace\thinspace & A_{21}B_{12}+A_{22}B_{22}
\end{array}\right).
\]
Observe that the action of a matrix on a vector, and the inner and outer products of vectors, are all just special cases of matrix multiplication – where a ket is a matrix with only one column, and a bra is a matrix with only one row!

Given a matrix $A$, if there exists another matrix $A^{-1}$ such that
\[
A^{-1}A=AA^{-1}=1,
\]
then the matrix $A$ is called \emph{invertible\index{Invertible matrix}} and $A^{-1}$ is called its \emph{inverse matrix}\index{Inverse matrix}. Note that $\left(A^{-1}\right)^{-1}=A$, so the operation of taking the inverse is an involution. Sometimes matrices do not have an inverse; such matrices are called \emph{singular}\index{Singular matrix}.
\begin{xca}\leavevmode
Calculate the products $AB$ and $BA$ where:
\[
A\equiv\left(\begin{array}{cc}
-1 & 3\\
-6\i & 2\i-1
\end{array}\right)\sp B\equiv\left(\begin{array}{cc}
9-8\i & 7\\
4\i & -2\i
\end{array}\right).
\]
\end{xca}

\begin{problem}\leavevmode
\label{prob:Find-a-general}Find a general formula for the inverse of a $2\xx2$ matrix by taking
\[
A\equiv\left(\begin{array}{cc}
a & b\\
c & d
\end{array}\right)\sp A^{-1}\equiv\left(\begin{array}{cc}
e & f\\
g & h
\end{array}\right),
\]
and solving for $e,f,g,h$ in terms of $a,b,c,d$.
\end{problem}

\begin{xca}\leavevmode
Find the inverse of the matrix
\[
A\equiv\left(\begin{array}{cc}
1 & 2-4\i\\
-\i & -2
\end{array}\right).
\]
You can use the formula you found in \zcref[S]{prob:Find-a-general}.
\end{xca}

\begin{problem}\leavevmode
\label{prob:adjoint-product}Show that $\left(AB\right)^{\dagger}=B^{\dagger}A^{\dagger}$ for any two matrices $A$ and $B$.
\end{problem}

\begin{problem}\leavevmode
Show that $\left(AB\right)^{-1}=B^{-1}A^{-1}$ for any two invertible matrices $A$ and $B$.
\end{problem}

\begin{problem}\leavevmode
\label{prob:Matrix-multiplication-is}Matrix multiplication is not \emph{commutative\index{Commutative operation}} in general. That is, for two arbitrary matrices $A$ and $B$, it is not in general true that $AB=BA$. Find an example of two matrices which commute, and an example of two matrices which do not commute. In each case, show that they indeed commute or don't commute.
\end{problem}

\begin{problem}\leavevmode
Show that multiplying by a scalar $\lambda\in\BBC$ is the same as multiplying by a matrix with all of its elements equal to zero except for the elements on the diagonal, which are all equal to $\lambda$:
\[
\lambda A=\left(\begin{array}{cc}
\lambda & 0\\
0 & \lambda
\end{array}\right)A.
\]
Such a matrix is known as a \emph{scalar matrix}\index{Scalar matrix}.
\end{problem}

\begin{problem}\leavevmode
Given two orthonormal bases $\left|B_{i}\right\rangle $ and $\left|C_{i}\right\rangle $, show that the change-of-basis matrix $P_{B\ot C}$ is the inverse of the change-of-basis matrix in the other direction, $P_{C\ot B}$.
\end{problem}

\subsubsection{Matrices Inside Inner Products}

Since $A\left|\Phi\right\rangle $ is itself a vector, we may calculate the inner product of that vector with the dual vector $\left\langle \Psi\right|$, which as usual gives us a complex number\index{Matrices inside inner products}:
\begin{align*}
\langle\Psi|A|\Phi\rangle & =\left(\begin{array}{cc}
\Psi^{*}_{1} & \Psi^{*}_{2}\end{array}\right)\left(\begin{array}{cc}
A_{11} & A_{12}\\
A_{21} & A_{22}
\end{array}\right)\left(\begin{array}{c}
\Phi_{1}\\
\Phi_{2}
\end{array}\right)\\
 & =\Psi^{*}_{1}A_{11}\Phi_{1}+\Psi^{*}_{2}A_{21}\Phi_{1}+\Psi^{*}_{1}A_{12}\Phi_{2}+\Psi^{*}_{2}A_{22}\Phi_{2}.
\end{align*}
If we take the dual of $A\left|\Phi\right\rangle $ we get $\left\langle \Phi\right|A^{\dagger}$, as you proved in \zcref[S]{prob:Show-that-the}. Thus, inverting the order of the inner product, we get
\begin{align*}
\langle\Phi|A^{\dagger}|\Psi\rangle & =\left(\begin{array}{cc}
\Phi^{*}_{1} & \Phi^{*}_{2}\end{array}\right)\left(\begin{array}{cc}
A^{*}_{11} & A^{*}_{21}\\
A^{*}_{12} & A^{*}_{22}
\end{array}\right)\left(\begin{array}{c}
\Psi_{1}\\
\Psi_{2}
\end{array}\right)\\
 & =\Psi_{1}A^{*}_{11}\Phi^{*}_{1}+\Psi_{2}A^{*}_{21}\Phi^{*}_{1}+\Psi_{1}A^{*}_{12}\Phi^{*}_{2}+\Psi_{2}A^{*}_{22}\Phi^{*}_{2}.
\end{align*}
This is, of course, the complex conjugate of $\langle\Psi|A|\Phi\rangle$, since inverting the order of the inner product results in the complex conjugate. In other words, we have the relation
\[
\langle\Psi|A|\Phi\rangle^{*}=\langle\Phi|A^{\dagger}|\Psi\rangle.
\]
Taking the complex conjugate reverses the order of the inner product, and also replaces the matrix with its adjoint.
\begin{xca}\leavevmode
Calculate the inner product $\langle\Psi|A|\Phi\rangle$ where
\[
\left|\Psi\right\rangle =\left(\begin{array}{c}
5+2\i\\
-3\i
\end{array}\right)\sp A=\left(\begin{array}{cc}
9 & 8\i\\
6\i & 5-4\i
\end{array}\right)\sp\left|\Phi\right\rangle =\left(\begin{array}{c}
3+4\i\\
2
\end{array}\right).
\]
\end{xca}

\subsubsection{Eigenvalues and Eigenvectors}

If the matrix $A$, acting on the (non-zero) vector $\left|\Psi\right\rangle $, results in a scalar multiple of $\left|\Psi\right\rangle $:
\[
A\left|\Psi\right\rangle =\lambda\left|\Psi\right\rangle \sp\lambda\in\BBC,
\]
then we call $\left|\Psi\right\rangle $ an \emph{eigenvector\index{Eigenvector} }of $A$ and $\lambda$ its \emph{eigenvalue}\index{Eigenvalue}. Note that $\left|\Psi\right\rangle $ cannot be the zero vector, but $\lambda$ can be zero.

For example, if
\[
A=\left(\begin{array}{cc}
1 & 0\\
0 & -1
\end{array}\right),
\]
then it's easy to see that
\[
\left|\Psi\right\rangle =\left(\begin{array}{c}
1\\
0
\end{array}\right)
\]
is an eigenvector with eigenvalue $+1$ and
\[
\left|\Phi\right\rangle =\left(\begin{array}{c}
0\\
1
\end{array}\right)
\]
is an eigenvector with eigenvalue $-1$:
\[
A\left|\Psi\right\rangle =\left|\Psi\right\rangle \sp A\left|\Phi\right\rangle =-\left|\Phi\right\rangle .
\]

\begin{xca}\leavevmode
The matrix
\[
A\equiv\left(\begin{array}{cc}
1 & 2\\
2 & 1
\end{array}\right)
\]
has two linearly independent eigenvectors. Find them and their corresponding eigenvalues.
\end{xca}

\begin{problem}\leavevmode
Prove that, if $\left|\Psi\right\rangle $ is an eigenvector of a matrix $A$, then $\alpha\left|\Psi\right\rangle $ is also an eigenvector of $A$ for any\footnote{Here, $\BBC\setminus\left\{ 0\right\} $ is the set of all complex numbers except zero. In general, $A\setminus B$ denotes the \emph{set difference}\index{Set difference}, that is, the set of elements in $A$ that are not in $B$.} $\alpha\in\BBC\setminus\left\{ 0\right\} $, and it has the same eigenvalue.
\end{problem}

\subsubsection{Hermitian Matrices}\label{subsec:Hermitian-Matrices}

A matrix $A$ is called \emph{Hermitian}\index{Hermitian matrix} if it's equal to its adjoint:
\[
A=A^{\dagger}.
\]
Thus, it is sometimes also referred to as a \emph{self-adjoint}\index{Self-adjoint matrix} matrix. For such a matrix, we have that
\begin{equation}
\langle\Psi|A|\Phi\rangle^{*}=\langle\Phi|A|\Psi\rangle.\label{eq:Hermitian}
\end{equation}
A Hermitian matrix is analogous to a real number, since $z=z^{*}$ implies that $z$ is real.

The eigenvalues of a Hermitian matrix must all be real. To see this, let $\lambda$ be an eigenvalue of the Hermitian matrix $A$ with the eigenvector $\left|\Psi\right\rangle $:
\[
A\left|\Psi\right\rangle =\lambda\left|\Psi\right\rangle .
\]
Then we can take the inner product of both sides with $\left\langle \Psi\right|$:
\[
\langle\Psi|A|\Psi\rangle=\langle\Psi|\lambda|\Psi\rangle=\lambda\langle\Psi|\Psi\rangle=\lambda\left\Vert \Psi\right\Vert ^{2},
\]
where we were able to move $\lambda$ out of the inner product because it's just a number. From \zcref[S]{eq:Hermitian}, we have:
\[
\langle\Psi|A|\Psi\rangle=\langle\Psi|A|\Psi\rangle^{*},
\]
so $\langle\Psi|A|\Psi\rangle$ is real. Since $\left\Vert \Psi\right\Vert ^{2}$ is also real – and non-zero, since $\left|\Psi\right\rangle $ is an eigenvector, so by definition it cannot be the zero vector – we conclude that $\lambda$ must be real.

Now, let $\left|\Psi\right\rangle $ and $\left|\Phi\right\rangle $ be two eigenvectors of $A$ corresponding to different eigenvalues $\lambda$ and $\mu$ respectively:
\[
A\left|\Psi\right\rangle =\lambda\left|\Psi\right\rangle \sp A\left|\Phi\right\rangle =\mu\left|\Phi\right\rangle \sp\lambda\ne\mu.
\]
Let us take the inner product of the first equation with $\left\langle \Phi\right|$ and of the second equation with $\left\langle \Psi\right|$:
\[
\langle\Phi|A|\Psi\rangle=\langle\Phi|\lambda|\Psi\rangle=\lambda\langle\Phi|\Psi\rangle,
\]
\[
\langle\Psi|A|\Phi\rangle=\langle\Psi|\mu|\Phi\rangle=\mu\langle\Psi|\Phi\rangle.
\]
From \zcref[S]{eq:Hermitian}, the first equation is the complex conjugate of the second equation. Since $\lambda$ must be real – as we just proved – we get
\[
\mu\langle\Psi|\Phi\rangle=\left(\lambda\langle\Phi|\Psi\rangle\right)^{*}=\lambda\langle\Psi|\Phi\rangle.
\]
Seeing that $\lambda\ne\mu$ by our assumption, this equation can only be true if
\[
\langle\Psi|\Phi\rangle=0.
\]
In other words, eigenvectors of a Hermitian matrix corresponding to different eigenvalues are orthogonal. Now, since an eigenvector multiplied by a non-zero scalar is still an eigenvector, the eigenvectors $\left|\Psi\right\rangle $ and $\left|\Phi\right\rangle $ can be divided by their norms, so that they are not only orthogonal but also orthonormal.

Moreover, one can prove that for any Hermitian matrix $A$ in $\BBC^{n}$, there is an orthonormal basis of $\BBC^{n}$ consisting of eigenvectors of $A$. Such a basis is called an \emph{orthonormal eigenbasis\index{Orthonormal eigenbasis}}. The proof requires some slightly more advanced tools from linear algebra, so we won't write it here. However, this theorem is extremely important in quantum theory. As we will see, Hermitian matrices represent physical observables in quantum theory, and their eigenvalues correspond to the possible values obtained by performing measurements on these observables. The fact that there is an orthonormal basis of eigenvectors will prove very useful for studying observables in quantum theory.
\begin{problem}\leavevmode
Let $A$ and $B$ be Hermitian matrices. Under what conditions is the product $AB$ Hermitian?
\end{problem}

\begin{xca}\leavevmode
Consider the matrix
\[
A\equiv\left(\begin{array}{cc}
0 & 2\i\\
c & 0
\end{array}\right).
\]
\begin{problemparts}
\item Find the value of $c$ for which $A$ is a Hermitian matrix.
\item Find the eigenvalues of $A$ with the value of $c$ that you found in (A).
\item Find an orthonormal eigenbasis of $A$ with the value of $c$ that you found in (A). Show that it is indeed orthonormal.
\end{problemparts}
\end{xca}

\begin{problem}\leavevmode
Find the most general $2\xx2$ Hermitian matrix by demanding that $A=A^{\dagger}$ and finding conditions on the components of $A$.
\end{problem}

\subsubsection{Unitary and Normal Matrices}\label{subsec:Unitary-and-Normal-Matrices}

A matrix $U$ is called \emph{unitary\index{Unitary matrix} }if its adjoint is also its inverse:
\[
U^{-1}=U^{\dagger}\soosp UU^{\dagger}=U^{\dagger}U=1.
\]
A unitary matrix is analogous to a complex number with magnitude $1$, since such a number satisfies $z^{*}z=zz^{*}=\left|z\right|^{2}=1$.

Acting with a unitary matrix on two vectors preserves their inner product. To see this, consider a unitary matrix $U$ and two vectors $\left|\Psi\right\rangle $ and $\left|\Phi\right\rangle $. If we act with $U$ on both vectors, we get $U\left|\Psi\right\rangle $ and $U\left|\Phi\right\rangle $. Taking the bra of the ket $U\left|\Psi\right\rangle $, we obtain
\[
\left(U\left|\Psi\right\rangle \right)^{\dagger}=\left\langle \Psi\right|U^{\dagger}.
\]
If we take the inner product of $U\left|\Psi\right\rangle $ and $U\left|\Phi\right\rangle $, we get
\[
\langle\Psi|U^{\dagger}U|\Phi\rangle=\langle\Psi|\Phi\rangle,
\]
since $U^{\dagger}U=1$. Therefore, the inner product of these two vectors is the same before and after acting on them with $U$.

Now, let $\lambda$ be an eigenvalue of the unitary matrix $U$ with the eigenvector $\left|\Psi\right\rangle $:
\[
U\left|\Psi\right\rangle =\lambda\left|\Psi\right\rangle .
\]
Taking the adjoint of both sides, we get
\[
\left\langle \Psi\right|U^{\dagger}=\left\langle \Psi\right|\lambda^{*}.
\]
Multiplying both equations together, we have
\[
\left\langle \Psi\right|U^{\dagger}U\left|\Psi\right\rangle =\left\langle \Psi\right|\lambda^{*}\lambda\left|\Psi\right\rangle .
\]
On the left-hand side $U^{\dagger}U=1$, and on the right-hand side $\lambda^{*}\lambda=\left|\lambda\right|^{2}$, so
\[
\langle\Psi|\Psi\rangle=\left|\lambda\right|^{2}\langle\Psi|\Psi\rangle\soosp\left\Vert \Psi\right\Vert ^{2}=\left|\lambda\right|^{2}\left\Vert \Psi\right\Vert ^{2}.
\]
Since $\left|\Psi\right\rangle $ is an eigenvector, it is not the zero vector, and $\left\Vert \Psi\right\Vert ^{2}\ne0$. Hence, we conclude that the eigenvalues of a unitary matrix must all have magnitude $1$. This means that they lie on the unit circle of the complex plane, and are of the form $z=\e^{\i\phi}$ for some $\phi\in\BBR$.

As with Hermitian matrices, eigenvectors of a unitary matrix corresponding to different eigenvalues are orthogonal (you will prove this in \zcref[S]{prob:Prove-that-eigenvectors}), and you can always find an orthonormal eigenbasis of a unitary matrix.

A \emph{normal matrix}\index{Normal matrix} is a matrix $A$ which satisfies $A^{\dagger}A=AA^{\dagger}$. Observe that a normal matrix is analogous to a complex number $z$, since such a number trivially satisfies $z^{*}z=zz^{*}$. It is easy to see that Hermitian matrices, which satisfy $A^{\dagger}=A$, and unitary matrices, which satisfy $A^{\dagger}=A^{-1}$, are both special cases of normal matrices.

If $A$ is normal and all of its eigenvalues are real, then it is Hermitian. If $A$ is normal and all of its eigenvalues have unit magnitude, then it is unitary. Furthermore, it turns out that the condition that the matrix has an orthonormal eigenbasis applies not just to Hermitian and unitary matrices, but in general to any normal matrix; in fact, it is true \textbf{if and only if }the matrix is normal.
\begin{problem}\leavevmode
Find the most general $2\xx2$ unitary matrix by demanding that $U^{-1}=U^{\dagger}$ or $UU^{\dagger}=U^{\dagger}U=1$ and finding conditions on the components of $U$.
\end{problem}

\begin{problem}\leavevmode
Find three $2\xx2$ matrices that are both Hermitian and unitary (other than the identity matrix).
\end{problem}

\begin{problem}\leavevmode
\label{prob:Prove-that-eigenvectors}Prove that eigenvectors of a unitary matrix corresponding to different eigenvalues are orthogonal.
\end{problem}

\begin{problem}\leavevmode
Prove that the columns of a unitary matrix, treated as kets, form an orthonormal basis of $\BBC^{n}$. Similarly, prove that the rows of a unitary matrix, treated as bras, form an orthonormal basis of the dual space $\left(\BBC^{n}\right)^{*}$.
\end{problem}

\begin{problem}\leavevmode
Let $A$ and $B$ be normal matrices such that $A^{\dagger}B=BA^{\dagger}$. Prove that both $AB$ and $A+B$ are also normal.
\end{problem}

\subsubsection{Representing Matrices in Different Bases}\label{subsec:Representing-Matrices-in}

In \zcref[S]{subsec:Representing-Vectors-in} we saw that vectors are abstract entities which can have different representations in different bases. The same is true for matrices.\index{Representing a matrix in a basis} Consider a matrix $A$ and a basis $\left|B_{i}\right\rangle $. Inserting the completeness relation in \zcref[S]{eq:completeness} \textbf{twice}, one time on each side of $A$, we get:
\begin{align*}
A & =\left(\sum^{n}_{i=1}|B_{i}\rangle\langle B_{i}|\right)A\left(\sum^{n}_{j=1}|B_{j}\rangle\langle B_{j}|\right)\\
 & =\sum^{n}_{i=1}\sum^{n}_{j=1}|B_{i}\rangle\langle B_{i}|A|B_{j}\rangle\langle B_{j}|\\
 & =\sum^{n}_{i=1}\sum^{n}_{j=1}\left(A_{ij}\right)_{B}|B_{i}\rangle\langle B_{j}|,
\end{align*}
where\footnote{Note that we could move $\left(A_{ij}\right)_{B}$ to the left since it is a scalar.}
\[
\left(A_{ij}\right)_{B}\equiv\langle B_{i}|A|B_{j}\rangle\in\BBC\sp i,j\in\left\{ 1,\ldots,n\right\} 
\]
are the \emph{coordinates}\index{Coordinates of a matrix in a basis} of $A$ in the basis $\left|B_{i}\right\rangle $.

We have obtained a sum over outer products of the form $|B_{i}\rangle\langle B_{j}|$. Recall that the outer product of two vectors is a matrix; thus $|B_{i}\rangle\langle B_{j}|$ can be thought of as “basis matrices”, in analogy with basis vectors. The representation of $A$ in terms of a linear combination of these “basis matrices” is called the \emph{outer product representation}\index{Outer product representation} of $A$, and it is very useful in quantum theory.

We can also write this representation in matrix form as
\[
\left(A\right)_{B}=\left(\begin{array}{ccc}
\langle B_{1}|A|B_{1}\rangle & \cdots & \langle B_{1}|A|B_{n}\rangle\\
\vdots & \ddots & \vdots\\
\langle B_{n}|A|B_{1}\rangle & \cdots & \langle B_{n}|A|B_{n}\rangle
\end{array}\right).
\]
In another basis $\left|C_{i}\right\rangle $, the matrix $A$ will have the representation
\[
\left(A\right)_{C}=\left(\begin{array}{ccc}
\langle C_{1}|A|C_{1}\rangle & \cdots & \langle C_{1}|A|C_{n}\rangle\\
\vdots & \ddots & \vdots\\
\langle C_{n}|A|C_{1}\rangle & \cdots & \langle C_{n}|A|C_{n}\rangle
\end{array}\right),
\]
where now the coordinates $\left(A_{ij}\right)_{C}$ are given by
\[
\left(A_{ij}\right)_{C}=\langle C_{i}|A|C_{j}\rangle\in\BBC\sp i,j\in\left\{ 1,\ldots,n\right\} .
\]
Inserting the completeness relation in \zcref[S]{eq:completeness} into the coordinates twice, we get
\begin{align*}
\langle C_{i}|A|C_{j}\rangle & =\langle C_{i}|\left(\sum^{n}_{k=1}|B_{k}\rangle\langle B_{k}|\right)A\left(\sum^{n}_{\ell=1}|B_{\ell}\rangle\langle B_{\ell}|\right)|C_{j}\rangle\\
 & =\sum^{n}_{k=1}\sum^{n}_{\ell=1}\langle C_{i}|B_{k}\rangle\langle B_{k}|A|B_{\ell}\rangle\langle B_{\ell}|C_{j}\rangle.
\end{align*}

\begin{problem}\leavevmode
Show that this relation can be written in matrix form as follows:
\begin{align*}
 & \left(\begin{array}{ccc}
\langle C_{1}|A|C_{1}\rangle & \cdots & \langle C_{1}|A|C_{n}\rangle\\
\vdots & \ddots & \vdots\\
\langle C_{n}|A|C_{1}\rangle & \cdots & \langle C_{n}|A|C_{n}\rangle
\end{array}\right)=\\
 & =\left(\begin{array}{ccc}
\langle C_{1}|B_{1}\rangle & \cdots & \langle C_{1}|B_{n}\rangle\\
\vdots & \ddots & \vdots\\
\langle C_{n}|B_{1}\rangle & \cdots & \langle C_{n}|B_{n}\rangle
\end{array}\right)\left(\begin{array}{ccc}
\langle B_{1}|A|B_{1}\rangle & \cdots & \langle B_{1}|A|B_{n}\rangle\\
\vdots & \ddots & \vdots\\
\langle B_{n}|A|B_{1}\rangle & \cdots & \langle B_{n}|A|B_{n}\rangle
\end{array}\right)\left(\begin{array}{ccc}
\langle B_{1}|C_{1}\rangle & \cdots & \langle B_{1}|C_{n}\rangle\\
\vdots & \ddots & \vdots\\
\langle B_{n}|C_{1}\rangle & \cdots & \langle B_{n}|C_{n}\rangle
\end{array}\right),
\end{align*}
and thus the relation between the representations of $A$ in different bases is given by
\[
\left(A\right)_{C}=P_{C\ot B}\left(A\right)_{B}P_{B\ot C},
\]
where
\[
P_{C\ot B}=\left(\begin{array}{ccc}
\langle C_{1}|B_{1}\rangle & \cdots & \langle C_{1}|B_{n}\rangle\\
\vdots & \ddots & \vdots\\
\langle C_{n}|B_{1}\rangle & \cdots & \langle C_{n}|B_{n}\rangle
\end{array}\right)
\]
is the change-of-basis matrix in \zcref[S]{eq:c-o-b}, and $P_{B\ot C}=P^{-1}_{C\ot B}$. This is analogous to the relation between vectors in different bases, $\left|\Psi\right\rangle \bll_{C}=P_{C\ot B}\left|\Psi\right\rangle \bll_{B}$. We can write this in simpler terms by defining $P\equiv P_{B\ot C}$:
\[
\left(A\right)_{C}=P^{-1}\left(A\right)_{B}P.
\]
\end{problem}

\begin{problem}\leavevmode
Let $U$ be a unitary matrix and let $\left|B_{i}\right\rangle $ be an orthonormal basis.
\begin{problemparts}
\item Prove that $\left|C_{i}\right\rangle \equiv U\left|B_{i}\right\rangle $ is also an orthonormal basis.
\item Prove that $U$ has the outer product representation
\begin{equation}
U=\sum^{n}_{i=1}\left|C_{i}\right\rangle \left\langle B_{i}\right|.\label{eq:unitary-outer}
\end{equation}
\item Conversely, prove that if $\left|B_{i}\right\rangle $ and $\left|C_{i}\right\rangle $ are two \textbf{arbitrary }orthonormal bases, then the matrix $U$ \textbf{defined }by \zcref[S]{eq:unitary-outer} is unitary.
\end{problemparts}
\end{problem}

\subsubsection{Diagonalizable Matrices}\label{subsec:Diagonalizable-Matrices}

A \emph{diagonal matrix}\index{Diagonal matrix} is a matrix with all of its elements equal to zero except for the elements on the diagonal, for example:
\[
D=\left(\begin{array}{cc}
D_{1} & 0\\
0 & D_{2}
\end{array}\right).
\]
A matrix $A$ is called \emph{diagonalizable\index{Diagonalizable matrix}} if there exists an invertible matrix $P$ such that the matrix $P^{-1}AP$ is diagonal. In quantum theory, we are mostly concerned with the case where $P$ is also a unitary matrix, such that $P^{\dagger}AP$ is diagonal. It turns out that a matrix $A$ is diagonalizable by a unitary matrix $P$ if and only if $A$ is normal. This means, in particular, that both Hermitian and unitary matrices are diagonalizable in such a way.

Let $A$ be a normal matrix with an orthonormal eigenbasis $\left|B_{i}\right\rangle $ with corresponding eigenvalues $\lambda_{i}$:
\begin{equation}
A\left|B_{i}\right\rangle =\lambda_{i}\left|B_{i}\right\rangle \sp\fa i.\label{eq:eigen-diag}
\end{equation}
Now, consider the change-of-basis matrix in \zcref[S]{eq:c-o-b}, this time from the eigenbasis $\left|B_{i}\right\rangle $ to the standard basis $\left|1_{i}\right\rangle $:
\begin{equation}
P_{1\ot B}=\left(\begin{array}{ccc}
\langle1_{1}|B_{1}\rangle & \cdots & \langle1_{1}|B_{n}\rangle\\
\vdots & \ddots & \vdots\\
\langle1_{n}|B_{1}\rangle & \cdots & \langle1_{n}|B_{n}\rangle
\end{array}\right).\label{eq:P-1-B}
\end{equation}
Note that each eigenvector $\left|B_{i}\right\rangle $ is represented in the standard basis $\left|1_{i}\right\rangle $ as follows:
\[
\left|B_{i}\right\rangle \bllll_{1}=\left(\begin{array}{c}
\langle1_{1}|B_{i}\rangle\\
\vdots\\
\langle1_{n}|B_{i}\rangle
\end{array}\right)\sp\fa i.
\]
Hence, the columns of $P_{1\ot B}$ are in fact the eigenvectors $\left|B_{i}\right\rangle $ themselves, as expressed in the standard basis:
\[
P_{1\ot B}=\left(\vphantom{\blll}\begin{array}{ccc}
\left|B_{1}\right\rangle  & \cdots & \left|B_{n}\right\rangle \end{array}\right).
\]
Let us denote $P\equiv P_{1\ot B}$ for short. Then
\begin{align*}
AP & =A\left(\vphantom{\blll}\begin{array}{ccc}
\left|B_{1}\right\rangle  & \cdots & \left|B_{n}\right\rangle \end{array}\right)\\
 & =\left(\vphantom{\blll}\begin{array}{ccc}
A\left|B_{1}\right\rangle  & \cdots & A\left|B_{n}\right\rangle \end{array}\right)\\
 & =\left(\vphantom{\blll}\begin{array}{ccc}
\lambda_{1}\left|B_{1}\right\rangle  & \cdots & \lambda_{n}\left|B_{n}\right\rangle \end{array}\right).
\end{align*}
How did we get this? Remember that in \zcref[S]{subsec:Multiplication-and-Inverse} we said that the element of the matrix $AP$ at row $i$, column $j$ is calculated by taking the dot product of row $i$ of $A$ with column $j$ of $P$. But column $j$ of $P$ is just $\left|B_{j}\right\rangle $. The product $A\left|B_{j}\right\rangle $ is another ket, whose components are obtained by taking the dot product of each row of $A$ with $\left|B_{j}\right\rangle $ respectively. The last equality follows from \zcref[S]{eq:eigen-diag}.

Next, we write the full matrix and decompose it into two matrices:
\begin{equation}
AP=\left(\begin{array}{ccc}
\lambda_{1}\langle1_{1}|B_{1}\rangle & \cdots & \lambda_{n}\langle1_{1}|B_{n}\rangle\\
\vdots & \ddots & \vdots\\
\lambda_{1}\langle1_{n}|B_{1}\rangle & \cdots & \lambda_{n}\langle1_{n}|B_{n}\rangle
\end{array}\right)=\left(\begin{array}{ccc}
\langle1_{1}|B_{1}\rangle & \cdots & \langle1_{1}|B_{n}\rangle\\
\vdots & \ddots & \vdots\\
\langle1_{n}|B_{1}\rangle & \cdots & \langle1_{n}|B_{n}\rangle
\end{array}\right)\left(\begin{array}{ccc}
\lambda_{1} & 0 & 0\\
0 & \ddots & 0\\
0 & 0 & \lambda_{n}
\end{array}\right).\label{eq:AP}
\end{equation}
You should calculate the product on the right-hand side (even just for $n=2$ or $n=3$) to convince yourself that this decomposition is indeed correct. Now, if we define a new diagonal matrix, with the eigenvalues on the diagonal:
\[
D\equiv\left(\begin{array}{ccc}
\lambda_{1} & 0 & 0\\
0 & \ddots & 0\\
0 & 0 & \lambda_{n}
\end{array}\right),
\]
then we can write \zcref[S]{eq:AP} as
\[
AP=PD.
\]
Finally, we multiply by $P^{-1}$ from the left to get
\[
P^{-1}AP=D.
\]

\begin{xca}\leavevmode
Diagonalize the following matrix:
\[
A=\left(\begin{array}{cc}
1 & 3\\
3 & 1
\end{array}\right).
\]
\end{xca}

\begin{problem}\leavevmode
Prove that the change-of-basis matrix $P=P_{1\ot B}$ in \zcref[S]{eq:P-1-B}, with $\left|B_{i}\right\rangle $ an orthonormal\textbf{ }eigenbasis, is unitary. This means that we can also write $P^{\dagger}AP=D$, since $P^{-1}=P^{\dagger}$ for unitary matrices.
\end{problem}

\subsubsection{Spectral Decomposition}\label{subsec:Spectral-Decomposition}

Recall from \zcref[S]{subsec:Unitary-and-Normal-Matrices} that a normal matrix $A$ is defined by the property $A^{\dagger}A=AA^{\dagger}$. If a matrix $A$ is normal, then it has an orthonormal eigenbasis $\left|B_{i}\right\rangle $ with corresponding eigenvalues $\lambda_{i}$. Multiplying $A$ by the completeness relation, \zcref[S]{eq:completeness}, we find
\[
A=\sum^{n}_{i=1}A\left|B_{i}\right\rangle \left\langle B_{i}\right|,
\]
and since $A\left|B_{i}\right\rangle =\lambda_{i}\left|B_{i}\right\rangle $, this becomes
\begin{equation}
A=\sum^{n}_{i=1}\lambda_{i}\left|B_{i}\right\rangle \left\langle B_{i}\right|.\label{eq:spectral-decomposition}
\end{equation}
This is called the \emph{spectral decomposition}\index{Spectral decomposition} of $A$: the normal matrix can be written as a sum of outer products $\left|B_{i}\right\rangle \left\langle B_{i}\right|$ for each element of the eigenbasis, weighted by the corresponding eigenvalues. Of course, since Hermitian and unitary matrices are also normal, this applies to them too.

Now, let $\lambda$ be an eigenvalue of $A$. The \emph{eigenspace}\index{Eigenspace} of $\lambda$ is the set of vectors $\left|\Psi\right\rangle \in\BBC^{n}$ that satisfy the eigenvalue equation $A\left|\Psi\right\rangle =\lambda\left|\Psi\right\rangle $. This means that it consists of all the eigenvectors with eigenvalue $\lambda$, together with the zero vector, since the zero vector satisfies this equation trivially. The eigenspace is itself a vector space, as you will prove in \zcref[S]{prob:eigen}.

An eigenvalue is called a \emph{degenerate eigenvalue}\index{Degenerate eigenvalue}\index{Eigenvalue!Degenerate} if it has at least two linearly independent eigenvectors. A simple example of a matrix with degenerate eigenvalues is the identity matrix 1, which has only one eigenvalue – namely, 1 – for which \textbf{every }non-zero vector is an eigenvector.

For a normal matrix $A$ with eigenvalues $\lambda_{i}$, since we can choose an orthonormal eigenbasis $\left|B_{i}\right\rangle $, which is linearly independent by definition, we can say that an eigenvalue is degenerate if two or more basis vectors have the same eigenvalue. Equivalently, the eigenvalue $\lambda_{i}$ is degenerate if $\lambda_{i}=\lambda_{j}$ for one or more $j\ne i$.

Given an eigenvalue $\lambda$, we can separate the eigenbasis of $n$ eigenvectors into $m$ vectors that have that eigenvalue and $n-m$ vectors that don't (rearranging them if needed):
\[
\underbrace{\left|B_{1}\right\rangle ,\ldots,\left|B_{m}\right\rangle }_{\text{eigenvalue }\lambda},\underbrace{\left|B_{m+1}\right\rangle ,\ldots,\left|B_{n}\right\rangle }_{\text{other eigenvalues}}.
\]
The number $m$ is called the \emph{multiplicity}\index{Eigenvalue!Multiplicity} of the eigenvalue. Now, let $V_{\lambda}$ be the $m$-dimensional vector space spanned by $\left|B_{1}\right\rangle ,\ldots,\left|B_{m}\right\rangle $. Any vector $\left|\Psi\right\rangle \in V_{\lambda}$ is clearly in the eigenspace of $\lambda$, since it's a linear combination of eigenvectors with eigenvalue $\lambda$. Conversely, let $\left|\Psi\right\rangle $ be any vector in the eigenspace of $\lambda$, and let $\left|B_{j}\right\rangle $ be any of the $n$ basis vectors. Then from \zcref[S]{eq:spectral-decomposition},
\[
\left\langle B_{j}\right|A=\left\langle B_{j}\right|\left(\sum^{n}_{i=1}\lambda_{i}\left|B_{i}\right\rangle \left\langle B_{i}\right|\right)=\sum^{n}_{i=1}\lambda_{i}\delta_{ij}\left\langle B_{i}\right|=\lambda_{j}\left\langle B_{j}\right|,
\]
since $\langle B_{j}|B_{i}\rangle=\delta_{ij}$. Therefore
\[
\langle B_{j}|A|\Psi\rangle=\lambda_{j}\langle B_{j}|\Psi\rangle,
\]
but we also have, since $A\left|\Psi\right\rangle =\lambda\left|\Psi\right\rangle $,
\[
\langle B_{j}|A|\Psi\rangle=\lambda\langle B_{j}|\Psi\rangle.
\]
Therefore
\[
\left(\lambda-\lambda_{j}\right)\langle B_{j}|\Psi\rangle=0,
\]
so if $\lambda_{j}\ne\lambda$, the component of $\left|\Psi\right\rangle $ along $\left|B_{j}\right\rangle $ vanishes. This means any vector in the eigenspace of $\lambda$ must be in $V_{\lambda}$. In other words, we have proven that the eigenspace of $\lambda$ is exactly $V_{\lambda}$. This also means the multiplicity of an eigenvalue is the dimension of its eigenspace.

We define the \emph{projector}\index{Projector} or \emph{projection operator}\index{Projection operator} onto the $\lambda$ eigenspace as the matrix
\[
P_{\lambda}\equiv\sum^{m}_{i=1}\left|B_{i}\right\rangle \left\langle B_{i}\right|,
\]
where the sum is over the basis vectors corresponding to $\lambda$. Given any vector $\left|\Psi\right\rangle \in\BBC^{n}$, we can multiply it by the projector $P_{\lambda}$ and get
\[
P_{\lambda}\left|\Psi\right\rangle =\sum^{m}_{i=1}\left|B_{i}\right\rangle \langle B_{i}|\Psi\rangle.
\]
Since this is a linear combination of the basis vectors of the eigenspace $V_{\lambda}$, we see that $P_{\lambda}\left|\Psi\right\rangle $ is in that eigenspace for any vector $\left|\Psi\right\rangle $. If we act with $P_{\lambda}$ on a vector $\left|\Psi_{\lambda}\right\rangle $ that's already in the eigenspace, we just get the same vector:
\begin{equation}
P_{\lambda}\left|\Psi_{\lambda}\right\rangle =\left|\Psi_{\lambda}\right\rangle .\label{eq:projector-already-in}
\end{equation}
This is because $\left|\Psi_{\lambda}\right\rangle $ is already written as a linear combination of the basis elements of the eigenspace, so the projector doesn't change anything (check that). In other words, $P_{\lambda}$ keeps the part of $\left|\Psi\right\rangle $ that's in $V_{\lambda}$, and removes the rest\footnote{For geometric intuition about projections, think about the real Euclidean space $\BBR^{3}$ with the standard basis
\[
\hat{\x}=\left(1,0,0\right)\sp\hat{\y}=\left(0,1,0\right)\sp\hat{\z}=\left(0,0,1\right).
\]
Let's define a 2-dimensional vector space $V$ as the subspace of $\BBR^{3}$ spanned by $\hat{\x}$ and $\hat{\y}$. This subspace is just the $xy$ plane, and it is isomorphic to $\BBR^{2}$. Given a real 3-vector $\v\equiv\left(x,y,z\right)$, its projection onto $V$ will simply be $\left(x,y,0\right)$.}.

The projector $P_{\lambda}$ is Hermitian:
\[
P^{\dagger}_{\lambda}=\left(\sum^{m}_{i=1}\left|B_{i}\right\rangle \left\langle B_{i}\right|\right)^{\dagger}=\sum^{m}_{i=1}\left|B_{i}\right\rangle \left\langle B_{i}\right|\soosp P^{\dagger}_{\lambda}=P_{\lambda},
\]
since $\left(\left|B_{i}\right\rangle \left\langle B_{i}\right|\right)^{\dagger}=\left|B_{i}\right\rangle \left\langle B_{i}\right|$ by \zcref[S]{prob:adjoint-outer}. Furthermore, we have for any vector $\left|\Psi\right\rangle \in\BBC^{n}$
\[
P^{2}_{\lambda}\left|\Psi\right\rangle =P_{\lambda}\left(P_{\lambda}\left|\Psi\right\rangle \right)=P_{\lambda}\left|\Psi\right\rangle ,
\]
since $P_{\lambda}\left|\Psi\right\rangle \in V_{\lambda}$ and by \zcref[S]{eq:projector-already-in} $P_{\lambda}$ acting on a vector already in $V_{\lambda}$ gives the same vector. Therefore,
\[
P^{2}_{\lambda}=P_{\lambda}.
\]
A matrix that is equal to its own square is called \emph{idempotent}\index{Idempotent}; hence, all projectors are idempotent.

If the multiplicity $m$ is greater than 1, then the eigenspace $V_{\lambda}$ has an infinite number of orthonormal bases. However, the projector $P_{\lambda}$ does not depend on which one we choose. To see this, let $P_{\lambda}$ and $Q_{\lambda}$ be two projectors onto the $\lambda$ eigenspace, defined by different bases. For any vector $\left|\Psi\right\rangle $, $P_{\lambda}\left|\Psi\right\rangle $ is in $V_{\lambda}$, and therefore
\[
Q_{\lambda}P_{\lambda}\left|\Psi\right\rangle =Q_{\lambda}\left(P_{\lambda}\left|\Psi\right\rangle \right)=P_{\lambda}\left|\Psi\right\rangle .
\]
Since this is correct for any vector $\left|\Psi\right\rangle $, it must be that
\[
Q_{\lambda}P_{\lambda}=P_{\lambda},
\]
and if we swap their roles, we see that
\[
P_{\lambda}Q_{\lambda}=Q_{\lambda}.
\]
On the other hand, using \zcref[S]{prob:adjoint-product} and the fact that both projectors are Hermitian, we have
\[
Q_{\lambda}P_{\lambda}=Q^{\dagger}_{\lambda}P^{\dagger}_{\lambda}=\left(P_{\lambda}Q_{\lambda}\right)^{\dagger}=Q^{\dagger}_{\lambda}=Q_{\lambda}.
\]
Comparing the two expressions for $Q_{\lambda}P_{\lambda}$, we conclude that $Q_{\lambda}=P_{\lambda}$.

Now, if we have a normal matrix $A$ with some degenerate eigenvalues, we can write down its spectral decomposition, \zcref[S]{eq:spectral-decomposition}, and then group together all the terms that correspond to the same eigenvalue. For example, consider the matrix
\[
A=\left(\begin{array}{cccc}
\lambda_{1} & 0 & 0 & 0\\
0 & \lambda_{1} & 0 & 0\\
0 & 0 & \lambda_{2} & 0\\
0 & 0 & 0 & \lambda_{2}
\end{array}\right),
\]
which has two eigenvalues, $\lambda_{1}$ and $\lambda_{2}$ (assuming $\lambda_{1}\ne\lambda_{2}$), each with multiplicity 2. Using the standard basis $\left|1_{i}\right\rangle $ of $\BBC^{4}$, which is given by
\[
\left|1_{1}\right\rangle =\left(\begin{array}{c}
1\\
0\\
0\\
0
\end{array}\right)\sp\left|1_{2}\right\rangle =\left(\begin{array}{c}
0\\
1\\
0\\
0
\end{array}\right)\sp\left|1_{3}\right\rangle =\left(\begin{array}{c}
0\\
0\\
1\\
0
\end{array}\right)\sp\left|1_{4}\right\rangle =\left(\begin{array}{c}
0\\
0\\
0\\
1
\end{array}\right),
\]
we can write
\[
A=\lambda_{1}\left|1_{1}\right\rangle \left\langle 1_{1}\right|+\lambda_{1}\left|1_{2}\right\rangle \left\langle 1_{2}\right|+\lambda_{2}\left|1_{3}\right\rangle \left\langle 1_{3}\right|+\lambda_{2}\left|1_{4}\right\rangle \left\langle 1_{4}\right|.
\]
We can then group the terms as follows:
\[
A=\lambda_{1}\left(\left|1_{1}\right\rangle \left\langle 1_{1}\right|+\left|1_{2}\right\rangle \left\langle 1_{2}\right|\right)+\lambda_{2}\left(\left|1_{3}\right\rangle \left\langle 1_{3}\right|+\left|1_{4}\right\rangle \left\langle 1_{4}\right|\right).
\]
Now, if we define the projectors onto each eigenspace:
\[
P_{1}\equiv\left|1_{1}\right\rangle \left\langle 1_{1}\right|+\left|1_{2}\right\rangle \left\langle 1_{2}\right|\sp P_{2}\equiv\left|1_{3}\right\rangle \left\langle 1_{3}\right|+\left|1_{4}\right\rangle \left\langle 1_{4}\right|,
\]
then we can write
\[
A=\sum^{2}_{i=1}\lambda_{i}P_{i}.
\]
Since the same procedure would work for any normal matrix, we conclude that if a normal $n\xx n$ matrix $A$ has $k$ distinct eigenvalues $\left\{ \lambda_{1},\ldots,\lambda_{k}\right\} $, we can write it down as a weighted sum of projectors onto the eigenspaces of each eigenvalue:
\[
A=\sum^{k}_{i=1}\lambda_{i}P_{i}.
\]
This is the spectral decomposition in terms of projectors. In the special case where none of the eigenvalues are degenerate, we simply have $P_{i}=\left|B_{i}\right\rangle \left\langle B_{i}\right|$ for $i\in\left\{ 1,\ldots,n\right\} $, and we recover \zcref[S]{eq:spectral-decomposition}.
\begin{problem}\leavevmode
\label{prob:eigen}Prove that an eigenspace is a vector space by showing that it satisfies the properties of a vector space as defined in \zcref[S]{subsec:Complex-Vector-Spaces}.
\end{problem}

\begin{problem}\leavevmode
Prove that the eigenvalues of a projector must be 0 or 1.
\end{problem}

\begin{problem}\leavevmode
For a normal matrix $A$ with $k$ distinct eigenvalues $\lambda_{i}$ and corresponding eigenspace projectors $P_{i}$, prove that
\[
P_{i}P_{j}=\delta_{ij}P_{i}\sp\sum^{k}_{i=1}P_{i}=1.
\]
\end{problem}

\subsubsection{The Matrix Commutator}\label{subsec:The-Matrix-Commutator}

In \zcref[S]{prob:Matrix-multiplication-is} we showed that matrix multiplication is not commutative in general; the order of multiplication matters. The \emph{commutator}\index{Matrix commutator}\index{Commutator} of two matrices $A$ and $B$ is:
\[
\left[A,B\right]\equiv AB-BA.
\]
The commutator can be used to check whether the matrices commute or not. If the matrices commute, then $AB=BA$ and thus the commutator vanishes: $\left[A,B\right]=0$. Otherwise, $AB\ne BA$ and the commutator is non-zero: $\left[A,B\right]\ne0$. The commutator thus tells us if the matrices commute or not. Note that any matrix commutes with itself: $\left[A,A\right]=0$ for any $A$.

The \emph{anti-commutator}\index{Matrix anti-commutator}\index{Anti-commutator} is defined as follows:
\[
\left\{ A,B\right\} \equiv AB+BA.
\]
It vanishes if the matrices \emph{anti-commute}\index{Anti-commuting matrices}: $AB=-BA$.
\begin{problem}\leavevmode
Prove that the commutator is anti-symmetric:
\[
\left[B,A\right]=-\left[A,B\right].
\]
\end{problem}

\begin{problem}\leavevmode
Prove that the commutator is linear:
\[
\left[A+B,C\right]=\left[A,C\right]+\left[B,C\right],
\]
\[
\left[A,B+C\right]=\left[A,B\right]+\left[A,C\right],
\]
\[
\left[\lambda A,B\right]=\left[A,\lambda B\right]=\lambda\left[A,B\right]\sp\lambda\in\BBC.
\]
\end{problem}

\begin{problem}\leavevmode
Prove that
\[
[A,B]^{\dagger}=[B^{\dagger},A^{\dagger}].
\]
\end{problem}

\begin{problem}\leavevmode
Prove the useful identities
\[
\left[AB,C\right]=A\left[B,C\right]+\left[A,C\right]B,
\]
\[
\left[A,BC\right]=B\left[A,C\right]+\left[A,B\right]C.
\]
\end{problem}

\begin{problem}\leavevmode
Prove the \emph{Jacobi identity}\index{Jacobi identity}:
\[
\left[A,\left[B,C\right]\right]+\left[B,\left[C,A\right]\right]+\left[C,\left[A,B\right]\right]=0.
\]
\end{problem}

\subsubsection{Matrix Exponentials and Logarithms}\label{subsec:Matrix-and-Operator}

To generalize the exponential to complex matrices $A$, we define the \emph{matrix exponential}\index{Matrix exponential}\index{Exponential!Of a matrix}:
\begin{equation}
\e^{A}\equiv\sum^{\infty}_{n=0}\frac{A^{n}}{n!}=1+A+\hf A^{2}+\frac{1}{3!}A^{3}+\cdots,\label{eq:matrix-power}
\end{equation}
where $1$ is now the identity matrix, and $A^{n}$ means the product of the matrix $A$ with itself $n$ times. Note that it satisfies
\begin{equation}
\e^{0}=1,\label{eq:exp0}
\end{equation}
that is, the exponential of the zero matrix is the identity matrix, in analogy with the fact that the exponential of zero is one. It also satisfies, as you will prove in \zcref[S]{prob:exp-adj},
\[
\left(\e^{A}\right)^{\dagger}=\e^{A^{\dagger}},
\]
in analogy with \zcref[S]{eq:exp-zconj}.

Things start to become more complicated when we consider the product of two exponentials, $\e^{A}\e^{B}$. For numbers (which can be considered $1\xx1$ matrices) we have $\e^{z}\e^{w}=\e^{z+w}$, but the proof of this identity relies on the fact that numbers commute. For arbitrary $n\xx n$ matrices $A$ and $B$, it is in general not true that $\e^{A}\e^{B}=\e^{A+B}$. However, in \zcref[S]{prob:The-identity-} you will prove that this identity is true if $\left[A,B\right]=0$, that is, if $A$ and $B$ commute.

The \emph{matrix logarithm}\index{Matrix logarithm}\index{Logarithm!Of a matrix} of a matrix $B$ is any matrix $A$ satisfying $\e^{A}=B$. If such a matrix exists, then we can write
\[
\e^{\log B}=B,
\]
in analogy with \zcref[S]{eq:log}. Just as with complex numbers, the logarithm is generally not unique; for example, both $0$ (the zero matrix) and $2\pi\i$ (the identity matrix times $2\pi\i$) exponentiate to $1$ (the identity matrix), in analogy with \zcref[S]{subsec:Exponentials-and-Logarithms}. However, although every non-zero complex number has a logarithm, a matrix has a logarithm if and only if it is invertible. One of the directions of this proof is easy: if a matrix $B$ has a logarithm $A=\log B$, then we can write $B=\e^{A}$, and then the inverse will be $B^{-1}=\e^{-A}$. We won't prove the other direction here.

Recall from \zcref[S]{subsec:Hermitian-Matrices,subsec:Unitary-and-Normal-Matrices} that Hermitian matrices are analogous to real numbers, and unitary matrices are analogous to complex numbers with unit magnitude. Since for real $\phi$ we have $\left|\e^{\i\phi}\right|=1$, we should expect that the exponential of $\i$ times a Hermitian matrix will be a unitary matrix. Let $H$ be a Hermitian matrix and let $t\in\BBR$ be a real number, and let us define\footnote{The minus sign here is a convention; the inverse matrix, $\e^{\i Ht}$, is of course unitary as well.}
\[
U\equiv\e^{-\i Ht}.
\]
To prove that $U$ is unitary, let us take its adjoint:
\[
U^{\dagger}=\left(\e^{-\i Ht}\right)^{\dagger}=\e^{-\i^{*}H^{\dagger}t^{*}}=\e^{\i Ht},
\]
since $H^{\dagger}=H$ due to $H$ being Hermitian, $t^{*}=t$ due to $t$ being real, and $\i^{*}=-\i$. Thus:
\[
UU^{\dagger}=\e^{-\i Ht}\e^{\i Ht}=\e^{0}=1,
\]
so $\e^{-\i Ht}$ is indeed unitary. Note that here we used the fact that $H$ commutes with itself, and therefore the product of the exponentials is the exponential of the sum, as you will prove in \zcref[S]{prob:The-identity-}. In fact, the converse is also true: any unitary matrix $U$ can be written as $\e^{-\i H}$ for some Hermitian matrix $H$. To show this, recall from \zcref[S]{subsec:Spectral-Decomposition} that a normal matrix $A$ can be written as a spectral decomposition:
\[
A=\sum^{n}_{i=1}\lambda_{i}\left|B_{i}\right\rangle \left\langle B_{i}\right|,
\]
where $\left|B_{i}\right\rangle $ is an orthonormal eigenbasis and $\lambda_{i}$ are the corresponding eigenvalues. A unitary matrix is normal, and its eigenvalues must have unit magnitude, so given a unitary matrix $U$, we can write
\[
U=\sum_{i}\e^{-\i\phi_{i}}\left|B_{i}\right\rangle \left\langle B_{i}\right|\sp\phi_{i}\in\BBR,
\]
where $\e^{-\i\phi_{i}}$ are the eigenvalues. Then we can define a corresponding matrix
\[
H=\sum_{i}\phi_{i}\left|B_{i}\right\rangle \left\langle B_{i}\right|.
\]
This matrix is Hermitian, since $H=H^{\dagger}$, and exponentiating $-\i H$ clearly produces the matrix $U$:
\[
U=\e^{-\i H}\soosp H=\i\log U.
\]
Of course, this $H$ is not unique, since replacing any $\phi_{i}$ by $\phi_{i}+2\pi k$ with integer $k$ will result in the same $U$.

Finally, \zcref[S]{eq:exp-lambda-t} generalizes to matrices as well:
\begin{equation}
\frac{\d}{\d t}\e^{At}=A\e^{At},\label{eq:exp-At}
\end{equation}
where $A$ is any \textbf{constant} complex matrix (i.e. independent of $t$).
\begin{problem}\leavevmode
\label{prob:exp-adj}Prove that $\left(\e^{A}\right)^{\dagger}=\e^{A^{\dagger}}$.
\end{problem}

\begin{problem}\leavevmode
\label{prob:The-identity-}Prove that if two matrices $A$ and $B$ commute, that is, $\left[A,B\right]=0$ (see \zcref[S]{subsec:The-Matrix-Commutator}), then
\[
\e^{A}\e^{B}=\e^{A+B}.
\]
One way to do this is using the power series definition in \zcref[S]{eq:matrix-power}.
\end{problem}

\begin{problem}\leavevmode
\begin{problemparts}
\item Prove that the exponential of a diagonal matrix with diagonal elements $\lambda_{i}$ is a diagonal matrix with diagonal elements $\e^{\lambda_{i}}$:
\[
\exp\left(\begin{array}{ccc}
\lambda_{1} & 0 & 0\\
0 & \ddots & 0\\
0 & 0 & \lambda_{n}
\end{array}\right)=\left(\begin{array}{ccc}
\e^{\lambda_{1}} & 0 & 0\\
0 & \ddots & 0\\
0 & 0 & \e^{\lambda_{n}}
\end{array}\right).
\]
\item Prove that if $B$ is an invertible matrix, then for any matrix $A$:
\[
\e^{BAB^{-1}}=B\e^{A}B^{-1}.
\]
\item Using (A) and (B), prove that if $A$ is diagonalizable, that is, $A=PDP^{-1}$ for some matrix $P$ and a diagonal matrix $D$ (recall \zcref[S]{subsec:Diagonalizable-Matrices}), then
\[
\e^{A}=P\e^{D}P^{-1}.
\]
This gives us a straightforward way to calculate the exponential of any diagonalizable matrix (and in particular every normal matrix, since they are always diagonalizable).
\end{problemparts}
\end{problem}

\begin{problem}\leavevmode
Find the matrix exponential $\e^{-\i\theta\sigma_{y}}$ where\footnote{$\sigma_{y}$ is one of the Pauli matrices; we will learn about them in \zcref[S]{subsec:The-Pauli-Matrices}.}
\[
\sigma_{y}\equiv\left(\begin{array}{cc}
0 & -\i\\
\i & 0
\end{array}\right).
\]
\end{problem}

\subsubsection{The Trace}\label{subsec:Trace}

The \emph{trace}\index{Trace} of a square matrix $A$, denoted $\tr A$, is the sum of elements on its diagonal. Let the components of the $n\times n$ matrix $A$ be $A_{ij}$ where $i,j\in\left\{ 1,\ldots,n\right\} $. Then we define
\[
\tr A\equiv\sum_{i}A_{ii}.
\]
The trace is clearly a linear operation:
\[
\tr\left(A+B\right)=\tr A+\tr B\sp\tr\left(\lambda A\right)=\lambda\tr A,
\]
for all square matrices $A$ and $B$ (of the same size) and all scalars $\lambda\in\BBC$. In \zcref[S]{prob:trace-invariant} you will prove that the trace satisfies the property
\[
\tr\left(AB\right)=\tr\left(BA\right).
\]
From this you will deduce that the trace is invariant under \emph{cyclic permutations}\index{Cyclic permutations}\footnote{These permutations are called “cyclic” because you can imagine the matrices as being placed on a circle. Rotating the circle will change which matrix goes first, but not their order. The identity is not necessarily true for \textbf{non-cyclic }permutations: $\tr\left(ABC\right)\ne\tr\left(ACB\right)$ in general, although equality does hold if all three matrices are symmetric (a matrix $M$ is symmetric if $M_{ij}=M_{ji}$ for all $i,j$).} of matrices:
\[
\tr\left(ABC\right)=\tr\left(BCA\right)=\tr\left(CAB\right),
\]
and similarly for any number $n\ge4$ of matrices:
\[
\tr\left(A_{1}A_{2}\cdots A_{n-1}A_{n}\right)=\tr\left(A_{2}A_{3}\cdots A_{n}A_{1}\right)=\cdots=\tr\left(A_{n}A_{1}\cdots A_{n-1}\right).
\]
Recall from \zcref[S]{subsec:Representing-Matrices-in} that the relation between the representations of $A$ in different bases is given by
\[
\left(A\right)_{C}=P^{-1}\left(A\right)_{B}P,
\]
where $P=P_{B\ot C}$ is the change-of-basis matrix in \zcref[S]{eq:c-o-b}. Since both sides are equal, we can take the trace of both:
\[
\tr\left(A\right)_{C}=\tr\left(P^{-1}\left(A\right)_{B}P\right)=\tr\left(\left(A\right)_{B}PP^{-1}\right)=\tr\left(A\right)_{B},
\]
where we used the cyclicity of the trace. We conclude that the trace of a matrix is actually independent of basis; no matter which basis you choose to represent the matrix in, the elements on the diagonal will magically sum to the same number.

Furthermore, recall from \zcref[S]{subsec:Diagonalizable-Matrices} that a matrix $A$ is diagonalizable if and only if
\[
P^{-1}AP=D,
\]
where $D$ is a diagonal matrix and $P$ is an invertible matrix. $D$ is just a representation of $A$ in a special basis that diagonalizes it, and $P=P_{1\ot B}$ is the change-of-basis matrix from that basis to the standard basis. As above, if we cycle the matrices on the left so that $P$ cancels out, we find that
\[
\tr A=\tr D.
\]
Therefore, the trace of a diagonalizable matrix is the same as the trace of the corresponding diagonal matrix. Since the elements on the diagonal are the eigenvalues $\lambda_{i}$ of the matrix, we conclude the trace of a diagonalizable matrix is the sum of its eigenvalues $\lambda_{i}$:
\[
\tr A=\sum_{i}\lambda_{i}.
\]
Finally, the trace of an outer product $\left|\Psi\right\rangle \left\langle \Phi\right|$ of two vectors is equal to their inner product:
\begin{equation}
\tr\left(\left|\Psi\right\rangle \left\langle \Phi\right|\right)=\langle\Phi|\Psi\rangle.\label{eq:trace-outer}
\end{equation}
This follows from the cyclicity of the trace. The trace is cyclic for any product of matrices, including non-square ones; we treat $\left|\Psi\right\rangle $ as an $n\times1$ matrix and $\left\langle \Phi\right|$ as a $1\times n$ matrix, so we can cycle $\left|\Psi\right\rangle \left\langle \Phi\right|$ (an $n\times n$ matrix) into $\langle\Phi|\Psi\rangle$ (a $1\times1$ matrix, that is, a scalar), and the trace will remain the same.
\begin{xca}\leavevmode
Show that, in general, $\tr\left(AB\right)\ne\tr\left(A\right)\tr\left(B\right)$.
\end{xca}

\begin{problem}\leavevmode
\label{prob:trace-invariant}
\begin{problemparts}
\item Prove that the trace satisfies the property $\tr\left(AB\right)=\tr\left(BA\right)$.
\item Prove that the trace is invariant under cyclic permutations of $n$ matrices.
\end{problemparts}
\end{problem}

\begin{problem}\leavevmode
\begin{problemparts}
\item Prove that any linear map $f:\BBC^{n\times n}\to\BBC$ that satisfies $f\left(AB\right)=f\left(BA\right)$ is proportional to the trace operation\footnote{Here, $\BBC^{n\xx n}$ denotes the space of $n\xx n$ complex matrices.}.
\item Prove that if the map also satisfies $f\left(1\right)=n$, where $1$ is the $n\times n$ identity matrix, then $f$ is equal to the trace operation.
\end{problemparts}
\end{problem}

\begin{problem}\leavevmode
\label{prob:Frobenius}We can use the trace to define an inner product on the space of matrices. This is called the \emph{Frobenius inner product}\index{Frobenius inner product}\index{Inner product!Frobenius}, and it is defined as follows:
\[
\langle A|B\rangle\equiv\tr\left(A^{\dagger}B\right)\sp A,B\in\BBC^{n\times n}.
\]
\begin{problemparts}
\item Find a general expression for the Frobenius inner product $\langle A|B\rangle$ of matrices with components $A_{ij}$ and $B_{ij}$.
\item Can the Frobenius inner product be used to define a norm?
\end{problemparts}
\end{problem}

\subsubsection{Positive-Definite and Positive Semi-Definite Matrices}\label{subsec:Positive-Definite-and-Positive}

A matrix $A$ is \emph{positive-definite}\index{Positive-definite matrix}\index{Matrix!Positive-definite} if and only if, for all $\left|\Psi\right\rangle \ne0$,
\[
\langle\Psi|A|\Psi\rangle>0.
\]
A matrix $A$ is \emph{positive semi-definite}\index{Positive semi-definite matrix}\index{Matrix!Positive semi-definite} if and only if, for all vectors $\left|\Psi\right\rangle $,
\[
\langle\Psi|A|\Psi\rangle\ge0.
\]
Note that a positive-definite matrix is automatically positive semi-definite, but not vice versa. \emph{Negative-definite}\index{Negative-definite matrix}\index{Matrix!Negative-definite} and \emph{negative semi-definite}\index{Negative semi-definite matrix}\index{Matrix!Negative semi-definite} matrices are defined analogously.
\begin{problem}\leavevmode
\label{prob:pos-hermitian}
\begin{problemparts}
\item Prove that any positive-definite (or semi-definite) matrix is Hermitian.
\item Show that not every Hermitian matrix is positive-definite (or semi-definite).
\end{problemparts}
\end{problem}

\subsubsection{Anti-Hermitian Matrices}\label{subsec:Anti-Hermitian-Matrices}

An \emph{anti-Hermitian matrix}\index{Anti-Hermitian matrix} $A$ is a matrix which satisfies
\[
A^{\dagger}=-A.
\]
Just as a Hermitian matrix is analogous to a real number, an anti-Hermitian matrix is analogous to an imaginary number $\i b$ where $b\in\BBR$, since such a number satisfies $\left(\i b\right)^{*}=-\i b$.
\begin{problem}\leavevmode
\begin{problemparts}
\item Prove that the eigenvalues of an anti-Hermitian matrix are all purely imaginary, as defined in \zcref[S]{prob:quadratic}.
\item Prove that an anti-Hermitian matrix is normal, and thus it has an orthonormal eigenbasis (see \zcref[S]{subsec:Unitary-and-Normal-Matrices}).
\item Prove that if $A$ and $B$ are Hermitian, then $\left[A,B\right]$ must be anti-Hermitian.
\end{problemparts}
\end{problem}

\subsubsection{The Cauchy-Schwarz Inequality}

The \emph{Cauchy-Schwarz inequality\index{Cauchy-Schwarz inequality}} states that for any two vectors $\left|\Psi\right\rangle $ and $\left|\Phi\right\rangle $, we have
\begin{equation}
\left|\langle\Psi|\Phi\rangle\right|\le\left\Vert \Psi\right\Vert \left\Vert \Phi\right\Vert .\label{eq:CSI}
\end{equation}
If $\left|\Phi\right\rangle =0$, both sides vanish, so the inequality holds. Otherwise, consider an orthonormal basis $\left|B_{i}\right\rangle $ such that\footnote{\label{fn:Such-a-basis}Such a basis can always be generated using a method called the \emph{Gram-Schmidt process}\index{Gram-Schmidt process}, which we will not describe here.}
\[
\left|B_{1}\right\rangle \equiv\frac{\left|\Phi\right\rangle }{\left\Vert \Phi\right\Vert }.
\]
Then, using the completeness relation in \zcref[S]{eq:completeness}, we find:
\begin{align*}
\left\Vert \Psi\right\Vert ^{2}\left\Vert \Phi\right\Vert ^{2} & =\langle\Psi|\Psi\rangle\left\Vert \Phi\right\Vert ^{2}\\
 & =\langle\Psi|\left(\sum^{n}_{i=1}|B_{i}\rangle\langle B_{i}|\right)|\Psi\rangle\left\Vert \Phi\right\Vert ^{2}\\
 & =\langle\Psi|\left(|B_{1}\rangle\langle B_{1}|+\sum^{n}_{i=2}|B_{i}\rangle\langle B_{i}|\right)|\Psi\rangle\left\Vert \Phi\right\Vert ^{2}\\
 & =\langle\Psi|\left(\frac{1}{\left\Vert \Phi\right\Vert ^{2}}\left|\Phi\right\rangle \left\langle \Phi\right|+\sum^{n}_{i=2}|B_{i}\rangle\langle B_{i}|\right)|\Psi\rangle\left\Vert \Phi\right\Vert ^{2}\\
 & =\left(\frac{1}{\left\Vert \Phi\right\Vert ^{2}}\langle\Psi|\Phi\rangle\langle\Phi|\Psi\rangle+\sum^{n}_{i=2}\langle\Psi|B_{i}\rangle\langle B_{i}|\Psi\rangle\right)\left\Vert \Phi\right\Vert ^{2}\\
 & =\left(\frac{1}{\left\Vert \Phi\right\Vert ^{2}}\left|\langle\Psi|\Phi\rangle\right|^{2}+\sum^{n}_{i=2}\left|\langle\Psi|B_{i}\rangle\right|^{2}\right)\left\Vert \Phi\right\Vert ^{2}\\
 & =\left|\langle\Psi|\Phi\rangle\right|^{2}+\sum^{n}_{i=2}\left|\langle\Psi|B_{i}\rangle\right|^{2}\left\Vert \Phi\right\Vert ^{2}\\
 & \ge\left|\langle\Psi|\Phi\rangle\right|^{2}.
\end{align*}
Taking the square root, we obtain \zcref[S]{eq:CSI}.
\begin{problem}\leavevmode
Explain each step in the proof above.
\end{problem}

\begin{xca}\leavevmode
Check this inequality explicitly for three pairs of vectors of your choice.
\end{xca}

\begin{problem}\leavevmode
Find a condition that is equivalent to an \textbf{equality }in the Cauchy-Schwarz inequality. That is, find an “if and only if” statement for $\left|\langle\Psi|\Phi\rangle\right|=\left\Vert \Psi\right\Vert \left\Vert \Phi\right\Vert $ involving properties of $\left|\Psi\right\rangle $ and $\left|\Phi\right\rangle $. Prove the statement in\textbf{ both directions}.
\end{problem}

\begin{problem}\leavevmode
You are told that a $3\xx3$ matrix $A$ satisfies for some non-zero vector $\left|\Psi\right\rangle $
\[
A\left|\Psi\right\rangle =2\left|\Psi\right\rangle .
\]
\begin{problemparts}
\item Can this matrix be Hermitian?
\item Can this matrix be unitary?
\item Can this matrix be anti-Hermitian (as defined in \zcref[S]{subsec:Anti-Hermitian-Matrices})?
\item You learn that this matrix also satisfies for two other non-zero vectors $\left|\Phi\right\rangle $ and $\left|\Theta\right\rangle $:
\[
A\left|\Phi\right\rangle =-2\i\left|\Phi\right\rangle \sp A\left|\Theta\right\rangle =2\left|\Theta\right\rangle .
\]
Can this matrix be Hermitian now?
\item Is the matrix diagonalizable now?
\item Finally, you find out that this matrix also satisfies for some non-zero vector $\left|\Xi\right\rangle $
\[
A\left|\Xi\right\rangle =\i\left|\Xi\right\rangle .
\]
Is the matrix diagonalizable now?
\end{problemparts}
\end{problem}

\subsection{Probability Theory}

\subsubsection{Random Variables and Probability Distributions}

Let's say we perform an experiment that can have multiple outcomes. The \emph{sample space}\index{Sample space} of the experiment, denoted $\Omega$, is the set of all possible outcomes. An \emph{event}\index{Event (probability)} is a subset of $\Omega$, or in other words, a set of outcomes. We say that an event “occurs” if the outcome of the experiment is an element of the event. For example, if we roll a 6-sided die, then the sample space is $\left\{ 1,2,3,4,5,6\right\} $. If the die lands on the outcome 5, then the event $\left\{ 5\right\} $ occurred, but the event $\left\{ 4,5,6\right\} $ (“the roll is at least 4”) also occurred.

A \emph{random variable\index{Random variable}} $X$ is a function which assigns a real value to each outcome. Sometimes these values will be the actual measured value in some way: for example, the value of the random variable $X$ for rolling a 6-sided die will simply be the number on the die. Other times, the value of the random variable will be just a numerical label assigned to each outcome: for example, for a coin toss we can assign 1 to heads and 0 to tails (but we can also assign any other numbers, if we want).

These examples were of \emph{discrete}\index{Random variable!Discrete} random variables, but we can also have \emph{continuous}\index{Random variable!Continuous} random variables, such as the position of a particle along a line, which in principle can take any real value. For simplicity, we will focus on discrete random variables here.

A (discrete) \emph{probability distribution\index{Probability distribution} }assigns a probability to each value of a random variable. We denote by $P\left(X=x\right)$ the probability that the random variable $X$ will have the value $x$. A \emph{probability\index{Probability}} is a number between $0$ and $1$, which denotes how likely it is (in percentage) for the value to occur, so $0$ means this value \textbf{never }occurs and $1$ ($=100\%$) means this value \textbf{always }occurs.

The probabilities for all the possible values must sum to 1, because if for example they only sum to $0.9$, this means that in 10\% of the cases the random variable has \textbf{no value}, which doesn't really make sense. Also, if $P\left(X=x\right)=0$ then there must be at least one other possible value that $X$ can take, since it will never evaluate to $x$, and if $P\left(X=x\right)=1$ then there cannot be any other possible values that $X$ can take, since it always evaluates to $x$.

For example, for the coin toss we have
\[
P\left(X=0\right)=\hf\sp P\left(X=1\right)=\hf,
\]
and for the 6-sided die roll we have
\[
P\left(X=1\right)=\sxt\sp P\left(X=2\right)=\sxt\sp P\left(X=3\right)=\sxt,
\]
\[
P\left(X=4\right)=\sxt\sp P\left(X=5\right)=\sxt\sp P\left(X=6\right)=\sxt.
\]
Note how the probabilities sum to $1$ in each case. Of course, we could also say that maybe the coin toss results in heads only 49.9\% of the time, and tails another 49.9\% of the time, and the remaining 0.2\% is the probability for the coin to balance perfectly on its edge... But usually we ignore subtleties like this and assume we have \textbf{idealized }coins. Similarly, we could also have a \emph{loaded coin}\index{Loaded coin or die} which lands on heads more or less frequently than it lands on tails, but usually we assume that the coins are fair\index{Fair coin or die} unless stated otherwise. The same discussion applies to dice, with any number of sides: they are, by default, assumed to be idealized and fair.

These probability distributions are \emph{uniform}\index{Uniform probability distribution}, since they assign the same probability to each value of $X$. However, probability distributions need not be uniform. A simple example is a loaded coin, which perhaps has
\[
P\left(X=0\right)=\frac{1}{3}\sp P\left(X=1\right)=\frac{2}{3}.
\]
As a more interesting example, if we toss \textbf{two }fair coins $X_{1}$ and $X_{2}$ and define a random variable to be the sum of the results, $X\equiv X_{1}+X_{2}$, then we can get any of the following 4 outcomes:
\[
0+0=0\sp0+1=1\sp1+0=1\sp1+1=2.
\]
The probability for each outcome is
\[
\hf\cdot\hf=\fr,
\]
but the value 1 appears twice; thus
\[
P\left(X=0\right)=\fr\sp P\left(X=1\right)=\hf\sp P\left(X=2\right)=\fr.
\]
Of course, the probabilities still sum to 1.
\begin{xca}\leavevmode
Calculate the probability distribution for the sum of two rolls of a 6-sided die. This is known to players of role-playing games (such as Dungeons \& Dragons) as a “$2\d6$”, where we define $n\d s$ to be the sum of $n$ rolls of an $s$-sided die.
\end{xca}

\subsubsection{Conditional Probability}\label{subsec:Conditional-Probability}

Consider two random variables, $X$ and $Y$. Let $X$ have $N$ possible values $x_{i}$, $i\in\left\{ 1,\ldots,N\right\} $ and let $Y$ have $M$ possible values $y_{j}$, $j\in\left\{ 1,\ldots,M\right\} $. Then the \emph{joint probability}\index{Joint probability} to get $X=x_{i}$ \textbf{and} $Y=y_{j}$ at the \textbf{same time}, for some specific choice of $i$ and $j$, is denoted
\[
P\left(X=x_{i}\cap Y=y_{j}\right),
\]
where $\cap$ means\footnote{More precisely, $\cap$ means the intersection of two sets, where one is the set of outcomes for which $X=x_{i}$ and the other is the set of outcomes for which $Y=y_{j}$.} “and”. Furthermore, we have
\begin{equation}
\sum^{M}_{j=1}P\left(X=x_{i}\cap Y=y_{j}\right)=P\left(X=x_{i}\right),\label{eq:PXY-PX}
\end{equation}
because the total probability to get $X=x_{i}$ is the sum of all the different probabilities that involve $X=x_{i}$ plus something else. To illustrate this, consider the following random variables:
\[
X=\textrm{whether a student passes a course},
\]
\[
Y=\textrm{whether the student did all the homework}.
\]
There are in total 4 different combinations, and their probabilities must sum to 1. Maybe the probabilities are as follows:
\[
P\left(\textrm{pass}\cap\textrm{did homework}\right)=40\%,
\]
\[
P\left(\textrm{pass}\cap\textrm{didn't do homework}\right)=20\%,
\]
\[
P\left(\textrm{didn't pass}\cap\textrm{did homework}\right)=10\%,
\]
\[
P\left(\textrm{didn't pass}\cap\textrm{didn't do homework}\right)=30\%.
\]

Then clearly the total probability that a student passes (whether or not they did the homework) is $40\%+20\%=60\%$, and the total probability that a student does not pass is $10\%+30\%=40\%$. This is exactly what \zcref[S]{eq:PXY-PX} means.

However, what the student probably wants to know is the probability that they pass given\textbf{ }that they did the homework vs. the probability that they pass given that they did not do the homework. This is called \emph{conditional probability\index{Conditional probability}}. Let $A$ and $B$ denote two events, such as “pass” and “did homework”. The probability for event $A$ \textbf{given} event $B$ is denoted $P\mr AB$, where $|$ is read as “given that”. It is related to $P\left(A\cap B\right)$ as follows:
\begin{equation}
P\mr AB=\frac{P\left(A\cap B\right)}{P\left(B\right)}\sp P\left(B\right)>0.\label{eq:conditional-joint}
\end{equation}
In other words, it is the probability that both $A$ and $B$ happened, divided by the probability for $B$ to happen. Let us calculate:
\[
P\mr{\textrm{pass}}{\textrm{did homework}}=\frac{40\%}{40\%+10\%}=80\%,
\]
\[
P\mr{\textrm{pass}}{\textrm{didn't do homework}}=\frac{20\%}{20\%+30\%}=40\%.
\]
In this example, students who did all the homework were twice as likely to pass the course.

Two events $A,B$ are \emph{independent}\index{Independent events} if their joint probability is just the product of the probabilities for each event:
\[
P\left(A\cap B\right)=P\left(A\right)P\left(B\right).
\]
By \zcref[S]{eq:conditional-joint}, this means that, if $P\left(A\right)>0$ and $P\left(B\right)>0$, we have
\[
P\mr AB=P\left(A\right)\sp P\mr BA=P\left(B\right).
\]
In other words, when two events are independent, the probability for one of them does not depend on whether or not the other happened.

Let us consider a finite or countably infinite set of events $B_{i}$ with $i\in\left\{ 1,2,\ldots\right\} $. We say that the events are \emph{mutually exclusive}\index{Mutually exclusive events} if $B_{i}\cap B_{j}=\emptyset$ for any $i\ne j$, where $\emptyset$ is the empty set. In other words, if one event occurs, then no other events can occur at the same time. We say that the events are \emph{collectively exhaustive}\index{Collectively exhaustive events} if at least one of the events must occur, or equivalently, if the sample space $\Omega$ is the union of all the events:
\[
\Omega=\bigcup_{i}B_{i}.
\]
The events form a \emph{partition}\index{Partition (probability)} of $\Omega$ if they are mutually exclusive and collectively exhaustive. For a 6-sided die, the set
\[
\left\{ \left\{ 1\right\} ,\left\{ 2\right\} ,\left\{ 3\right\} ,\left\{ 4\right\} ,\left\{ 5\right\} ,\left\{ 6\right\} \right\} ,
\]
that is, the set of events $B_{i}=\left\{ i\right\} $ for $i\in\left\{ 1,2,3,4,5,6\right\} $, is a partition, but so is the set
\[
\left\{ \left\{ 1,2\right\} ,\left\{ 3,4\right\} ,\left\{ 5,6\right\} \right\} ,
\]
that is, the set of events $B_{1}=\left\{ 1,2\right\} $, $B_{2}=\left\{ 3,4\right\} $, and $B_{3}=\left\{ 5,6\right\} $.

Given a partition $B_{i}$, for any event $A$ we have
\[
A=A\cap\Omega=A\cap\bigcup_{i}B_{i}=\bigcup_{i}\left(A\cap B_{i}\right).
\]
In the first equality, we used the fact that by definition $A\subseteq\Omega$ (an event is a subset of the sample space) and therefore $A=A\cap\Omega$. In the second equality, we used that the events are collectively exhaustive, so their union is the sample space. To understand the third equality, first note that an outcome is in $A\cap\bigcup_{i}B_{i}$ exactly when it's both in $A$ and in $\bigcup_{i}B_{i}$ (that's just the definition of the set intersection $\cap$), and it's in $\bigcup_{i}B_{i}$ exactly when there's an index $i$ such that the outcome is in the set $B_{i}$ (that's just the definition of the set union $\cup$). Together, these two conditions hold exactly when there's an index $i$ such that the outcome is in both $A$ and $B_{i}$, which is the same as saying it's in $\bigcup_{i}\left(A\cap B_{i}\right)$.

Because the sets $B_{i}$ are mutually exclusive, the sets $A\cap B_{i}$ are also mutually exclusive, that is,
\[
\left(A\cap B_{i}\right)\cap\left(A\cap B_{j}\right)=\emptyset
\]
for all $i\ne j$. Since the probability of a countable union of mutually exclusive events is the sum of their probabilities, we get
\[
P\left(A\right)=\sum_{i}P\left(A\cap B_{i}\right).
\]
This is the \emph{law of total probability}\index{Law of total probability}: the probability of event $A$ happening is the sum of the probabilities for event $A$ happening together with each event $B_{i}$ in the partition. Using \zcref[S]{eq:conditional-joint} we can also write this as
\[
P\left(A\right)=\sum_{i}P\mr A{B_{i}}P\left(B_{i}\right),
\]
where we omit\footnote{If $P\left(B_{i}\right)=0$ then $P\mr A{B_{i}}$ is undefined, but these terms don't contribute to the sum anyway since $P\left(B_{i}\right)=0$ means $P\left(A\cap B_{i}\right)=0$ for any $A$.} any terms with $P\left(B_{i}\right)=0$. This means that the total probability of event $A$ happening is the sum of the probabilities for event $A$ conditional on each event $B_{i}$, weighted by the probabilities to get the events $B_{i}$ themselves.
\begin{xca}\leavevmode
There are six more conditional probabilities from the pass/homework example that we did not calculate here. Calculate them. What do you learn from the results?
\end{xca}

\begin{xca}\leavevmode
A test for a certain disease has a 1\% chance of a \textbf{false positive} for a healthy patient, i.e. the result is positive but the patient \textbf{isn't }actually sick, and a 1\% chance of a \textbf{false negative} for a sick patient, i.e. the result is negative but the patient \textbf{is }actually sick. Assume that 0.1\% of the population is actually sick.

\[
\begin{array}{c|c|c|c}
 & \textrm{Sick} & \textrm{Healthy} & \textrm{Total}\\
\hline \textrm{Actual status} &  & 99.9\% & 100\%\\
\hline \textrm{Positive test} & 0.099\% &  & 1.098\%\\
\hline \textrm{Negative test} & 0.001\% & 98.901\%
\end{array}
\]
\begin{problemparts}
\item Fill in the blanks in the table above.
\item Given that you tested positive, what is the conditional probability that you actually have the disease?
\item Given that you tested negative, what is the conditional probability that you actually don't have the disease?
\item Which result should you trust, a positive one or a negative one?
\end{problemparts}
\end{xca}

\begin{problem}\leavevmode
Prove that the law of total probability also applies when all the terms are conditional on some other event $C$ with $P\left(C\right)>0$, that is,
\begin{equation}
P\mr AC=\sum_{i}P\mr A{B_{i}\cap C}P\mr{B_{i}}C,\label{eq:total-probability-conditional}
\end{equation}
where we omit any terms with $P\left(B_{i}\cap C\right)=0$.
\end{problem}

\subsubsection{Expected Values}

The \emph{expected value} \index{Expected (or expectation) value}(or \emph{expectation value} or \emph{mean}\index{Mean}) $\left\langle X\right\rangle $ of a random variable $X$ is the average over all the possible values $X$ can take, weighted by their assigned probabilities:
\begin{equation}
\left\langle X\right\rangle \equiv\sum^{N}_{i=1}P\left(X=x_{i}\right)x_{i},\label{eq:expected}
\end{equation}
where $N$ is the total number of possible values of $X$, $x_{i}$ is the $i$-th possible value, and $P\left(X=x_{i}\right)$ is the probability to get $x_{i}$. In the example of the coin toss, we have:
\[
\left\langle X\right\rangle =\hf\cdot0+\hf\cdot1=\hf=0.5,
\]
and for the 6-sided die roll, we have:
\[
\left\langle X\right\rangle =\sxt\cdot1+\sxt\cdot2+\sxt\cdot3+\sxt\cdot4+\sxt\cdot5+\sxt\cdot6=\frac{7}{2}=3.5.
\]
Observe that the expected value in both cases is not an actual value the random variable can take! This is often the case with discrete random variables.

We will now prove that the expected value is linear:
\begin{equation}
\left\langle \alpha X+\beta Y\right\rangle =\alpha\left\langle X\right\rangle +\beta\left\langle Y\right\rangle \sp\alpha,\beta\in\BBR.\label{eq:exp-val-lin}
\end{equation}
This can be broken down into two rules:
\[
\left\langle \alpha X\right\rangle =\alpha\left\langle X\right\rangle \sp\left\langle X+Y\right\rangle =\left\langle X\right\rangle +\left\langle Y\right\rangle .
\]
The first rule is easy to prove:
\begin{align*}
\left\langle \alpha X\right\rangle  & =\sum^{N}_{i=1}P\left(\alpha X=\alpha x_{i}\right)\left(\alpha x_{i}\right)\\
 & =\alpha\sum^{N}_{i=1}P\left(X=x_{i}\right)x_{i}=\alpha\left\langle X\right\rangle .
\end{align*}
To prove the second part, let $X$ have $N$ possible values $x_{i}$ and let $Y$ have $M$ possible values $y_{j}$, as in \zcref[S]{subsec:Conditional-Probability}. Then in calculating $\left\langle X+Y\right\rangle $ we need to sum over both $i$ and $j$, to ensure we take all possible combinations of $X$ and $Y$ into account. Using \zcref[S]{eq:PXY-PX} (and the analogous relation with the roles of $X$ and $Y$ exchanged), we get:
\begin{align*}
\left\langle X+Y\right\rangle  & =\sum^{N}_{i=1}\sum^{M}_{j=1}P\left(X=x_{i}\cap Y=y_{j}\right)\left(x_{i}+y_{j}\right)\\
 & =\sum^{N}_{i=1}\left(\sum^{M}_{j=1}P\left(X=x_{i}\cap Y=y_{j}\right)\right)x_{i}+\sum^{M}_{j=1}\left(\sum^{N}_{i=1}P\left(X=x_{i}\cap Y=y_{j}\right)\right)y_{j}\\
 & =\sum^{N}_{i=1}P\left(X=x_{i}\right)x_{i}+\sum^{M}_{j=1}P\left(Y=y_{j}\right)y_{j}\\
 & =\left\langle X\right\rangle +\left\langle Y\right\rangle ,
\end{align*}
as we wanted to prove.
\begin{xca}\leavevmode
Calculate the expected value for the sum of two coin tosses and for a $2\d6$ roll (the sum of two 6-sided dice). First, do it by defining one random variable for the sum, calculating the probabilities, and then using the definition of the expected value. Then, do it by considering just one coin or one 6-sided die respectively, and use \zcref[S]{eq:exp-val-lin}. Compare your results.
\end{xca}

\subsubsection{Standard Deviation}\label{subsec:Standard-Deviation}

The \emph{standard deviation\index{Standard deviation}}\footnote{By the way, the square of the standard deviation is called the \emph{variance}\index{Variance}, but it will not interest us in these notes.}\emph{ }measures how far the outcomes are expected to be from the expected value. To calculate the standard deviation, we take the expected value of $\left(X-\left\langle X\right\rangle \right)^{2}$, that is, the square of the difference between the \textbf{actual }value of $X$ and its \textbf{expected }value $\left\langle X\right\rangle $. Then, we take the square root of the result to obtain the standard deviation $\Delta X$:
\[
\Delta X\equiv\sqrt{\left\langle \left(X-\left\langle X\right\rangle \right)^{2}\right\rangle }.
\]
To simplify this, first we note that
\[
\left(X-\left\langle X\right\rangle \right)^{2}=X^{2}-2X\left\langle X\right\rangle +\left\langle X\right\rangle ^{2}.
\]
In this formula, $X^{2}$ is a random variable (whose values are the squares of the values of $X$), but $\left\langle X\right\rangle $ is just a \textbf{number}, not a random variable. Since it is a number, we can treat it as a random variable that only returns one value with 100\% probability, which means that
\[
\left\langle \left\langle X\right\rangle \right\rangle =\left\langle X\right\rangle .
\]
So, by \zcref[S]{eq:exp-val-lin}, we have:
\begin{align*}
\left\langle \left(X-\left\langle X\right\rangle \right)^{2}\right\rangle  & =\left\langle X^{2}\right\rangle -2\left\langle X\right\rangle \left\langle X\right\rangle +\left\langle X\right\rangle ^{2}\\
 & =\left\langle X^{2}\right\rangle -\left\langle X\right\rangle ^{2}.
\end{align*}
Therefore, the standard deviation can be written as follows:
\[
\Delta X=\sqrt{\left\langle X^{2}\right\rangle -\left\langle X\right\rangle ^{2}}.
\]
This form is easier to do calculations with. For example, for the coin toss we have from before
\[
\left\langle X\right\rangle =\hf,
\]
and we also calculate:
\[
\left\langle X^{2}\right\rangle =\hf\cdot0^{2}+\hf\cdot1^{2}=\hf,
\]
which gives us
\[
\Delta X=\sqrt{\hf-\fr}=\hf.
\]
This makes sense, as the two actual values of the outcomes, 0 and 1, lie exactly $1/2$ away from the expected value $\left\langle X\right\rangle =1/2$ in each direction. So they each “deviate” from it by $1/2$.

For the die roll, we have from before
\[
\left\langle X\right\rangle =\frac{7}{2},
\]
and we also calculate:
\[
\left\langle X^{2}\right\rangle =\sxt\left(1^{2}+2^{2}+3^{2}+4^{2}+5^{2}+6^{2}\right)=\frac{91}{6},
\]
which gives us
\[
\Delta X=\sqrt{\frac{91}{6}-\frac{49}{4}}=\sqrt{\frac{35}{12}}\ap1.7.
\]

\begin{xca}\leavevmode
Calculate the standard deviation for the sum of two coin tosses and for a $2\d6$ roll.
\end{xca}

\begin{problem}\leavevmode
A wizard can cast one of two spells on a dragon.
\begin{itemize}
\item For \textbf{Psychic Ray}, the dragon makes a saving throw: it rolls a d20, and if it gets 8 or higher, it succeeds. If the saving throw fails, the spell does 10d8 points of psychic damage. If the saving throw succeeds, the damage is reduced by 50\%.
\item For \textbf{Fiery Missile}, the wizard rolls a d20, and if the result is 9 or higher, the attack hits. If the attack misses, no damage will be done. If the attack hits, the spell will do 6d6+8d10 points of fire damage. However, the wizard is not sure if the dragon is resistant to fire or not; there is a probability of 30\% that it is resistant, in which case the damage is reduced by 50\%.
\end{itemize}
Which spell will do more damage on average?
\end{problem}

\subsubsection{Information and Entropy}\label{subsec:Information-and-Entropy}

Consider a random variable $X$ which has $n$ possible outcomes $x_{1},\ldots,x_{n}$ with associated probabilities $p_{1},\ldots,p_{n}$. If we measured the variable and found that the outcome is $x_{i}$ for some $i$, how much information have we gained from that? One way to think about this is to consider how much that outcome “surprised” us. If we were not surprised at all, then what we measured is exactly what we expected, so we have gained no information. If we were extremely surprised, then clearly we have gained a lot of information.

When defining the information content of $X$, we shouldn't use the \textbf{values }$x_{i}$, only the \textbf{probabilities }$p_{i}$ to measure each value. This is because the values are up to our interpretation, but the probabilities are fixed. For example, I can define the two sides of a fair coin to mean 0 and 1, or 1 and 2, or $-1$ and $+1$, or just as “heads” and “tails”; in each case, the values are up to me to choose, but the probability for that coin to land on each side is fixed at 1/2. So the information content of the coin should only be based on the probabilities, not the values.

To relate information to probabilities, note that the lower the probability of the outcome, the higher the information learned from it. If we measure two independent outcomes $x_{i},x_{j}$ with probabilities $p_{i}$ and $p_{j}$, the joint probability to measure both is the product $p_{i}p_{j}$ (as we saw in \zcref[S]{subsec:Conditional-Probability}), but the total information we gain should be the sum of the information gained by measuring each outcome separately, not their product (if we learned 3 bits of information from $x_{i}$ and $4$ bits from $x_{j}$, then we learned a total of 7 bits, not 12). In other words, we \textbf{multiply} probabilities, but we \textbf{add} information. This suggests that the information should be the \textbf{logarithm} of the probability (as defined in \zcref[S]{subsec:Exponentials-and-Logarithms}). More precisely, we define the \emph{information content}\index{Information content} or \emph{surprisal}\index{Surprisal} $I\left(p\right)$ of an outcome with probability $p$ as:
\[
I\left(p\right)\equiv-\log p.
\]
Here $\log$ is the \textbf{base-2} logarithm, that is, $\log\left(2^{x}\right)=2^{\log x}=x$, since we want to measure information in \textbf{bits}. The minus sign in this formula may be a bit confusing. In fact, information is always \textbf{non-negative}, because $p\in\left[0,1\right]$, and this means that $\log p\le0$, so $-\log p\ge0$. Note that $I\left(1\right)=0$, so an outcome with probability 1 gives us no information. Conversely, $I\left(p\right)$ tends to infinity as $p$ tends to zero, so the smaller the probability, the more information we gain. If we find an outcome with probability zero, then
\[
I\left(0\right)=-\lim_{p\to0}\log p=+\infty,
\]
so we will be infinitely surprised!

The \emph{Shannon entropy}\index{Shannon entropy} or \emph{information entropy}\index{Information entropy} of a random variable $X$ quantifies the following two related notions of information:
\begin{enumerate}
\item How much \textbf{uncertainty }we have \textbf{before }we learn the value of $X$.
\item How much \textbf{information }we gain, on average, \textbf{after }we learn the value of $X$.
\end{enumerate}
At first glance, these two notions may seem to be contradictory, but they are actually complementary once you take into account the “before” and “after” parts: the more uncertainty we had \textbf{before }learning the value of $X$, the more information we have obtained \textbf{after }learning that value. In the extreme cases, if we had maximum uncertainty about $X$, then learning the value of $X$ provided us with the maximum amount of information; and if we had no uncertainty at all about $X$, then learning the value of $X$ provided us with no new information. Therefore, this makes sense.

As above, let the possible measured values of $X$ be $x_{1},\ldots,x_{n}$, and let the probability to measure the value $x_{i}$ be $p_{i}$, that is,
\[
p_{i}\equiv P\left(X=x_{i}\right)\sp i\in\left\{ 1,\ldots,n\right\} .
\]
We define the \emph{Shannon entropy}\index{Shannon entropy} of $X$, denoted $H\left(X\right)$, as the expected value of the information content:
\[
H\left(X\right)\equiv\left\langle I\left(X\right)\right\rangle =\left\langle -\log P\left(X\right)\right\rangle =\sum_{i}p_{i}I\left(p_{i}\right)=-\sum_{i}p_{i}\log p_{i}.
\]
Here, $I\left(X\right)$ is the random variable that equals $I\left(p_{i}\right)=-\log p_{i}$ when $X=x_{i}$, and $P\left(X\right)$ is the random variable that equals $p_{i}$ when $X=x_{i}$.

If $p_{i}=0$ for some $i$ then $\log p_{i}$ is undefined, but note that even though $\lim_{x\to0^{+}}\left(\log x\right)=-\infty$, we have that $\lim_{x\to0^{+}}\left(x\log x\right)=0$. Therefore, if $p_{i}=0$, we simply have $p_{i}\log p_{i}=0$ and that particular value does not contribute to the entropy\footnote{Technically we should have defined $H\left(X\right)\equiv-\sum_{i}\lim_{x\to p_{i}}\left(x\log x\right)$ for this to work, but it's simpler to just have a convention that $0\log0\equiv0$.}. This makes sense, because if a certain value can never happen (it has probability zero), then it should not contribute to the entropy; it does not add uncertainty (because it is irrelevant) and we will never measure it and thus never gain information from it. Similarly, if $p_{i}=1$ for some $i$ (which necessarily means that all the other probabilities are zero), then the entropy is $-1\log1=0$ because we know exactly what value $X$ is with 100\% probability, so we have zero uncertainty about it and gain no information from measuring it, thus zero entropy.

Just like information, entropy is always non-negative, due to the minus sign. We have $p_{i}\in\left[0,1\right]$ for all $i$, so $\log p_{i}\le0$, therefore $-\log p_{i}\ge0$, and since $p_{i}\ge0$, we get that $-p_{i}\log p_{i}\ge0$ for all $i$, with equality only if $p_{i}=0$ or $p_{i}=1$.

Let us consider some simple examples. First, a fair coin will have $p_{1}=p_{2}=1/2$, so:
\[
I\left(p_{i}\right)=-\log\hf=\log2=1\sp i\in\left\{ 1,2\right\} .
\]
Therefore, we learn 1 bit of information regardless of the outcome, and the entropy is
\[
H=-\hf\log\hf-\hf\log\hf=1.
\]
If the coin is loaded, for example if it lands on one side 1/3 of the time and on the other side 2/3 of the time, we get
\[
I\left(\trd\right)=-\log\trd\ap1.585\sp I\left(\frac{2}{3}\right)=-\log\frac{2}{3}\ap0.585,
\]
so we would be more surprised if we get the less likely side, and
\[
H=-\trd\log\trd-\frac{2}{3}\log\frac{2}{3}\ap0.918,
\]
so a loaded coin has a lower entropy than a fair one. More generally, a coin that lands on one side with probability $p$ and the other side with probability $1-p$ will have entropy
\[
H\left(p\right)=-p\log p-\left(1-p\right)\log\left(1-p\right),
\]
and we can see that the maximum entropy of 1 is reached when the coin is fair ($p=1/2$) and the minimum entropy of 0 is reached when the coin is maximally loaded ($p=0$ or $p=1$, i.e. it only lands on one side). This makes sense, because if the coin is fair then no side is preferred over the other, and therefore we have maximum uncertainty and learn the maximum amount of information by tossing the coin, while if it is maximally loaded then we can be certain it will land on the loaded side and therefore we have no uncertainty and gain no information by tossing it.

In the case of a fair die with $s$ sides, each side will have probability $1/s$, so we will have
\[
I=-\log\frac{1}{s}=\log s,
\]
\[
H=s\cdot\left(-\frac{1}{s}\log\frac{1}{s}\right)=\log s.
\]
If $s=1$, the entropy is zero, because obviously there is no uncertainty about a die that only has one side. If $s=2$, we reproduce the entropy 1 of a fair coin. If $s>2$, the entropy will increase beyond 1, although it won't get too far as the increase is only logarithmic. For example, the entropy of a d6 is $\log6\ap2.58$ and the entropy of a d20 is $\log20\ap4.32$. Again, this makes sense because the more sides the die has, the more uncertainty we will have before we roll it, and the more information we will gain after we see the result of the roll.

\section{The Foundations of Quantum Theory}

Now that we have obtained the required mathematical tools, we can finally present quantum theory! This theory provides the correct fundamental framework for virtually all of known physics. We will see that its fundamental ingredients are Hilbert spaces with states and operators. These universal ingredients are then used to create particular models describing specific physical systems.

In this chapter, we will work exclusively with \textbf{discrete }quantum systems and \textbf{finite-dimensional }Hilbert spaces. These are much simpler than \textbf{continuous }quantum systems and/or \textbf{infinite-dimensional }Hilbert spaces. In particular, the math is much simpler – just linear algebra, without any calculus. However, it turns out that finite-dimensional Hilbert spaces are sufficient to define all of the fundamental concepts in quantum theory, and derive almost all of the most important results.

\subsection{Axiomatic Definition}

\subsubsection{Dimensionless and Dimensionful Constants}\label{subsec:Dimensionless-and-Dimensionful}

Consider the \emph{fine-structure constant}\index{Fine-structure constant}, which represents the strength of the electromagnetic interaction:
\[
\alpha\ap0.0073.
\]
This constant is not specified in any particular units, such as meters or seconds; it is a \textbf{pure number}. We call such a constant \emph{dimensionless}\index{Dimensionless constants}.

In contrast, some constants in physics are \emph{dimensionful}\index{Dimensionful constants}. This means that their numerical value depends on the system of units we use. For example, the speed of light $c$ has the following values in different systems of units:
\begin{align*}
c & \ap3.0\xx10^{8}\thinspace\textrm{meters/second}\\
 & \ap1.1\xx10^{7}\thinspace\textrm{miles/minute}\\
 & \ap170\thinspace\textrm{astronomical units/day}\\
 & \ap3.5\xx10^{-5}\thinspace\textrm{parsecs/hour}\\
 & \ap1\thinspace\textrm{light year/year}.
\end{align*}
What this means is that the numerical value of the speed of light does not have any physical meaning whatsoever\footnote{Indeed, in modern SI units the speed of light is \textbf{defined }to be 299,792,458 meters per second, and this definition is used to measure the length of a meter – not the other way around.}! It is merely a consequence of choosing to work with one system of units and not the other. But units are human constructs; the universe could not care less what units humans choose to measure things with. Therefore, none of the numbers written above have any actual meaning.

The numerical values of \textbf{dimensionless }constants are the only numbers that have a physical meaning, as they do not depend on the system of units. However, keep in mind that they are still, unavoidably, just parameters that are defined by humans in a certain way. There's nothing special about the number $\alpha\ap0.0073$ itself; we could also define another parameter $\beta\equiv2\alpha$ and use that in our equations instead. So don't try to do numerology\footnote{Interestingly, in the past some physicists tried to claim that $\alpha$ equals exactly $1/137$, but more precise measurements revealed that this is not actually the case. Still, you will often see it written as $1/137$ for that historical reason.} with the specific value of $\alpha$! What is important here is not the numerical value itself, but the fact that it is independent of the choice of units.

For this reason, it is most natural to work in \emph{Planck units}\index{Planck units}, where:
\begin{equation}
c=G=\hbar=\frac{1}{4\pi\varepsilon_{0}}=k_{B}=1.\label{eq:Planck-units}
\end{equation}
Here $c$ is the speed of light, $G$ is the \emph{gravitational constant}\index{Gravitational constant} used in Newtonian gravity, $\hbar$ is the \emph{(reduced) Planck constant\index{Planck constant}} used in quantum mechanics, $1/4\pi\varepsilon_{0}$ is the \emph{Coulomb constant\index{Coulomb constant} }used in electromagnetism, and $k_{B}$ is the \emph{Boltzmann constant\index{Boltzmann constant} }used in statistical mechanics.

All of these are \textbf{dimensionful }constants, which means we don't really care about their numerical values – so we might as well just set them to 1. This allows us to simply remove them from our equations. For example, instead of writing $\sqrt{\hbar G/c^{3}}$ – also known as the \emph{Planck length}\index{Planck length} – we just write 1. Much simpler, right?

Planck units are commonly used when doing research in theoretical physics, because they make equations simpler, more elegant, and less cluttered. However, sometimes we get numerical results that we wish to convert to real-world units such as kilograms and meters. To do this, all we need to do is to find the combination of the constants in \zcref[S]{eq:Planck-units} that has the desired units. For example, if we know that our pure number represents length, then we can multiply it by the Planck length $\sqrt{\hbar G/c^{3}}\ap1.6\xx10^{-35}\thinspace\mathrm{meters}$ to find its value in meters.

Since these notes were written by a theorist, we will use Planck units exclusively. This means that unlike in traditional quantum mechanics textbooks, $\hbar$ will not appear in any of our equations!
\begin{xca}\leavevmode
Calculate your age, height, mass, and body temperature in Planck units. For this, you will have to find combinations of the dimensionful constants we set to 1 in \zcref[S]{eq:Planck-units} that give you the desired units, as we did for the Planck length.
\end{xca}

\subsubsection{Hilbert Spaces, States, and Operators}

Recall that in \zcref[S]{subsec:Dual-Vectors} we defined a Hilbert space as a vector space with an inner product that is also a complete metric space with respect to that inner product. Quantum theory can be defined \textbf{axiomatically} using the theory of Hilbert spaces. In this chapter we will list a total of seven fundamental axioms.
\begin{quote}
\textbf{The System Axiom}\index{Axioms of quantum mechanics!System Axiom}\index{System Axiom}\textbf{:} A \emph{system\index{System}\index{Quantum system} }in quantum theory is the mathematical representation of a physical system (such as a particle) as a Hilbert space. The type and dimension of the Hilbert space will depend on the particular system. Note that the dimension of the Hilbert space is \textbf{unrelated} to the dimension of spacetime.
\end{quote}
In the finite-dimensional case, for example when the system involves spin, the Hilbert space will usually be $\BBC^{n}$ for some $n$, such as $\BBC^{2}$, which was used in most of the examples above and will continue to be used below. In the infinite-dimensional case, for example when the system involves position and momentum (which are, in general, continuous and not discrete), the Hilbert space will usually be a space of functions, which is much more complicated.
\begin{quote}
\textbf{The State Axiom}\index{Axioms of quantum mechanics!State Axiom}\index{State Axiom}\textbf{:} A \emph{state\index{State}\index{Quantum state} }of a quantum system is a vector with unit norm in the system's Hilbert space, that is, a vector $\left|\Psi\right\rangle $ which satisfies
\[
\left\Vert \Psi\right\Vert =\sqrt{\langle\Psi|\Psi\rangle}=1.
\]
\end{quote}
States represent the different configurations the system can have. It is important to stress that \textbf{only }unit vectors can represent states. If for some reason we have a (non-zero) vector with non-unit norm, we must normalize it (divide it by its norm) to obtain a unit vector, which can then represent a state.

Another important aspect of states is that they are only defined \textbf{up to a phase factor}. This means that, if the vector $\left|\Psi\right\rangle $ represents a state, then all vectors of the form $\e^{\i\phi}\left|\Psi\right\rangle $ for $\phi\in\BBR$ represent the \textbf{same}\footnote{\label{fn:Equiv}Actually, the more precise definition is that a state is a \emph{ray} in a Hilbert space\index{Ray in a Hilbert space}. Rays are defined as \emph{equivalence classes\index{Equivalence class of vectors}} of vectors such that a non-zero vector $\left|\Psi\right\rangle $ is equivalent to $\lambda\left|\Psi\right\rangle $ for any scalar $\lambda\in\BBC\setminus\left\{ 0\right\} $. The scalar can be separated into a polar representation, $\lambda=r\e^{\i\phi}$, as we discussed in \zcref[S]{subsec:Polar-Coordinates-and}. The $r$ part stretches the magnitude of the vector by a factor of $r$, and the $\e^{\i\phi}$ part (the phase factor) rotates it by $\phi$ radians. Any vector in the same equivalence class represents the \textbf{same }state, so multiplying the vector by a scalar will not change the state it represents, whatever the magnitude and phase are. However, it is conventional to choose states to be represented specifically by a unit vector from the equivalence class, since otherwise we would have to normalize vectors to 1 all the time. This is also the reason we prefer to use orthonormal bases in quantum theory.}\textbf{ }state as $\left|\Psi\right\rangle $. Note that adding a phase to a vector does not change the norm, since
\[
\left\Vert \e^{\i\phi}\Psi\right\Vert =\sqrt{\left(\e^{\i\phi}\left|\Psi\right\rangle \right)^{\dagger}\e^{\i\phi}\left|\Psi\right\rangle }=\sqrt{\left\langle \Psi\right|\e^{-\i\phi}\e^{\i\phi}\left|\Psi\right\rangle }=\sqrt{\langle\Psi|\Psi\rangle}=\left\Vert \Psi\right\Vert .
\]

\begin{quote}
\textbf{The Operator Axiom}\index{Axioms of quantum mechanics!Operator Axiom}\index{Operator Axiom}\textbf{:} An \emph{operator\index{Operator}\index{Quantum operator}} on a Hilbert space is a linear transformation which takes a vector and outputs another vector. By “linear” we mean that the operator $A$ satisfies
\[
A\left(\left|\Psi\right\rangle +\left|\Phi\right\rangle \right)=A\left|\Psi\right\rangle +A\left|\Phi\right\rangle \sp A\left(\lambda\left|\Psi\right\rangle \right)=\lambda A\left|\Psi\right\rangle ,
\]
where $\left|\Psi\right\rangle $ and $\left|\Phi\right\rangle $ are vectors and $\lambda$ is a scalar. In a finite-dimensional Hilbert space, operators are represented by $n\xx n$ complex matrices.
\end{quote}

\subsubsection{Hermitian Operators and Observables}

An operator corresponding to a Hermitian matrix\footnote{In an infinite-dimensional Hilbert space, where we don't necessarily have a matrix representation, a Hermitian operator is defined using the property that it is self-adjoint.} is a \emph{Hermitian operator}\index{Hermitian operator}. In \zcref[S]{subsec:Hermitian-Matrices} we discussed some interesting properties of these operators. In particular, their eigenvalues are real, and there is an orthonormal basis consisting of their eigenvectors (a.k.a. an eigenbasis).
\begin{quote}
\textbf{The Observable Axiom}\index{Axioms of quantum mechanics!Observable Axiom}\index{Observable Axiom}\textbf{:} In quantum theory, Hermitian operators correspond to \emph{observables}\index{Observable}\index{Quantum observable}, that is, properties of the system that can be measured. The eigenvalues of these operators, which are real as for all Hermitian operators, exactly correspond to all the different possible outcomes of the measurement.
\end{quote}
The mapping is one-to-one and onto (a.k.a. a \emph{bijection}\index{Bijection}), meaning that each eigenvalue corresponds to exactly one measurement outcome and vice versa. This makes sense because we always measure real numbers; there are no measurement devices that measure complex numbers!

Examples of observables are position, momentum, angular momentum, energy, and spin (which is intrinsic angular momentum). All of these may be represented as Hermitian operators on an appropriate Hilbert space.

\subsubsection{Probability Amplitudes}\label{subsec:Probability-Amplitudes}

Let the state of a quantum system be $\left|\Psi\right\rangle $. Once we have chosen a Hermitian operator to represent our observable, we may obtain an orthonormal basis of states $\left|B_{i}\right\rangle $ corresponding to the eigenvectors of that operator. In quantum mechanics, these eigenvectors are called \emph{eigenstates}\index{Eigenstate}. For now, we assume that the operator only has non-degenerate eigenvalues (as defined in \zcref[S]{subsec:Spectral-Decomposition}); we will deal with the case of degenerate eigenvalues in \zcref[S]{subsec:The-Measurement-Axiom-Projective}. Having non-degenerate eigenvalues simplifies things, since once we choose an eigenbasis $\left|B_{i}\right\rangle $, each eigenvalue corresponds to exactly one eigenstate in that basis.
\begin{quote}
\textbf{The Probability Axiom}\index{Axioms of quantum mechanics!Probability Axiom}\index{Probability Axiom}\textbf{:} For an observable with non-degenerate eigenvalues, the inner product $\langle B_{i}|\Psi\rangle$ is called the \emph{probability amplitude\index{Probability amplitude}} to measure the eigenvalue $\lambda_{i}$ corresponding to the eigenstate $\left|B_{i}\right\rangle $, given the state $\left|\Psi\right\rangle $. When we take the magnitude-squared of a probability amplitude, we get the corresponding probability. Thus
\[
\left|\langle B_{i}|\Psi\rangle\right|^{2}
\]
is the probability to measure the eigenvalue $\lambda_{i}$ corresponding to the eigenstate $\left|B_{i}\right\rangle $, given the state $\left|\Psi\right\rangle $. This is also known as the \emph{Born rule}\index{Born rule}.
\end{quote}
The first four axioms that we presented here simply defined the meaning of systems, states, operators, and observables in mathematical terms. The Probability Axiom, on the other hand, has to do with the relations between these mathematical structures. One can thus justifiably ask: why would this be a probability in the first place?

Unfortunately, since this is an axiom, it cannot be derived from anything more fundamental, such as other axioms. However, at the very least, we can verify that it indeed behaves exactly like a probability is expected to. This follows from the fact that
\begin{align*}
\sum^{n}_{i=1}\left|\langle B_{i}|\Psi\rangle\right|^{2} & =\sum^{n}_{i=1}\langle B_{i}|\Psi\rangle^{*}\langle B_{i}|\Psi\rangle\\
 & =\sum^{n}_{i=1}\langle\Psi|B_{i}\rangle\langle B_{i}|\Psi\rangle\\
 & =\langle\Psi|\left(\sum^{n}_{i=1}|B_{i}\rangle\langle B_{i}|\right)|\Psi\rangle\\
 & =\langle\Psi|\Psi\rangle\\
 & =\left\Vert \Psi\right\Vert ^{2}\\
 & =1,
\end{align*}
where we used the following:
\begin{itemize}
\item Taking the complex conjugate of an inner product switches the order of the vectors,
\item The completeness relation in \zcref[S]{eq:completeness},
\item All quantum states have a norm (and thus also norm-squared) of 1.
\end{itemize}
What does this mean? The number $\left|\langle B_{i}|\Psi\rangle\right|^{2}$ for each value of $i$ from $1$ to $n$ corresponds to each of the $n$ possible outcomes of a measurement. We know that it must be non-negative, because it is the magnitude-squared of a complex number. Also, when taking the sum of all such numbers, we get 1. In other words, the numbers $\left|\langle B_{i}|\Psi\rangle\right|^{2}$ behave like probabilities: they are real numbers between 0 and 1, which always sum to 1. Why they actually represent probabilities is a question that seems to have no good answer except that this is just how quantum theory works, and it is mathematically consistent and experimentally verified.

One might wonder why we are interested in the probability amplitudes $\langle B_{i}|\Psi\rangle$, which are complex numbers, instead of just the probabilities $\left|\langle B_{i}|\Psi\rangle\right|^{2}$. It turns out that the fact that probability amplitudes are complex numbers is an essential part of what makes quantum mechanics different from classical mechanics. In fact, you can even think of quantum mechanics as a generalization of classical probability theory where the “probabilities” are allowed to be complex numbers. This means that probability amplitudes can \textbf{interfere} with one another: add up to increase the probability, or cancel each other to decrease the probability. We will see examples of this later, for example in \zcref[S]{subsec:The-Deutsch-Jozsa-Algorithm}.
\begin{xca}\leavevmode
\label{exer:sigma-x}In the Hilbert space $\BBC^{2}$, consider the Hermitian operator
\[
\sigma_{x}\equiv\left(\begin{array}{cc}
0 & 1\\
1 & 0
\end{array}\right).
\]
Find its eigenstates (make sure they are normalized to 1!) and eigenvalues. Then, calculate the probability to measure each of the eigenvalues given that the system is in the state
\[
\left|\Psi\right\rangle \equiv\frac{1}{\sqrt{10}}\left(\begin{array}{c}
1\\
3
\end{array}\right).
\]
Verify that the probabilities sum to 1.
\end{xca}

\subsubsection{Superposition}\label{subsec:Superposition}

Consider an observable represented by a Hermitian operator with an orthonormal basis of eigenstates $\left|B_{i}\right\rangle $. As with any basis, we may write the state vector $\left|\Psi\right\rangle $ as a linear combination of the basis eigenstates $\left|B_{i}\right\rangle $:
\[
\left|\Psi\right\rangle =\sum^{n}_{i=1}|B_{i}\rangle\langle B_{i}|\Psi\rangle.
\]
Remember that each coefficient $\langle B_{i}|\Psi\rangle$ is the probability amplitude to measure the eigenvalue corresponding to the eigenstate $\left|B_{i}\right\rangle $ given that the system is in the state $\left|\Psi\right\rangle $. So this is a sum over the basis states $\left|B_{i}\right\rangle $, corresponding to the possible measurement outcomes, with a probability amplitude attached to each of these outcomes, which depends on the state $\left|\Psi\right\rangle $. Such a linear combination of states\footnote{More generally, a superposition is any linear combination of states. The states don't have to be basis eigenstates and the coefficients don't have to be probability amplitudes – but they usually are.} is called a \emph{superposition}\index{Superposition}.

The concept of superposition is responsible for many of the weird properties of quantum mechanics, as we will soon see. Importantly, superposition is \textbf{not }an axiom, but simply an (almost trivial) mathematical property of vectors in Hilbert spaces. This means that superposition follows automatically from the previous axioms; it is not something that needs to be introduced separately.

You will often hear people (including physicists, if they are being sloppy) say that superposition means that “the system is in multiple states at the same time”. For example, it is frequently said about particles – which can be in a superposition of eigenstates corresponding to different outcomes for the measurement of their position – that “the particle is in multiple places at the same time”. However, this is a common misconception – or at the very least, an overly literal interpretation of the math.

The fact that a state $\left|\Psi\right\rangle $ can be written in a superposition of eigenstates $\left|B_{i}\right\rangle $ doesn't mean that the system is actually “in” all of these different states at once. The system is, in fact, in only \textbf{one} state: the state $\left|\Psi\right\rangle $. This state can be represented in the eigenbasis $\left|B_{i}\right\rangle $, and doing this reveals the probability to measure each of the eigenvalues. However, one can always find\footnote{Using the Gram-Schmidt process mentioned in \zcref[S]{fn:Such-a-basis}.} an orthonormal basis where $\left|\Psi\right\rangle $ itself is one of the basis states – and often, this can be an eigenbasis corresponding to another observable of the system. In that basis, the system will \textbf{not} be in a superposition – it will just be in the state $\left|\Psi\right\rangle $, with a probability amplitude of $\langle\Psi|\Psi\rangle=1$!

So instead of saying that “the system is in all of the states $\left|B_{1}\right\rangle ,\ldots,\left|B_{n}\right\rangle $ at once”, it is more precise to say that the system is currently in the state $\left|\Psi\right\rangle $, and a measurement of the observable with the eigenbasis $\left|B_{i}\right\rangle $ could yield different outcomes, with the probability amplitude for outcome number $i$ given by the \emph{scalar projection}\footnote{In $\BBR^{n}$, given a vector $\v$ and a unit vector $\hat{\u}$, the scalar projection of $\v$ in the direction of $\hat{\u}$ is given by the dot product $\v\cdot\hat{\u}$. Scalar projections in $\BBC^{n}$ generalize this concept, with the inner product replacing the dot product.}\index{Scalar projection} of $\left|\Psi\right\rangle $ in the direction of the unit vector $\left|B_{i}\right\rangle $, calculated by taking the inner product $\langle B_{i}|\Psi\rangle$. It sounds less cool and mysterious, but it is more accurate and less prone to confusion and misinterpretation.

Of course, this description is too technical for the average person, which is why physicists usually choose to just say, incorrectly, that “the system is in multiple states at the same time”. But now that you actually know the math of quantum theory, you should be able to understand the correct definition of superposition! I will let you digest all of this for now, and in \zcref[S]{subsec:The-Meaning-of} we will discuss an analogy, using a concrete quantum system, that should help you understand this better.
\begin{xca}\leavevmode
Consider again the Hermitian operator from \zcref[S]{exer:sigma-x},
\[
\sigma_{x}=\left(\begin{array}{cc}
0 & 1\\
1 & 0
\end{array}\right).
\]
Write down an example of a state $\left|\Psi\right\rangle $ which corresponds to a probability of $1/3$ to measure the eigenvalue $+1$ and a probability of $2/3$ to measure the eigenvalue $-1$.
\end{xca}

\begin{xca}\leavevmode
\label{exer:A-quantum-system}A quantum system described by the Hilbert space $\BBC^{3}$ has an observable corresponding to a Hermitian operator $A$ with the matrix representation
\[
A\equiv\left(\begin{array}{ccc}
0 & 1 & 0\\
1 & 0 & 0\\
0 & 0 & 2
\end{array}\right).
\]
\begin{problemparts}
\item Find its eigenvalues and their corresponding eigenstates. Make sure the states are normalized to 1.
\item Find three \textbf{different }states such that a measurement of the observable $A$ will produce the lowest eigenvalue with probability $1/7$, the highest eigenvalue with probability $2/7$, and the middle eigenvalue with probability $4/7$. When we say different states, we mean that the vectors that represent them cannot be scalar multiples of each other; recall from \zcref[S]{fn:Equiv} that such vectors are in the same equivalence class, and thus represent the same state. Make sure the states are normalized to 1.
\item Write the state
\[
\left|\Psi\right\rangle \equiv\frac{1}{\sqrt{15}}\left(\begin{array}{c}
1\\
-2\\
3-\i
\end{array}\right)
\]
as a superposition of eigenstates of $A$, and calculate the probabilities to measure each eigenvalue of $A$ given that the system is in the state $\left|\Psi\right\rangle $. Verify that the probabilities sum to 1.
\end{problemparts}
\end{xca}

\subsubsection{Inner Products with Matrices, and the Expectation Value}

Consider a Hermitian operator $A$ with an orthonormal basis of $n$ eigenstates $\left|B_{i}\right\rangle $ and $n$ eigenvalues $\lambda_{i}$. To remind you, this means that
\[
A\left|B_{i}\right\rangle =\lambda_{i}\left|B_{i}\right\rangle \sp\langle B_{i}|B_{j}\rangle=\delta_{ij},
\]
where $\delta_{ij}$ is the Kronecker delta, which we defined in \zcref[S]{eq:Kron}:
\[
\delta_{ij}=\begin{cases}
0 & \textrm{if }i\ne j,\\
1 & \textrm{if }i=j.
\end{cases}
\]
Then we have, for any $i,j\in\left\{ 1,\ldots,n\right\} $:
\begin{align*}
\langle B_{i}|A|B_{j}\rangle & =\langle B_{i}|\left(A|B_{j}\rangle\right)\\
 & =\langle B_{i}|\lambda_{j}|B_{j}\rangle\\
 & =\lambda_{j}\langle B_{i}|B_{j}\rangle\\
 & =\lambda_{j}\delta_{ij}.
\end{align*}
Let us also recall the completeness relation in \zcref[S]{eq:completeness}:
\[
\sum^{n}_{i=1}|B_{i}\rangle\langle B_{i}|=1.
\]
Now, let $\left|\Psi\right\rangle $ be the state of the system. Then:
\begin{align*}
\langle\Psi|A|\Psi\rangle & =\langle\Psi|\left(\sum^{n}_{i=1}|B_{i}\rangle\langle B_{i}|\right)A\left(\sum^{n}_{j=1}|B_{j}\rangle\langle B_{j}|\right)|\Psi\rangle\\
 & =\sum^{n}_{i=1}\sum^{n}_{j=1}\langle\Psi|B_{i}\rangle\langle B_{i}|A|B_{j}\rangle\langle B_{j}|\Psi\rangle\\
 & =\sum^{n}_{i=1}\sum^{n}_{j=1}\lambda_{j}\delta_{ij}\langle\Psi|B_{i}\rangle\langle B_{j}|\Psi\rangle.
\end{align*}
When taking the sum over $j$, the Kronecker delta $\delta_{ij}$ is always 0 except when $j=i$. Therefore the sum over $j$ always reduces to just one element, the one where $j=i$. We get:
\begin{align*}
\langle\Psi|A|\Psi\rangle & =\sum^{n}_{i=1}\lambda_{i}\langle\Psi|B_{i}\rangle\langle B_{i}|\Psi\rangle\\
 & =\sum^{n}_{i=1}\lambda_{i}\langle\Psi|B_{i}\rangle\langle\Psi|B_{i}\rangle^{*}\\
 & =\sum^{n}_{i=1}\lambda_{i}\left|\langle\Psi|B_{i}\rangle\right|^{2},
\end{align*}
where in the second line we used the fact that switching the order of the vectors in the inner product is equivalent to taking the complex conjugate.

Recall that $\left|\langle\Psi|B_{i}\rangle\right|^{2}$ is the probability to measure the eigenvalue $\lambda_{i}$ associated with the eigenstate $\left|B_{i}\right\rangle $ given the state $\left|\Psi\right\rangle $. Therefore, this is a sum of the possible values of the measurement of $A$, weighted by their probabilities. But this is exactly the expected value\index{Expected (or expectation) value!Of a quantum observable} for the measurement of $A$, as we defined in \zcref[S]{eq:expected}. For this reason, we sometimes simply write $\left\langle A\right\rangle $ (the usual notation for the expected value) instead of $\langle\Psi|A|\Psi\rangle$, as long as it is clear that the expected value is taken with respect to the state $\left|\Psi\right\rangle $. If we want to specify the state explicitly, we can also use the notation
\[
\left\langle A\right\rangle _{\Psi}\equiv\langle\Psi|A|\Psi\rangle.
\]
Note that the terms “expected value” and “expectation value” are often used interchangeably, but the former seems to be more popular in classical probability theory while the latter is more popular in quantum theory.
\begin{xca}\leavevmode
Calculate $\left\langle A\right\rangle _{\Psi}$ where
\[
A\equiv\left(\begin{array}{cc}
1 & 0\\
0 & -1
\end{array}\right),
\]
for the following three states:
\[
\left|\Psi_{1}\right\rangle =\frac{1}{\sqrt{2}}\left(\begin{array}{c}
1\\
1
\end{array}\right),
\]
\[
\left|\Psi_{2}\right\rangle =\frac{1}{\sqrt{5}}\left(\begin{array}{c}
1\\
2
\end{array}\right),
\]
\[
\left|\Psi_{3}\right\rangle =\frac{1}{\sqrt{13}}\left(\begin{array}{c}
3\\
2
\end{array}\right).
\]
\end{xca}

\begin{xca}\leavevmode
Calculate $\left\langle A\right\rangle _{\Psi}$ for $A$ and $\left|\Psi\right\rangle $ as defined in \zcref[S]{exer:A-quantum-system}:
\[
A=\left(\begin{array}{ccc}
0 & 1 & 0\\
1 & 0 & 0\\
0 & 0 & 2
\end{array}\right)\sp\left|\Psi\right\rangle =\frac{1}{\sqrt{15}}\left(\begin{array}{c}
1\\
-2\\
3-\i
\end{array}\right).
\]
Then, calculate the expected value explicitly as defined in \zcref[S]{eq:expected}, using the probabilities you calculated in part (C) of \zcref[S]{exer:A-quantum-system}, and verify that you get the same result.
\end{xca}

\subsubsection{Summary for Discrete Systems}

To summarize, here are the axioms of quantum theory we formulated so far. Here we formulate them specifically for discrete systems with finite-dimensional Hilbert spaces:
\begin{enumerate}
\item \textbf{The System Axiom: }Physical systems are represented by complex $n$-dimensional Hilbert spaces $\BBC^{n}$, where $n$ depends on the specific system.
\item \textbf{The State Axiom: }The states of the system are represented by unit $n$-vectors in the system's Hilbert space, up to a phase factor.
\item \textbf{The Operator Axiom: }The operators on the system, which act on states, are represented by $n\xx n$ matrices in the system's Hilbert space.
\item \textbf{The Observable Axiom: }Physical observables in the system are represented by Hermitian operators on the system's Hilbert space. The eigenvalues of the observable (which are always real, since it's Hermitian) represent its possible measured values. The eigenstates of the observable can be used to form an orthonormal eigenbasis of the Hilbert space.
\item \textbf{The Probability Axiom: }For an observable with non-degenerate eigenvalues, the probability amplitude to measure the eigenvalue corresponding to the eigenstate $\left|B_{i}\right\rangle $, given that the system is in the state $\left|\Psi\right\rangle $, is the inner product $\langle B_{i}|\Psi\rangle$. The probability is given by the magnitude-squared of the amplitude, $\left|\langle B_{i}|\Psi\rangle\right|^{2}$.
\end{enumerate}
We also discussed two important consequences of these axioms:
\begin{itemize}
\item \textbf{Superposition:} Any state $\left|\Psi\right\rangle $ can be written as a linear combination of the eigenstates $\left|B_{i}\right\rangle $ of an observable, with the probability amplitudes $\langle B_{i}|\Psi\rangle$ as coefficients:
\[
\left|\Psi\right\rangle =\sum^{n}_{i=1}|B_{i}\rangle\langle B_{i}|\Psi\rangle.
\]
\item \textbf{Expectation Value:} If the system is in the state $\left|\Psi\right\rangle $, the expectation value for the measurement of the observable $A$ is given by $\left\langle A\right\rangle _{\Psi}=\langle\Psi|A|\Psi\rangle$.
\end{itemize}
There are some more axioms that we will add later, but first let us discuss a concrete example of a physical quantum system and see these axioms in action.
\begin{problem}\leavevmode
Are these axioms enough to actually do physics? If not, what do you think is missing and why?
\end{problem}

\subsection{Two-State Systems, Spin $1/2$, and Qubits}

So far in this chapter, we discussed quantum theory in an abstract way. However, a theory of physics is useless without a concrete mapping between the theory and reality. The simplest non-trivial\footnote{1-dimensional Hilbert spaces are of course simpler, but they are trivial, since there is only one state the system can be in, with probability 1.} quantum system is described by a 2-dimensional Hilbert space, and is thus called a \emph{two-state system}\index{Two-state system}. All such systems can also be used as \emph{qubits}\index{Qubit}, or quantum bits – where one state (doesn't matter which one) corresponds to 0 and the other state corresponds to 1. Let us now describe such systems in detail.

\subsubsection{The Pauli Matrices}\label{subsec:The-Pauli-Matrices}

Let us introduce the \emph{Pauli matrices}\index{Pauli matrices}:
\[
\sigma_{x}=\left(\begin{array}{cc}
0 & 1\\
1 & 0
\end{array}\right)\sp\sigma_{y}\equiv\left(\begin{array}{cc}
0 & -\i\\
\i & 0
\end{array}\right)\sp\sigma_{z}\equiv\left(\begin{array}{cc}
1 & 0\\
0 & -1
\end{array}\right).
\]
As the notation suggests, each matrix is associated with a spatial axis: $x$, $y$, and $z$. These three matrices have the following properties (here $i$ stands for $x$, $y$, or $z$):
\begin{itemize}
\item They are Hermitian: $\sigma^{\dagger}_{i}=\sigma_{i}$. This means they can represent observables.
\item They are unitary: $\sigma^{\dagger}_{i}=\sigma^{-1}_{i}$. This means they can represent transformations.
\begin{itemize}
\item Since they are both Hermitian and unitary, they are their own inverse: $\sigma_{i}=\sigma^{\dagger}_{i}=\sigma^{-1}_{i}$. This means that $\sigma^{2}_{i}=1$. A matrix which is its own inverse is called \emph{involutory}\index{Involutory matrix}.
\end{itemize}
\item They are traceless: $\tr\sigma_{i}=0$. (We defined the trace operation in \zcref[S]{subsec:Trace}.)
\item They have two eigenvalues: $+1$ and $-1$.
\begin{itemize}
\item The eigenstates of $\sigma_{x}$ are:
\[
\left|+x\right\rangle \equiv\left|+\right\rangle \equiv\frac{1}{\sqrt{2}}\left(\begin{array}{c}
1\\
1
\end{array}\right)\sp\left|-x\right\rangle \equiv\left|-\right\rangle \equiv\frac{1}{\sqrt{2}}\left(\begin{array}{c}
1\\
-1
\end{array}\right).
\]
\item The eigenstates of $\sigma_{y}$ are:
\[
\left|+y\right\rangle \equiv\left|+\i\right\rangle \equiv\frac{1}{\sqrt{2}}\left(\begin{array}{c}
1\\
\i
\end{array}\right)\sp\left|-y\right\rangle \equiv\left|-\i\right\rangle \equiv\frac{1}{\sqrt{2}}\left(\begin{array}{c}
1\\
-\i
\end{array}\right).
\]
\item The eigenstates of $\sigma_{z}$ are:
\[
\left|+z\right\rangle \equiv\left|0\right\rangle \equiv\left(\begin{array}{c}
1\\
0
\end{array}\right)\sp\left|-z\right\rangle \equiv\left|1\right\rangle \equiv\left(\begin{array}{c}
0\\
1
\end{array}\right),
\]
where, confusingly, $\left|0\right\rangle $ corresponds to the eigenvalue $+1$ and $\left|1\right\rangle $ corresponds to the eigenvalue $-1$ (but that is the standard convention).
\end{itemize}
\item Since the Pauli matrices are normal, the eigenstates of each matrix form an orthonormal eigenbasis of $\BBC^{2}$. As you can see, the eigenstates of $\sigma_{z}$ are just the standard basis.
\item The eigenstates of $\sigma_{x}$ and $\sigma_{z}$ are related to each other as follows:
\[
\left|+\right\rangle =\frac{1}{\sqrt{2}}\left(\left|0\right\rangle +\left|1\right\rangle \right)\sp\left|-\right\rangle =\frac{1}{\sqrt{2}}\left(\left|0\right\rangle -\left|1\right\rangle \right),
\]
\[
\left|0\right\rangle =\frac{1}{\sqrt{2}}\left(\left|+\right\rangle +\left|-\right\rangle \right)\sp\left|1\right\rangle =\frac{1}{\sqrt{2}}\left(\left|+\right\rangle -\left|-\right\rangle \right).
\]
\item A useful mnemonic is as follows: for each $n$, the state $\left|n\right\rangle $ has the number $n$ in the bottom component (excluding the factor of $1/\sqrt{2}$ if it exists). So the states $\left|\pm\right\rangle $ have $\pm1$ in the bottom component, the states $\left|\pm\i\right\rangle $ have $\pm\i$ in the bottom component, and the states $\left|0\right\rangle $ and $\left|1\right\rangle $ have $0$ and $1$ in the bottom component respectively.
\end{itemize}
\begin{problem}\leavevmode
Prove that $\sigma_{x}$, $\sigma_{y}$, and $\sigma_{z}$ are Hermitian.
\end{problem}

\begin{problem}\leavevmode
Prove that $\sigma_{x}$, $\sigma_{y}$, and $\sigma_{z}$ are unitary.
\end{problem}

\begin{problem}\leavevmode
\label{prob:Consider-the-real}Consider the real vector space of $2\xx2$ Hermitian matrices. This is a vector space where \textbf{the vectors are Hermitian matrices} and the scalars are real numbers. Don't get confused: in an \textbf{abstract }vector space, anything can be a “vector” – including numbers, matrices, tensors of higher rank, functions, and even weirder stuff.
\begin{problemparts}
\item Show that the real vector space of $2\xx2$ Hermitian matrices satisfies all of the conditions in our definition of a vector space in \zcref[S]{subsec:Complex-Vector-Spaces}.
\item Show that the set $\left\{ 1,\sigma_{x},\sigma_{y},\sigma_{z}\right\} $, composed of the identity matrix $1$ and the three Pauli matrices, is a basis of the real vector space of $2\xx2$ Hermitian matrices.
\item Show that this basis is orthogonal with respect to the Frobenius inner product defined in \zcref[S]{prob:Frobenius}, and normalize its elements to obtain an orthonormal basis.
\end{problemparts}
\end{problem}

\subsubsection{Spin $1/2$}

In quantum theory, every particle has a property called \emph{spin}\index{Spin} $s$, which is a multiple of $1/2$:
\[
s\in\left\{ 0,\hf,1,\frac{3}{2},2,\ldots\right\} .
\]
Spin can be \emph{integer}\index{Integer spin} ($0$, $1$, $2$, and so on) or \emph{half-integer}\index{Half-integer spin} ($1/2$, $3/2$, $5/2$, and so on). Particles with integer spin are called \emph{bosons}\index{Boson}, and particles with half-integer spin are called \emph{fermions}\index{Fermion}.

Spin is \textbf{intrinsic} angular momentum. In the rest frame of a massive particle of spin $s$ in three spatial dimensions, the magnitude of spin is $\sqrt{s\left(s+1\right)}$. However, a measurement of a single spin component along any chosen axis returns one of the results in the set
\[
\left\{ -s,-s+1,\ldots,s-1,s\right\} .
\]
Note that this set always contains $2s+1$ values. For example:
\begin{itemize}
\item The measured spin components of a massive particle of spin $0$ will always be $0$;
\item The measured spin components of a massive particle of spin $1/2$ can be $-1/2$ or $+1/2$;
\item The measured spin components of a massive particle of spin $1$ can be $-1$, $0$, or $+1$;
\item The measured spin components of a massive particle of spin $3/2$ can be $-3/2$, $-1/2$, $+1/2$, or $+3/2$;
\item and so on.
\end{itemize}
For massless particles, spin behaves a bit differently. For example, photons are massless spin-1 particles. Their spin component along the direction of motion is called \emph{helicity}\index{Helicity}; it only has two possible measured values, $+1$ and $-1$, corresponding to right-handed and left-handed circular \emph{polarization}\index{Polarization} respectively. Unlike massive spin-1 particles, a photon has no helicity-zero state. This is because electromagnetic waves are \emph{transverse}\index{Transverse wave}: their electric field is perpendicular to their direction of propagation, which leaves only two independent polarizations.

For a massive particle with spin $1/2$, a measurement of intrinsic angular momentum along a chosen axis has two possible outcomes: $+1/2$, called “spin up”, and $-1/2$, called “spin down”. We can thus describe the particle's spin using a two-state quantum system. The spin-up and spin-down states will be eigenstates of the corresponding spin operator, with eigenvalues $+1/2$ and $-1/2$ respectively. We construct the spin operators from the Pauli matrices $\sigma_{i}$, where $i\in\left\{ x,y,z\right\} $. The Pauli matrices have eigenvalues $\pm1$, so they represent \textbf{twice }the spin component along the corresponding axis. We therefore define
\begin{equation}
S_{x}\equiv\hf\sigma_{x}\sp S_{y}\equiv\hf\sigma_{y}\sp S_{z}\equiv\hf\sigma_{z},\label{eq:Ssigma}
\end{equation}
such that $S_{i}$ is a Hermitian operator corresponding to spin $\pm1/2$ along the $i$ direction. You can check that each $S_{i}$ has the same eigenstates as the corresponding $\sigma_{i}$, but they correspond to the eigenvalues $\pm1/2$ instead of $\pm1$.

In \zcref[S]{prob:Consider-the-real} you proved that the set $\left\{ 1,\sigma_{x},\sigma_{y},\sigma_{z}\right\} $ forms a basis for the real vector space of $2\xx2$ Hermitian matrices. This means that any Hermitian operator on the Hilbert space $\BBC^{2}$ can be written as a linear combination of these 4 matrices. Since Hermitian operators correspond to observables, this means that \textbf{every possible observable} in $\BBC^{2}$ can be written in terms of the Pauli matrices and the identity matrix. Note that if we omit the identity matrix, and use a linear combination of just the three Pauli matrices, then we can only generate traceless Hermitian operators, since the Pauli matrices are all traceless.

In particular, given a unit vector $\v\in\BBR^{3}$ pointing in an arbitrary direction in space (the real space, not the Hilbert space!)
\[
\v\equiv\left(x,y,z\right)\sp\sqrt{x^{2}+y^{2}+z^{2}}=1,
\]
we can represent the measurement of intrinsic angular momentum along that direction as the Hermitian operator (on the Hilbert space, $\BBC^{2}$)
\begin{equation}
S_{\v}\equiv xS_{x}+yS_{y}+zS_{z}=\hf\left(\begin{array}{cc}
z & x-\i y\\
x+\i y & -z
\end{array}\right).\label{eq:spin-xyz}
\end{equation}
For $\v=\left(0,0,1\right)$, that is, spin along the $z$ direction, we get
\[
S_{\v}=S_{z}=\hf\sigma_{z}=\hf\left(\begin{array}{cc}
1 & 0\\
0 & -1
\end{array}\right),
\]
and the spin-up and spin-down eigenstates are the $+1/2$ and $-1/2$ eigenstates respectively:
\[
\left|\up\right\rangle =\left(\begin{array}{c}
1\\
0
\end{array}\right)\sp\left|\dn\right\rangle =\left(\begin{array}{c}
0\\
1
\end{array}\right).
\]

For $\v=\left(0,0,-1\right)$, that is, spin along the negative $z$ direction, we similarly get
\[
S_{\v}=-S_{z}=-\hf\sigma_{z}=\hf\left(\begin{array}{cc}
-1 & 0\\
0 & 1
\end{array}\right),
\]
and the spin-up and spin-down eigenstates are just the previous eigenstates flipped along the $z$ axis:
\[
\left|\up\right\rangle =\left(\begin{array}{c}
0\\
1
\end{array}\right)\sp\left|\dn\right\rangle =\left(\begin{array}{c}
1\\
0
\end{array}\right).
\]
In all other cases, we have $z\notin\left\{ -1,+1\right\} $, so the spin operator is given by \zcref[S]{eq:spin-xyz}, and the spin-up and spin-down eigenstates are
\begin{equation}
\left|\up\right\rangle =\frac{1}{\sqrt{2\left(1+z\right)}}\left(\begin{array}{c}
1+z\\
x+\i y
\end{array}\right)\sp\left|\dn\right\rangle =\frac{1}{\sqrt{2\left(1-z\right)}}\left(\begin{array}{c}
1-z\\
-x-\i y
\end{array}\right).\label{eq:spin-xyz-eigenstates}
\end{equation}
So we learn that, for a spin-$1/2$ particle, the measurement of intrinsic angular momentum along \textbf{any} direction $\v$ in space always yields one of exactly two possible results: spin up, eigenvalue $+1/2$, or spin down, eigenvalue $-1/2$, with the eigenstates determined by the Hermitian operator $S_{\v}$.
\begin{xca}\leavevmode
Show that the eigenstates $\left|\up\right\rangle $ and $\left|\dn\right\rangle $ in \zcref[S]{eq:spin-xyz-eigenstates} are indeed the eigenstates of the operator in \zcref[S]{eq:spin-xyz}.
\end{xca}

\begin{xca}\leavevmode
A spin-$1/2$ particle is in the state
\[
\left|\Psi\right\rangle \equiv\frac{1}{\sqrt{10}}\left(\begin{array}{c}
1\\
3
\end{array}\right).
\]
\begin{problemparts}
\item What are the probabilities to measure spin up or down in the $x$ direction?
\item What are the probabilities to measure spin up or down in the $y$ direction?
\item What are the probabilities to measure spin up or down in the $z$ direction?
\item What are the probabilities to measure spin up or down in the direction of the unit vector $\left(\frac{1}{3},\frac{2}{3},\frac{2}{3}\right)$?
\item What are the expectation values for a measurement of spin in each of the directions specified in (A)–(D)? (Make sure you are using $S_{i}$ and not $\sigma_{i}$ for this calculation!)
\end{problemparts}
\end{xca}

\begin{problem}\leavevmode
\begin{problemparts}
\item In \zcref[S]{subsec:The-Matrix-Commutator} we defined the \emph{matrix commutator}\index{Matrix commutator}\index{Commutator}:
\[
\left[A,B\right]\equiv AB-BA.
\]
Show that the spin-$1/2$ operators $S_{i}$ have the \emph{commutation relations}\index{Commutation relation!Of spin matrices}:
\begin{equation}
\left[S_{x},S_{y}\right]=\i S_{z}\sp\left[S_{y},S_{z}\right]=\i S_{x}\sp\left[S_{z},S_{x}\right]=\i S_{y}.\label{eq:commute-Pauli}
\end{equation}
\item Show that the commutation relations in \zcref[S]{eq:commute-Pauli} can be written compactly as
\[
\left[S_{i},S_{j}\right]=\i\sum^{3}_{k=1}\dui{\epsilon}{ij}kS_{k},
\]
where the indices $i,j,k$ take the values $\left\{ 1,2,3\right\} $ corresponding to $\left\{ x,y,z\right\} $, and $\dui{\epsilon}{ij}k$ is the \emph{Levi-Civita symbol}\index{Levi-Civita symbol}, defined as
\[
\dui{\epsilon}{ij}k\equiv\begin{cases}
+1 & \textrm{if }\left(i,j,k\right)\textrm{ is an even permutation of }\left(1,2,3\right),\\
-1 & \textrm{if }\left(i,j,k\right)\textrm{ is an odd permutation of }\left(1,2,3\right),\\
0 & \textrm{otherwise}.
\end{cases}
\]
By \emph{even permutation}\index{Even permutation} or \emph{odd permutation}\index{Odd permutation} we mean that the permutation\index{Permutation} involves exchanging elements an even or odd number of times. For example, $\left(1,3,2\right)$ is an odd permutation, because we exchanged elements once: $2\tot3$. However, $\left(3,1,2\right)$ is an even permutation, because we exchanged elements twice: $2\tot3$ and then $1\tot3$.
\item In \zcref[S]{subsec:The-Matrix-Commutator} we also defined the \emph{matrix anti-commutator}\index{Matrix anti-commutator}\index{Anti-commutator}:
\[
\left\{ A,B\right\} \equiv AB+BA.
\]
Show that the spin-$1/2$ operators $S_{i}$ have the \emph{anti-commutation relation}\index{Anti-commutation relation}
\[
\left\{ S_{i},S_{j}\right\} =\hf\delta_{ij},
\]
where $\delta_{ij}$ is the Kronecker delta (times the identity matrix $1$).
\end{problemparts}
\end{problem}

\subsubsection{Qubits}

A \emph{classical bit}\index{Classical bit} can be in one of two states: 0 or 1. A \emph{quantum bit}, or \emph{qubit}\index{Qubit} for short, is instead in a \textbf{superposition }of two states, denoted $\left|0\right\rangle $ and $\left|1\right\rangle $:
\[
\left|\Psi\right\rangle =a\left|0\right\rangle +b\left|1\right\rangle \sp\left|a\right|^{2}+\left|b\right|^{2}=1,
\]
where $a,b\in\BBC$ are the probability amplitudes:
\[
a\equiv\langle0|\Psi\rangle\sp b\equiv\langle1|\Psi\rangle.
\]
Since the system has two basis states, it can be represented by the Hilbert space $\BBC^{2}$, and it is conventional to choose $\left|0\right\rangle $ and $\left|1\right\rangle $ to be the vectors in the standard basis, which in this case is called the \emph{computational basis}\index{Computational basis}:
\[
\left|0\right\rangle =\left(\begin{array}{c}
1\\
0
\end{array}\right)\sp\left|1\right\rangle =\left(\begin{array}{c}
0\\
1
\end{array}\right).
\]
\textbf{Any} two-state quantum system can serve as a qubit. In fact, even systems with more than two states can be used, as long as two of these states can be decoupled (separated) from the rest. Some examples include:
\begin{itemize}
\item Any spin-$1/2$ particle, such as an electron, where $\left|0\right\rangle $ and $\left|1\right\rangle $ are the eigenstates of the spin operator along the $z$ direction, $S_{z}$, so they represent spin up and spin down, respectively, along that direction.
\item The number of particles (for example, photons) in a system, where $\left|0\right\rangle $ corresponds to a state with no particles (a \emph{vacuum}\index{Vacuum state}) and $\left|1\right\rangle $ corresponds to a state with exactly one particle.
\item The \emph{polarization}\index{Polarization} of a photon, where $\left|0\right\rangle $ and $\left|1\right\rangle $ correspond to right-handed and left-handed circular polarization respectively.
\end{itemize}
Qubits are used in quantum computers as the basic units of computation, just like bits in classical computers. Since so many different systems can be represented mathematically in the same way, we can build quantum computers in many different ways. We will discuss quantum computers (from the theoretical point of view) in more detail later.

\subsubsection{The Meaning of Superposition}\label{subsec:The-Meaning-of}

In \zcref[S]{subsec:Superposition} we discussed the concept of superposition, and we emphasized that it is inaccurate to describe a system in a superposition of two states as being “in both states at once”. Similarly, it is a common misconception that quantum computers\index{Quantum computer} are powerful because qubits, which are in a superposition of $\left|0\right\rangle $ and $\left|1\right\rangle $, are in some way “both 0 and 1 at the same time”, and this allows the quantum computer to “calculate all the possibilities at once”. That would have been awesome, but unfortunately that is \textbf{not }how quantum computers work! We will see how they really work later in these notes.

\begin{figure}[tbph]
\centering
\begin{centering}
\includegraphics[width=0.9\textwidth]{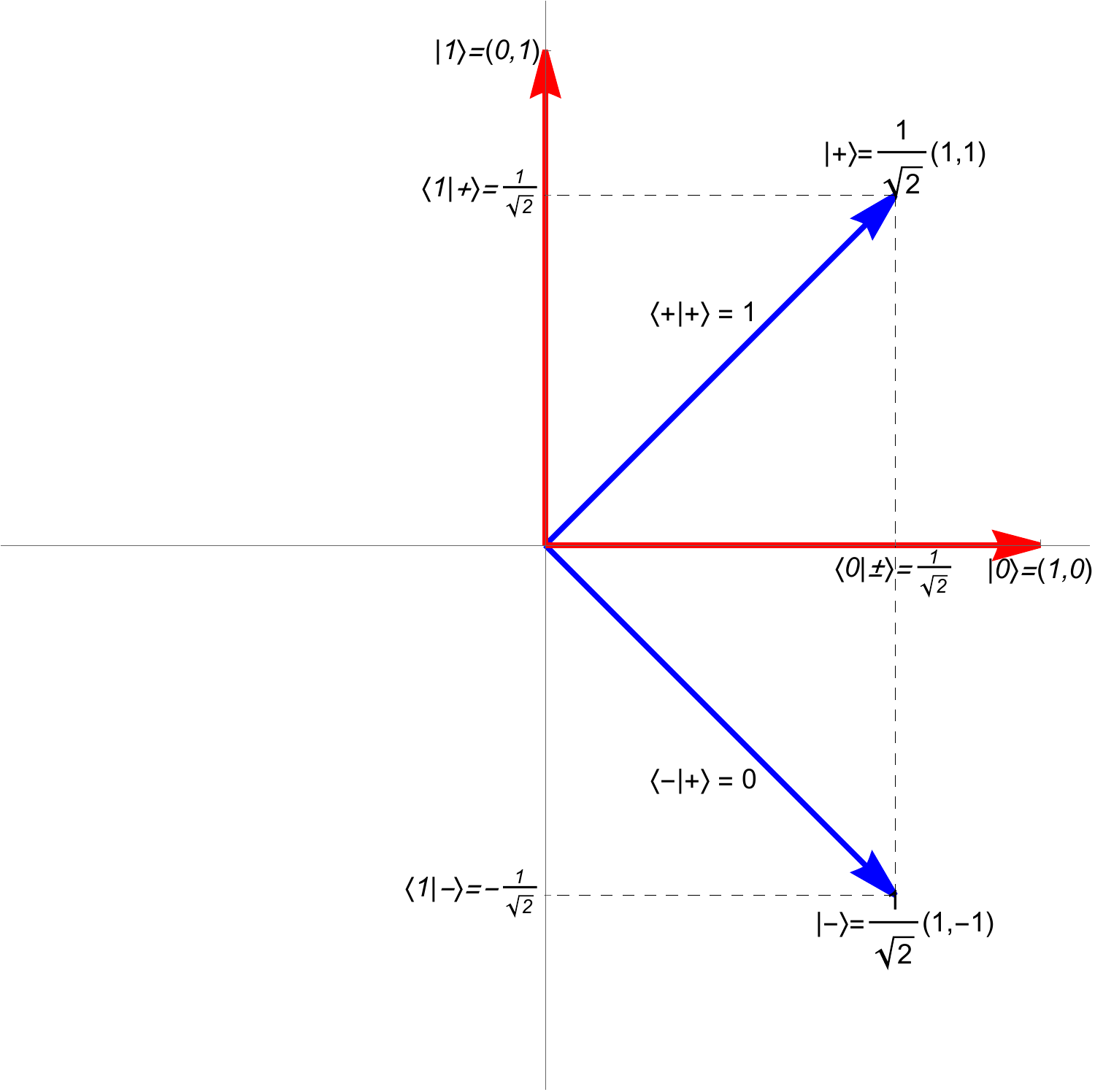}
\par\end{centering}
\caption[A qubit in a superposition of $\left|0\right\rangle $ and $\left|1\right\rangle $]{The eigenbasis $\left\{ \left|0\right\rangle ,\left|1\right\rangle \right\} $, in red, and the eigenbasis $\left\{ \left|+\right\rangle ,\left|-\right\rangle \right\} $, in blue. A qubit in the state $\left|+\right\rangle $ is in a superposition of $\left|0\right\rangle $ and $\left|1\right\rangle $, but this does not mean it is in the states $\left|0\right\rangle $ and $\left|1\right\rangle $ “at the same time” – it is only in one state, $\left|+\right\rangle $.}\label{fig:A-qubit-in}
\end{figure}

\index{Superposition!Meaning of}Now that we are familiar with a concrete quantum system, we can use it to illustrate further the meaning of superposition. Let us consider, as a simple example, a qubit in the state $\left|+\right\rangle $ (the eigenstate of spin $+1/2$ in the $x$ direction):
\[
\left|+\right\rangle =\frac{1}{\sqrt{2}}\left(\left|0\right\rangle +\left|1\right\rangle \right)=\frac{1}{\sqrt{2}}\left(\left(\begin{array}{c}
1\\
0
\end{array}\right)+\left(\begin{array}{c}
0\\
1
\end{array}\right)\right)=\frac{1}{\sqrt{2}}\left(\begin{array}{c}
1\\
1
\end{array}\right).
\]
For simplicity, let us forget for a second that we are dealing with complex vectors, and imagine that they are vectors in $\BBR^{2}$, since that is much easier to visualize; see \zcref[S]{fig:A-qubit-in}. As vectors in $\BBR^{2}$, the state $\left|0\right\rangle =\left(1,0\right)$ points east and the state $\left|1\right\rangle =\left(0,1\right)$ points north. This does \textbf{not }mean that $\left|+\right\rangle $ is “pointing both north and east at the same time”. It does not, in fact, point in either of these directions; instead, it points in a third direction, namely north-east.

In other words, if we just look at the vector represented by $\left|+\right\rangle $, without considering any particular basis, it is just a vector pointing in \textbf{one }particular direction, and in this direction only. The superposition only exists if we insist on representing $\left|+\right\rangle $ in this particular eigenbasis, but there can be another eigenbasis, e.g. the eigenbasis composed of $\left|+\right\rangle $ itself along with $\left|-\right\rangle $, in which $\left|+\right\rangle $ is \textbf{not }in a superposition.

What this means is that a state only appears to be in a superposition when we choose a particular observable and represent that state as a superposition of eigenstates with respect to that observable. But the system itself is still in the same state, regardless of which eigenbasis we choose. The scalar projections of the state of the system in the directions of the basis eigenstates give us the probability amplitudes relevant to that measurement; for example, in \zcref[S]{fig:A-qubit-in} we see that the probability amplitudes to measure $\left|0\right\rangle $ or $\left|1\right\rangle $ are both $1/\sqrt{2}$. However, in the basis consisting of $\left|+\right\rangle $ and $\left|-\right\rangle $, we instead have that the probability amplitude to measure $\left|+\right\rangle $ is 1 and the probability amplitude to measure $\left|-\right\rangle $ is 0.

In the specific case where the qubit is the spin of a spin-$1/2$ particle, we know that if the qubit is in the state $\left|+\right\rangle $, this means that a measurement of spin along the $x$ axis will yield spin up with probability 1. We can say, if we want, that the system is in a state of spin up along the $x$ axis, and this defines the state uniquely. We also see that, in this basis, the system is not in a superposition; it is just one state.

However, in the basis corresponding to measurement of spin along the $z$ axis, we may write the state as a superposition, $\left|+\right\rangle =\left(\left|0\right\rangle +\left|1\right\rangle \right)/\sqrt{2}$. This doesn't mean that the qubit is in both the states $\left|0\right\rangle $ and $\left|1\right\rangle $ “at the same time”; it means that it is in a state where a measurement of spin along the $z$ axis will yield spin up or spin down with equal probability.

If being in the superposition $\left(\left|0\right\rangle +\left|1\right\rangle \right)/\sqrt{2}$ doesn't mean that the qubit is \textbf{both} $\left|0\right\rangle $ \textbf{and }$\left|1\right\rangle $ at the same time, perhaps it could mean that the qubit is \textbf{either} $\left|0\right\rangle $ \textbf{or }$\left|1\right\rangle $, but we just don't know which one it is, and when we perform a measurement we will discover which state it was in all along? Unfortunately, that interpretation doesn't work either. To see why, let's assume that someone secretly tossed a fair coin and then prepared the qubit in the state $\left|0\right\rangle $ for heads or $\left|1\right\rangle $ for tails. If we measure spin along the $z$ axis, then we will indeed find spin up or down with probability 1/2 each, just as if the qubit was in an equal superposition of $\left|0\right\rangle $ and $\left|1\right\rangle $. However, if we measure spin along the $x$ axis, then because
\[
\left|0\right\rangle =\frac{1}{\sqrt{2}}\left(\left|+\right\rangle +\left|-\right\rangle \right)\sp\left|1\right\rangle =\frac{1}{\sqrt{2}}\left(\left|+\right\rangle -\left|-\right\rangle \right),
\]
as explained in \zcref[S]{subsec:The-Pauli-Matrices}, we will get either spin up or down with equal probability (the relative sign does not affect probabilities). This is not the same as what we would get from the qubit being in the superposition $\left|+\right\rangle =\left(\left|0\right\rangle +\left|1\right\rangle \right)/\sqrt{2}$, which is spin up along the $x$ axis with probability 1. Thus, having a qubit in a superposition of $\left|0\right\rangle $ and $\left|1\right\rangle $ is not the same as having a qubit either in the state $\left|0\right\rangle $ or $\left|1\right\rangle $ and simply not knowing which state it has until we measure it\footnote{In \zcref[S]{subsec:Density-Operators-and} we will learn about density operators, which will allow us to define the latter notion more precisely.}.

So in conclusion, being in a superposition of two states doesn't mean being in \textbf{both} the first state \textbf{and }the second state, but also doesn't mean being in \textbf{either} the first state \textbf{or }the second state. It's a secret third option, something “in between”, like the vector pointing north-east. Compare this with the familiar (and also confusing) notion of \emph{wave-particle duality}\index{Wave-particle duality}. This duality doesn't mean that light is “\textbf{both }a wave \textbf{and }a particle”, and it also doesn't mean that light is “\textbf{either }a wave \textbf{or }a particle”. What it really means is that the classical concepts of “wave” and “particle” are not the proper way to describe reality at the quantum level.

\subsection{Composite Systems and Quantum Entanglement}

\subsubsection{The Tensor Product}\label{subsec:The-Tensor-Product}

So far, we have only considered single, isolated physical systems, described by a single Hilbert space. What if we have more than one system, such as a collection of particles? This calls for a new axiom:
\begin{quote}
\textbf{The Composite System Axiom}\index{Axioms of quantum mechanics!Composite System Axiom}\index{Composite System Axiom}\textbf{:} Given two physical quantum systems represented by two Hilbert spaces $\HH_{A}$ and $\HH_{B}$ respectively, the \emph{tensor product\index{Tensor product}} of the two spaces, denoted
\[
\HH_{A}\otimes\HH_{B},
\]
is another Hilbert space, representing the \emph{composite system}\index{Composite system} which combines the two original systems. The dimension of the composite Hilbert space is the product of the dimensions of the individual spaces:
\[
\dim\left(\HH_{A}\otimes\HH_{B}\right)=\dim\HH_{A}\cdot\dim\HH_{B}.
\]
For example, the dimension of $\BBC^{m}\otimes\BBC^{n}$ is $mn$.
\end{quote}
Given a state $\left|\Psi_{A}\right\rangle $ in $\HH_{A}$ and a state $\left|\Psi_{B}\right\rangle $ in $\HH_{B}$, we can use the tensor product to form a new state in $\HH_{A}\otimes\HH_{B}$:
\[
\left|\Psi_{A}\right\rangle \otimes\left|\Psi_{B}\right\rangle \in\HH_{A}\otimes\HH_{B}.
\]
However, \textbf{not all states} in $\HH_{A}\otimes\HH_{B}$ are necessarily of this form; this fact will prove essential soon, when we discuss entanglement. Furthermore, if $\left|A_{i}\right\rangle $, $i\in\left\{ 1,\ldots,m\right\} $ is an orthonormal basis of $\HH_{A}$ and $\left|B_{j}\right\rangle $, $j\in\left\{ 1,\ldots,n\right\} $ is an orthonormal basis of $\HH_{B}$, then
\[
\left|A_{i}\right\rangle \otimes\left|B_{j}\right\rangle \sp i\in\left\{ 1,\ldots,m\right\} \sp j\in\left\{ 1,\ldots,n\right\} ,
\]
is an orthonormal basis of $\HH_{A}\otimes\HH_{B}$. Note that there are $mn$ basis states in total, since the dimension of the composite Hilbert space is $mn$.

The tensor product is \emph{bilinear}\index{Tensor product!Bilinearity}, that is, linear in each argument separately. This means that for $\lambda\in\BBC$, $\left|\Psi_{A}\right\rangle \in\HH_{A}$, and $\left|\Psi_{B}\right\rangle \in\HH_{B}$ we have
\begin{equation}
\lambda\left(\left|\Psi_{A}\right\rangle \otimes\left|\Psi_{B}\right\rangle \right)=\left(\vphantom{\blll}\lambda\left|\Psi_{A}\right\rangle \right)\otimes\left|\Psi_{B}\right\rangle =\left|\Psi_{A}\right\rangle \otimes\left(\vphantom{\blll}\lambda\left|\Psi_{B}\right\rangle \right),\label{eq:tensor-linear}
\end{equation}
for $\left|\Psi_{A}\right\rangle ,\left|\Phi_{A}\right\rangle \in\HH_{A}$ and $\left|\Theta_{B}\right\rangle \in\HH_{B}$ we have
\[
\left(\vphantom{\blll}\left|\Psi_{A}\right\rangle +\left|\Phi_{A}\right\rangle \right)\otimes\left|\Theta_{B}\right\rangle =\left|\Psi_{A}\right\rangle \otimes\left|\Theta_{B}\right\rangle +\left|\Phi_{A}\right\rangle \otimes\left|\Theta_{B}\right\rangle ,
\]
and for $\left|\Psi_{A}\right\rangle \in\HH_{A}$ and $\left|\Theta_{B}\right\rangle ,\left|\Omega_{B}\right\rangle \in\HH_{B}$ we have
\[
\left|\Psi_{A}\right\rangle \otimes\left(\vphantom{\blll}\left|\Theta_{B}\right\rangle +\left|\Omega_{B}\right\rangle \right)=\left|\Psi_{A}\right\rangle \otimes\left|\Theta_{B}\right\rangle +\left|\Psi_{A}\right\rangle \otimes\left|\Omega_{B}\right\rangle .
\]
In particular, notice from \zcref[S]{eq:tensor-linear} that scalars commute with the tensor product, so we can move them in or out of the product as we see fit – just as, until now, we have been moving scalars in and out of inner and outer products. Importantly, the tensor product itself is \textbf{not }commutative:
\[
\left|\Psi_{A}\right\rangle \otimes\left|\Psi_{B}\right\rangle \ne\left|\Psi_{B}\right\rangle \otimes\left|\Psi_{A}\right\rangle .
\]
The order matters, since the \textbf{first} state must come from the \textbf{first} Hilbert space, and the \textbf{second} state must come from the \textbf{second} Hilbert space – which may be a completely different space with completely different states. For example, in the tensor product $\BBC^{2}\otimes\BBC^{3}$ the first state must be represented by a 2-vector while the second state must be represented by a 3-vector – so they cannot be interchanged.

Now, if $O_{A}$ is an operator on $\HH_{A}$ and $O_{B}$ is an operator on $\HH_{B}$, then $O_{A}\otimes O_{B}$ is an operator on $\HH_{A}\otimes\HH_{B}$, which is defined such that each operator acts only on the state coming from the same space as that operator:
\[
\left(\vphantom{\blll}O_{A}\otimes O_{B}\right)\left(\vphantom{\blll}\left|\Psi_{A}\right\rangle \otimes\left|\Psi_{B}\right\rangle \right)=\left(\vphantom{\blll}O_{A}\left|\Psi_{A}\right\rangle \right)\otimes\left(\vphantom{\blll}O_{B}\left|\Psi_{B}\right\rangle \right).
\]
In other words, the \textbf{first }operator in the product $O_{A}\otimes O_{B}$ acts only on the \textbf{first }state in the product $\left|\Psi_{A}\right\rangle \otimes\left|\Psi_{B}\right\rangle $, and the \textbf{second }operator in the product $O_{A}\otimes O_{B}$ acts only on the \textbf{second }state in the product $\left|\Psi_{A}\right\rangle \otimes\left|\Psi_{B}\right\rangle $. This \textbf{has }to be the case, since e.g. in the tensor product $\BBC^{2}\otimes\BBC^{3}$ the first operator must be represented by a $2\xx2$ matrix and act on 2-vectors while the second operator must be represented by a $3\xx3$ matrix and act on 3-vectors. Note that, as for the tensor product of states, not all operators in $\HH_{A}\otimes\HH_{B}$ are necessarily of this form.

If we have two bras $\left\langle \Psi_{A}\right|\in\HH^{*}_{A}$ and $\left\langle \Psi_{B}\right|\in\HH^{*}_{B}$, their tensor product $\left\langle \Psi_{A}\right|\otimes\left\langle \Psi_{B}\right|$ is a bra in $\left(\HH_{A}\otimes\HH_{B}\right)^{*}$, and the inner product of this bra with a ket of the form $\left|\Phi_{A}\right\rangle \otimes\left|\Phi_{B}\right\rangle $ in $\HH_{A}\otimes\HH_{B}$ is defined by taking the inner products of each ket with the bra from its respective dual space:
\begin{equation}
\left(\vphantom{\blll}\left\langle \Psi_{A}\right|\otimes\left\langle \Psi_{B}\right|\right)\left(\vphantom{\blll}\left|\Phi_{A}\right\rangle \otimes\left|\Phi_{B}\right\rangle \right)=\langle\Psi_{A}|\Phi_{A}\rangle\langle\Psi_{B}|\Phi_{B}\rangle.\label{eq:inner-tensor}
\end{equation}
The \textbf{first }bra acts only on the \textbf{first }ket and the \textbf{second }bra acts only on the \textbf{second }ket. Once again, the inner product \textbf{must} work this way, since for example in $\BBC^{2}\otimes\BBC^{3}$ we can only take the inner product of 2-vectors with 2-vectors and 3-vectors with 3-vectors – the inner product of a 2-vector with a 3-vector is undefined.

Similarly, if we have two operators $O_{A},P_{A}:\HH_{A}\to\HH_{A}$ and two operators $O_{B},P_{B}:\HH_{B}\to\HH_{B}$, then the composite operator $O_{A}\otimes O_{B}:\HH_{A}\otimes\HH_{B}\to\HH_{A}\otimes\HH_{B}$ multiplies the composite operator $P_{A}\otimes P_{B}:\HH_{A}\otimes\HH_{B}\to\HH_{A}\otimes\HH_{B}$ in the only way that makes sense, with each operator multiplying the operator from the same space:
\[
\left(\vphantom{\blll}O_{A}\otimes O_{B}\right)\left(\vphantom{\blll}P_{A}\otimes P_{B}\right)=O_{A}P_{A}\otimes O_{B}P_{B}.
\]
Outer products also work the way you would expect. For vectors $\left|\Psi_{A}\right\rangle ,\left|\Phi_{A}\right\rangle \in\HH_{A}$ and $\left|\Psi_{B}\right\rangle ,\left|\Phi_{B}\right\rangle \in\HH_{B}$, we have
\begin{equation}
\left(\left|\Psi_{A}\right\rangle \otimes\left|\Psi_{B}\right\rangle \right)\left(\left\langle \Phi_{A}\right|\otimes\left\langle \Phi_{B}\right|\right)=\left|\Psi_{A}\right\rangle \left\langle \Phi_{A}\right|\otimes\left|\Psi_{B}\right\rangle \left\langle \Phi_{B}\right|.\label{eq:outer-tensor}
\end{equation}

Finally, above we stated the Composite System Axiom for two quantum systems, but we can use it recursively to define the composite Hilbert space of any number of systems: just take the tensor product of all the Hilbert spaces together,
\[
\HH_{A}\otimes\HH_{B}\otimes\HH_{C}\otimes\cdots.
\]
Everything we defined above still applies, with the obvious generalizations.
\begin{problem}\leavevmode
Let $\HH_{A}$ and $\HH_{B}$ be two Hilbert spaces. Find an isomorphism between the composite Hilbert spaces $\HH_{A}\otimes\HH_{B}$ and $\HH_{B}\otimes\HH_{A}$.
\end{problem}

\begin{problem}\leavevmode
Let $\HH_{A}$ and $\HH_{B}$ be two Hilbert spaces, let $A$ be an operator on $\HH_{A}$, and let $B$ be an operator on $\HH_{B}$.
\begin{problemparts}
\item Construct an operator on $\HH_{A}\otimes\HH_{B}$ which acts as $A$ does on the states of $\HH_{A}$, but leaves the states of $\HH_{B}$ unchanged.
\item Construct an operator on $\HH_{A}\otimes\HH_{B}$ which acts as $B$ does on the states of $\HH_{B}$, but leaves the states of $\HH_{A}$ unchanged.
\item Show that the two operators you constructed commute by calculating their commutator as defined in \zcref[S]{subsec:The-Matrix-Commutator}.
\end{problemparts}
\end{problem}

\subsubsection{Vectors and Matrices in the Composite Hilbert Space}

Consider the tensor product $\BBC^{m}\otimes\BBC^{n}$. Since the dimension of this Hilbert space is $mn$, and since in any finite-dimensional Hilbert space we know how to represent states as vectors and operators as matrices of the same dimension as the Hilbert space (as discussed in \zcref[S]{subsec:Representing-Vectors-in,subsec:Representing-Matrices-in} respectively), we conclude that states in $\BBC^{m}\otimes\BBC^{n}$ can be represented as $mn$-vectors and operators in $\BBC^{m}\otimes\BBC^{n}$ can be represented as $mn\xx mn$ matrices. In other words, $\BBC^{m}\otimes\BBC^{n}$ is isomorphic to $\BBC^{mn}$.

Explicitly, for two states represented by the vectors\footnote{Note that before, we used the subscript to indicate which space the state belongs to, but now the subscript is instead a vector index.}
\[
\left|\Psi\right\rangle \equiv\left(\begin{array}{c}
\Psi_{1}\\
\vdots\\
\Psi_{m}
\end{array}\right)\in\BBC^{m}\sp\left|\Phi\right\rangle \equiv\left(\begin{array}{c}
\Phi_{1}\\
\vdots\\
\Phi_{n}
\end{array}\right)\in\BBC^{n},
\]
we define the tensor product\index{Tensor product} as follows:
\[
\left|\Psi\right\rangle \otimes\left|\Phi\right\rangle \equiv\left(\begin{array}{c}
\Psi_{1}\left|\Phi\right\rangle \\
\vdots\\
\Psi_{m}\left|\Phi\right\rangle 
\end{array}\right)=\left(\begin{array}{c}
\Psi_{1}\left(\begin{array}{c}
\Phi_{1}\\
\vdots\\
\Phi_{n}
\end{array}\right)\\
\vdots\\
\Psi_{m}\left(\begin{array}{c}
\Phi_{1}\\
\vdots\\
\Phi_{n}
\end{array}\right)
\end{array}\right)=\left(\begin{array}{c}
\Psi_{1}\Phi_{1}\\
\vdots\\
\Psi_{1}\Phi_{n}\\
\vdots\\
\Psi_{m}\Phi_{1}\\
\vdots\\
\Psi_{m}\Phi_{n}
\end{array}\right)\in\BBC^{mn}.
\]
For example:
\[
\left(\begin{array}{c}
1\\
2
\end{array}\right)\otimes\left(\begin{array}{c}
3\\
4
\end{array}\right)=\left(\begin{array}{c}
1\cdot\left(\begin{array}{c}
3\\
4
\end{array}\right)\\
2\cdot\left(\begin{array}{c}
3\\
4
\end{array}\right)
\end{array}\right)=\left(\begin{array}{c}
1\cdot3\\
1\cdot4\\
2\cdot3\\
2\cdot4
\end{array}\right)=\left(\begin{array}{c}
3\\
4\\
6\\
8
\end{array}\right).
\]
Similarly, for two operators represented by the matrices
\[
A\equiv\left(\begin{array}{ccc}
A_{11} & \cdots & A_{1m}\\
\vdots & \ddots & \vdots\\
A_{m1} & \cdots & A_{mm}
\end{array}\right)\in\BBC^{m\xx m}\sp B\equiv\left(\begin{array}{ccc}
B_{11} & \cdots & B_{1n}\\
\vdots & \ddots & \vdots\\
B_{n1} & \cdots & B_{nn}
\end{array}\right)\in\BBC^{n\xx n},
\]
we define the tensor product\index{Tensor product} as follows\footnote{Note that the tensor product of vectors is a special case of the tensor product of matrices, with the vectors treated as single-column matrices.}:
\begin{align*}
A\otimes B & \equiv\left(\begin{array}{ccc}
A_{11}B & \cdots & A_{1m}B\\
\vdots & \ddots & \vdots\\
A_{m1}B & \cdots & A_{mm}B
\end{array}\right)\\
 & =\left(\begin{array}{ccc}
A_{11}\left(\begin{array}{ccc}
B_{11} & \cdots & B_{1n}\\
\vdots & \ddots & \vdots\\
B_{n1} & \cdots & B_{nn}
\end{array}\right) & \cdots & A_{1m}\left(\begin{array}{ccc}
B_{11} & \cdots & B_{1n}\\
\vdots & \ddots & \vdots\\
B_{n1} & \cdots & B_{nn}
\end{array}\right)\\
\vdots & \ddots & \vdots\\
A_{m1}\left(\begin{array}{ccc}
B_{11} & \cdots & B_{1n}\\
\vdots & \ddots & \vdots\\
B_{n1} & \cdots & B_{nn}
\end{array}\right) & \cdots & A_{mm}\left(\begin{array}{ccc}
B_{11} & \cdots & B_{1n}\\
\vdots & \ddots & \vdots\\
B_{n1} & \cdots & B_{nn}
\end{array}\right)
\end{array}\right)\\
 & =\left(\begin{array}{ccccccc}
A_{11}B_{11} & \cdots & A_{11}B_{1n} & \cdots & A_{1m}B_{11} & \cdots & A_{1m}B_{1n}\\
\vdots & \ddots & \vdots &  & \vdots & \ddots & \vdots\\
A_{11}B_{n1} & \cdots & A_{11}B_{nn} & \cdots & A_{1m}B_{n1} & \cdots & A_{1m}B_{nn}\\
\vdots &  & \vdots & \ddots & \vdots &  & \vdots\\
A_{m1}B_{11} & \cdots & A_{m1}B_{1n} & \cdots & A_{mm}B_{11} & \cdots & A_{mm}B_{1n}\\
\vdots & \ddots & \vdots &  & \vdots & \ddots & \vdots\\
A_{m1}B_{n1} & \cdots & A_{m1}B_{nn} & \cdots & A_{mm}B_{n1} & \cdots & A_{mm}B_{nn}
\end{array}\right)\in\BBC^{mn\xx mn}.
\end{align*}
For example:
\begin{align*}
\left(\begin{array}{cc}
0 & 1\\
2 & 0
\end{array}\right)\otimes\left(\begin{array}{cc}
3 & 0\\
0 & 4
\end{array}\right) & =\left(\begin{array}{cc}
0\cdot\left(\begin{array}{cc}
3 & 0\\
0 & 4
\end{array}\right) & 1\cdot\left(\begin{array}{cc}
3 & 0\\
0 & 4
\end{array}\right)\\
2\cdot\left(\begin{array}{cc}
3 & 0\\
0 & 4
\end{array}\right) & 0\cdot\left(\begin{array}{cc}
3 & 0\\
0 & 4
\end{array}\right)
\end{array}\right)\\
 & =\left(\begin{array}{cccc}
0\cdot3 & 0\cdot0 & 1\cdot3 & 1\cdot0\\
0\cdot0 & 0\cdot4 & 1\cdot0 & 1\cdot4\\
2\cdot3 & 2\cdot0 & 0\cdot3 & 0\cdot0\\
2\cdot0 & 2\cdot4 & 0\cdot0 & 0\cdot4
\end{array}\right)\\
 & =\left(\begin{array}{cccc}
0 & 0 & 3 & 0\\
0 & 0 & 0 & 4\\
6 & 0 & 0 & 0\\
0 & 8 & 0 & 0
\end{array}\right).
\end{align*}

\begin{xca}\leavevmode
For the specific $\left|\Psi\right\rangle $, $\left|\Phi\right\rangle $, $A$, and $B$ we used above:
\[
\left|\Psi\right\rangle \equiv\left(\begin{array}{c}
1\\
2
\end{array}\right)\sp\left|\Phi\right\rangle \equiv\left(\begin{array}{c}
3\\
4
\end{array}\right),
\]
\[
A\equiv\left(\begin{array}{cc}
0 & 1\\
2 & 0
\end{array}\right)\sp B\equiv\left(\begin{array}{cc}
3 & 0\\
0 & 4
\end{array}\right),
\]
calculate
\[
\left(\vphantom{\bll}A\otimes B\right)\left(\vphantom{\bll}\left|\Psi\right\rangle \otimes\left|\Phi\right\rangle \right).
\]
Do so in two ways:
\begin{enumerate}
\item Directly in the composite Hilbert space\footnote{Here, $\simeq$ means “isomorphic to”.} $\BBC^{2}\otimes\BBC^{2}\simeq\BBC^{4}$ using the $4\xx4$ matrix and 4-vector found above.
\item Separately in each of the component spaces using the two $2\xx2$ matrices and the two 2-vectors (acting with the first matrix on the first vector and the second matrix on the second vector), and then calculating the tensor product of the results.
\end{enumerate}
Then compare your results and verify that they are the same.
\end{xca}

\begin{problem}\leavevmode
\begin{problemparts}
\item Prove that the tensor product preserves the adjoint operation on both vectors and matrices. That is,
\[
\left(\vphantom{\bll}\left|\Psi\right\rangle \otimes\left|\Phi\right\rangle \right)^{\dagger}=\left\langle \Psi\right|\otimes\left\langle \Phi\right|\sp\left(A\otimes B\right)^{\dagger}=A^{\dagger}\otimes B^{\dagger}.
\]
\item Prove that the tensor product of two Hermitian operators is Hermitian, and the tensor product of two unitary operators is unitary.
\end{problemparts}
\end{problem}

\begin{problem}\leavevmode
Consider the tensor product $\BBC^{m}\otimes\BBC^{n}$ for arbitrary $m$ and $n$. Show that the standard basis of $\BBC^{m}\otimes\BBC^{n}$ is obtained by taking the tensor products of the standard basis states of $\BBC^{m}$ and $\BBC^{n}$.
\end{problem}

\begin{xca}\leavevmode
Calculate the tensor product
\[
\left|+\right\rangle \otimes\left|-\right\rangle \otimes\left|0\right\rangle ,
\]
where $\left|+\right\rangle $ and $\left|-\right\rangle $ are the $+1$ and $-1$ eigenstates of $\sigma_{x}$ respectively, and $\left|0\right\rangle $ is the $+1$ eigenstate of $\sigma_{z}$ (see \zcref[S]{subsec:The-Pauli-Matrices}).
\end{xca}

\begin{xca}\leavevmode
\begin{problemparts}
\item Calculate the tensor product operator
\[
A\equiv S_{x}\otimes S_{z},
\]
where $S_{x}$ and $S_{z}$ were defined in \zcref[S]{eq:Ssigma}.
\item Calculate the tensor product state
\[
\left|\Psi\right\rangle \equiv\left|+\right\rangle \otimes\left|1\right\rangle ,
\]
where $\left|+\right\rangle $ and $\left|1\right\rangle $ were defined in \zcref[S]{subsec:The-Pauli-Matrices}.
\item Calculate the expectation value $\langle A\rangle_{\Psi}$.
\end{problemparts}
\end{xca}

\subsubsection{Quantum Entanglement}

Consider a composite system of two qubits. Let us name the first qubit $A$ and the second qubit $B$. In the composite Hilbert space $\HH_{A}\otimes\HH_{B}$, the computational basis has four eigenstates:
\[
\left|0\right\rangle \otimes\left|0\right\rangle \sp\left|0\right\rangle \otimes\left|1\right\rangle \sp\left|1\right\rangle \otimes\left|0\right\rangle \sp\left|1\right\rangle \otimes\left|1\right\rangle ,
\]
where in each of these, the first state is the state of qubit $A$ and the second is the state of qubit $B$. Thus $\left|0\right\rangle \otimes\left|0\right\rangle $ corresponds to $\left|0\right\rangle $ for both qubits, $\left|0\right\rangle \otimes\left|1\right\rangle $ corresponds to $\left|0\right\rangle $ for qubit $A$ and $\left|1\right\rangle $ for qubit $B$, $\left|1\right\rangle \otimes\left|0\right\rangle $ corresponds to $\left|1\right\rangle $ for qubit $A$ and $\left|0\right\rangle $ for qubit $B$, and $\left|1\right\rangle \otimes\left|1\right\rangle $ corresponds to $\left|1\right\rangle $ for both qubits.

These four eigenstates have the following representations in terms of vectors in $\BBC^{4}$:
\[
\left|0\right\rangle \otimes\left|0\right\rangle =\left(\begin{array}{c}
1\\
0
\end{array}\right)\otimes\left(\begin{array}{c}
1\\
0
\end{array}\right)=\left(\begin{array}{c}
1\\
0\\
0\\
0
\end{array}\right)\sp\left|0\right\rangle \otimes\left|1\right\rangle =\left(\begin{array}{c}
1\\
0
\end{array}\right)\otimes\left(\begin{array}{c}
0\\
1
\end{array}\right)=\left(\begin{array}{c}
0\\
1\\
0\\
0
\end{array}\right),
\]
\[
\left|1\right\rangle \otimes\left|0\right\rangle =\left(\begin{array}{c}
0\\
1
\end{array}\right)\otimes\left(\begin{array}{c}
1\\
0
\end{array}\right)=\left(\begin{array}{c}
0\\
0\\
1\\
0
\end{array}\right)\sp\left|1\right\rangle \otimes\left|1\right\rangle =\left(\begin{array}{c}
0\\
1
\end{array}\right)\otimes\left(\begin{array}{c}
0\\
1
\end{array}\right)=\left(\begin{array}{c}
0\\
0\\
0\\
1
\end{array}\right).
\]
So we see that they are, in fact, just the standard basis of $\BBC^{4}$.

The most general state of both qubits is described as a superposition of all possible combinations:
\begin{equation}
\left|\Psi\right\rangle =\alpha_{00}\left|0\right\rangle \otimes\left|0\right\rangle +\alpha_{01}\left|0\right\rangle \otimes\left|1\right\rangle +\alpha_{10}\left|1\right\rangle \otimes\left|0\right\rangle +\alpha_{11}\left|1\right\rangle \otimes\left|1\right\rangle =\left(\begin{array}{c}
\alpha_{00}\\
\alpha_{01}\\
\alpha_{10}\\
\alpha_{11}
\end{array}\right),\label{eq:general-two-qubit}
\end{equation}
where $\alpha_{00},\alpha_{01},\alpha_{10},\alpha_{11}\in\BBC$ and, of course, the coefficients should be chosen such that the state is normalized to 1:
\[
\left|\alpha_{00}\right|^{2}+\left|\alpha_{01}\right|^{2}+\left|\alpha_{10}\right|^{2}+\left|\alpha_{11}\right|^{2}=1.
\]
We would now like to ask: when do the two qubits depend on each other? More precisely, under what conditions can qubit $A$ be $\left|0\right\rangle $ or $\left|1\right\rangle $ \textbf{independently} of the state of qubit $B$, and vice versa? As we will now see, this depends on the coefficients $\alpha_{ij}$.

A \emph{separable state\index{Separable state}} is a state which can be written as just \textbf{one} tensor product instead of a \textbf{sum} of tensor products, that is, a state of the form
\[
\left|\Psi\right\rangle =\left|\Psi_{A}\right\rangle \otimes\left|\Psi_{B}\right\rangle ,
\]
where $\left|\Psi_{A}\right\rangle $ is the state of qubit $A$ and $\left|\Psi_{B}\right\rangle $ is the state of qubit $B$. If we can write the state in this way, then we have \textbf{separated} the states from one another, in the sense that whatever value $\left|\Psi_{A}\right\rangle $ has is completely independent of the value of $\left|\Psi_{B}\right\rangle $ (and vice versa). In other words, the overall state of the composite system is just the tensor product of the independent states of the individual systems.

A simple example of a separable state would be:
\begin{equation}
\left|\Psi\right\rangle =\left|0\right\rangle \otimes\left|0\right\rangle .\label{eq:sep-state}
\end{equation}
This just means that both qubits are, with 100\% probability, in the state $\left|0\right\rangle $:
\[
\left|\Psi_{A}\right\rangle =\left|0\right\rangle \sp\left|\Psi_{B}\right\rangle =\left|0\right\rangle .
\]
A more interesting separable state is:
\begin{equation}
\left|\Psi\right\rangle =\hf\left(\left|0\right\rangle \otimes\left|0\right\rangle +\left|0\right\rangle \otimes\left|1\right\rangle +\left|1\right\rangle \otimes\left|0\right\rangle +\left|1\right\rangle \otimes\left|1\right\rangle \right).\label{eq:sep-state2}
\end{equation}
To see that it is separable, we simplify it using the distributive property, and get:
\[
\left|\Psi\right\rangle =\frac{1}{\sqrt{2}}\left(\left|0\right\rangle +\left|1\right\rangle \right)\otimes\frac{1}{\sqrt{2}}\left(\left|0\right\rangle +\left|1\right\rangle \right).
\]
In other words, both qubits are in a state where either 0 or 1 is possible with 50\% probability, that is:
\[
\left|\Psi_{A}\right\rangle =\frac{1}{\sqrt{2}}\left(\left|0\right\rangle +\left|1\right\rangle \right)\sp\left|\Psi_{B}\right\rangle =\frac{1}{\sqrt{2}}\left(\left|0\right\rangle +\left|1\right\rangle \right).
\]

A state which is \textbf{not} separable is called an \emph{entangled state}\index{Entangled state}\index{Quantum entanglement}. Here is an example of an entangled state:
\begin{equation}
\left|\Psi\right\rangle =\frac{1}{\sqrt{2}}\left(\left|0\right\rangle \otimes\left|1\right\rangle +\left|1\right\rangle \otimes\left|0\right\rangle \right).\label{eq:ent-state}
\end{equation}
No matter how much we try, we can never write it as just one tensor product – it is always going to be the sum of two tensor products! This means that the state of each qubit is no longer independent of the state of the other qubit; the measurement results are correlated. Indeed, if a measurement of qubit $A$ yields the state $\left|0\right\rangle $ then a measurement of qubit $B$ must yield the state $\left|1\right\rangle $ (due to the first term), and if a measurement of qubit $A$ yields the state $\left|1\right\rangle $ then a measurement of qubit $B$ must yield the state $\left|0\right\rangle $ (due to the second term).

More generally, consider again a composite system in the state
\[
\left|\Psi\right\rangle =\alpha_{00}\left|0\right\rangle \otimes\left|0\right\rangle +\alpha_{01}\left|0\right\rangle \otimes\left|1\right\rangle +\alpha_{10}\left|1\right\rangle \otimes\left|0\right\rangle +\alpha_{11}\left|1\right\rangle \otimes\left|1\right\rangle =\left(\begin{array}{c}
\alpha_{00}\\
\alpha_{01}\\
\alpha_{10}\\
\alpha_{11}
\end{array}\right),
\]
where $\alpha_{00},\alpha_{01},\alpha_{10},\alpha_{11}\in\BBC$. If it is separable, then we should be able to write it in the form
\[
\left|\Psi\right\rangle =\left(\beta_{0}\left|0\right\rangle +\beta_{1}\left|1\right\rangle \right)\otimes\left(\gamma_{0}\left|0\right\rangle +\gamma_{1}\left|1\right\rangle \right),
\]
where $\beta_{0},\beta_{1},\gamma_{0},\gamma_{1}\in\BBC$. Expanding the last equation, we get
\[
\left|\Psi\right\rangle =\beta_{0}\gamma_{0}\left|0\right\rangle \otimes\left|0\right\rangle +\beta_{0}\gamma_{1}\left|0\right\rangle \otimes\left|1\right\rangle +\beta_{1}\gamma_{0}\left|1\right\rangle \otimes\left|0\right\rangle +\beta_{1}\gamma_{1}\left|1\right\rangle \otimes\left|1\right\rangle .
\]
So we should have:
\[
\alpha_{ij}=\beta_{i}\gamma_{j}\sp i,j\in\left\{ 0,1\right\} ,
\]
or explicitly:
\[
\alpha_{00}=\beta_{0}\gamma_{0}\sp\alpha_{01}=\beta_{0}\gamma_{1}\sp\alpha_{10}=\beta_{1}\gamma_{0}\sp\alpha_{11}=\beta_{1}\gamma_{1}.
\]
If this is true, then in particular
\begin{align*}
\alpha_{00}\alpha_{11}-\alpha_{01}\alpha_{10} & =\left(\beta_{0}\gamma_{0}\right)\left(\beta_{1}\gamma_{1}\right)-\left(\beta_{0}\gamma_{1}\right)\left(\beta_{1}\gamma_{0}\right)\\
 & =\beta_{0}\beta_{1}\gamma_{0}\gamma_{1}-\beta_{0}\beta_{1}\gamma_{0}\gamma_{1}\\
 & =0.
\end{align*}
Now, if $\alpha_{ij}$ are the components of a matrix,
\[
\alpha=\left(\begin{array}{cc}
\alpha_{00} & \alpha_{01}\\
\alpha_{10} & \alpha_{11}
\end{array}\right),
\]
then the quantity $\alpha_{00}\alpha_{11}-\alpha_{01}\alpha_{10}$ is called the \emph{determinant}\index{Determinant}\index{Matrix determinant} of the matrix, denoted $\det\alpha$:
\[
\det\alpha\equiv\alpha_{00}\alpha_{11}-\alpha_{01}\alpha_{10}.
\]
We have proven that, \textbf{if }the composite state is separable (not entangled), \textbf{then }the matrix of the coefficients has vanishing determinant. In \zcref[S]{prob:det-separable} you will prove that this also works in the opposite direction; thus, a composite state of two qubits is separable \textbf{if and only if} $\det\alpha=0$.

Let us check this. The state in \zcref[S]{eq:sep-state} is separable, since it has
\[
\det\alpha=1\cdot0-0\cdot0=0.
\]
The state in \zcref[S]{eq:sep-state2} is also separable, since it has
\[
\det\alpha=\hf\cdot\hf-\hf\cdot\hf=0.
\]
However, the state in \zcref[S]{eq:ent-state} is entangled, since it has
\[
\det\alpha=0\cdot0-\frac{1}{\sqrt{2}}\cdot\frac{1}{\sqrt{2}}=-\hf\ne0.
\]
Unfortunately, this simple rule only works for a composite system of two qubits. The problem of finding whether a given state of a composite system is separable or entangled is called the \emph{separability problem}\index{Separability problem}, and it is, for general states, a difficult problem to solve!
\begin{problem}\leavevmode
\label{prob:det-separable}Prove that, for a composite state of two qubits given by
\[
\left|\Psi\right\rangle =\alpha_{00}\left|0\right\rangle \otimes\left|0\right\rangle +\alpha_{01}\left|0\right\rangle \otimes\left|1\right\rangle +\alpha_{10}\left|1\right\rangle \otimes\left|0\right\rangle +\alpha_{11}\left|1\right\rangle \otimes\left|1\right\rangle ,
\]
the state is separable if
\[
\det\left(\begin{array}{cc}
\alpha_{00} & \alpha_{01}\\
\alpha_{10} & \alpha_{11}
\end{array}\right)=\alpha_{00}\alpha_{11}-\alpha_{01}\alpha_{10}=0.
\]
This is the opposite direction to what we proved above, which is that \textbf{if} the state is separable, \textbf{then} the determinant is zero.
\end{problem}

\begin{problem}\leavevmode
Find two separable states and two entangled states of \textbf{three }qubits, and prove that they are separable/entangled.
\end{problem}

\subsubsection{The Bell States}\label{subsec:The-Bell-States}

Let us define the \emph{Bell states}\index{Bell states}, also known as\footnote{EPR stands for Einstein, Podolsky, and Rosen.} \emph{EPR states}\index{EPR states}:
\[
\left|\beta_{xy}\right\rangle \equiv\frac{1}{\sqrt{2}}\left(\left|0\right\rangle \otimes\left|y\right\rangle +\left(-1\right)^{x}\left|1\right\rangle \otimes\left|1-y\right\rangle \right)\sp x,y\in\left\{ 0,1\right\} .
\]
Explicitly, the four choices for $x$ and $y$ give:
\[
\left|\beta_{00}\right\rangle =\frac{1}{\sqrt{2}}\left(\left|0\right\rangle \otimes\left|0\right\rangle +\left|1\right\rangle \otimes\left|1\right\rangle \right),
\]
\[
\left|\beta_{01}\right\rangle =\frac{1}{\sqrt{2}}\left(\left|0\right\rangle \otimes\left|1\right\rangle +\left|1\right\rangle \otimes\left|0\right\rangle \right),
\]
\[
\left|\beta_{10}\right\rangle =\frac{1}{\sqrt{2}}\left(\left|0\right\rangle \otimes\left|0\right\rangle -\left|1\right\rangle \otimes\left|1\right\rangle \right),
\]
\[
\left|\beta_{11}\right\rangle =\frac{1}{\sqrt{2}}\left(\left|0\right\rangle \otimes\left|1\right\rangle -\left|1\right\rangle \otimes\left|0\right\rangle \right).
\]
It is useful to adopt a shorthand notation where we write
\[
\left|xy\right\rangle \equiv\left|x\right\rangle \otimes\left|y\right\rangle ,
\]
so
\[
\left|00\right\rangle =\left|0\right\rangle \otimes\left|0\right\rangle \sp\left|01\right\rangle =\left|0\right\rangle \otimes\left|1\right\rangle \sp\left|10\right\rangle =\left|1\right\rangle \otimes\left|0\right\rangle \sp\left|11\right\rangle =\left|1\right\rangle \otimes\left|1\right\rangle .
\]
In this notation, the Bell states are
\begin{equation}
\left|\beta_{00}\right\rangle =\frac{1}{\sqrt{2}}\left(\left|00\right\rangle +\left|11\right\rangle \right),\label{eq:Bell00-short}
\end{equation}
\begin{equation}
\left|\beta_{01}\right\rangle =\frac{1}{\sqrt{2}}\left(\left|01\right\rangle +\left|10\right\rangle \right),\label{eq:Bell01-short}
\end{equation}
\[
\left|\beta_{10}\right\rangle =\frac{1}{\sqrt{2}}\left(\left|00\right\rangle -\left|11\right\rangle \right),
\]
\[
\left|\beta_{11}\right\rangle =\frac{1}{\sqrt{2}}\left(\left|01\right\rangle -\left|10\right\rangle \right).
\]
The Bell states have important applications in quantum information and computation, as we will see below.
\begin{xca}\leavevmode
Write down the representations of the four Bell states as 4-vectors in $\BBC^{2}\otimes\BBC^{2}\simeq\BBC^{4}$.
\end{xca}

\begin{problem}\leavevmode
Prove that the four Bell states form an orthonormal basis for the composite Hilbert space of two qubits, by showing that they span that space, are linearly independent, are orthogonal, and are normalized to 1.
\end{problem}

\begin{problem}\leavevmode
Prove that each of the four Bell states is entangled.
\end{problem}

\begin{xca}\leavevmode
Write down the four Bell states in terms of $\left|+\right\rangle $ and $\left|-\right\rangle $, the eigenstates of $\sigma_{x}$. You may wish to use the shorthand notation $\left|\pm\pm\right\rangle \equiv\left|\pm\right\rangle \otimes\left|\pm\right\rangle $.
\end{xca}

\subsubsection{Entanglement Does Not Transmit Information}\label{subsec:Entanglement-Does-Not}

Now that we have rigorously defined quantum entanglement, let us debunk the most common misconception associated with it: that quantum entanglement allows us to transmit information, and in particular, that it allows us to do so faster than the speed of light (or even instantaneously), in violation of relativity. This is, in fact, not true.

To illustrate this, imagine the following scenario. Alice and Bob create an entangled pair of qubits, for example in the Bell state
\[
\left|\beta_{01}\right\rangle =\frac{1}{\sqrt{2}}\left(\left|01\right\rangle +\left|10\right\rangle \right).
\]
Alice takes the first qubit in the pair, and Bob takes the second qubit. Alice then stays on Earth, while Bob embarks on a long journey to Alpha Centauri, about 4.4 light years away. When Bob gets there, he measures his qubit. He has a 50\% chance to observe 0 and a 50\% chance to observe 1. However, if he observes 0 he knows that Alice will surely observe 1 whenever she measures her qubit, and if he observes 1 he knows that Alice will surely observe 0, since the qubits must have opposite values.

So it seems that Bob now knows something about Alice's qubit that he did not know before. Furthermore, he knows that \textbf{instantly} – even though Alice is 4.4 light years away, and thus according to relativity, that information should have taken at least 4.4 years to travel between them. But has any information actually been transferred between them?

The answer is no! All Bob did was observe a \textbf{random} event. Bob cannot \textbf{control} which value he observes when he measures the qubit, 0 or 1; he can only observe it, and randomly get whatever he gets. He gains information about Alice's qubit, which is completely random, but he does not receive any specific message from Alice, nor can he transmit any specific information to Alice by observing his qubit.

In fact, there is a theorem called the \emph{no-communication theorem\index{No-communication theorem} }which rigorously proves that no information can be transmitted using quantum entanglement alone, whether faster than light or otherwise.

The fact that a measurement of one qubit determines the measurement of another qubit might seem like it indicates that some information must be transmitted between the qubits themselves, so that they “know” about each other's states. However, there isn't any actual need to transmit information between the two entangled qubits in order for them to match their measurements! After all, the entangled state does \textbf{not }depend on the distance between the qubits, whether in time or in space; it is simply the combined state of the two qubits, wherever or whenever they might be.

Consider now the following completely classical scenario. Let's say I write 0 on one piece of paper and 1 on another piece of paper – essentially, two classical bits. I then put each piece of paper in a separate sealed envelope, and randomly give one envelope to Alice and the other to Bob. When Bob gets to Alpha Centauri, he opens his envelope. If he sees 0 he knows that Alice's envelope says 1, and if he sees 1 he knows that Alice's envelope says 0.

Obviously, this does not allow any information to be transmitted between Alice and Bob, nor does each envelope need to “know” what's inside the other envelope in order for the measurements to match. If Bob sees 0, then the piece of paper saying 0 was inside the envelope \textbf{all along}, and the piece of paper saying 1 was inside Alice's envelope all along – and vice versa. The envelopes are \textbf{classically correlated}, and nothing weird is going on. What, then, is the difference between this classical correlation and quantum entanglement? The answer to this question can be made precise using Bell's theorem, which we will formulate in \zcref[S]{subsec:Bell}.

\subsection{Non-Commuting Observables and the Uncertainty Principle}

\subsubsection{The Uncertainty Principle}

When two quantum observables do not commute, we get an \emph{uncertainty relation}\index{Uncertainty principle}. \emph{Uncertainty\index{Uncertainty}} is just another name for \emph{standard deviation}\index{Standard deviation}, which we defined in \zcref[S]{subsec:Standard-Deviation}. The most well-known such relation is the \emph{position-momentum uncertainty relation}\index{Position-momentum uncertainty relation}\footnote{Recall that we are using units where $\hbar=1$!}:
\[
\Delta x\Delta p\ge\hf.
\]
Here, $x$ and $p$ are two Hermitian operators, corresponding to the observables position and momentum respectively. This inequality means that the product of the uncertainty in position $\Delta x$ and the uncertainty in momentum $\Delta p$ cannot go below $1/2$; it follows that $\Delta x$ and $\Delta p$ cannot both be zero at the same time, so we can never know both the position and momentum with arbitrarily high certainty.

Let us prove this relation for the general case of any two observables represented by Hermitian operators, $A$ and $B$, which do not commute:
\[
\left[A,B\right]\ne0.
\]
Recall that the (square of the) standard deviation $\Delta A$ of $A$ is given by
\[
\left(\Delta A\right)^{2}=\left\langle \left(A-\left\langle A\right\rangle \right)^{2}\right\rangle .
\]
We have seen that expectation values in quantum theory are calculated using the inner product “sandwich”
\[
\left\langle A\right\rangle =\langle\Psi|A|\Psi\rangle,
\]
where $\left|\Psi\right\rangle $ is the state with respect to which the expectation value is calculated. The (square of the) standard deviation is thus
\begin{align*}
\left(\Delta A\right)^{2} & =\langle\Psi|\left(A-\left\langle A\right\rangle \right)^{2}|\Psi\rangle\\
 & =\langle\Psi|\left(A-\left\langle A\right\rangle \right)\left(A-\left\langle A\right\rangle \right)|\Psi\rangle.
\end{align*}
Let us now define a new vector:
\[
\left|a\right\rangle =\left(A-\left\langle A\right\rangle \right)|\Psi\rangle.
\]
Then we simply have\footnote{$A-\left\langle A\right\rangle $ is the \textbf{operator }$A$ minus the \textbf{real number }$\left\langle A\right\rangle $ times the identity operator $1$ (the identity operator is implied). Thus $A-\left\langle A\right\rangle $ is Hermitian, and the bra of $\left(A-\left\langle A\right\rangle \right)|\Psi\rangle$ is $\langle\Psi|\left(A-\left\langle A\right\rangle \right)$.}
\[
\left(\Delta A\right)^{2}=\langle a|a\rangle=\left\Vert a\right\Vert ^{2}.
\]
Similarly, for $B$ we define
\[
\left|b\right\rangle =\left(B-\left\langle B\right\rangle \right)|\Psi\rangle,
\]
and get
\[
\left(\Delta B\right)^{2}=\langle b|b\rangle=\left\Vert b\right\Vert ^{2}.
\]
The product of the (squares of the) standard deviations in $A$ and $B$ is thus
\[
\left(\Delta A\right)^{2}\left(\Delta B\right)^{2}=\left\Vert a\right\Vert ^{2}\left\Vert b\right\Vert ^{2}.
\]
Using the Cauchy-Schwarz inequality in \zcref[S]{eq:CSI}, we have
\begin{align*}
\left(\Delta A\right)^{2}\left(\Delta B\right)^{2} & =\left\Vert a\right\Vert ^{2}\left\Vert b\right\Vert ^{2}\\
 & \ge\left|\langle a|b\rangle\right|^{2}\\
\left(*\right) & =\left(\re\langle a|b\rangle\right)^{2}+\left(\im\langle a|b\rangle\right)^{2}\\
\left(**\right) & \ge\left(\im\langle a|b\rangle\right)^{2}\\
\left(***\right) & =\left(\frac{\langle a|b\rangle-\langle b|a\rangle}{2\i}\right)^{2},
\end{align*}
where in $\left(*\right)$ we used \zcref[S]{eq:mag-re-im}, in $\left(**\right)$ we used the fact that $\left(\re\langle a|b\rangle\right)^{2}\ge0$ since it's the square of a real number, and in $\left(***\right)$ we used \zcref[S]{eq:re-im} and the fact that $\langle b|a\rangle=\langle a|b\rangle^{*}$.

Next, we note that
\begin{align*}
\langle a|b\rangle & =\langle\Psi|\left(A-\langle A\rangle\right)\left(B-\langle B\rangle\right)|\Psi\rangle\\
 & =\langle\left(A-\langle A\rangle\right)\left(B-\langle B\rangle\right)\rangle\\
 & =\langle AB-A\langle B\rangle-\langle A\rangle B+\langle A\rangle\langle B\rangle\rangle\\
 & =\langle AB\rangle-\langle A\langle B\rangle\rangle-\langle\langle A\rangle B\rangle+\langle\langle A\rangle\langle B\rangle\rangle\\
 & =\langle AB\rangle-\langle A\rangle\langle B\rangle-\langle A\rangle\langle B\rangle+\langle A\rangle\langle B\rangle\\
 & =\langle AB\rangle-\langle A\rangle\langle B\rangle,
\end{align*}
where we used the linearity of the inner product (see \zcref[S]{prob:inner3}) and the fact that $\langle\Psi|\Psi\rangle=1$. Similarly,
\[
\langle b|a\rangle=\langle BA\rangle-\langle A\rangle\langle B\rangle.
\]
Thus
\begin{align*}
\langle a|b\rangle-\langle b|a\rangle & =\left(\langle AB\rangle-\langle A\rangle\langle B\rangle\right)-\left(\langle BA\rangle-\langle A\rangle\langle B\rangle\right)\\
 & =\langle AB\rangle-\langle BA\rangle\\
 & =\langle\left[A,B\right]\rangle,
\end{align*}
and so we get
\begin{equation}
\left(\Delta A\right)^{2}\left(\Delta B\right)^{2}\ge\left\langle \frac{1}{2\i}\left[A,B\right]\right\rangle ^{2}.\label{eq:uncertainty-squared}
\end{equation}
Now, by definition, $\Delta A$ and $\Delta B$ are real and non-negative. If $\left\langle \frac{1}{2\i}\left[A,B\right]\right\rangle $ is also real, we could take the square root (but we have to add an absolute value because it could actually be negative):
\begin{equation}
\Delta A\Delta B\ge\hf\left|\left\langle \left[A,B\right]\right\rangle \right|.\label{eq:uncertainty}
\end{equation}
You will show in \zcref[S]{prob:Inequalities-are-only} that it is indeed always real. Note that the uncertainty relation we found still depends on the choice of state $\left|\Psi\right\rangle $ with which to calculate the expected values and standard deviations, but sometimes, as in the position-momentum uncertainty relation, the same relation applies to all states.

As we will explain in more detail later, when we discuss continuous systems, the operators $x$ and $p$ have the commutator
\[
\left[x,p\right]=\i.
\]
By plugging this commutator into the uncertainty relation in \zcref[S]{eq:uncertainty}, we indeed get the familiar result
\[
\Delta x\Delta p\ge\hf.
\]

\begin{problem}\leavevmode
\label{prob:Inequalities-are-only}Inequalities are only defined for real numbers, not complex numbers. Prove that if $A$ and $B$ are Hermitian, then $\left\langle \left[A,B\right]\right\rangle $ must always be an imaginary number, and thus $\left\langle \frac{1}{2\i}\left[A,B\right]\right\rangle $ is always real, so the inequality in \zcref[S]{eq:uncertainty-squared} is well-defined. You may wish to recall the properties of anti-Hermitian matrices from \zcref[S]{subsec:Anti-Hermitian-Matrices}.
\end{problem}

\begin{xca}\leavevmode
Calculate the uncertainty relation for $\sigma_{x}$ and $\sigma_{y}$ given an arbitrary qubit:
\[
\left|\Psi\right\rangle =a\left|0\right\rangle +b\left|1\right\rangle \sp\left|a\right|^{2}+\left|b\right|^{2}=1.
\]
That is, find the right-hand side of
\[
\Delta\sigma_{x}\Delta\sigma_{y}\ge\left(?\right).
\]
Comment on the consequences of the relation you found for choices of different states, that is, different values of $a$ and $b$.
\end{xca}

\subsubsection{Simultaneous Diagonalization}

Why is there uncertainty when two observables don't commute? Some insight may be gained from the fact that two Hermitian operators may be \emph{simultaneously diagonalizable}\index{Simultaneous diagonalization} if and only if they commute\footnote{This is a special case of a more general theorem: a set of diagonalizable\textbf{ }matrices commute if and only if they are simultaneously diagonalizable. Of course, here we are dealing specifically with Hermitian matrices, and such matrices are always diagonalizable; furthermore, for our purposes it is enough to talk about two matrices rather than a larger set.}.

Recall that in \zcref[S]{subsec:Diagonalizable-Matrices} we saw that for any Hermitian matrix\footnote{Or more generally for any normal matrix, which satisfies $A^{\dagger}A=AA^{\dagger}$. As we mentioned in \zcref[S]{subsec:Unitary-and-Normal-Matrices}, both Hermitian and unitary matrices are special cases of normal matrices.} $A$ there exists a unitary matrix $P$ such that
\[
P^{\dagger}AP=D,
\]
where $D$ is a diagonal matrix. Furthermore, the elements on the diagonal are none other than the eigenvalues of $A$. This is called \emph{diagonalizing}\index{Diagonalizable matrix} the matrix $A$.

Now, let $A_{1}$ and $A_{2}$ be two Hermitian matrices. We say that $A_{1}$ and $A_{2}$ are \emph{simultaneously diagonalizable}\index{Simultaneous diagonalization} if both matrices are diagonalizable using the \textbf{same }unitary matrix $P$:
\begin{equation}
P^{\dagger}A_{1}P=D_{1}\sp P^{\dagger}A_{2}P=D_{2},\label{eq:sim-diag}
\end{equation}
where $D_{1}$ and $D_{2}$ are two diagonal matrices.

If $A_{1}$ and $A_{2}$ are simultaneously diagonalizable, we can invert \zcref[S]{eq:sim-diag} (by multiplying both sides by $P$ from the left and $P^{\dagger}$ from the right) to find:
\[
A_{1}=PD_{1}P^{\dagger}\sp A_{2}=PD_{2}P^{\dagger}.
\]
Then the commutator of the two matrices is
\begin{align*}
\left[A_{1},A_{2}\right] & =A_{1}A_{2}-A_{2}A_{1}\\
 & =\left(PD_{1}P^{\dagger}\right)\left(PD_{2}P^{\dagger}\right)-\left(PD_{2}P^{\dagger}\right)\left(PD_{1}P^{\dagger}\right)\\
 & =PD_{1}\left(P^{\dagger}P\right)D_{2}P^{\dagger}-PD_{2}\left(P^{\dagger}P\right)D_{1}P^{\dagger}\\
 & =PD_{1}D_{2}P^{\dagger}-PD_{2}D_{1}P^{\dagger}\\
 & =P\left(D_{1}D_{2}-D_{2}D_{1}\right)P^{\dagger}\\
 & =P\left[D_{1},D_{2}\right]P^{\dagger}.
\end{align*}
However, any two diagonal matrices commute with each other. Indeed, if
\begin{equation}
D_{1}\equiv\left(\begin{array}{ccc}
\lambda_{1} & 0 & 0\\
0 & \ddots & 0\\
0 & 0 & \lambda_{n}
\end{array}\right)\sp D_{2}\equiv\left(\begin{array}{ccc}
\mu_{1} & 0 & 0\\
0 & \ddots & 0\\
0 & 0 & \mu_{n}
\end{array}\right),\label{eq:D1D2}
\end{equation}
then it is easy to see that
\[
D_{1}D_{2}=D_{2}D_{1}=\left(\begin{array}{ccc}
\lambda_{1}\mu_{1} & 0 & 0\\
0 & \ddots & 0\\
0 & 0 & \lambda_{n}\mu_{n}
\end{array}\right).
\]
Therefore $\left[D_{1},D_{2}\right]=0$, and we conclude that $A_{1}$ and $A_{2}$ commute:
\[
\left[A_{1},A_{2}\right]=0.
\]
It is possible to prove the opposite direction as well: \textbf{if} $A_{1}$ and $A_{2}$ commute, \textbf{then} they are simultaneously diagonalizable. However, we won't do this here.

So what does this mean? Let $A_{1}$ and $A_{2}$ be two commuting observables, represented by Hermitian operators. Then they are simultaneously diagonalizable. Now, remember that in \zcref[S]{subsec:Diagonalizable-Matrices} we said that the unitary matrix $P$, which in this case diagonalizes both matrices, has for its columns an orthonormal eigenbasis $\left|B_{i}\right\rangle $:
\[
P=\left(\vphantom{\blll}\begin{array}{ccc}
\left|B_{1}\right\rangle  & \cdots & \left|B_{n}\right\rangle \end{array}\right).
\]
By inspecting \zcref[S]{eq:sim-diag,eq:D1D2}, we see that the basis states $\left|B_{i}\right\rangle $ are eigenstates of \textbf{both }$A_{1}$ and $A_{2}$, with the eigenvalues:
\[
A_{1}\left|B_{i}\right\rangle =\lambda_{i}\left|B_{i}\right\rangle \sp A_{2}\left|B_{i}\right\rangle =\mu_{i}\left|B_{i}\right\rangle .
\]
This means that the eigenstates $\left|B_{i}\right\rangle $ are states where the system \textbf{simultaneously} has the exact value $\lambda_{i}$ for the observable $A_{1}$ and the exact value $\mu_{i}$ for the observable $A_{2}$.

Conversely, since this is an if-and-only-if relationship, if $A_{1}$ and $A_{2}$ \textbf{don't }commute, then one \textbf{cannot }find a basis of eigenstates of both observables simultaneously; if we found such a basis, then they would be simultaneously diagonalizable, in contradiction. Some eigenstates might still be shared between the two, but there must be some eigenstates of $A_{1}$ that are not also eigenstates of $A_{2}$. If the system is in such an eigenstate of $A_{1}$, then it must be in a \textbf{superposition} of eigenstates of $A_{2}$ with at least two distinct eigenvalues\footnote{This is because a superposition of eigenstates with the same eigenvalue is itself an eigenstate, so at least two of the eigenvalues must be distinct.}. Thus, the value of $A_{1}$ is certain, while a measurement of $A_{2}$ has multiple distinct outcomes with non-zero probabilities. This is, essentially, why non-commuting observables satisfy the uncertainty principle.
\begin{xca}\leavevmode
\begin{problemparts}
\item Show that the following Hermitian operator commutes with the Pauli operator $\sigma_{x}$:
\[
A\equiv\left(\begin{array}{cc}
1 & -3\\
-3 & 1
\end{array}\right)\sp\sigma_{x}=\left(\begin{array}{cc}
0 & 1\\
1 & 0
\end{array}\right).
\]
Therefore, they are simultaneously diagonalizable.
\item Show that the eigenstates of $\sigma_{x}$ (see \zcref[S]{subsec:The-Pauli-Matrices}) are also eigenstates of $A$, and find their eigenvalues.
\item Find a unitary matrix $P$ which diagonalizes both $A$ and $\sigma_{x}$, and find the resulting diagonal matrices.
\end{problemparts}
\end{xca}

\subsection{Dynamics, Transformations, and Measurements}

\subsubsection{Unitary Transformations and Evolution}\label{subsec:Unitary-Transformations-and}

We have covered almost all of the basic properties of quantum theory. However, notice that so far we only talked about quantum systems that are in one given state, and never change. In real life, physical systems change all the time, whether it's because some transformation was explicitly done to the system, or simply because time has passed. To account for that in the mathematical framework of quantum theory, let us introduce a new axiom:
\begin{quote}
\textbf{The Evolution Axiom}\index{Axioms of quantum mechanics!Evolution Axiom}\index{Evolution Axiom}\textbf{:} Isolated quantum systems evolve in time via the action of a unitary operator. More precisely, given two points in time $t_{1}$ and $t_{2}$, there is a unitary operator $U$ such that any state $\left|\Psi\right\rangle $ at $t_{1}$ evolves to the state $U\left|\Psi\right\rangle $ at time $t_{2}$.
\end{quote}
This is called \emph{unitary evolution}\index{Unitary evolution} or a \emph{unitary transformation}\index{Unitary transformation}. The exact form of $U$ is determined by the specific quantum system in question and the specific transformation performed, for example rotating the system, moving it in space, or letting it “move itself” in time (i.e. just waiting for time to pass). All the Evolution Axiom tells us is that $U$ must be a unitary operator – just like the Observable Axiom tells us that an observable must be represented by a Hermitian operator, but the exact form of the Hermitian operator depends on the specific system and the specific observable.

Note that the Evolution Axiom is formulated specifically for \emph{isolated systems}\index{Isolated system}. By this we mean that the system is not entangled with any other system and is not measured by any observer or apparatus (which would itself be a separate system). The isolated system may still be acted upon in order to perform certain operations or transformations, like rotating the system, and we assume that such operations do not entangle the system with any other system, at least to a good approximation. These external effects are encoded in the choice of $U$.

Now, in \zcref[S]{subsec:Unitary-and-Normal-Matrices} we proved that unitary operators preserve the inner product between two states. This means that if we have two states $\left|\Psi_{1}\right\rangle $ and $\left|\Phi_{1}\right\rangle $ at one time, and they evolve to $\left|\Psi_{2}\right\rangle =U\left|\Psi_{1}\right\rangle $ and $\left|\Phi_{2}\right\rangle =U\left|\Phi_{1}\right\rangle $ at another time, then the inner product of the new states $\langle\Psi_{2}|\Phi_{2}\rangle$ is equal to the inner product of the old states $\langle\Psi_{1}|\Phi_{1}\rangle$, because $U^{\dagger}U=1$:
\[
\langle\Psi_{2}|\Phi_{2}\rangle=\left(\langle\Psi_{1}|U^{\dagger}\right)\left(U|\Phi_{1}\rangle\right)=\langle\Psi_{1}|U^{\dagger}U|\Phi_{1}\rangle=\langle\Psi_{1}|\Phi_{1}\rangle.
\]
As a corollary, unitary evolution also preserves the norm of a vector – that is, $\left|\Psi\right\rangle $ and $U\left|\Psi\right\rangle $ have the same norm:
\[
\left\Vert U\Psi\right\Vert =\sqrt{\langle\Psi|U^{\dagger}U|\Psi\rangle}=\sqrt{\langle\Psi|\Psi\rangle}=\left\Vert \Psi\right\Vert .
\]
This has to be the case, since quantum states must have norm 1! So if we start with a properly normalized quantum state, we end up with another properly normalized quantum state.

Furthermore, recall that probabilities must sum to one. Indeed, for an orthonormal eigenbasis $\left|B_{i}\right\rangle $, we have
\[
\sum^{n}_{i=1}\left|\langle B_{i}|\Psi\rangle\right|^{2}=\left\Vert \Psi\right\Vert ^{2}=1,
\]
as we proved in \zcref[S]{subsec:Probability-Amplitudes}. Since unitary evolution preserves the norm of the state, we are guaranteed that $\left\Vert U\Psi\right\Vert =\left\Vert \Psi\right\Vert =1$, so the probabilities still sum to 1 after the state has evolved.

Lastly, observe that since any unitary operator is invertible (with the inverse of $U$ being $U^{-1}=U^{\dagger}$), any unitary transformation has an inverse transformation. This means that unitary evolution is \emph{reversible}\index{Unitary evolution!Reversibility}: the same evolution operator that evolved the system forward in time can also be used (via its adjoint) to evolve the system backward in time. More concretely, if at time $t_{1}$ the system is in the state $\left|\Psi_{1}\right\rangle $ and at time $t_{2}>t_{1}$ the system is in the state $\left|\Psi_{2}\right\rangle $, then they are either related by $\left|\Psi_{2}\right\rangle =U\left|\Psi_{1}\right\rangle $, evolving \textbf{forward }in time, or $\left|\Psi_{1}\right\rangle =U^{\dagger}\left|\Psi_{2}\right\rangle $ for the same $U$, evolving \textbf{backward }in time.
\begin{problem}\leavevmode
The system was previously in the state
\[
\left|\Psi_{1}\right\rangle =\frac{1}{\sqrt{5}}\left(\begin{array}{c}
1\\
2\i
\end{array}\right).
\]
Now, it is in the state
\[
\left|\Psi_{2}\right\rangle =\frac{1}{\sqrt{5}}\left(\begin{array}{c}
2\i\\
-1
\end{array}\right).
\]
Find the complete set of unitary operators $U$ that could have been responsible for this evolution, that is, all unitary operators satisfying $\left|\Psi_{2}\right\rangle =U\left|\Psi_{1}\right\rangle $.
\end{problem}

\subsubsection{Quantum Logic Gates}\label{subsec:Quantum-Logic-Gates}

In a classical computer, bits are manipulated using \emph{logic gates}\index{Logic gate!Classical}\index{Classical logic gate}. In logic terms, these gates treat 0 as “false” and 1 as “true”. Let us list some examples of logic gates.

\emph{NOT}\index{NOT gate!Classical}\index{Logic gate!Classical NOT}\index{Classical logic gate!NOT} gets a single bit as input, and outputs 1 minus that bit. In logic terms, it outputs “true” if it gets “false” and vice versa, so the output is the negation of the input:
\begin{center}
\begin{tabular}{|c|c|}
\hline 
Input & NOT\tabularnewline
\hline 
\hline 
0 & 1\tabularnewline
\hline 
1 & 0\tabularnewline
\hline 
\end{tabular}
\par\end{center}

\emph{AND}\index{AND gate}\index{Logic gate!Classical AND}\index{Classical logic gate!AND} gets two bits as input, and outputs 1 if both bits are 1, otherwise it outputs 0. In logic terms, it outputs “true” only if both bit A \textbf{and }bit B are “true”:
\begin{center}
\begin{tabular}{|c|c|c|}
\hline 
Input A & Input B & AND\tabularnewline
\hline 
\hline 
0 & 0 & 0\tabularnewline
\hline 
0 & 1 & 0\tabularnewline
\hline 
1 & 0 & 0\tabularnewline
\hline 
1 & 1 & 1\tabularnewline
\hline 
\end{tabular}
\par\end{center}

\emph{OR}\index{OR gate}\index{Logic gate!Classical OR}\index{Classical logic gate!OR} gets two bits as input, and outputs 1 if at least one of the bits is 1, otherwise it outputs 0. In logic terms, it outputs “true” if either bit A \textbf{or }bit B \textbf{or both} are “true”:
\begin{center}
\begin{tabular}{|c|c|c|}
\hline 
Input A & Input B & OR\tabularnewline
\hline 
\hline 
0 & 0 & 0\tabularnewline
\hline 
0 & 1 & 1\tabularnewline
\hline 
1 & 0 & 1\tabularnewline
\hline 
1 & 1 & 1\tabularnewline
\hline 
\end{tabular}
\par\end{center}

\emph{XOR}\index{XOR gate}\index{Logic gate!Classical XOR}\index{Classical logic gate!XOR} (\emph{eXclusive OR}\index{Exclusive OR}, pronounced “ex or”) gets two bits as input, and outputs 1 if exactly one of the bits is 1, otherwise it outputs 0. In logic terms, it outputs “true” if either bit A \textbf{or }bit B, but \textbf{not both}, is “true”:
\begin{center}
\begin{tabular}{|c|c|c|}
\hline 
Input A & Input B & XOR\tabularnewline
\hline 
\hline 
0 & 0 & 0\tabularnewline
\hline 
0 & 1 & 1\tabularnewline
\hline 
1 & 0 & 1\tabularnewline
\hline 
1 & 1 & 0\tabularnewline
\hline 
\end{tabular}
\par\end{center}

In quantum computers\index{Quantum computer} we have qubits instead of classical bits, and thus we must use \emph{quantum logic gates}, or \emph{quantum gates}\index{Logic gate!Quantum}\index{Quantum logic gate} for short. Since they transform qubits from one state to another, quantum gates must take the form of unitary operators, by the Evolution Axiom.

As a simple example, let us define the \emph{quantum NOT gate}\index{Logic gate!Quantum NOT (X)}\index{Quantum logic gate!NOT (X)}\index{NOT gate!Quantum}\index{NOT (X) gate}\index{X (NOT) gate}, which flips $\left|0\right\rangle \tot\left|1\right\rangle $, just like a classical NOT gate flips $0\tot1$. This gate is none other than the Pauli matrix $\sigma_{x}$, which is of course unitary:
\[
\textrm{NOT}\equiv X\equiv\sigma_{x}=\left(\begin{array}{cc}
0 & 1\\
1 & 0
\end{array}\right).
\]
(The notation $X$ for the NOT gate is common in quantum computing.) Indeed, we have
\[
\textrm{NOT}\left|0\right\rangle =\left(\begin{array}{cc}
0 & 1\\
1 & 0
\end{array}\right)\left(\begin{array}{c}
1\\
0
\end{array}\right)=\left(\begin{array}{c}
0\\
1
\end{array}\right)=\left|1\right\rangle ,
\]
\[
\textrm{NOT}\left|1\right\rangle =\left(\begin{array}{cc}
0 & 1\\
1 & 0
\end{array}\right)\left(\begin{array}{c}
0\\
1
\end{array}\right)=\left(\begin{array}{c}
1\\
0
\end{array}\right)=\left|0\right\rangle .
\]
Since unitary transformations are linear, this means that for a general qubit state we have
\[
\textrm{NOT}\left(a\left|0\right\rangle +b\left|1\right\rangle \right)=a\left|1\right\rangle +b\left|0\right\rangle ,
\]
where $\left|a\right|^{2}+\left|b\right|^{2}=1$.

In classical computers there is only one non-trivial single-bit gate, the NOT gate\footnote{There are exactly three other single-bit gates: the identity gate is trivial because it doesn't do anything, and the gates that take everything to 0 or everything to 1 are trivial because their output is fixed and does not depend on the input.}. However, in quantum computers, since qubits are in a superposition of $\left|0\right\rangle $ and $\left|1\right\rangle $, we have more options; in fact, we have an \textbf{infinite }number of possible single-qubit gates, since any $2\xx2$ unitary operator can be a single-qubit gate.

One example of a useful quantum gate is the \emph{$Z$ gate}, which is just the Pauli matrix $\sigma_{z}$:\index{Logic gate!Quantum Z}\index{Quantum logic gate!Z}\index{Z gate}
\[
Z\equiv\sigma_{z}=\left(\begin{array}{cc}
1 & 0\\
0 & -1
\end{array}\right),
\]
and has the action
\[
Z\left|0\right\rangle =\left(\begin{array}{cc}
1 & 0\\
0 & -1
\end{array}\right)\left(\begin{array}{c}
1\\
0
\end{array}\right)=\left(\begin{array}{c}
1\\
0
\end{array}\right)=\left|0\right\rangle ,
\]
\[
Z\left|1\right\rangle =\left(\begin{array}{cc}
1 & 0\\
0 & -1
\end{array}\right)\left(\begin{array}{c}
0\\
1
\end{array}\right)=\left(\begin{array}{c}
0\\
-1
\end{array}\right)=-\left|1\right\rangle ,
\]
so it leaves $\left|0\right\rangle $ unchanged but flips the phase of $\left|1\right\rangle $.

Another example is the \emph{Hadamard gate}\index{Logic gate!Quantum Hadamard}\index{Quantum logic gate!Hadamard}\index{Hadamard gate}:
\begin{equation}
H\equiv\frac{1}{\sqrt{2}}\left(\begin{array}{cc}
1 & 1\\
1 & -1
\end{array}\right),\label{eq:Hadamard}
\end{equation}
which turns $\left|0\right\rangle $ and $\left|1\right\rangle $ (the eigenstates of $\sigma_{z}$) into $\left|+\right\rangle $ and $\left|-\right\rangle $ respectively (the eigenstates of $\sigma_{x}$):
\[
H\left|0\right\rangle =\frac{1}{\sqrt{2}}\left(\begin{array}{cc}
1 & 1\\
1 & -1
\end{array}\right)\left(\begin{array}{c}
1\\
0
\end{array}\right)=\frac{1}{\sqrt{2}}\left(\left|0\right\rangle +\left|1\right\rangle \right)=\left|+\right\rangle ,
\]
\[
H\left|1\right\rangle =\frac{1}{\sqrt{2}}\left(\begin{array}{cc}
1 & 1\\
1 & -1
\end{array}\right)\left(\begin{array}{c}
0\\
1
\end{array}\right)=\frac{1}{\sqrt{2}}\left(\left|0\right\rangle -\left|1\right\rangle \right)=\left|-\right\rangle .
\]
What about two-qubit gates? Notice that classical two-bit gates such as AND, OR, and XOR are \textbf{irreversible}, since if we are given the single output bit of any of these gates, we cannot in general reconstruct the two input bits. For example, if AND outputs 0, then the inputs could have been any of 00, 01, or 10. In contrast, quantum gates must be represented by unitary operators, and as we saw in \zcref[S]{subsec:Unitary-Transformations-and}, unitary transformations are \textbf{reversible}. Thus we cannot use AND, OR, XOR, and other irreversible logic gates in quantum computing.

We can, however, define other two-qubit quantum gates. A very useful example is the \emph{controlled-NOT}\index{Controlled-NOT gate} or CNOT\index{CNOT gate}\index{Logic gate!Quantum CNOT}\index{Quantum logic gate!CNOT} gate. Here, the first qubit controls whether the second qubit gets flipped or not. If the first qubit is $\left|0\right\rangle $, then the second qubit is unchanged; if the first qubit is $\left|1\right\rangle $, then the second qubit is flipped $\left|0\right\rangle \tot\left|1\right\rangle $. So, given an input state of two qubits, we have:
\begin{equation}
\textrm{CNOT}\left(\left|0\right\rangle \otimes\left|0\right\rangle \right)=\left|0\right\rangle \otimes\left|0\right\rangle ,\label{eq:CNOT00}
\end{equation}
\begin{equation}
\textrm{CNOT}\left(\left|0\right\rangle \otimes\left|1\right\rangle \right)=\left|0\right\rangle \otimes\left|1\right\rangle ,\label{eq:CNOT01}
\end{equation}
\begin{equation}
\textrm{CNOT}\left(\left|1\right\rangle \otimes\left|0\right\rangle \right)=\left|1\right\rangle \otimes\left|1\right\rangle ,\label{eq:CNOT10}
\end{equation}
\begin{equation}
\textrm{CNOT}\left(\left|1\right\rangle \otimes\left|1\right\rangle \right)=\left|1\right\rangle \otimes\left|0\right\rangle .\label{eq:CNOT11}
\end{equation}
As you will verify in \zcref[S]{exer:CNOT-matrix}, the CNOT gate can be represented by the unitary matrix
\begin{equation}
\textrm{CNOT}=\left(\begin{array}{cccc}
1 & 0 & 0 & 0\\
0 & 1 & 0 & 0\\
0 & 0 & 0 & 1\\
0 & 0 & 1 & 0
\end{array}\right).\label{eq:CNOT}
\end{equation}
Alternatively, as you will verify in \zcref[S]{exer:CNOT-outer}, the CNOT gate can be represented by a sum of tensor products of outer products:
\[
\textrm{CNOT}=\left|0\right\rangle \left\langle 0\right|\otimes\left(\vphantom{\bll}\left|0\right\rangle \left\langle 0\right|+\left|1\right\rangle \left\langle 1\right|\right)+\left|1\right\rangle \left\langle 1\right|\otimes\left(\vphantom{\bll}\left|0\right\rangle \left\langle 1\right|+\left|1\right\rangle \left\langle 0\right|\right).
\]
Note that, in the computational basis, the CNOT gate is similar to a classical XOR gate: the second qubit of the result is the XOR of the two original qubits. However, unlike XOR, the CNOT gate is reversible because the result retains the first qubit, which allows us to reconstruct the original state.
\begin{xca}\leavevmode
\label{exer:CNOT-matrix}Verify that the matrix definition of the CNOT operator given in \zcref[S]{eq:CNOT} indeed has the action described in \zcref[S]{eq:CNOT00,eq:CNOT01,eq:CNOT10,eq:CNOT11}.
\end{xca}

\begin{xca}\leavevmode
\label{exer:CNOT-outer}Verify that the CNOT operator has the outer product representation
\[
\textrm{CNOT}=\left|0\right\rangle \left\langle 0\right|\otimes\left(\vphantom{\bll}\left|0\right\rangle \left\langle 0\right|+\left|1\right\rangle \left\langle 1\right|\right)+\left|1\right\rangle \left\langle 1\right|\otimes\left(\vphantom{\bll}\left|0\right\rangle \left\langle 1\right|+\left|1\right\rangle \left\langle 0\right|\right).
\]
You can either do so by explicitly calculating the matrix representations of the outer products and tensor products and adding them up to get the matrix in \zcref[S]{eq:CNOT}, or by showing that this operator has the required action on the two-qubit basis states.
\end{xca}

\begin{xca}\leavevmode
Show that the Hadamard gate turns $\left|+\right\rangle $ back into $\left|0\right\rangle $ and $\left|-\right\rangle $ back into $\left|1\right\rangle $. \textbf{Note:} You don't actually have to do an explicit calculation; you can simply use a certain property of the matrix $H$ itself.
\end{xca}

\begin{problem}\leavevmode
Find an outer product representation for the Hadamard operator in \zcref[S]{eq:Hadamard}.
\end{problem}

\begin{problem}\leavevmode
Show how you can generate each of the four entangled Bell states by acting on the separable state $\left|0\right\rangle \otimes\left|0\right\rangle $ with various quantum gates. This means that quantum gates can be used to generate entanglement if it's not already there.
\end{problem}

\subsubsection{The Measurement Axiom (Projective)}\label{subsec:The-Measurement-Axiom-Projective}

In \zcref[S]{subsec:Probability-Amplitudes}, we formulated the Probability Axiom: for an observable with non-degenerate eigenvalues, if the system is in the state $\left|\Psi\right\rangle $, then the probability to measure the eigenvalue $\lambda_{i}$ corresponding to the eigenstate $\left|B_{i}\right\rangle $ is given by $\left|\langle B_{i}|\Psi\rangle\right|^{2}$. This axiom was good enough at the time, but after all that we have learned since we formulated this axiom, we can now see that it is missing three important things:
\begin{enumerate}
\item It doesn't deal with the case of degenerate eigenvalues.
\item It doesn't tell us what happens if we measure just one part of a composite system.
\item It doesn't take into account dynamics: what happens to the system \textbf{after} we perform the measurement.
\end{enumerate}
To correct that, we now replace the Probability Axiom with a new and improved axiom, which we call the Measurement Axiom. In order to formulate it, let us recall from \zcref[S]{subsec:Spectral-Decomposition} that if $A$ is a normal operator (so in particular, if it is Hermitian and thus an observable), then it has the spectral decomposition
\[
A=\sum^{k}_{i=1}\lambda_{i}P_{i},
\]
where $\lambda_{i}$ are the $k$ distinct eigenvalues (which may be degenerate) and $P_{i}$ are the projectors onto the eigenspaces of these eigenvalues. Using this, we can now define a very general Measurement Axiom, which employs so-called \emph{projective measurements}\index{Projective measurement}.
\begin{quote}
\textbf{The Measurement Axiom (Projective)}\index{Axioms of quantum mechanics!Measurement Axiom (Projective)}\index{Measurement Axiom!Projective}\textbf{:} Consider an observable $A$ on an $n$-dimensional Hilbert space, with $k\le n$ distinct eigenvalues $\left\{ \lambda_{1},\ldots,\lambda_{k}\right\} $, corresponding projectors $\left\{ P_{1},\ldots,P_{k}\right\} $, and spectral decomposition
\[
A=\sum^{k}_{i=1}\lambda_{i}P_{i}.
\]
If the system is in the state $\left|\Psi\right\rangle $, then the probability to measure the eigenvalue $\lambda_{i}$ is given by
\[
\langle\Psi|P_{i}|\Psi\rangle.
\]
The measurement yields exactly one of the eigenvalues $\lambda_{i}$, and after the measurement, the system \emph{collapses}\index{Collapse}\index{State collapse} to the state\footnote{\label{fn:phase}When the projective measurement is applied in the non-degenerate case, where we can also use probability amplitudes via the Probability Axiom, notice that the square root of the probability, $\sqrt{\langle\Psi|P_{i}|\Psi\rangle}$, is not necessarily the probability amplitude. For example, if the amplitude is $\i/2$ then the probability is $1/4$, but the square root of that is $1/2$, which is not the amplitude we started with! However, recall that the two vectors $\left|\Psi\right\rangle $ and $\e^{\i\phi}\left|\Psi\right\rangle $, which differ by an overall phase factor $\e^{\i\phi}$, represent the same state. Since the square root of the probability is the same as the amplitude up to a phase factor, dividing by $\i/2$ or $1/2$ both result in the \textbf{same }state.}
\[
\left|\Psi\right\rangle \mt\frac{P_{i}\left|\Psi\right\rangle }{\sqrt{\langle\Psi|P_{i}|\Psi\rangle}}.
\]
\end{quote}
What does this mean? Since $P_{i}$ is a projector, it satisfies $P_{i}=P^{2}_{i}$ and $P_{i}=P^{\dagger}_{i}$, and therefore $P_{i}=P^{\dagger}_{i}P_{i}$. Therefore the probability can be written as follows:
\[
\langle\Psi|P_{i}|\Psi\rangle=\langle\Psi|P^{\dagger}_{i}P_{i}|\Psi\rangle=\left\Vert P_{i}\Psi\right\Vert ^{2}.
\]
In other words, the probability to measure an eigenvalue is given by projecting the state onto that eigenvalue's eigenspace and taking the norm-squared. Since the projectors satisfy the completeness relation
\[
\sum_{i}P_{i}=1,
\]
we can split the state into the sum of its projections onto each eigenspace:
\[
\left|\Psi\right\rangle =\sum_{i}P_{i}\left|\Psi\right\rangle .
\]
Then, the norm-squared of each term in the sum tells us the corresponding probability. Intuitively, how much of the state vector “fits” in each eigenspace tells us about the probability to measure the corresponding eigenvalue. After the measurement, the system collapses onto the eigenspace corresponding to the eigenvalue that was measured. The denominator simply normalizes that projection to 1:
\[
\left|\Psi\right\rangle \mt\frac{P_{i}\left|\Psi\right\rangle }{\left\Vert P_{i}\Psi\right\Vert }.
\]

\begin{xca}\leavevmode
Find the eigenvalues of the CNOT operator in \zcref[S]{eq:CNOT} and their corresponding eigenvectors and eigenspaces.
\end{xca}

\subsubsection{Applications of the Measurement Axiom}\label{subsec:Applications-of-the-Measurement-Axiom}

Let us now see some examples of the Measurement Axiom in action. First of all, consider a qubit in the general state
\[
\left|\Psi\right\rangle =a\left|0\right\rangle +b\left|1\right\rangle \sp\left|a\right|^{2}+\left|b\right|^{2}=1.
\]
The observable corresponding to the eigenbasis $\left|0\right\rangle ,\left|1\right\rangle $ is the Pauli matrix $\sigma_{z}$, whose spectral decomposition is
\[
\sigma_{z}=\left(\begin{array}{cc}
1 & 0\\
0 & -1
\end{array}\right)=\left|0\right\rangle \left\langle 0\right|-\left|1\right\rangle \left\langle 1\right|.
\]
This means that we have\footnote{Note that I decided to start counting $i$ from 0 to 1 instead of from 1 to 2, so that the subscript of $\lambda_{i}$ will correspond to the value of the qubit. Also, recall that the eigenvalue of $\left|0\right\rangle $ is \textbf{not} 0, it's $+1$, and the eigenvalue of $\left|1\right\rangle $ is \textbf{not} 1, it's $-1$; this is confusing, but unfortunately it's standard notation, since qubits are analogous to classical bits which have the values 0 and 1.}
\[
\lambda_{0}=+1\sp P_{0}=\left|0\right\rangle \left\langle 0\right|,
\]
\[
\lambda_{1}=-1\sp P_{1}=\left|1\right\rangle \left\langle 1\right|.
\]
The probability to measure the eigenvalue $+1$ (corresponding to a value of 0 for the qubit) is thus
\[
\langle\Psi|P_{0}|\Psi\rangle=\langle\Psi|\left(\vphantom{\bll}|0\rangle\langle0|\right)|\Psi\rangle=\langle\Psi|0\rangle\langle0|\Psi\rangle=\left|\langle0|\Psi\rangle\right|^{2}=\left|a\right|^{2},
\]
and the probability to measure the eigenvalue $-1$ (corresponding to a value of 1 for the qubit) is
\[
\langle\Psi|P_{1}|\Psi\rangle=\langle\Psi|\left(\vphantom{\bll}|1\rangle\langle1|\right)|\Psi\rangle=\langle\Psi|1\rangle\langle1|\Psi\rangle=\left|\langle1|\Psi\rangle\right|^{2}=\left|b\right|^{2}.
\]
This indeed matches the old Probability Axiom. The new part is that after the measurement, if we measured 0, then the system will collapse to the state
\[
\left|\Psi\right\rangle \mt\frac{P_{0}\left|\Psi\right\rangle }{\sqrt{\langle\Psi|P_{0}|\Psi\rangle}}=\frac{|0\rangle\langle0|\Psi\rangle}{\left|a\right|}=\frac{a}{\left|a\right|}|0\rangle\simeq\left|0\right\rangle ,
\]
where by $\simeq$ we mean that $\frac{a}{\left|a\right|}|0\rangle$ and $\left|0\right\rangle $ are the same state, since they only differ by a phase factor (see \zcref[S]{fn:phase}; in polar coordinates we have $a=\left|a\right|\e^{\i\phi}$ where $\e^{\i\phi}$ is the phase factor of $a$, so if we divide $a$ by $\left|a\right|$ we are left with just the phase factor). Similarly, if we measured 1 then the system will collapse to the state
\[
\left|\Psi\right\rangle \mt\frac{P_{1}\left|\Psi\right\rangle }{\sqrt{\langle\Psi|P_{1}|\Psi\rangle}}=\frac{|1\rangle\langle1|\Psi\rangle}{\left|b\right|}=\frac{b}{\left|b\right|}|1\rangle\simeq\left|1\right\rangle .
\]

Consider now the general composite state of two qubits given in \zcref[S]{eq:general-two-qubit}:
\[
\left|\Psi\right\rangle =\alpha_{00}\left|0\right\rangle \otimes\left|0\right\rangle +\alpha_{01}\left|0\right\rangle \otimes\left|1\right\rangle +\alpha_{10}\left|1\right\rangle \otimes\left|0\right\rangle +\alpha_{11}\left|1\right\rangle \otimes\left|1\right\rangle ,
\]
which is of course normalized such that
\[
\left\Vert \Psi\right\Vert ^{2}=\left|\alpha_{00}\right|^{2}+\left|\alpha_{01}\right|^{2}+\left|\alpha_{10}\right|^{2}+\left|\alpha_{11}\right|^{2}=1.
\]
We can define an observable corresponding to a measurement of only the \textbf{first} qubit as follows:
\[
\sigma_{z}\otimes1=\left(\vphantom{\bll}\left|0\right\rangle \left\langle 0\right|-\left|1\right\rangle \left\langle 1\right|\right)\otimes1,
\]
where $1$ is the identity operator. So we have
\[
P_{0}=\left|0\right\rangle \left\langle 0\right|\otimes1\sp P_{1}=\left|1\right\rangle \left\langle 1\right|\otimes1.
\]

Note that any observable of the form $A\otimes1$ (or more generally, any tensor product of operators in which one of the factors is the identity operator) always has degenerate eigenvalues, since if
\[
A\left|B_{i}\right\rangle =\lambda_{i}\left|B_{i}\right\rangle ,
\]
then $\left|B_{i}\right\rangle \otimes\left|\Phi\right\rangle $ is an eigenstate of $A\otimes1$ with the eigenvalue $\lambda_{i}$ for any state $\left|\Phi\right\rangle $ of the second system, so the multiplicity of the eigenvalue is multiplied by the dimension of that system. For example, in the case of the operator $\sigma_{z}\otimes1$, both $\left|0\right\rangle \otimes\left|0\right\rangle $ and $\left|0\right\rangle \otimes\left|1\right\rangle $ are eigenstates with the eigenvalue $+1$. This is exactly why we need projectors to measure only part of a composite system.

From the Measurement Axiom, the probability to measure 0 for the first qubit is
\[
\langle\Psi|P_{0}|\Psi\rangle=\langle\Psi|\left(\vphantom{\bll}|0\rangle\langle0|\otimes1\right)|\Psi\rangle.
\]
Let us first calculate\footnote{To understand this calculation, you might want to review how tensor products work, which we discussed in \zcref[S]{subsec:The-Tensor-Product}.} the action of the operator $P_{0}=|0\rangle\langle0|\otimes1$ on the ket $\left|\Psi\right\rangle $:
\begin{align*}
P_{0}\left|\Psi\right\rangle  & =\left(\vphantom{\bll}|0\rangle\langle0|\otimes1\right)|\Psi\rangle\\
 & =\left(\vphantom{\bll}|0\rangle\langle0|\otimes1\right)\left(\vphantom{\bll}\alpha_{00}\left|0\right\rangle \otimes\left|0\right\rangle +\alpha_{01}\left|0\right\rangle \otimes\left|1\right\rangle +\alpha_{10}\left|1\right\rangle \otimes\left|0\right\rangle +\alpha_{11}\left|1\right\rangle \otimes\left|1\right\rangle \right)\\
 & =|0\rangle\otimes\left(\vphantom{\bll}\alpha_{00}\langle0|0\rangle\left|0\right\rangle +\alpha_{01}\langle0|0\rangle\left|1\right\rangle +\alpha_{10}\langle0|1\rangle\left|0\right\rangle +\alpha_{11}\langle0|1\rangle\left|1\right\rangle \right)\\
 & =|0\rangle\otimes\left(\vphantom{\bll}\alpha_{00}\left|0\right\rangle +\alpha_{01}\left|1\right\rangle \right),
\end{align*}
since $\left|0\right\rangle $ and $\left|1\right\rangle $ form an orthonormal basis, so $\langle0|0\rangle=\langle1|1\rangle=1$ and $\langle0|1\rangle=\langle1|0\rangle=0$. Then we act with the bra $\left\langle \Psi\right|$ from the left:
\begin{align*}
\langle\Psi|P_{0}|\Psi\rangle & =\langle\Psi|\left(\vphantom{\bll}|0\rangle\langle0|\otimes1\right)|\Psi\rangle\\
 & =\langle\Psi|\left(\vphantom{\bll}|0\rangle\otimes\left(\alpha_{00}\left|0\right\rangle +\alpha_{01}\left|1\right\rangle \right)\right)\\
 & =\left(\vphantom{\bll}\alpha^{*}_{00}\left\langle 0\right|\otimes\left\langle 0\right|+\alpha^{*}_{01}\left\langle 0\right|\otimes\left\langle 1\right|+\alpha^{*}_{10}\left\langle 1\right|\otimes\left\langle 0\right|+\alpha^{*}_{11}\left\langle 1\right|\otimes\left\langle 1\right|\right)\cdot\\
 & \qquad\cdot\left(\vphantom{\bll}|0\rangle\otimes\left(\vphantom{\bll}\alpha_{00}\left|0\right\rangle +\alpha_{01}\left|1\right\rangle \right)\right)\\
 & =\alpha^{*}_{00}\left\langle 0\right|\left(\vphantom{\bll}\alpha_{00}\left|0\right\rangle +\alpha_{01}\left|1\right\rangle \right)+\alpha^{*}_{01}\left\langle 1\right|\left(\vphantom{\bll}\alpha_{00}\left|0\right\rangle +\alpha_{01}\left|1\right\rangle \right)\\
 & =\left|\alpha_{00}\right|^{2}+\left|\alpha_{01}\right|^{2}.
\end{align*}
Similarly, we also find that the probability to measure 1 for the first qubit is
\[
\langle\Psi|P_{1}|\Psi\rangle=\langle\Psi|\left(\vphantom{\bll}|1\rangle\langle1|\otimes1\right)|\Psi\rangle=\left|\alpha_{10}\right|^{2}+\left|\alpha_{11}\right|^{2}.
\]
These very complicated calculations tell us what we could have just guessed from common sense: the total probability to measure $\left|0\right\rangle $ is the sum of the probabilities to measure all the composite states which have $\left|0\right\rangle $ as the state of the first qubit, and similarly for $\left|1\right\rangle $.

What about collapse? If we measured 0, then the system will collapse to the state
\[
\frac{P_{0}\left|\Psi\right\rangle }{\sqrt{\langle\Psi|P_{0}|\Psi\rangle}}=\frac{|0\rangle\otimes\left(\alpha_{00}|0\rangle+\alpha_{01}|1\rangle\right)}{\sqrt{\left|\alpha_{00}\right|^{2}+\left|\alpha_{01}\right|^{2}}},
\]
and if we measured 1, it will collapse to the state
\[
\frac{P_{1}\left|\Psi\right\rangle }{\sqrt{\langle\Psi|P_{1}|\Psi\rangle}}=\frac{|1\rangle\otimes\left(\alpha_{10}|0\rangle+\alpha_{11}|1\rangle\right)}{\sqrt{\left|\alpha_{10}\right|^{2}+\left|\alpha_{11}\right|^{2}}}.
\]
Again, we could have just guessed the result: the qubit that we measured collapses into either $\left|0\right\rangle $ or $\left|1\right\rangle $, while the other qubit is left in a superposition (assuming both coefficients are non-zero). The denominator is there simply to normalize the vector so it has norm 1, and can thus represent a state.
\begin{problem}\leavevmode
\label{prob:Consider-a-composite}Consider a composite system of three qubits. Which projectors will you use to measure only the state of the middle qubit in the $\left|+\right\rangle ,\left|-\right\rangle $ eigenbasis? Which projectors will you use to measure only the state of the first two qubits in the $\left|0\right\rangle ,\left|1\right\rangle $ eigenbasis?
\end{problem}

\subsubsection{The Measurement Axiom (Simplified)}

Now that we understand how projective measurements work, we can formulate a simpler version of the Measurement Axiom, which does not explicitly use projectors, and will be sufficient for most of our purposes in the rest of these notes. In the simple case where there are no degenerate eigenvalues, if the eigenvalues are $\lambda_{i}$ and the eigenbasis is $\left|B_{i}\right\rangle $, the projectors are simply
\[
P_{i}=\left|B_{i}\right\rangle \left\langle B_{i}\right|.
\]
If the system is in the state $\left|\Psi\right\rangle $, then the probability to measure the eigenvalue $\lambda_{i}$ is given by
\[
\langle\Psi|P_{i}|\Psi\rangle=\langle\Psi|B_{i}\rangle\langle B_{i}|\Psi\rangle=\left|\langle B_{i}|\Psi\rangle\right|^{2},
\]
so we recover the probability amplitudes of the Probability Axiom. The measurement yields exactly one of the eigenvalues $\lambda_{i}$, and after the measurement, the system collapses to
\[
\left|\Psi\right\rangle \mt\frac{P_{i}\left|\Psi\right\rangle }{\sqrt{\langle\Psi|P_{i}|\Psi\rangle}}=\frac{|B_{i}\rangle\langle B_{i}|\Psi\rangle}{\left|\langle B_{i}|\Psi\rangle\right|},
\]
but the fraction $\langle B_{i}|\Psi\rangle/\left|\langle B_{i}|\Psi\rangle\right|$ is just some phase factor $\e^{\i\phi}$ (since a complex number divided by its magnitude is just its phase factor), and states are defined only up to a phase factor, so we can simply say that $\left|\Psi\right\rangle \mt\left|B_{i}\right\rangle $. The simplified Measurement Axiom can therefore be formulated as follows:
\begin{quote}
\textbf{The Measurement Axiom (Simplified)}\index{Axioms of quantum mechanics!Measurement Axiom (Simplified)}\index{Measurement Axiom!Simplified}\textbf{:}
\begin{itemize}
\item Consider an observable with an eigenbasis of eigenstates $\left|B_{i}\right\rangle $ corresponding to non-degenerate eigenvalues $\lambda_{i}$. If the system is in the state $\left|\Psi\right\rangle $, then the probability to measure the eigenvalue $\lambda_{i}$ corresponding to the eigenstate $\left|B_{i}\right\rangle $ is given by
\[
\left|\langle B_{i}|\Psi\rangle\right|^{2}.
\]
After the measurement, if the eigenvalue $\lambda_{i}$ was measured, then the system will collapse to the eigenstate $\left|B_{i}\right\rangle $:
\[
\left|\Psi\right\rangle \mt\left|B_{i}\right\rangle .
\]
This works the same whether the system in question is composite or not, provided that the measurement is performed on the entire system at once.
\item Now consider a composite system and an observable defined only on part of that system, with eigenstates $\left|B_{i}\right\rangle $ corresponding to non-degenerate eigenvalues $\lambda_{i}$. The total probability to measure the eigenvalue $\lambda_{i}$ is the sum of the probabilities for all the possible ways in which this eigenvalue can be measured – that is, the sum of the magnitude-squared of the probability amplitudes of all the composite states where the part being measured is in the eigenstate $\left|B_{i}\right\rangle $. After the measurement, if the eigenvalue $\lambda_{i}$ was measured, then only the system we measured will collapse to the eigenstate $\left|B_{i}\right\rangle $, while the other systems may still be in a superposition.
\end{itemize}
\end{quote}
The process described by the Measurement Axiom, where the state of the system changes after a measurement, is what people mean when they talk about \emph{wavefunction collapse}\index{Wavefunction collapse}. However, we haven't yet defined what a “wavefunction” is. This is because in the modern abstract formulation of quantum mechanics, which is what we have been studying so far, \textbf{states} are the fundamental entities, not wavefunctions. We will explain this in more detail when we define wavefunctions in \zcref[S]{subsec:Wavefunctions}.
\begin{problem}\leavevmode
A composite system of two qubits is in the state
\[
\left|\Psi\right\rangle =\frac{1}{\sqrt{14}}\left(2\left|00\right\rangle -\i\left|10\right\rangle +3\left|11\right\rangle \right).
\]
A measurement is performed on only the first qubit, in the $\left|+\right\rangle ,\left|-\right\rangle $ eigenbasis. For each of the two possible outcomes, what is the probability to measure that outcome and what will be the state of the system after the measurement?
\end{problem}

\begin{problem}\leavevmode
A composite system of three qubits is in the state
\[
\left|\Psi\right\rangle =\frac{1}{\sqrt{35}}\left(\left|000\right\rangle +2\left|010\right\rangle -3\i\left|011\right\rangle -4\left|101\right\rangle +\i\left|110\right\rangle +2\i\left|111\right\rangle \right),
\]
where $\left|000\right\rangle \equiv\left|0\right\rangle \otimes\left|0\right\rangle \otimes\left|0\right\rangle $ and so on. A measurement is performed on only the \textbf{first two} qubits in the $\left|0\right\rangle ,\left|1\right\rangle $ basis. For each of the \textbf{four} possible outcomes, what is the probability to measure that outcome and what will be the state of the system after the measurement? You can either solve this problem by inspection using the simplified axiom, or by explicit calculation using the projectors you found in \zcref[S]{prob:Consider-a-composite}.
\end{problem}

\subsection{Bell's Theorem and Interpretations of Quantum Mechanics}\label{subsec:Bell}

\subsubsection{Setting Up the Experiment}\label{subsec:Setting-Up-the-Experiment}

Consider the following experiment. There is a machine which, when a certain button is pressed, produces two sealed boxes. No one knows what's going on inside the machine, and no one can open the boxes. Alice and Bob press the button, and each of them takes one of the sealed boxes to their respective labs. At the labs, they each have an apparatus that can probe the contents of the box using two different methods; they choose which method to use via a switch that has two settings, 0 or 1. The apparatus also has a screen that displays one of two possible measurement outcomes: $-1$ or $+1$. Alice chooses a setting bit $x\in\left\{ 0,1\right\} $ at random (for example by a coin flip) and records an outcome $a\in\left\{ -1,+1\right\} $, while Bob chooses a setting bit $y\in\left\{ 0,1\right\} $ at random and records an outcome $b\in\left\{ -1,+1\right\} $. Each of them performs only one measurement.

Alice and Bob repeat this experiment many times, each time with a fresh pair of boxes, one set of random choices of $x$ and $y$, and one set of outcomes $a,b$. This allows them to define random variables $X,Y$ for the settings and $A,B$ for the outcomes, for Alice and Bob respectively. We assume that each of the four combinations of $x,y\in\left\{ 0,1\right\} $ was chosen at least once\footnote{If this is not the case, they can just continue doing more experiments until this assumption is satisfied.}, so
\[
p\left(x,y\right)\equiv P\left(X=x\cap Y=y\right)>0\sp\fa x,y\in\left\{ 0,1\right\} ,
\]
with\footnote{Throughout this section, we will always use uppercase $P$ for the full probability specification in terms of the random variables and their values, for example $P\left(X=x\cap Y=y\right)$ here, and lowercase $p$ as a shorthand in terms of the values only, for example $p\left(x,y\right)$ here. When lowercase $p$ is used, the random variables involved can always be inferred from the values, as we will consistently use $A=a$, $B=b$, $X=x$, and $Y=y$.}
\[
\sum_{x,y\in\left\{ 0,1\right\} }p\left(x,y\right)=1.
\]

Once Alice and Bob have enough data, they can calculate the conditional probabilities (as defined in \zcref[S]{subsec:Conditional-Probability}) to obtain the outcomes $a$ and $b$ given the settings $x$ and $y$:
\[
p\mr{a,b}{x,y}\equiv P\mr{A=a\cap B=b}{X=x\cap Y=y}\ge0,
\]
for all 16 combinations of $a,b\in\left\{ -1,+1\right\} $ and $x,y\in\left\{ 0,1\right\} $. These probabilities satisfy
\[
\sum_{a,b\in\left\{ -1,+1\right\} }p\mr{a,b}{x,y}=1\sp\fa x,y\in\left\{ 0,1\right\} .
\]
That is, for each choice of $x$ and $y$, the sum of the probabilities for all four possible outcomes is 1. They also calculate the quantities
\begin{equation}
\left\langle A_{x}B_{y}\right\rangle \equiv\sum_{a,b\in\left\{ -1,+1\right\} }ab\thinspace p\mr{a,b}{x,y}\sp\fa x,y\in\left\{ 0,1\right\} .\label{eq:mean-of-products}
\end{equation}
$\left\langle A_{x}B_{y}\right\rangle $ is just the mean of the products of the two outcomes $a$ and $b$, conditional on the settings $x$ and $y$. Since $ab=+1$ for equal outcomes and $ab=-1$ for opposite outcomes, $\left\langle A_{x}B_{y}\right\rangle $ is also the total probability of equal outcomes minus the total probability of opposite outcomes:
\[
\left\langle A_{x}B_{y}\right\rangle =p\mr{a=b}{x,y}-p\mr{a\ne b}{x,y}.
\]
The largest possible value of $\left\langle A_{x}B_{y}\right\rangle $ is $1$, obtained when $p\mr{a\ne b}{x,y}=0$, in which case we must have $p\mr{a=b}{x,y}=1$. Conversely, the smallest possible value of $\left\langle A_{x}B_{y}\right\rangle $ is $-1$. In other words, $\left\langle A_{x}B_{y}\right\rangle $ is equal to $+1$ when the two outcomes $a$ and $b$ always agree for the settings $x$ and $y$, $-1$ when they always disagree, and $0$ if they agree exactly half the time.

Finally, Alice and Bob calculate the \emph{CHSH expression}\index{CHSH expression}\footnote{CHSH is named after Clauser, Horne, Shimony, and Holt, who first derived the CHSH inequality, which we will prove in \zcref[S]{subsec:Correlations-and-Hidden}.}
\begin{equation}
S\equiv\left\langle A_{0}B_{0}\right\rangle +\left\langle A_{0}B_{1}\right\rangle +\left\langle A_{1}B_{0}\right\rangle -\left\langle A_{1}B_{1}\right\rangle .\label{eq:CHSH-expression}
\end{equation}
Since each term is in the range $\left[-1,+1\right]$, the largest possible value of $S$ is $4$, obtained when the first three terms are $+1$ and the last term is $-1$, and conversely the smallest possible value of $S$ is $-4$, so
\[
\left|S\right|\le4.
\]
Alternatively, we can also derive this using the triangle inequality, \zcref[S]{eq:triangle-inequality-general}:
\begin{align*}
\left|S\right| & \le\left|\left\langle A_{0}B_{0}\right\rangle \right|+\left|\left\langle A_{0}B_{1}\right\rangle \right|+\left|\left\langle A_{1}B_{0}\right\rangle \right|+\left|-\left\langle A_{1}B_{1}\right\rangle \right|\\
 & \le1+1+1+1=4.
\end{align*}

\begin{xca}\leavevmode
Find an example of conditional probabilities $p\mr{a,b}{x,y}$ for which $S=4$. In your example, does the probability for Bob's outcome depend on Alice's setting?
\end{xca}

\subsubsection{Correlations and Hidden Variables}\label{subsec:Correlations-and-Hidden}

Let us now assume that the experimental data shows
\[
p\mr{a,b}{x,y}\ne p\mr a{x,y}p\mr b{x,y},
\]
meaning that Alice's and Bob's measurements are correlated. They wonder if this might be because the two boxes are secretly communicating with each other. They repeat the experiment, but this time ensure the measurements are causally disconnected, for example by choosing the settings and performing the measurements at the exact same time but in two distant locations. However, they find that the correlations still exist. This can no longer be explained by a causal relationship between the boxes, because that would require information to travel faster than light, violating relativity. They conclude that the causal relationship must instead be between each of the boxes and the original machine.

A simple example of such correlations happens when the machine flips a fair coin, with heads corresponding to $+1$ and tails to $-1$. It then prepares both boxes so that any measurement, regardless of the settings $x$ and $y$, produces the result of the coin. In this example, there is some extra information that is “hidden” from us, because we can't look into the machine, namely the result of the coin flip. In general, with any correlation, there must be similar extra information that is responsible for the correlation.

We are therefore motivated to introduce a \emph{hidden variable theory}\index{Hidden variable theory}: a theory containing some random variable $\Lambda$, with value\footnote{For simplicity, we will assume that $\lambda$ is taken from a finite or countable set. If $\lambda$ can take continuous values, the sums over $\lambda$ below should be replaced with integrals, but the proof proceeds similarly.} $\lambda$, that may be hidden from us, and affects the results of the measurements. Therefore, the measurement outcomes are now conditional not only on the settings $x$ and $y$, but also on the hidden variable $\lambda$, and we need to look at probabilities of the form
\[
p\mr{a,b}{x,y,\lambda}\equiv P\mr{A=a\cap B=b}{X=x\cap Y=y\cap\Lambda=\lambda}.
\]

We make two crucial assumptions about the hidden variable:
\begin{enumerate}
\item \textbf{Measurement independence.}\index{Measurement independence} The hidden variable is statistically independent of the measurement settings:
\[
p\mr{\lambda}{x,y}=p\left(\lambda\right)\sp\fa x,y,\lambda.
\]
Note that statistical independence from each setting separately is not enough. As an example, consider the case where the machine imposes the relationship $\Lambda=X\oplus Y$, where $\oplus$ means addition modulo 2, that is,
\[
0\oplus0=0\sp0\oplus1=1\sp1\oplus0=1\sp1\oplus1=0.
\]
If we fix one of the setting bits at either 0 or 1, but keep the other bit random and equally likely to be 0 or 1, then you can check that
\[
p\mr{\lambda}x=p\mr{\lambda}y=p\left(\lambda\right)=\hf.
\]
However, if we fix both settings, then $\lambda=x\oplus y$ is determined uniquely. Therefore, the hidden variable can be independent from each setting separately while still being dependent on both together.
\item \textbf{Local causality (or factorization).}\index{Local causality} Conditional on the hidden variable, the joint outcome probabilities factorize to the product of an Alice-only probability and a Bob-only probability:
\[
p\mr{a,b}{x,y,\lambda}=p\mr a{x,\lambda}p\mr b{y,\lambda}.
\]
This means that $\lambda$, on its own, is sufficient to explain any correlations; both Alice's and Bob's measurements are completely independent of the other person's outcome or setting, but they can both depend on $\lambda$. Local causality can be split into two sub-assumptions:
\begin{enumerate}
\item \textbf{Outcome independence.}\index{Outcome independence} Alice's outcome does not depend on Bob's outcome, and vice versa:
\[
p\mr{a,b}{x,y,\lambda}=p\mr a{x,y,\lambda}p\mr b{x,y,\lambda}.
\]
\item \textbf{Setting independence.}\index{Setting independence} Alice's outcome does not depend on Bob's setting, and vice versa:
\[
p\mr a{x,y,\lambda}=p\mr a{x,\lambda},
\]
\[
p\mr b{x,y,\lambda}=p\mr b{y,\lambda}.
\]
\end{enumerate}
\end{enumerate}
A hidden variable theory that satisfies measurement independence and local causality is called a \emph{local hidden variable theory}\index{Local hidden variable theory}\index{Hidden variable theory!Local}.

With these assumptions, we can split the outcome probability into a sum over the possible values of the hidden variable as follows:
\begin{align*}
p\mr{a,b}{x,y} & =\sum_{\lambda}p\mr{\lambda}{x,y}p\mr{a,b}{x,y,\lambda}\\
 & =\sum_{\lambda}p\left(\lambda\right)p\mr a{x,\lambda}p\mr b{y,\lambda}.
\end{align*}
In the first line we used the law of total probability in its conditional form, \zcref[S]{eq:total-probability-conditional}. In the second line we used measurement independence on the first factor and local causality on the second. Substituting this into the means defined in \zcref[S]{eq:mean-of-products} gives:
\begin{align*}
\left\langle A_{x}B_{y}\right\rangle  & =\sum_{a,b}ab\thinspace p\mr{a,b}{x,y}\\
 & =\sum_{a,b}ab\sum_{\lambda}p\left(\lambda\right)p\mr a{x,\lambda}p\mr b{y,\lambda}\\
 & =\sum_{\lambda}p\left(\lambda\right)\sum_{a}a\thinspace p\mr a{x,\lambda}\sum_{b}b\thinspace p\mr b{y,\lambda}.
\end{align*}
Let us define the quantities
\[
\alpha_{x}\left(\lambda\right)\equiv\sum_{a}a\thinspace p\mr a{x,\lambda},
\]
\[
\beta_{y}\left(\lambda\right)\equiv\sum_{b}b\thinspace p\mr b{y,\lambda},
\]
which are the mean outcomes of Alice and Bob respectively for the settings $x,y$ and the hidden variable $\lambda$. Note that
\[
\alpha_{x}\left(\lambda\right)=p\mr{a=+1}{x,\lambda}-p\mr{a=-1}{x,\lambda},
\]
hence $\left|\alpha_{x}\left(\lambda\right)\right|\le1$, and similarly $\left|\beta_{y}\left(\lambda\right)\right|\le1$. Substituting these into $\left\langle A_{x}B_{y}\right\rangle $, we get
\[
\left\langle A_{x}B_{y}\right\rangle =\sum_{\lambda}p\left(\lambda\right)\alpha_{x}\left(\lambda\right)\beta_{y}\left(\lambda\right).
\]
Plugging this into the CHSH expression, \zcref[S]{eq:CHSH-expression}, we get
\[
S=\sum_{\lambda}p\left(\lambda\right)\left(\alpha_{0}\left(\lambda\right)\beta_{0}\left(\lambda\right)+\alpha_{0}\left(\lambda\right)\beta_{1}\left(\lambda\right)+\alpha_{1}\left(\lambda\right)\beta_{0}\left(\lambda\right)-\alpha_{1}\left(\lambda\right)\beta_{1}\left(\lambda\right)\right).
\]
Let us define for brevity
\[
S=\sum_{\lambda}p\left(\lambda\right)s\left(\lambda\right),
\]
where
\[
s\equiv\alpha_{0}\beta_{0}+\alpha_{0}\beta_{1}+\alpha_{1}\beta_{0}-\alpha_{1}\beta_{1}
\]
and we suppressed the argument $\lambda$ for further brevity. We can factorize $s$ as follows:
\begin{align*}
s & =\alpha_{0}\beta_{0}+\alpha_{0}\beta_{1}+\alpha_{1}\beta_{0}-\alpha_{1}\beta_{1}\\
 & =\alpha_{0}\left(\beta_{0}+\beta_{1}\right)+\alpha_{1}\left(\beta_{0}-\beta_{1}\right).
\end{align*}
Using the triangle inequality, \zcref[S]{eq:triangle-inequality-general}, we find
\begin{align*}
\left|s\right| & \le\left|\alpha_{0}\right|\left|\beta_{0}+\beta_{1}\right|+\left|\alpha_{1}\right|\left|\beta_{0}-\beta_{1}\right|\\
 & \le\left|\beta_{0}+\beta_{1}\right|+\left|\beta_{0}-\beta_{1}\right|,
\end{align*}
where we used the fact that $\left|\alpha_{x}\right|\le1$. Assume without loss of generality that $\left|\beta_{0}\right|\ge\left|\beta_{1}\right|$ (exchanging them leaves the upper bound unchanged). Then $\beta_{0}+\beta_{1}$ and $\beta_{0}-\beta_{1}$ cannot have opposite signs, so the sum of their absolute values is the absolute value of their sum:
\[
\left|s\right|\le\left|\beta_{0}+\beta_{1}\right|+\left|\beta_{0}-\beta_{1}\right|=\left|\beta_{0}+\beta_{1}+\beta_{0}-\beta_{1}\right|=2\left|\beta_{0}\right|.
\]
Since $\left|\beta_{0}\right|\le1$, we conclude that
\[
\left|s\right|\le2.
\]
Lastly, we have from the triangle inequality
\[
\left|S\right|\le\sum_{\lambda}p\left(\lambda\right)\left|s\left(\lambda\right)\right|\le2\sum_{\lambda}p\left(\lambda\right)=2,
\]
since the probabilities sum to 1. In conclusion, in \zcref[S]{subsec:Setting-Up-the-Experiment} we showed that $\left|S\right|\le4$, but in this section we see that with the assumptions of measurement independence and local causality, we get a much stricter bound of
\begin{equation}
\left|S\right|\le2.\label{eq:CHSH-inequality}
\end{equation}
This is known as the \emph{CHSH inequality}\index{CHSH inequality}.
\begin{problem}\leavevmode
\label{prob:Prove-that-local}Prove that local causality is equivalent to outcome independence and setting independence together. That is, prove that local causality implies both outcome independence and setting independence, and conversely, that outcome independence and setting independence together imply local causality.
\end{problem}

\subsubsection{The Quantum Experiment}

What if the mysterious machine creates quantum systems, not classical ones? Let us assume that this machine generates a pair of qubits, and each qubit is placed in a sealed box. Alice and Bob take one box each, which means each of them now has one of the two qubits. Let the Hilbert spaces of Alice's and Bob's qubits be $\HH_{A}$ and $\HH_{B}$ respectively, and let their joint state be
\[
\left|\Psi\right\rangle \in\HH_{A}\otimes\HH_{B}.
\]
The setting on each apparatus selects which observable will be measured for each qubit. Alice can select one of the Hermitian operators $A_{0}$ or $A_{1}$ on $\HH_{A}$, which we denote $A_{x}$ where $x\in\left\{ 0,1\right\} $, and Bob can select one of the Hermitian operators $B_{0}$ or $B_{1}$ on $\HH_{B}$, which we denote $B_{y}$ where $y\in\left\{ 0,1\right\} $. The Hermitian operators are chosen such that their eigenvalues are $-1$ and $+1$. We define the corresponding eigenstates as follows:
\[
A_{x}\left|a_{x}\right\rangle =a\left|a_{x}\right\rangle \sp B_{y}\left|b_{y}\right\rangle =b\left|b_{y}\right\rangle ,
\]
where
\[
x,y\in\left\{ 0,1\right\} \sp a,b\in\left\{ -1,+1\right\} ,
\]
so that Alice's measurement result is $a$ and Bob's measurement result is $b$. These observables have the spectral decompositions (recall \zcref[S]{subsec:Spectral-Decomposition})
\[
A_{x}=\sum_{a\in\left\{ -1,+1\right\} }a\left|a_{x}\right\rangle \left\langle a_{x}\right|\sp B_{y}=\sum_{b\in\left\{ -1,+1\right\} }b\left|b_{y}\right\rangle \left\langle b_{y}\right|.
\]
This notation may be a bit confusing, so let us write down all the states explicitly:
\begin{itemize}
\item Alice's first observable $A_{0}$ has the eigenstates $\left|+1_{0}\right\rangle $ and $\left|-1_{0}\right\rangle $ and her second observable $A_{1}$ has the eigenstates $\left|+1_{1}\right\rangle $ and $\left|-1_{1}\right\rangle $.
\item Bob's first observable $B_{0}$ has the eigenstates $\left|+1_{0}\right\rangle $ and $\left|-1_{0}\right\rangle $ and his second observable $B_{1}$ has the eigenstates $\left|+1_{1}\right\rangle $ and $\left|-1_{1}\right\rangle $.
\end{itemize}
We use the same notation for Alice's and Bob's eigenstates because their eigenvalues are the same, but the difference is that Alice's eigenstates live in $\HH_{A}$ (on the left of the tensor product) and Bob's eigenstates live in $\HH_{B}$ (on the right of the tensor product). This works exactly the same as in systems of multiple qubits, where each qubit can have states $\left|0\right\rangle $ and $\left|1\right\rangle $ but in a different Hilbert space.

Alice measures only her own qubit, so her measurement uses the observable $A_{x}\otimes1$ on the composite Hilbert space (recall \zcref[S]{subsec:Applications-of-the-Measurement-Axiom}). Note that the eigenvalues of this observable are degenerate, with multiplicity 2; each eigenvalue $a\in\left\{ -1,+1\right\} $ has an eigenspace spanned by the basis
\[
\left|a_{x}\right\rangle \otimes\left|0\right\rangle \sp\left|a_{x}\right\rangle \otimes\left|1\right\rangle .
\]
The projectors onto these eigenspaces are
\[
P_{a}\equiv\left|a_{x}\right\rangle \left\langle a_{x}\right|\otimes1,
\]
and therefore the observable can be written as a spectral decomposition in terms of these projectors:
\[
A_{x}\otimes1=\sum_{a\in\left\{ -1,+1\right\} }a\left(\left|a_{x}\right\rangle \left\langle a_{x}\right|\otimes1\right).
\]
Similarly, Bob measures only his own qubit, using the observable
\[
1\otimes B_{y}=\sum_{b\in\left\{ -1,+1\right\} }b\left(1\otimes\left|b_{y}\right\rangle \left\langle b_{y}\right|\right).
\]

Let us assume that Alice measures her qubit first. By the projective Measurement Axiom (see \zcref[S]{subsec:The-Measurement-Axiom-Projective}), the probability that she obtains the outcome $a$ is
\[
p\mr a{x,y}=\left\langle \Psi\right|\left(\left|a_{x}\right\rangle \left\langle a_{x}\right|\otimes1\right)\left|\Psi\right\rangle ,
\]
and if that outcome is obtained, the state of the two qubits collapses onto the corresponding eigenspace:
\[
\left|\Psi\right\rangle \mt\left|\Psi_{a}\right\rangle \equiv\frac{\left(\left|a_{x}\right\rangle \left\langle a_{x}\right|\otimes1\right)\left|\Psi\right\rangle }{\sqrt{p\mr a{x,y}}}.
\]
Bob measures his qubit next, so his measurement is applied to the collapsed state $\left|\Psi_{a}\right\rangle $. By the projective Measurement Axiom, the probability that he obtains the outcome $b$, given that Alice obtained the outcome $a$, is
\begin{align*}
p\mr b{a,x,y} & =\left\langle \Psi_{a}\right|\left(1\otimes\left|b_{y}\right\rangle \left\langle b_{y}\right|\right)\left|\Psi_{a}\right\rangle \\
 & =\frac{\left\langle \Psi\right|\left(\left|a_{x}\right\rangle \left\langle a_{x}\right|\otimes1\right)}{\sqrt{p\mr a{x,y}}}\left(1\otimes\left|b_{y}\right\rangle \left\langle b_{y}\right|\right)\frac{\left(\left|a_{x}\right\rangle \left\langle a_{x}\right|\otimes1\right)\left|\Psi\right\rangle }{\sqrt{p\mr a{x,y}}}\\
 & =\frac{\left\langle \Psi\right|\left(\left|a_{x}\right\rangle \left\langle a_{x}\right|\otimes\left|b_{y}\right\rangle \left\langle b_{y}\right|\right)\left|\Psi\right\rangle }{p\mr a{x,y}},
\end{align*}
where we used the fact that
\[
\left(1\otimes\left|b_{y}\right\rangle \left\langle b_{y}\right|\right)\left(\left|a_{x}\right\rangle \left\langle a_{x}\right|\otimes1\right)=\left|a_{x}\right\rangle \left\langle a_{x}\right|\otimes\left|b_{y}\right\rangle \left\langle b_{y}\right|,
\]
because each operator acts only on its own Hilbert space, and
\[
\left(\left|a_{x}\right\rangle \left\langle a_{x}\right|\right)\left(\left|a_{x}\right\rangle \left\langle a_{x}\right|\right)=\left|a_{x}\right\rangle \left\langle a_{x}\right|.
\]
If we multiply both sides by $p\mr a{x,y}$, we get
\begin{align*}
p\mr a{x,y}p\mr b{a,x,y} & =\left\langle \Psi\right|\left(\left|a_{x}\right\rangle \left\langle a_{x}\right|\otimes\left|b_{y}\right\rangle \left\langle b_{y}\right|\right)\left|\Psi\right\rangle \\
 & =\left\langle \Psi\right|\left(\left|a_{x}\right\rangle \otimes\left|b_{y}\right\rangle \right)\left(\left\langle a_{x}\right|\otimes\left\langle b_{y}\right|\right)\left|\Psi\right\rangle \\
 & =\langle\Psi|a_{x}b_{y}\rangle\langle a_{x}b_{y}|\Psi\rangle\\
 & =\left|\langle a_{x}b_{y}|\Psi\rangle\right|^{2},
\end{align*}
where to get to the second line we used \zcref[S]{eq:outer-tensor}, and we defined
\[
\left|a_{x}b_{y}\right\rangle \equiv\left|a_{x}\right\rangle \otimes\left|b_{y}\right\rangle .
\]
From \zcref[S]{eq:conditional-joint} (making all the probabilities conditional on $x,y$), we get that
\[
p\mr{a,b}{x,y}=p\mr b{a,x,y}p\mr a{x,y},
\]
but that's the expression we already calculated. We conclude that the probability that Alice obtains the outcome $a$ and Bob obtains the outcome $b$ is:
\[
p\mr{a,b}{x,y}=\left|\langle a_{x}b_{y}|\Psi\rangle\right|^{2}.
\]
Note that the result we got is just the usual Born rule, but we did it carefully step by step because we had degenerate eigenvalues and a multi-step measurement. If we repeat the proof with the roles of Alice and Bob exchanged, we see that if Bob measures first and Alice second, we will get the same result. This means that either Alice can measure first, Bob can measure first, or they can both measure at the same time\footnote{In fact, according to special relativity, simultaneity is relative, so who went first depends on the observer: some observers see Alice go first, some see Bob go first, and others see them go at the same time. We have therefore proved that all observers will agree on the probabilities, which is important. In relativistic terms, Alice's and Bob's measurement events are assumed to be separated by a spacelike interval, which means no information can pass between the measurements, and this fact is also observer-independent. We won't go into more details, since relativity is out of scope for these notes.}, without changing the result. The reason is that Alice's and Bob's observables commute, so they can act in either order without changing the outcome:
\begin{align*}
\left[A_{x}\otimes1,1\otimes B_{y}\right] & =\left(A_{x}\otimes1\right)\left(1\otimes B_{y}\right)-\left(1\otimes B_{y}\right)\left(A_{x}\otimes1\right)\\
 & =A_{x}\otimes B_{y}-A_{x}\otimes B_{y}\\
 & =0.
\end{align*}

The expressions $\left\langle A_{x}B_{y}\right\rangle $ defined in \zcref[S]{eq:mean-of-products} are the expectation values of the product of outcomes $ab$, which corresponds to the product of Alice's and Bob's observables:
\[
\left(A_{x}\otimes1\right)\left(1\otimes B_{y}\right)=A_{x}\otimes B_{y}.
\]
Therefore, we can calculate them as follows:
\[
\left\langle A_{x}B_{y}\right\rangle =\langle\Psi|A_{x}\otimes B_{y}|\Psi\rangle=\sum_{a,b\in\left\{ -1,+1\right\} }ab\thinspace p\mr{a,b}{x,y},
\]
for each choice of $x,y$.

To do a concrete calculation, we must choose the operators $A_{x}$ and $B_{y}$. Let us define them as follows\footnote{Do not confuse the Pauli matrix $X$, which is used here to define the observables, with the random variable $X$, which was used in \zcref[S]{subsec:Setting-Up-the-Experiment,subsec:Correlations-and-Hidden} to denote Alice's measurement setting.}:
\[
A_{0}\equiv Z\sp A_{1}\equiv X\sp B_{0}\equiv\frac{1}{\sqrt{2}}\left(Z+X\right)\sp B_{1}\equiv\frac{1}{\sqrt{2}}\left(Z-X\right),
\]
where $X$ and $Z$ are Pauli matrices, in the notation introduced in \zcref[S]{subsec:Quantum-Logic-Gates}. Note that all four observables have eigenvalues $\pm1$, as you should check.

For the composite state $\left|\Psi\right\rangle $, we use the Bell state $\left|\beta_{00}\right\rangle $ defined in \zcref[S]{subsec:The-Bell-States}:
\[
\left|\Psi\right\rangle \equiv\frac{1}{\sqrt{2}}\left(\left|00\right\rangle +\left|11\right\rangle \right).
\]
This is an entangled state; if both qubits are measured with $Z$, the outcomes will always match. As a reminder, we use 0 and 1 as the qubit labels in analogy with classical bits, but the $Z$ eigenvalues are confusingly $+1$ for the eigenstate $\left|0\right\rangle $ and $-1$ for the eigenstate $\left|1\right\rangle $, so this does match the required eigenvalues in our experiment. Of course, we could also get correlation between Alice's and Bob's measurements with classical randomness alone, as in the coin example of \zcref[S]{subsec:Correlations-and-Hidden}. However, we will now see that quantum entanglement produces different predictions.

In \zcref[S]{exer:Prove-CHSH-sqrt2} you will prove that
\begin{equation}
\left\langle A_{0}B_{0}\right\rangle =\left\langle A_{0}B_{1}\right\rangle =\left\langle A_{1}B_{0}\right\rangle =\frac{1}{\sqrt{2}}\sp\left\langle A_{1}B_{1}\right\rangle =-\frac{1}{\sqrt{2}}.\label{eq:CHSH-sqrt2}
\end{equation}
\zcref[S]{eq:CHSH-expression} for the quantum state is therefore (note how the minus sign in front of the $1,1$ term cancels out the minus sign in the expectation value):
\begin{equation}
S=\left\langle A_{0}B_{0}\right\rangle +\left\langle A_{0}B_{1}\right\rangle +\left\langle A_{1}B_{0}\right\rangle -\left\langle A_{1}B_{1}\right\rangle =\frac{4}{\sqrt{2}}=2\sqrt{2}>2.\label{eq:Bell-quantum}
\end{equation}
The quantum experiment produces a value of $2\sqrt{2}$ for $S$, which is greater than the local hidden variable experiment's $\left|S\right|\le2$.
\begin{xca}\leavevmode
\label{exer:Prove-CHSH-sqrt2}Prove \zcref[S]{eq:CHSH-sqrt2}.
\end{xca}

\subsubsection{Interpretations of Quantum Mechanics}

In the previous sections, we showed that any local hidden variable theory must obey the CHSH inequality $\left|S\right|\le2$, \zcref[S]{eq:CHSH-inequality}, but quantum mechanics with a suitable choice of state and observables predicts \zcref[S]{eq:Bell-quantum}, $S=2\sqrt{2}$. This proves \emph{Bell's theorem}\index{Bell's theorem}, which states that no local hidden variable theory can reproduce all the predictions of quantum mechanics.

At the end of \zcref[S]{subsec:Entanglement-Does-Not}, we asked: how does giving Alice and Bob correlated classical bits – two sealed envelopes, one with a piece of paper that says 0 and another with a piece of paper that says 1 – differ from giving them entangled qubits? Now we can answer that question: the envelopes describe a local hidden variable theory, where the hidden variable $\lambda$ determines which piece of paper was placed in which envelope (for example, $\lambda$ could indicate Alice's bit, and then $1-\lambda$ automatically indicates Bob's bit). This kind of classical correlation will always satisfy the CHSH inequality, no matter how Alice and Bob choose to probe it. However, entangled qubits can violate this inequality, so the correlation created by entanglement simply cannot be explained in terms of some unknown quantity $\lambda$ in a local hidden variable theory.

Importantly, this is not just a mathematical theorem; it has been \textbf{verified experimentally}, so this is how nature actually works. Therefore, the consequence of Bell's theorem is that any theory that correctly describes nature must violate one of the assumptions we used to derive the CHSH inequality. Which assumptions can we give up? Let's consider our options:
\begin{itemize}
\item \textbf{Give up measurement independence.} This means that we could have a different $p\mr{\lambda}{x,y}$ for each choice of $x$ and $y$, so the hidden variable must somehow “know” in advance which settings Alice and Bob will choose, before they choose them. One way this could work is if Alice and Bob were already correlated with the hidden variable before the experiment even started; this is called \emph{superdeterminism}\index{Superdeterminism}. Hypothetically, Alice and Bob could base their choices on, say, the color of a photon emitted billions of years ago from a galaxy near the edge of the observable universe, so superdeterminism actually requires that everything in the universe is predetermined in just the right way to produce these correlations – potentially all the way back to the Big Bang. Another option is that the choice of settings somehow influenced the hidden variable back in time; this is called \emph{retrocausality}\index{Retrocausality}. Both options allow the hidden variable theory to remain local (that is, satisfy local causality) and still violate the CHSH inequality.
\item \textbf{Give up local causality.} A theory that violates local causality is known as a \emph{non-local hidden variable theory}\index{Non-local hidden variable theory}\index{Hidden variable theory!Non-local}. As you proved in \zcref[S]{prob:Prove-that-local}, local causality is equivalent to setting independence and outcome independence together, so each one could be violated separately:
\begin{itemize}
\item \textbf{Give up setting independence:} This means the probabilities for Alice's outcomes depend on Bob's choice of setting (or vice versa) – no matter how far they are from each other. The most well-known theory of this kind is \emph{de Broglie-Bohm theory}\index{de Broglie-Bohm theory}, also known as \emph{Bohmian mechanics}\index{Bohmian mechanics}. In this theory, each particle has a definite position at all times, and that information is essentially the hidden variable. The quantum state then acts as a \emph{pilot wave}\index{Pilot wave} that guides the motion of the particles. When particles are entangled, the velocity of each particle depends on the positions of the other particles, and on which observable is being used to measure them, no matter how far apart they are; a particle in the Andromeda Galaxy can instantaneously influence its entangled particle on Earth.
\item \textbf{Give up outcome independence:} This means Alice's outcome depends on Bob's outcome. One example of such a theory is... quantum mechanics itself! We can think of the quantum state $\left|\Psi\right\rangle $ as the “hidden variable” $\lambda$, even though it is not really “hidden”. This theory clearly satisfies measurement independence and setting independence, but it doesn't satisfy outcome independence, because we know that if Alice and Bob have the state $\left|\beta_{00}\right\rangle $, and both measure $Z$, their outcomes will always be the same. This means that the probability for both of them to get $+1$ is $1/2$, while outcome independence requires it to be $1/2\cdot1/2=1/4$, so outcome independence is violated. This means local causality is also violated, so quantum mechanics is non-local in that sense, although as we have seen in \zcref[S]{subsec:Entanglement-Does-Not}, this non-locality cannot be used to transmit information.
\end{itemize}
\item \textbf{Give up single outcomes.} In our analysis, we implicitly assumed that each measurement has a single, definite outcome, so it makes sense to talk about “the” outcomes $a$ and $b$. This seems so obvious that we didn't even bother to state it explicitly. However, there are theories in which this assumption actually fails – most famously, the \emph{many-worlds interpretation}\index{Many-worlds interpretation}, also known as the \emph{Everett interpretation}\index{Everett interpretation}. In this interpretation, when Alice measures a qubit in a superposition of 0 and 1, she gets entangled with the qubit, and “branches” into two versions of herself, “Alice who measured 0” and “Alice who measured 1”. Schematically, we have:
\begin{equation}
\begin{gathered}\phantomsection\left|\text{Alice ready}\right\rangle \otimes\left(\alpha\left|0\right\rangle +\beta\left|1\right\rangle \right)\\
\mapsdown\\
\alpha\left|\text{Alice measured 0}\right\rangle \otimes\left|0\right\rangle +\beta\left|\text{Alice measured 1}\right\rangle \otimes\left|1\right\rangle .
\end{gathered}
\label{eq:many-worlds-measurement}
\end{equation}
The same thing happens when Alice and Bob have an entangled pair of qubits: the measurement entangles them with the qubits, and they “branch” into different versions who measured different outcomes. In standard quantum mechanics, Alice's measurement instantly collapses the state of both qubits, no matter how far away Bob is. In the many-worlds interpretation, Alice's measurement is just a local interaction between her and her qubit, which does not affect anything at Bob's location. The correlation between them only becomes meaningful when they later meet and share their results – which cannot happen faster than light. Therefore, the CHSH inequality can be violated without giving up measurement independence, and without any non-local action. (The probabilities are still the same as in standard quantum mechanics, so they do not factorize the way local causality requires; but since local causality assumed single outcomes, this does not indicate non-locality.)
\end{itemize}
The violation of the CHSH inequality is not the only feature of quantum mechanics that raises important interpretational and philosophical issues. According to the Evolution Axiom, an isolated system evolves deterministically via a unitary operator. If the measurement device and the object being measured form an isolated system, we would expect that system to evolve in the same way. A measurement would be an interaction of the device with the object, and if the measurement device changed its state based on the eigenvalue that was measured, then whenever the object was in a superposition of eigenstates, the device would become entangled with the object, as in \zcref[S]{eq:many-worlds-measurement}. However, according to the Measurement Axiom, a measurement instead collapses the state. The collapse process is neither deterministic (the final state is chosen at random) nor unitary (the collapse is not invertible). Therefore, the Measurement Axiom is incompatible with the Evolution Axiom. So how exactly does a measurement differ from a normal interaction, and why do we always observe only one outcome? This is known as the \emph{measurement problem}\index{Measurement problem}.

This problem is famously illustrated by a thought experiment known as \emph{Schrödinger's cat}\index{Schrödinger's cat}. A cat is placed in a sealed box, together with a qubit in a superposition of $\left|0\right\rangle $ and $\left|1\right\rangle $ with equal probability. The qubit is measured; if the result is 0, nothing happens, but if it's 1, a soothing lullaby is played, causing the cat to fall asleep. If the box is isolated, then according to the Evolution Axiom, the qubit and the cat become entangled, and the state of the box is a superposition of two states, one where the cat is awake and another where it is asleep; see \zcref[S]{fig:cat}. However, when we open the box, we will obviously find the cat either awake or asleep. What happened to the superposition?

\begin{figure}[tbph]
\centering
\includegraphics[width=0.8\textwidth]{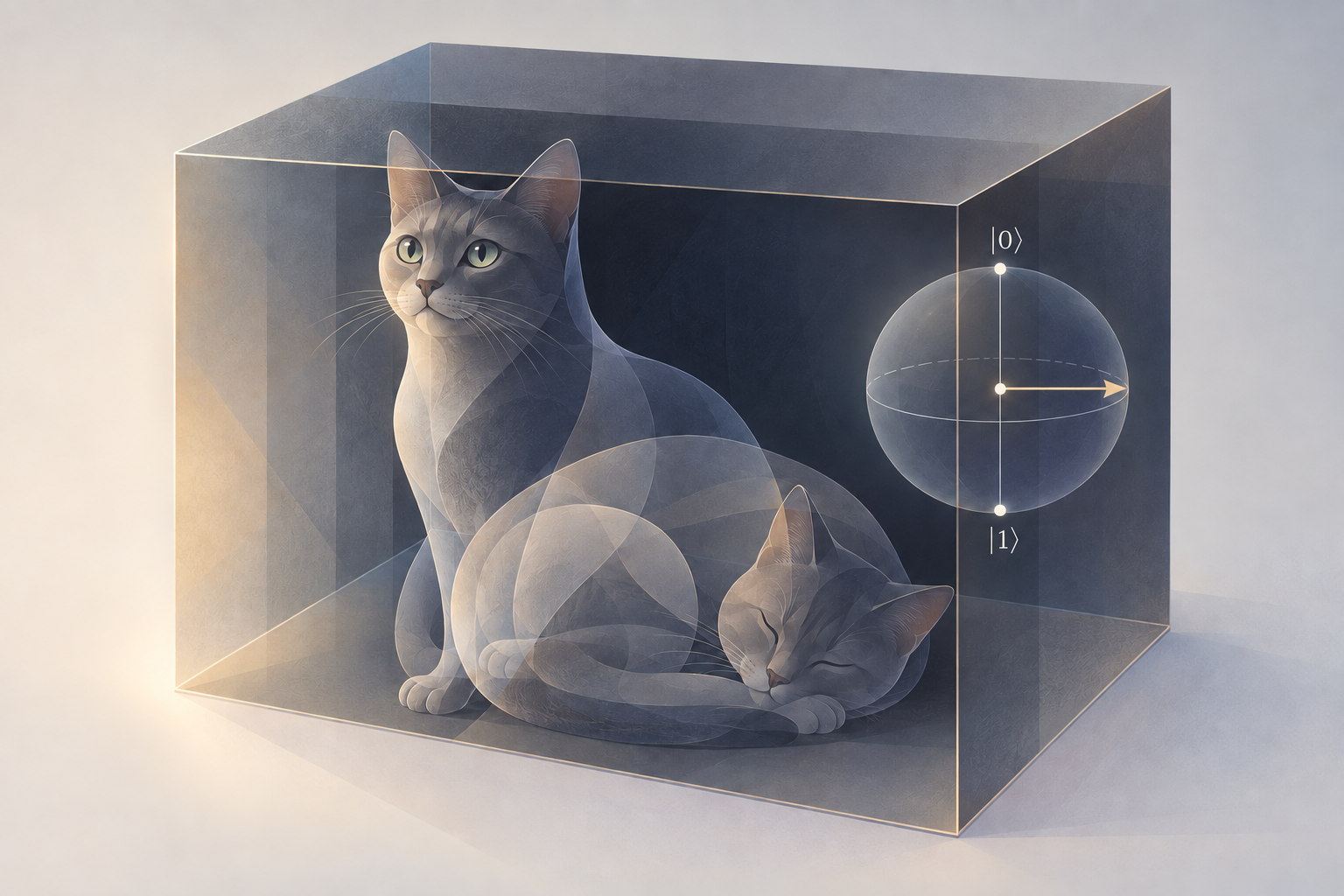}

\caption[Schrödinger's cat]{Schrödinger's cat. The qubit is represented by the Bloch sphere, which we will learn about in \zcref[S]{subsec:The-Bloch-Sphere}.}\label{fig:cat}

\end{figure}

To resolve the measurement problem, various \emph{interpretations of quantum mechanics}\index{Interpretations of quantum mechanics} have been proposed. Here are some of the most common interpretations, some of which we've already mentioned above:
\begin{itemize}
\item \textbf{The many-worlds interpretation.} In this interpretation, as we have seen, there is no single outcome; all outcomes are obtained, but in different branches. Therefore, there is no need for collapse. In \zcref[S]{eq:many-worlds-measurement}, Alice will branch into two versions of herself, one where she perceives the qubit as having collapsed to 0, and another where she perceives the qubit as having collapsed to 1. In reality, no collapse occurred, and the evolution is perfectly unitary throughout.
\item \textbf{De Broglie-Bohm theory.} Here there is no collapse either; since the particles have definite positions, the outcome is predetermined instead of being chosen at random. The theory is fully deterministic, not probabilistic like quantum mechanics; for a chosen measurement, the qubit was “always going to have” whatever value we measured, with 100\% probability – it's just that we did not know that in advance.
\item \textbf{The Copenhagen interpretation.}\index{Copenhagen interpretation} The Measurement Axiom works exactly as we've formulated it, including the collapse to a single outcome chosen at random based on the probability amplitudes. Measurements simply work differently from other interactions; but where exactly the boundary between the quantum system and the classical measurement device lies is never explained, so the measurement problem is essentially ignored. This approach is the one usually presented in textbooks and is the most popular among physicists, as it views quantum mechanics as a tool to calculate probabilities, without having to worry about any deeper meaning; it is therefore sometimes also referred to as “shut up and calculate”.
\item \textbf{Objective collapse models.}\index{Objective collapse model} These models promote the collapse to a real physical process; in fact, collapses happen all the time, spontaneously and at random, even if no one is measuring anything. The best-known example is the \emph{GRW model}\index{GRW model}\footnote{Named after Ghirardi, Rimini, and Weber.}. Each particle undergoes a spontaneous collapse very rarely. However, upon measurement, the huge number of particles in the measuring device are all entangled with the measured object. Since there are so many of them, it's essentially guaranteed that one of the particles will undergo a spontaneous collapse almost immediately – and this collapses both the device and the system into a single outcome. These models make slightly different predictions than standard quantum mechanics, so they can be tested experimentally; some of them have already been ruled out.
\item \textbf{Information-based interpretations.} In interpretations such as \emph{QBism}\index{QBism} (originally short for “quantum Bayesianism”) and \emph{relational quantum mechanics}\index{Relational quantum mechanics}, the quantum state does not describe physical reality itself; it simply indicates an observer's information about the outcomes of future measurements. This means that collapse is not a physical process – it's just the update of information that occurs when the observer learns the measurement outcome.
\end{itemize}
\begin{problem}\leavevmode
Find a unitary operator that could be responsible for the process described in \zcref[S]{eq:many-worlds-measurement}. Note that Alice's Hilbert space is 3-dimensional.
\end{problem}

\subsection{The Foundations of Quantum Theory: Summary}\label{subsec:The-Foundations-of}

Quantum theory is a fundamental mathematical framework for describing physical systems in our universe. For discrete systems with finite-dimensional Hilbert spaces, we defined this framework using a set of seven axioms:
\begin{enumerate}
\item \textbf{The System Axiom: }Physical systems are represented by complex $n$-dimensional Hilbert spaces $\BBC^{n}$, where $n$ depends on the specific system.
\item \textbf{The State Axiom: }The states of the system are represented by unit $n$-vectors in the system's Hilbert space, up to a phase factor.
\item \textbf{The Operator Axiom: }The operators on the system, which act on states, are represented by $n\xx n$ matrices in the system's Hilbert space.
\item \textbf{The Observable Axiom: }Physical observables in the system are represented by Hermitian operators on the system's Hilbert space. The eigenvalues of the observable (which are always real, since it's Hermitian) represent its possible measured values. The eigenstates of the observable can be used to form an orthonormal eigenbasis of the Hilbert space.
\begin{itemize}
\item \textbf{Superposition:} Any state $\left|\Psi\right\rangle $ can be written as a linear combination of the eigenstates $\left|B_{i}\right\rangle $ of an observable:
\[
\left|\Psi\right\rangle =\sum^{n}_{i=1}|B_{i}\rangle\langle B_{i}|\Psi\rangle.
\]
\end{itemize}
\item \textbf{The Composite System Axiom:} The Hilbert space of a composite system is represented by the tensor product of the Hilbert spaces of the individual systems.
\begin{itemize}
\item \textbf{Entanglement:} A state of a composite system that cannot be written as a single tensor product of states of the individual systems is entangled. Quantum entanglement is a form of correlation between systems, and by Bell's theorem, it is stronger than classical correlation.
\end{itemize}
\item \textbf{The Evolution Axiom:} Isolated quantum systems evolve in time via the action of a unitary operator. More precisely, given two points in time $t_{1}$ and $t_{2}$, there is a unitary operator $U$ such that any state $\left|\Psi\right\rangle $ at $t_{1}$ evolves to the state $U\left|\Psi\right\rangle $ at time $t_{2}$.
\item \textbf{The Measurement Axiom:} Consider an observable $A$ on an $n$-dimensional Hilbert space, with $k\le n$ distinct eigenvalues $\left\{ \lambda_{1},\ldots,\lambda_{k}\right\} $, corresponding projectors $\left\{ P_{1},\ldots,P_{k}\right\} $, and spectral decomposition
\[
A=\sum^{k}_{i=1}\lambda_{i}P_{i}.
\]
If the system is in the state $\left|\Psi\right\rangle $, then the probability to measure the eigenvalue $\lambda_{i}$ is given by
\[
\langle\Psi|P_{i}|\Psi\rangle.
\]
The measurement yields exactly one of the eigenvalues $\lambda_{i}$, and after the measurement, the system collapses to the state
\[
\left|\Psi\right\rangle \mt\frac{P_{i}\left|\Psi\right\rangle }{\sqrt{\langle\Psi|P_{i}|\Psi\rangle}}.
\]

\begin{itemize}
\item \textbf{The Simplified Measurement Axiom:} Consider an observable with an eigenbasis of eigenstates $\left|B_{i}\right\rangle $ corresponding to non-degenerate eigenvalues $\lambda_{i}$. If the system is in the state $\left|\Psi\right\rangle $, then the probability to measure the eigenvalue $\lambda_{i}$ corresponding to the eigenstate $\left|B_{i}\right\rangle $ is given by
\[
\left|\langle B_{i}|\Psi\rangle\right|^{2}.
\]
After the measurement, if the eigenvalue $\lambda_{i}$ was measured, then the system will collapse to the eigenstate $\left|B_{i}\right\rangle $:
\[
\left|\Psi\right\rangle \mt\left|B_{i}\right\rangle .
\]
If a measurement is performed only on part of a composite system, the total probability to measure the eigenvalue $\lambda_{i}$ is the sum of the probabilities for all the possible ways in which this eigenvalue can be measured. After the measurement, if the eigenvalue $\lambda_{i}$ was measured, then only the system we measured will collapse to the eigenstate $\left|B_{i}\right\rangle $, while the other systems may still be in a superposition.
\item \textbf{Expectation Value:} If the system is in the state $\left|\Psi\right\rangle $, the expectation value for the measurement of the observable $A$ is given by $\left\langle A\right\rangle _{\Psi}=\langle\Psi|A|\Psi\rangle$.
\item \textbf{Uncertainty Principle:} If two observables $A$ and $B$ don't commute, the standard deviations of their measurements satisfy the uncertainty relation
\[
\Delta A\Delta B\ge\hf\left|\left\langle \left[A,B\right]\right\rangle \right|.
\]
\end{itemize}
\end{enumerate}
The mathematical framework we have defined here is not enough on its own; one must use the framework to define different \emph{models}\index{Model}, which map the framework to specific physical systems. A model is a specific choice of the following ingredients:
\begin{itemize}
\item A Hilbert space describing a specific physical system,
\item Hermitian operators corresponding to specific physical observables that may be measured for the system,
\item Unitary operators corresponding to the time evolution and other possible transformations of the system,
\item The states on which these operators act, which correspond to different configurations of the system.
\end{itemize}
In the simple case of a qubit, we saw that the Hilbert space is $\BBC^{2}$, the Hermitian operators corresponding to observables are linear combinations of the Pauli matrices and the identity matrix, the unitary operators corresponding to transformations are the quantum gates, and the states are the two possible values of the qubit, 0 and 1 (and superpositions thereof).

Of course, not every possible model we can make will actually correspond to a physical system that we can find in nature. However, amazingly, the opposite statement does seem to be true: every physical system that we find in nature\footnote{Except perhaps general relativity, but we are pretty sure that there is a quantum theory of general relativity; we just don't have a consistent formulation of it yet. This theory is known as \emph{quantum gravity}\index{Quantum gravity}.} can be precisely described by a model built using the ingredients of quantum theory.

We can think of quantum theory as a sort of \textbf{language}. Just like English is a language with rules such as grammar and spelling, so is quantum theory a language with its own rules: observables must be Hermitian operators, possible measurement results are given by the eigenvalues of these operators, and so on. And just like we can use English to make any sentence we want, both true and false, we can use quantum theory to make any model we want, both models that correspond to real physical systems and those that do not.

\section{Quantum Information and Computation}

Now that we have successfully formulated the mathematical basis of quantum theory, I would like to discuss two of its modern applications: \emph{quantum information}\index{Quantum information} and \emph{quantum computation}\index{Quantum computation}. We will only present a few basic concepts and examples from these cutting-edge areas of research, but I encourage you to look them up and read more about them.

\subsection{The No-Cloning Theorem and Quantum Teleportation}

\subsubsection{The No-Cloning Theorem}

The \emph{no-cloning theorem\index{No-cloning theorem}} states that it is impossible to make a copy of an arbitrary unknown quantum state. Note that it is possible, in principle, to generate a \textbf{known }quantum state as many times as we want; all we need to do is repeat whatever process is known to generate that state. However, if someone gives you an \textbf{unknown }quantum state $\left|\Psi\right\rangle $ and doesn't tell you anything about it, the no-cloning theorem states that you will never be able to make another copy of $\left|\Psi\right\rangle $.

To prove the theorem, let us assume that we have a “copying operator” $U$ which gets a tensor product of two states as input, and copies the state from the first slot into the second slot:
\[
U\left(\vphantom{\bll}\left|\Psi\right\rangle \otimes\left|?\right\rangle \right)=\left|\Psi\right\rangle \otimes\left|\Psi\right\rangle .
\]
By the Evolution Axiom, $U$ must be a unitary operator. The state $\left|?\right\rangle $ is a “blank state” that will be overwritten with the state $\left|\Psi\right\rangle $ that we are copying. Since $\left|?\right\rangle $ only exists to be overwritten, any state would work, but $\left|?\right\rangle $ cannot depend on $\left|\Psi\right\rangle $, because we don't know $\left|\Psi\right\rangle $ in advance, and a given $U$ must always use the same $\left|?\right\rangle $, otherwise the process is not invertible and therefore cannot be unitary. We also assume that this process is completely isolated from the environment.

We are looking for a \textbf{universal }copying operator, so the same $U$ must be able to copy \textbf{any }state $\left|\Psi\right\rangle $, even if we don't know in advance what the state is. Let us use $U$ to copy two states, $\left|\Psi_{1}\right\rangle $ and $\left|\Psi_{2}\right\rangle $:
\[
U\left(\vphantom{\bll}\left|\Psi_{1}\right\rangle \otimes\left|?\right\rangle \right)=\left|\Psi_{1}\right\rangle \otimes\left|\Psi_{1}\right\rangle ,
\]
\[
U\left(\vphantom{\bll}\left|\Psi_{2}\right\rangle \otimes\left|?\right\rangle \right)=\left|\Psi_{2}\right\rangle \otimes\left|\Psi_{2}\right\rangle .
\]
We can take the inner product of the last two equations by turning the second equation into a bra:
\[
\left(\vphantom{\bll}\left\langle \Psi_{2}\right|\otimes\left\langle ?\right|\right)U^{\dagger}U\left(\vphantom{\bll}\left|\Psi_{1}\right\rangle \otimes\left|?\right\rangle \right)=\left(\vphantom{\bll}\left\langle \Psi_{2}\right|\otimes\left\langle \Psi_{2}\right|\right)\left(\vphantom{\bll}\left|\Psi_{1}\right\rangle \otimes\left|\Psi_{1}\right\rangle \right).
\]
Since $U$ is unitary, we have $U^{\dagger}U=1$:
\[
\left(\vphantom{\bll}\left\langle \Psi_{2}\right|\otimes\left\langle ?\right|\right)\left(\vphantom{\bll}\left|\Psi_{1}\right\rangle \otimes\left|?\right\rangle \right)=\left(\vphantom{\bll}\left\langle \Psi_{2}\right|\otimes\left\langle \Psi_{2}\right|\right)\left(\vphantom{\bll}\left|\Psi_{1}\right\rangle \otimes\left|\Psi_{1}\right\rangle \right).
\]
The inner products can be calculated using \zcref[S]{eq:inner-tensor}:
\[
\langle\Psi_{2}|\Psi_{1}\rangle\langle?|?\rangle=\langle\Psi_{2}|\Psi_{1}\rangle\langle\Psi_{2}|\Psi_{1}\rangle.
\]
On the right-hand side, we have $\langle\Psi_{2}|\Psi_{1}\rangle\langle\Psi_{2}|\Psi_{1}\rangle=\langle\Psi_{2}|\Psi_{1}\rangle^{2}$:
\[
\langle\Psi_{2}|\Psi_{1}\rangle\langle?|?\rangle=\langle\Psi_{2}|\Psi_{1}\rangle^{2}.
\]

Finally, even though we haven't specified the state $\left|?\right\rangle $ (since we don't care what it is), we still know it must be normalized such that $\langle?|?\rangle=1$, since otherwise it won't be a proper state. Therefore, we obtain:
\[
\langle\Psi_{2}|\Psi_{1}\rangle=\langle\Psi_{2}|\Psi_{1}\rangle^{2}.
\]
This is a quadratic equation, so it has two solutions:
\begin{itemize}
\item The first solution is $\langle\Psi_{2}|\Psi_{1}\rangle=1$, in which case the states must be the same state: $\left|\Psi_{1}\right\rangle =\left|\Psi_{2}\right\rangle $. But this means $U$ is a copying operator that can only copy one specific state, in contradiction with our requirement above that $U$ is universal.
\item The second solution is $\langle\Psi_{2}|\Psi_{1}\rangle=0$, in which case $\left|\Psi_{1}\right\rangle $ and $\left|\Psi_{2}\right\rangle $ must be orthogonal. Again, this means that $U$ cannot be universal, since it can only copy states that are orthogonal to a specific state, and thus we cannot clone an unknown quantum state.
\end{itemize}
In conclusion, we have proven that it is impossible to find a unitary operator $U$ that can clone any arbitrary state $\left|\Psi\right\rangle $.

By the way, this is one of the reasons quantum computers\index{Quantum computer} are so hard to build. In a classical computer, we can just make several copies of each bit, and use that for error correction in case the bit gets corrupted. In a quantum computer, we cannot do that, since we cannot make copies of a qubit, due to the no-cloning theorem. Still, quantum error correction is possible – but it is much more complicated.

The opposite of the no-cloning theorem is the \emph{no-deleting theorem}\index{No-deleting theorem}: given two copies of an arbitrary unknown state, $\left|\Psi\right\rangle \otimes\left|\Psi\right\rangle $, there is no unitary operator $U$ that “deletes” one of them by turning it into a fixed blank state:
\[
U\left(\left|\Psi\right\rangle \otimes\left|\Psi\right\rangle \right)=\left|\Psi\right\rangle \otimes\left|?\right\rangle .
\]
The proof is trivial: if such a $U$ existed, then $U^{\dagger}$ would be a copying operator, contradicting the no-cloning theorem.
\begin{problem}\leavevmode
Remarkably, if cloning a quantum state \textbf{were }possible, it would have allowed faster-than-light communication! Assuming Alice and Bob each have one qubit of an entangled Bell state, and Bob can make as many copies of his qubit as he wants, show that it is possible for Alice to send Bob a message instantaneously, regardless of the distance between them. This result is especially noteworthy due to the fact that we can freely copy classical bits, and yet classical correlation definitely does not allow instantaneous communication. This demonstrates something very special about quantum entanglement, which does not apply to classical correlation.
\end{problem}

\subsubsection{Quantum Teleportation}

We discovered that it is impossible to copy a quantum state, which is quite surprising. \emph{Quantum teleportation\index{Quantum teleportation}} is another surprising discovery, which also serves to illustrate the powerful consequences of entanglement. We begin with the Bell state in \zcref[S]{eq:Bell00-short}:
\[
\left|\beta_{00}\right\rangle \equiv\frac{1}{\sqrt{2}}\left(\left|00\right\rangle +\left|11\right\rangle \right),
\]
where we again used the shorthand notation $\left|xy\right\rangle \equiv\left|x\right\rangle \otimes\left|y\right\rangle $. Alice takes the first qubit, Bob takes the second, and they go their separate ways. In this entangled state, if Alice measures 0, Bob will also measure 0, and if Alice measures 1, Bob will also measure 1.

Later, Alice receives an arbitrary qubit
\[
\left|\Psi\right\rangle =a\left|0\right\rangle +b\left|1\right\rangle \sp\left|a\right|^{2}+\left|b\right|^{2}=1,
\]
but she does \textbf{not }know the state of the qubit, that is, the coefficients $a$ and $b$. Alice needs to transfer this unknown qubit in its entirety to Bob using only two \textbf{classical }bits. This seems impossible, for two different reasons:
\begin{enumerate}
\item The exact state of the qubit is determined by the two arbitrary complex numbers $a$ and $b$. Even if Alice did know the values of these numbers, transferring that information requires much more than two classical bits – in fact, to transmit the \textbf{precise }value of an arbitrary complex (or even real) number, an \textbf{infinite }number of bits is required.
\item Even if Alice were somehow able to magically transmit two complex numbers using only two classical bits, there is no way she could determine the values of $a$ and $b$ in the first place. Any measurement that Alice makes on her qubit will simply result in either 0 or 1; it does not tell Alice anything about the probabilities, not to mention the probability amplitudes, except that the probability of the result is not zero. To estimate the probabilities, Alice must make a large number of measurements (in fact, an infinite number of them, if she wants to know the precise values of the probabilities). However, this is impossible due to the no-cloning theorem; Alice can only measure the qubit once, and that's it.
\end{enumerate}
To make the impossible possible, Alice can use the fact that her half of the Bell state is entangled with Bob's half. All three qubits can be represented together by the composite state
\begin{align*}
\left|\gamma\right\rangle  & \equiv\left|\Psi\right\rangle \otimes\left|\beta_{00}\right\rangle \\
 & =\frac{1}{\sqrt{2}}\left(a\left|0\right\rangle +b\left|1\right\rangle \right)\otimes\left(\left|00\right\rangle +\left|11\right\rangle \right)\\
 & =\frac{1}{\sqrt{2}}\left(\vphantom{\bll}a\left(\left|000\right\rangle +\left|011\right\rangle \right)+b\left(\left|100\right\rangle +\left|111\right\rangle \right)\right),
\end{align*}
where we used the shorthand notation
\[
\left|xyz\right\rangle \equiv\left|x\right\rangle \otimes\left|y\right\rangle \otimes\left|z\right\rangle .
\]
Here the first qubit is the one that is to be teleported from Alice to Bob, the second is Alice's half of the Bell state, and the third is Bob's half.

First, Alice sends the first qubit (the unknown qubit $\left|\Psi\right\rangle $) and the second qubit (her half of the Bell state) through a CNOT gate, which, as you recall, flips the second qubit only if the first qubit is $\left|1\right\rangle $:
\begin{align*}
\textrm{CNOT}_{1,2}\left|\gamma\right\rangle  & =\frac{1}{\sqrt{2}}\left(\vphantom{\bll}a\left(\left|000\right\rangle +\left|011\right\rangle \right)+b\left(\left|110\right\rangle +\left|101\right\rangle \right)\right)\\
 & =\frac{1}{\sqrt{2}}\left(\vphantom{\bll}a\left|0\right\rangle \otimes\left(\left|00\right\rangle +\left|11\right\rangle \right)+b\left|1\right\rangle \otimes\left(\left|10\right\rangle +\left|01\right\rangle \right)\right).
\end{align*}
Here we used the notation $\textrm{CNOT}_{1,2}$ to indicate that the gate only acts on qubits 1 and 2 out of the three qubits. Explicitly, this would be the tensor product of the CNOT gate on the left with the $2\xx2$ identity matrix on the right:
\[
\textrm{CNOT}_{1,2}\equiv\textrm{CNOT}\otimes1=\left(\begin{array}{cccc}
1 & 0 & 0 & 0\\
0 & 1 & 0 & 0\\
0 & 0 & 0 & 1\\
0 & 0 & 1 & 0
\end{array}\right)\otimes\left(\begin{array}{cc}
1 & 0\\
0 & 1
\end{array}\right).
\]
Next, she sends the first qubit through the Hadamard gate, which, as you recall, takes $\left|0\right\rangle $ to $\left|+\right\rangle \equiv\left(\left|0\right\rangle +\left|1\right\rangle \right)/\sqrt{2}$ and $\left|1\right\rangle $ to $\left|-\right\rangle \equiv\left(\left|0\right\rangle -\left|1\right\rangle \right)/\sqrt{2}$:
\begin{align*}
H_{1}\cdot\textrm{CNOT}_{1,2}\left|\gamma\right\rangle  & =\frac{1}{\sqrt{2}}\left(\vphantom{\bll}a\left|+\right\rangle \otimes\left(\left|00\right\rangle +\left|11\right\rangle \right)+b\left|-\right\rangle \otimes\left(\left|10\right\rangle +\left|01\right\rangle \right)\right)\\
 & =\hf\left(\vphantom{\bll}a\left(\left|0\right\rangle +\left|1\right\rangle \right)\otimes\left(\left|00\right\rangle +\left|11\right\rangle \right)+b\left(\left|0\right\rangle -\left|1\right\rangle \right)\otimes\left(\left|10\right\rangle +\left|01\right\rangle \right)\right)\\
 & =\hf\left(\vphantom{\bll}a\left(\left|000\right\rangle +\left|011\right\rangle +\left|100\right\rangle +\left|111\right\rangle \right)+b\left(\left|010\right\rangle +\left|001\right\rangle -\left|110\right\rangle -\left|101\right\rangle \right)\right)\\
 & =\hf a\left(\left|00\right\rangle \otimes\left|0\right\rangle +\left|01\right\rangle \otimes\left|1\right\rangle +\left|10\right\rangle \otimes\left|0\right\rangle +\left|11\right\rangle \otimes\left|1\right\rangle \right)+\\
 & \qquad+\hf b\left(\left|01\right\rangle \otimes\left|0\right\rangle +\left|00\right\rangle \otimes\left|1\right\rangle -\left|11\right\rangle \otimes\left|0\right\rangle -\left|10\right\rangle \otimes\left|1\right\rangle \right).
\end{align*}
Again, the notation $H_{1}$ means we act with the Hadamard gate only on the first qubit:
\[
H_{1}\equiv H\otimes1\otimes1=\frac{1}{\sqrt{2}}\left(\begin{array}{cc}
1 & 1\\
1 & -1
\end{array}\right)\otimes\left(\begin{array}{cc}
1 & 0\\
0 & 1
\end{array}\right)\otimes\left(\begin{array}{cc}
1 & 0\\
0 & 1
\end{array}\right).
\]
We can rearrange the transformed state as follows:
\begin{align*}
H_{1}\cdot\textrm{CNOT}_{1,2}\left|\gamma\right\rangle  & =\hf\left|00\right\rangle \otimes\left(a\left|0\right\rangle +b\left|1\right\rangle \right)+\\
 & \qquad+\hf\left|01\right\rangle \otimes\left(a\left|1\right\rangle +b\left|0\right\rangle \right)+\\
 & \qquad+\hf\left|10\right\rangle \otimes\left(a\left|0\right\rangle -b\left|1\right\rangle \right)+\\
 & \qquad+\hf\left|11\right\rangle \otimes\left(a\left|1\right\rangle -b\left|0\right\rangle \right).
\end{align*}
Finally, Alice performs a measurement on the first two qubits (the one to be teleported, and her half of the Bell state), and obtains one of four results: 00, 01, 10, or 11. These are \textbf{two classical bits}, which she can then send to Bob. With this information, Bob can read from the last equation exactly which operations he has to perform on his qubit (which you will determine in \zcref[S]{prob:For-each-of}) in order to obtain the original qubit $\left|\Psi\right\rangle =a\left|0\right\rangle +b\left|1\right\rangle $. The qubit has been successfully teleported from Alice to Bob!

Note that since Alice measured the original qubit, it collapsed and its quantum state has been destroyed. Therefore, quantum teleportation does not violate the no-cloning theorem; the state of the qubit was not cloned or copied, it was just moved from one qubit to another. Also, since Alice had to send two classical bits to Bob – for example through a cable or radio waves – the speed of teleportation is limited by the speed of light, and there is no violation of relativity.

Finally, since quantum teleportation requires Alice and Bob to already have one half of an entangled pair each, and the entanglement is destroyed in the process due to Alice's measurement, the number of qubits they can teleport is limited by the number of entangled pairs they have. Once they run out of entangled pairs, they can no longer teleport any qubits until they physically exchange more entangled pairs. This means that you can't just establish two teleportation stations on two planets and teleport qubits between them forever; you will have to distribute a fresh supply of entangled pairs every once in a while, for example by transmitting entangled photons between the planets.
\begin{problem}\leavevmode
Quantum teleportation has been demonstrated in many different experiments, over distances of up to 1,400 km, and not just with qubits but even with more complicated systems. Whenever a new quantum teleportation experiment happens, articles appear in the media with sensationalist headlines such as “scientists demonstrate teleportation is possible!” or “is teleportation closer than we think?”, where by “teleportation” they actually mean the \textbf{science-fiction }concept of “teleportation”, where a macroscopic object is sent from one place to another without going through the space in between. Is the word “teleportation” in “quantum teleportation” indeed justified? In what ways is quantum teleportation the same as science-fiction teleportation, and in what ways is it different?
\end{problem}

\begin{problem}\leavevmode
\label{prob:For-each-of}For each of the four results of Alice's measurement, 00, 01, 10, and 11, determine which unitary transformations Bob must perform on his qubit in order to obtain the original $\left|\Psi\right\rangle =a\left|0\right\rangle +b\left|1\right\rangle $.
\end{problem}

\begin{problem}\leavevmode
\label{prob:Write-a-computer-qubits}Write a computer program, in any language of your choice, that gets an arbitrary composite state of $n$ qubits as input and allows the user to perform the following actions:
\begin{itemize}
\item Act on one or more of the qubits with a quantum gate; for example, act with Hadamard on one qubit or with CNOT on two qubits.
\item Simulate a measurement of one or more of the qubits as dictated by the projective Measurement Axiom, with the result determined randomly according to the appropriate probability distribution, and the state collapsing after the measurement according to the value that was measured.
\end{itemize}
Use your program to simulate quantum teleportation, and show that it indeed works.
\end{problem}

\subsection{Quantum Algorithms}

\subsubsection{Quantum Parallelism}

It's a common misconception that quantum computers\index{Quantum computer} work by using superposition to “calculate the answer for every possible combination of qubits in parallel”. As the claim goes, if you have a quantum computer with $n$ qubits, then because each qubit can be “0 and 1 at the same time”, you can operate on all $2^{n}$ possible combinations at once, and then simply pick out the correct answer.

You've probably already heard this incorrect claim before – perhaps even from someone with a PhD in physics! This misconception stems from the more general misconception about the meaning of superposition which we discussed in \zcref[S]{subsec:The-Meaning-of}. Sure, it would have been great if this kind of parallelism were actually possible... but unfortunately, it's not.

So how do quantum computers actually work? Generally, they make clever use of properties of quantum states, such as superposition, entanglement, and interference, to solve certain problems. Just as classical computers operate by sending one or more classical bits through classical logic gates\index{Logic gate!Classical}\index{Classical logic gate}, quantum computers operate by sending one or more qubits through quantum logic gates\index{Logic gate!Quantum}\index{Quantum logic gate} (recall \zcref[S]{subsec:Quantum-Logic-Gates}). The arrangement of gates is called a \emph{quantum circuit}\index{Quantum circuit}, and algorithms which make use of qubits and quantum gates are called \emph{quantum algorithms}\index{Quantum algorithm}.

Even though quantum computers don't really “calculate everything in parallel”, there is still a concept called \emph{quantum parallelism}\index{Quantum parallelism} which is used in most quantum algorithms. Let us demonstrate it with a simple example. Consider a function $f:\left\{ 0,1\right\} \to\left\{ 0,1\right\} $ which takes one bit as input and gives one bit as output. There are, in fact, exactly 4 such functions, because each of the input bits 0 or 1 can be sent to either 0 or 1 as output:
\[
f\left(x\right)=0\sp f\left(x\right)=1\sp f\left(x\right)=x\sp f\left(x\right)=1-x.
\]
Say we have a quantum computer which can manipulate two qubits. Let $U_{f}$ be a unitary operator which transforms any composite two-qubit state
\[
\left|x,y\right\rangle \equiv\left|x\right\rangle \otimes\left|y\right\rangle ,
\]
where $x,y\in\left\{ 0,1\right\} $, as follows:
\begin{equation}
U_{f}\left|x,y\right\rangle =\left|x,y\oplus f\left(x\right)\right\rangle ,\label{eq:Uf}
\end{equation}
where $\oplus$ means addition modulo 2 (recall \zcref[S]{subsec:Correlations-and-Hidden}), that is,
\[
0\oplus0=0\sp0\oplus1=1\sp1\oplus0=1\sp1\oplus1=0.
\]
The form of $U_{f}$ depends on the choice of $f\left(x\right)$. For example, if $f\left(x\right)=0$, then $U_{f}$ is simply the identity operator. In \zcref[S]{prob:Find-the-form} you will find the exact form of $U_{f}$ for each choice of $f\left(x\right)$ and see that it is indeed unitary.

We assume that initially, the state of the two qubits in our quantum computer is
\[
\left|\Psi\right\rangle =\left|0\right\rangle \otimes\left|0\right\rangle =\left|00\right\rangle .
\]
The first qubit, which will store the input, is called the \emph{data register}\index{Register!Data}, and the second qubit, which will store the output, is called the \emph{target register}\index{Register!Target}. We now build a quantum circuit as follows. First, recall the Hadamard gate\index{Logic gate!Quantum Hadamard}\index{Quantum logic gate!Hadamard}\index{Hadamard gate} in \zcref[S]{eq:Hadamard}:
\[
H\equiv\frac{1}{\sqrt{2}}\left(\begin{array}{cc}
1 & 1\\
1 & -1
\end{array}\right),
\]
which acts as follows:
\[
H\left|0\right\rangle =\frac{1}{\sqrt{2}}\left(\left|0\right\rangle +\left|1\right\rangle \right),
\]
\[
H\left|1\right\rangle =\frac{1}{\sqrt{2}}\left(\left|0\right\rangle -\left|1\right\rangle \right).
\]
Let us apply it to the first qubit of $\left|\Psi\right\rangle $:
\begin{align*}
H_{1}\left|\Psi\right\rangle  & =\frac{1}{\sqrt{2}}\left(\left|0\right\rangle +\left|1\right\rangle \right)\otimes\left|0\right\rangle \\
 & =\frac{1}{\sqrt{2}}\left(\left|00\right\rangle +\left|10\right\rangle \right).
\end{align*}
We now apply $U_{f}$ to this state. Since $U_{f}$ is a linear operator, it acts on both terms in the superposition independently:
\[
U_{f}H_{1}\left|\Psi\right\rangle =\frac{1}{\sqrt{2}}\left(\left|0,f\left(0\right)\right\rangle +\left|1,f\left(1\right)\right\rangle \right).
\]
Even though we applied $U_{f}$ only once, the state now contains information about the value of $f\left(x\right)$ for both 0 and 1! This is an example of quantum parallelism. Does this actually mean we “calculated all the possibilities in parallel”, as the common misconception says? Not exactly. The state is still in a superposition; there is no way to tell the values of $f\left(x\right)$ from the state, since \textbf{we don't know the state}; we can only \textbf{measure }the state, and the measurement will collapse it to either $\left|0,f\left(0\right)\right\rangle $ or $\left|1,f\left(1\right)\right\rangle $, each with probability $1/2$. Thus we can only know \textbf{one} of the values of $f\left(x\right)$, chosen at random.
\begin{problem}\leavevmode
\label{prob:Find-the-form}Find the form of the operator $U_{f}$ defined in \zcref[S]{eq:Uf} for each possible choice of $f\left(x\right)$ and prove that it is indeed unitary for each choice.
\end{problem}

\begin{xca}\leavevmode
For each of the functions $f\left(x\right)=x$ and $f\left(x\right)=1-x$, find the matrix representation of $U_{f}H_{1}$ and show that it takes the vector $\left|\Psi\right\rangle =\left|00\right\rangle $ to the Bell states $\left|\beta_{00}\right\rangle $ and $\left|\beta_{01}\right\rangle $ respectively, as given in \zcref[S]{eq:Bell00-short,eq:Bell01-short}.
\end{xca}

\subsubsection{Deutsch's Algorithm}

If we can only know one of the values of $f\left(x\right)$, and not all of them, then what's the point of quantum parallelism? Well, the point isn't to know all the values at once; it's to take advantage of the parallelism to find \textbf{relations }between the different values. Consider, for example, the problem of determining whether $f\left(x\right)$ is \textbf{constant }or not.

Classically, it is clear that we must evaluate both $f\left(0\right)$ and $f\left(1\right)$. The function is constant if $f\left(0\right)=f\left(1\right)$ and not constant if $f\left(0\right)\ne f\left(1\right)$, but we must know \textbf{both }values to find the answer. However, with a quantum computer, that is not the case. We can determine if the function is constant \textbf{without }knowing both of its values, and in fact, without knowing \textbf{any }of its values!

This can be done using \emph{Deutsch's algorithm}\index{Deutsch's algorithm}. In this algorithm, we start with the state
\[
\left|\Psi\right\rangle =\left|0\right\rangle \otimes\left|1\right\rangle =\left|01\right\rangle ,
\]
and then act on \textbf{both }qubits with a Hadamard gate:
\[
H_{12}\left|\Psi\right\rangle =\frac{1}{\sqrt{2}}\left(\left|0\right\rangle +\left|1\right\rangle \right)\otimes\frac{1}{\sqrt{2}}\left(\left|0\right\rangle -\left|1\right\rangle \right).
\]
Next, we notice that
\[
U_{f}\left(\left|x\right\rangle \otimes\frac{1}{\sqrt{2}}\left(\left|0\right\rangle -\left|1\right\rangle \right)\right)=\left|x\right\rangle \otimes\frac{1}{\sqrt{2}}\left(\left|f\left(x\right)\right\rangle -\left|1\oplus f\left(x\right)\right\rangle \right).
\]
On the right-hand side, we have
\begin{align*}
\frac{1}{\sqrt{2}}\left(\left|f\left(x\right)\right\rangle -\left|1\oplus f\left(x\right)\right\rangle \right) & =\begin{cases}
\frac{1}{\sqrt{2}}\left(\left|0\right\rangle -\left|1\right\rangle \right) & f\left(x\right)=0,\\
\frac{1}{\sqrt{2}}\left(\left|1\right\rangle -\left|0\right\rangle \right) & f\left(x\right)=1,
\end{cases}\\
 & =\left(-1\right)^{f\left(x\right)}\frac{1}{\sqrt{2}}\left(\left|0\right\rangle -\left|1\right\rangle \right),
\end{align*}
which means that
\begin{equation}
U_{f}\left(\left|x\right\rangle \otimes\frac{1}{\sqrt{2}}\left(\left|0\right\rangle -\left|1\right\rangle \right)\right)=\left(-1\right)^{f\left(x\right)}\left|x\right\rangle \otimes\frac{1}{\sqrt{2}}\left(\left|0\right\rangle -\left|1\right\rangle \right).\label{eq:Uf2}
\end{equation}
Using this information, we can act with $U_{f}$ on the full state:
\[
U_{f}H_{12}\left|\Psi\right\rangle =\frac{1}{\sqrt{2}}\left(\left(-1\right)^{f\left(0\right)}\left|0\right\rangle +\left(-1\right)^{f\left(1\right)}\left|1\right\rangle \right)\otimes\frac{1}{\sqrt{2}}\left(\left|0\right\rangle -\left|1\right\rangle \right).
\]
If $f\left(x\right)$ is constant, then $f\left(0\right)=f\left(1\right)$, and both terms in the first qubit have the same sign, either both $+$ or both $-$. However, if $f\left(x\right)$ is not constant, then the terms will have opposite signs, one $+$ and one $-$. In other words:
\[
U_{f}H_{12}\left|\Psi\right\rangle =\begin{cases}
\pm\frac{1}{\sqrt{2}}\left(\left|0\right\rangle +\left|1\right\rangle \right)\otimes\frac{1}{\sqrt{2}}\left(\left|0\right\rangle -\left|1\right\rangle \right) & f\left(0\right)=f\left(1\right),\\
\pm\frac{1}{\sqrt{2}}\left(\left|0\right\rangle -\left|1\right\rangle \right)\otimes\frac{1}{\sqrt{2}}\left(\left|0\right\rangle -\left|1\right\rangle \right) & f\left(0\right)\ne f\left(1\right).
\end{cases}
\]
Next, we pass the first qubit through a Hadamard gate again:
\[
H_{1}U_{f}H_{12}\left|\Psi\right\rangle =\begin{cases}
\pm\left|0\right\rangle \otimes\frac{1}{\sqrt{2}}\left(\left|0\right\rangle -\left|1\right\rangle \right) & f\left(0\right)=f\left(1\right),\\
\pm\left|1\right\rangle \otimes\frac{1}{\sqrt{2}}\left(\left|0\right\rangle -\left|1\right\rangle \right) & f\left(0\right)\ne f\left(1\right),
\end{cases}
\]
where we used the fact that the Hadamard gate is its own inverse:
\[
H\left(\frac{1}{\sqrt{2}}\left(\left|0\right\rangle +\left|1\right\rangle \right)\right)=\frac{1}{\sqrt{2}}\left(\frac{1}{\sqrt{2}}\left(\left|0\right\rangle +\left|1\right\rangle \right)+\frac{1}{\sqrt{2}}\left(\left|0\right\rangle -\left|1\right\rangle \right)\right)=\left|0\right\rangle ,
\]
\[
H\left(\frac{1}{\sqrt{2}}\left(\left|0\right\rangle -\left|1\right\rangle \right)\right)=\frac{1}{\sqrt{2}}\left(\frac{1}{\sqrt{2}}\left(\left|0\right\rangle +\left|1\right\rangle \right)-\frac{1}{\sqrt{2}}\left(\left|0\right\rangle -\left|1\right\rangle \right)\right)=\left|1\right\rangle .
\]
Finally, we note that
\[
f\left(0\right)\oplus f\left(1\right)=\begin{cases}
0 & f\left(0\right)=f\left(1\right),\\
1 & f\left(0\right)\ne f\left(1\right),
\end{cases}
\]
so we can write concisely
\begin{equation}
H_{1}U_{f}H_{12}\left|\Psi\right\rangle =\pm\left|f\left(0\right)\oplus f\left(1\right)\right\rangle \otimes\frac{1}{\sqrt{2}}\left(\left|0\right\rangle -\left|1\right\rangle \right).\label{eq:f0f1}
\end{equation}
To determine if $f\left(x\right)$ is constant or not, all we need to do is simply to measure the first qubit. If it's 0, then we know $f\left(x\right)$ is constant, and if it's 1, then we know $f\left(x\right)$ is not constant!
\begin{xca}\leavevmode
For each of the functions $f\left(x\right)=0$ (constant) and $f\left(x\right)=x$ (not constant), find the matrix representation of $H_{1}U_{f}H_{12}$ and demonstrate by explicit matrix multiplication that it takes the vector $\left|\Psi\right\rangle =\left|01\right\rangle $ to a state of the form given in \zcref[S]{eq:f0f1}.
\end{xca}

\subsubsection{The Deutsch-Jozsa Algorithm}\label{subsec:The-Deutsch-Jozsa-Algorithm}

You may be asking yourself: okay, we can determine if the function is constant, but so what? We could also just calculate both its values. Well, imagine now that $f\left(x\right)$ is much more complicated. Instead of getting just one bit as input, it gets $n$ bits. This means that the input can be any integer between 0 and $2^{n}-1$, as can be seen from the following table:
\begin{center}
\begin{tabular}{|c|c|}
\hline 
Binary ($n$ bits) & Decimal\tabularnewline
\hline 
\hline 
$000\ldots000$ & $0$\tabularnewline
\hline 
$000\ldots001$ & $1$\tabularnewline
\hline 
$000\ldots010$ & $2$\tabularnewline
\hline 
$000\ldots011$ & $3$\tabularnewline
\hline 
$\vdots$ & $\vdots$\tabularnewline
\hline 
$111\ldots100$ & $2^{n}-4$\tabularnewline
\hline 
$111\ldots101$ & $2^{n}-3$\tabularnewline
\hline 
$111\ldots110$ & $2^{n}-2$\tabularnewline
\hline 
$111\ldots111$ & $2^{n}-1$\tabularnewline
\hline 
\end{tabular}
\par\end{center}

For example, for the case of $n=3$ we have:
\begin{center}
\begin{tabular}{|c|c|}
\hline 
Binary ($3$ bits) & Decimal\tabularnewline
\hline 
\hline 
$000$ & $0$\tabularnewline
\hline 
$001$ & $1$\tabularnewline
\hline 
$010$ & $2$\tabularnewline
\hline 
$011$ & $3$\tabularnewline
\hline 
$100$ & $4$\tabularnewline
\hline 
$101$ & $5$\tabularnewline
\hline 
$110$ & $6$\tabularnewline
\hline 
$111$ & $7$\tabularnewline
\hline 
\end{tabular}
\par\end{center}

In this case $2^{n}=8$, so $2^{n}-1=7$. For a large number of bits, the number of possible input values increases \textbf{exponentially}. For example, if we have $n=64$ bits, then the maximum input value is $2^{64}-1\ap1.8\xx10^{19}$.

Assume that we are given the information that $f\left(x\right)$ is one of two kinds of functions:
\begin{enumerate}
\item A \textbf{constant }function, that is, $f\left(x\right)$ \textbf{always} gives the same value regardless of the input; either $f\left(x\right)=0$ for all $x$ or $f\left(x\right)=1$ for all $x$.
\item A \textbf{balanced }function, that is, $f\left(x\right)=0$ for \textbf{exactly half }of all possible values of $x$ and $f\left(x\right)=1$ for the other half.
\end{enumerate}
$f\left(x\right)$ can never be anything in between, e.g. equal to 0 for all values of $x$ except one; it must be one of these two options. The problem is now to determine whether $f\left(x\right)$ is constant or balanced.

With a classical computer, in the best-case scenario, we only need to calculate $f\left(x\right)$ for two different values of $x$. For example, if $f\left(0\right)=0$ and $f\left(1\right)=1$, then we immediately know that $f\left(x\right)$ cannot be constant, and thus it must be balanced. However, in the worst-case scenario we will have to calculate $f\left(x\right)$ for $2^{n-1}+1$ different values of $x$.

The worst-case scenario occurs when the first $2^{n-1}$ values (i.e. half of all the possible values) all turn out to be the same. That is still not enough to know if the function is constant or balanced; it could still be either of the two. We must calculate one more value, and if that is \textbf{also }the same, then we know the function must be constant, otherwise it must be balanced. Hence the total number of calculations required is $2^{n-1}+1$.

This means that we may need to calculate $f\left(x\right)$ for a very large number of values. For example, if $n=64$, then we may need to make up to $2^{63}+1\ap9.2\xx10^{18}$ calculations. Even if each calculation only takes the classical computer one microsecond (a millionth of a second), it would still take around \textbf{300,000 years }to finish all the required calculations in the worst-case scenario\footnote{Note that this only applies if we want to know for sure. If a probabilistic answer is acceptable, we could just choose $k\ge2$ distinct values at random; if they're all the same, we guess the function is constant. If it's actually balanced, it can be shown that the probability that all $k$ values are the same is less than $1/2^{k-1}$, which decreases exponentially with $k$.}!

Things become much easier if we happen to have a quantum computer with $n+1$ qubits. The \emph{Deutsch-Jozsa algorithm}\index{Deutsch-Jozsa algorithm} is a straightforward generalization of Deutsch's algorithm, replacing the single input qubit with $n$ input qubits. We start in the initial state
\[
\left|\Psi\right\rangle =\left|0\right\rangle ^{\otimes n}\otimes\left|1\right\rangle ,
\]
where the notation $\left|0\right\rangle ^{\otimes n}$, the order-$n$ \emph{tensor power}\index{Tensor power} of $\left|0\right\rangle $, means “the tensor product of $\left|0\right\rangle $ with itself $n$ times”, that is,
\[
\left|0\right\rangle ^{\otimes n}\equiv\underbrace{\left|0\right\rangle \otimes\cdots\otimes\left|0\right\rangle }_{n\textrm{ copies}}.
\]
Let us send each of the first $n$ qubits through a Hadamard gate. This is called a \emph{Hadamard transform}\index{Hadamard transform}. We get:
\[
H_{1,\ldots,n}\left|0\right\rangle ^{\otimes n}=\underbrace{\frac{1}{\sqrt{2}}\left(\left|0\right\rangle +\left|1\right\rangle \right)\otimes\cdots\otimes\frac{1}{\sqrt{2}}\left(\left|0\right\rangle +\left|1\right\rangle \right)}_{n\textrm{ copies}}.
\]
Consider, for clarity, the case of $n=3$. Then we have:
\begin{align*}
H_{123}\left|0\right\rangle ^{\otimes3} & =\frac{1}{\sqrt{2}}\left(\left|0\right\rangle +\left|1\right\rangle \right)\otimes\frac{1}{\sqrt{2}}\left(\left|0\right\rangle +\left|1\right\rangle \right)\otimes\frac{1}{\sqrt{2}}\left(\left|0\right\rangle +\left|1\right\rangle \right)\\
 & =\frac{1}{\sqrt{2^{3}}}\left(\left|000\right\rangle +\left|001\right\rangle +\left|010\right\rangle +\left|011\right\rangle +\left|100\right\rangle +\left|101\right\rangle +\left|110\right\rangle +\left|111\right\rangle \right),
\end{align*}
or in other words, $H_{123}\left|0\right\rangle ^{\otimes3}$ is a superposition of all the possible combinations of 3 qubits, with equal probabilities. We can write this in a compact notation as follows:
\[
H_{123}\left|0\right\rangle ^{\otimes3}=\frac{1}{\sqrt{2^{3}}}\sum_{x\in\left\{ 0,1\right\} ^{3}}\left|x\right\rangle ,
\]
where the notation $\left\{ 0,1\right\} ^{3}$ means “all the possible combinations of 3 bits”, and as we have seen in the table above, is equivalent to a sum over the integers from 0 to $2^{3}-1=7$. Generalizing to $n$ bits, we see that:
\[
H_{1,\ldots,n}\left|0\right\rangle ^{\otimes n}=\frac{1}{\sqrt{2^{n}}}\sum_{x\in\left\{ 0,1\right\} ^{n}}\left|x\right\rangle ,
\]
where now the sum on $x$ is over all $n$-bit integers, so from 0 to $2^{n}-1$. The Hadamard transform thus takes $n$ qubits in the state $\left|0\right\rangle $ and transforms them into a superposition of every possible combination of $n$ qubits with equal probabilities. In conclusion, applying this to the first $n$ qubits of the full state $\left|\Psi\right\rangle =\left|0\right\rangle ^{\otimes n}\otimes\left|1\right\rangle $, we get:
\[
H_{1,\ldots,n}\left|\Psi\right\rangle =\frac{1}{\sqrt{2^{n}}}\sum_{x\in\left\{ 0,1\right\} ^{n}}\left|x\right\rangle \otimes\left|1\right\rangle .
\]
Let us also send the target qubit, number $n+1$, which started in the state $\left|1\right\rangle $, through the Hadamard gate:
\[
H_{n+1}H_{1,\ldots,n}\left|\Psi\right\rangle =\frac{1}{\sqrt{2^{n}}}\sum_{x\in\left\{ 0,1\right\} ^{n}}\left|x\right\rangle \otimes\frac{1}{\sqrt{2}}\left(\left|0\right\rangle -\left|1\right\rangle \right).
\]
We now use the same $U_{f}$ from \zcref[S]{eq:Uf}. In our discussion of Deutsch's algorithm, we found in \zcref[S]{eq:Uf2} that
\[
U_{f}\left(\left|x\right\rangle \otimes\frac{1}{\sqrt{2}}\left(\left|0\right\rangle -\left|1\right\rangle \right)\right)=\left(-1\right)^{f\left(x\right)}\left|x\right\rangle \otimes\frac{1}{\sqrt{2}}\left(\left|0\right\rangle -\left|1\right\rangle \right).
\]
Therefore
\begin{equation}
U_{f}H_{n+1}H_{1,\ldots,n}\left|\Psi\right\rangle =\frac{1}{\sqrt{2^{n}}}\sum_{x\in\left\{ 0,1\right\} ^{n}}\left(-1\right)^{f\left(x\right)}\left|x\right\rangle \otimes\frac{1}{\sqrt{2}}\left(\left|0\right\rangle -\left|1\right\rangle \right).\label{eq:UHH}
\end{equation}
Notice that we now automatically get the results of calculating $f\left(x\right)$, all $2^{n}$ of them, in the \textbf{amplitudes} of each state in the superposition, even though we never actually calculated the function for each input individually!

However, as I stressed before, this information cannot actually be obtained, since we need to perform a \textbf{measurement}, and we will get only one value of $x$ chosen at random; and even then, we won't actually know the value of $f\left(x\right)$ for that $x$, since it's hidden in the amplitude. The amplitude determines the probability, but we cannot measure it directly, and even the probability is actually determined by the \textbf{magnitude-squared }of the amplitude, which means we lose the phase factor $\left(-1\right)^{f\left(x\right)}$ in any case.

Nonetheless, there is still a clever way to extract the information we want – whether $f\left(x\right)$ is constant or balanced – by taking advantage of \textbf{quantum interference}. That is, we make all the different amplitudes interfere with each other constructively or destructively in a specific way. Let us see how this works.

The action of the Hadamard gate on one qubit can be written succinctly as follows:
\[
H\left|x\right\rangle =\frac{1}{\sqrt{2}}\left(\left|0\right\rangle +\left(-1\right)^{x}\left|1\right\rangle \right)\sp x\in\left\{ 0,1\right\} .
\]
You can check that this works for each of the options $x=0$ and $x=1$. We can write it even more compactly as follows:
\[
H\left|x\right\rangle =\frac{1}{\sqrt{2}}\sum_{z\in\left\{ 0,1\right\} }\left(-1\right)^{xz}\left|z\right\rangle \sp x\in\left\{ 0,1\right\} .
\]
Let us consider how the Hadamard transform acts on an $n$-qubit state $\left|x\right\rangle $ for some $x\in\left\{ 0,1\right\} ^{n}$. Explicitly, the state is
\[
\left|x\right\rangle \equiv\left|x_{1},\ldots,x_{n}\right\rangle \equiv\left|x_{1}\right\rangle \otimes\cdots\otimes\left|x_{n}\right\rangle \sp x_{i}\in\left\{ 0,1\right\} \textrm{ for all }i\in\left\{ 1,\ldots,n\right\} .
\]
Thus:
\begin{align*}
H_{1,\ldots,n}\left|x\right\rangle  & =\frac{1}{\sqrt{2}}\sum_{z_{1}\in\left\{ 0,1\right\} }\left(-1\right)^{x_{1}z_{1}}\left|z_{1}\right\rangle \otimes\cdots\otimes\frac{1}{\sqrt{2}}\sum_{z_{n}\in\left\{ 0,1\right\} }\left(-1\right)^{x_{n}z_{n}}\left|z_{n}\right\rangle \\
 & =\frac{1}{\sqrt{2^{n}}}\sum_{z_{1},\ldots,z_{n}\in\left\{ 0,1\right\} }\left(-1\right)^{x_{1}z_{1}+\cdots+x_{n}z_{n}}\left|z_{1}\right\rangle \otimes\cdots\otimes\left|z_{n}\right\rangle ,
\end{align*}
where we simply combined all the phase factors $\left(-1\right)^{x_{i}z_{i}}$ into one.

Just as we treat $x_{1},\ldots,x_{n}$ as the bits of an integer $x$ between 0 and $2^{n}-1$, we can also treat $z_{1},\ldots,z_{n}$ as the bits of an integer $z$. We can then define a \emph{bitwise inner product}\index{Bitwise inner product}:
\[
x\odot z\equiv x_{1}z_{1}+\cdots+x_{n}z_{n}.
\]
This allows us to write the result of the Hadamard transform simply as:
\[
H_{1,\ldots,n}\left|x\right\rangle =\frac{1}{\sqrt{2^{n}}}\sum_{z\in\left\{ 0,1\right\} ^{n}}\left(-1\right)^{x\odot z}\left|z\right\rangle .
\]
Using this equation, we can now calculate the Hadamard transform of the first $n$ qubits of the state in \zcref[S]{eq:UHH}:
\begin{align*}
 & H_{1,\ldots,n}U_{f}H_{n+1}H_{1,\ldots,n}\left|\Psi\right\rangle =\\
 & \qquad=\frac{1}{\sqrt{2^{n}}}\sum_{x\in\left\{ 0,1\right\} ^{n}}\left(-1\right)^{f\left(x\right)}\left(\vphantom{\blll}H_{1,\ldots,n}\left|x\right\rangle \right)\otimes\frac{1}{\sqrt{2}}\left(\left|0\right\rangle -\left|1\right\rangle \right)\\
 & \qquad=\frac{1}{\sqrt{2^{n}}}\sum_{x\in\left\{ 0,1\right\} ^{n}}\left(-1\right)^{f\left(x\right)}\left(\frac{1}{\sqrt{2^{n}}}\sum_{z\in\left\{ 0,1\right\} ^{n}}\left(-1\right)^{x\odot z}\left|z\right\rangle \right)\otimes\frac{1}{\sqrt{2}}\left(\left|0\right\rangle -\left|1\right\rangle \right)\\
 & \qquad=\frac{1}{2^{n}}\sum_{x,z\in\left\{ 0,1\right\} ^{n}}\left(-1\right)^{x\odot z+f\left(x\right)}\left|z\right\rangle \otimes\frac{1}{\sqrt{2}}\left(\left|0\right\rangle -\left|1\right\rangle \right).
\end{align*}
We can write this in a more suggestive way as follows:
\[
H_{1,\ldots,n}U_{f}H_{n+1}H_{1,\ldots,n}\left|\Psi\right\rangle =\sum_{z\in\left\{ 0,1\right\} ^{n}}\left(\frac{1}{2^{n}}\sum_{x\in\left\{ 0,1\right\} ^{n}}\left(-1\right)^{x\odot z+f\left(x\right)}\right)\left|z\right\rangle \otimes\frac{1}{\sqrt{2}}\left(\left|0\right\rangle -\left|1\right\rangle \right).
\]
This means that, when we measure the first $n$ qubits, the amplitude to measure the integer $z$ is given by the \textbf{sum}:
\[
A\left(z\right)\equiv\frac{1}{2^{n}}\sum_{x\in\left\{ 0,1\right\} ^{n}}\left(-1\right)^{x\odot z+f\left(x\right)}.
\]
Consider specifically the amplitude to measure $z=0$, that is, the term in the superposition where all $n$ qubits $z_{1},\ldots,z_{n}$ are zero. In that case $x\odot z=0$, so the amplitude will simply be:
\[
A\left(0\right)=\frac{1}{2^{n}}\sum_{x\in\left\{ 0,1\right\} ^{n}}\left(-1\right)^{f\left(x\right)}.
\]
The sum will depend on which of the two possible kinds of $f\left(x\right)$ we have:
\begin{itemize}
\item If $f\left(x\right)$ is \textbf{constant}, then we have
\[
A\left(0\right)=\begin{cases}
\frac{1}{2^{n}}\sum_{x\in\left\{ 0,1\right\} ^{n}}\left(+1\right)=+1 & f\left(x\right)=0,\\
\frac{1}{2^{n}}\sum_{x\in\left\{ 0,1\right\} ^{n}}\left(-1\right)=-1 & f\left(x\right)=1.
\end{cases}
\]
In each case, we are summing over $2^{n}$ terms which are all the same (either $+1$ or $-1$), and then dividing by $2^{n}$, so the result is just $+1$ or $-1$ depending on the constant value of $f\left(x\right)$. But this means that the amplitude has magnitude 1, so the probability to measure $z=0$ is $\left|A\left(0\right)\right|^{2}=1$. In other words, if $f\left(x\right)$ is \textbf{constant}, then we will \textbf{always }measure $z=0$, with 100\% probability, and we can never measure any other values for $z$, since they must have a 0\% probability. Notice that in this case we got \textbf{constructive interference}\index{Interference} between the amplitudes.
\item If $f\left(x\right)$ is \textbf{balanced}, then it is equal to 0 for exactly half the values of $x$ and to 1 for the other half. Therefore, in the sum $\sum_{x\in\left\{ 0,1\right\} ^{n}}\left(-1\right)^{f\left(x\right)}$, half the terms will be $+1$ and half will be $-1$. The terms will exactly cancel each other, and the amplitude will be zero, so the probability to measure $z=0$ is $\left|A\left(0\right)\right|^{2}=0$. In other words, if $f\left(x\right)$ is \textbf{balanced}, then we will \textbf{never }measure $z=0$, since that outcome has 0\% probability. Notice that in this case we got \textbf{destructive interference} between the amplitudes.
\end{itemize}
In conclusion, if we measure $z=0$ (all qubits are 0) then we know for sure that $f\left(x\right)$ is constant, and if we measure $z\ne0$ (at least one qubit is 1) then we know for sure that $f\left(x\right)$ is balanced. We found this just by passing the qubits through a few gates, without having to actually calculate $2^{n-1}+1$ values of $f\left(x\right)$ as in the classical case. Going back to our example of $n=64$, it could take a classical computer up to 300,000 years to find the answer, while a quantum computer can theoretically find it in the time it takes to pass the qubits through the gates.

This result is very impressive, but we should also mention that an algorithm for determining whether a function is constant or balanced is not something you would ever realistically need. This problem was specifically designed to show the potential benefits of quantum computers, but it has no known applications.

A more useful quantum algorithm, which you may have heard of, is \emph{Shor's algorithm}\index{Shor's algorithm}, which is used to factorize integers, and can do so much faster than any known classical algorithm. Factorizing integers is a very common problem, especially in cryptography. This algorithm is a bit more complicated, so we will not discuss it here, but you should look up this and other quantum algorithms if you're interested.
\begin{problem}\leavevmode
You meet a layperson with no knowledge of the math of quantum mechanics. They tell you about an amazing YouTube video with 100 million likes, where a famous YouTuber with 10 million subscribers talked about quantum computers. According to the famous YouTuber, who has a master's degree in physics, quantum computers are so immensely powerful because they can apply any calculation to all the possible inputs in parallel, while a classical computer can only calculate them one at a time. Explain to the layperson, in simple terms and without using any math, what the YouTuber got wrong. Identify the concepts on which the YouTuber's misconception is most likely based, and explain how the YouTuber applied these concepts incorrectly. Then clarify how quantum computers really work.
\end{problem}

\begin{problem}\leavevmode
\begin{problemparts}
\item In \zcref[S]{prob:Write-a-computer-qubits} you wrote a computer program which allows the user to pass qubits through quantum gates and perform measurements. Your program should therefore be able to do anything a quantum computer can. Use it to perform the Deutsch-Jozsa algorithm.
\item Your program is running on a classical computer, so obviously, it shouldn't be able to reproduce the speedup of a quantum computer. How can you reconcile this with the fact that the program is able to perform the Deutsch-Jozsa algorithm? You should be able to answer this part of the question even if you didn't do \zcref[S]{prob:Write-a-computer-qubits} or part A, but writing the program should give you some insight into the answer.
\end{problemparts}
\end{problem}

\section{Density Operators and Mixed States}

\subsection{Introduction to Density Operators}

\subsubsection{Density Operators and Matrices}\label{subsec:Density-Operators-and}

So far, we have described quantum states as vectors in a Hilbert space. We will refer to states described by a vector $\left|\Psi\right\rangle $ as \emph{pure states}\index{Pure state}\index{Quantum state!Pure}. However, this is not the most general way to define quantum states. Consider a quantum system which can be in any of several different states $\left|\Psi_{i}\right\rangle $ with corresponding \textbf{classical} probabilities $p_{i}$, such that
\[
\sum_{i}p_{i}=1.
\]
We call this a \emph{statistical ensemble of pure states}\index{Ensemble of pure states}\index{Statistical ensemble of pure states}. The system is then described not by a vector, but by a \emph{density operator}\index{Density operator}, usually denoted by $\rho$:
\[
\rho\equiv\sum_{i}p_{i}\left|\Psi_{i}\right\rangle \left\langle \Psi_{i}\right|.
\]
The density operator is simply the sum of outer products of the pure states, each weighted by its probability. Since it is a sum of outer products, it is also an operator on the Hilbert space, but don't get confused – it does not represent an observable or transformation, like other operators; it represents a \textbf{state}.

The density operator can be represented as a matrix given a suitable choice of basis, as we learned in \zcref[S]{subsec:Representing-Matrices-in}. Therefore, a density operator is also often referred to as a \emph{density matrix}\index{Density matrix}.

Let's consider some examples using qubits in $\BBC^{2}$. Recall that the outer products of the qubit computational basis states are represented by the following matrices:
\[
\left|0\right\rangle \left\langle 0\right|=\left(\begin{array}{cc}
1 & 0\\
0 & 0
\end{array}\right)\sp\left|0\right\rangle \left\langle 1\right|=\left(\begin{array}{cc}
0 & 1\\
0 & 0
\end{array}\right),
\]
\[
\left|1\right\rangle \left\langle 0\right|=\left(\begin{array}{cc}
0 & 0\\
1 & 0
\end{array}\right)\sp\left|1\right\rangle \left\langle 1\right|=\left(\begin{array}{cc}
0 & 0\\
0 & 1
\end{array}\right).
\]
Also, for the states
\[
\left|+\right\rangle =\frac{1}{\sqrt{2}}\left(\left|0\right\rangle +\left|1\right\rangle \right)\sp\left|-\right\rangle =\frac{1}{\sqrt{2}}\left(\left|0\right\rangle -\left|1\right\rangle \right),
\]
we find that
\begin{align*}
\left|+\right\rangle \left\langle +\right| & =\frac{1}{\sqrt{2}}\left(\left|0\right\rangle +\left|1\right\rangle \right)\frac{1}{\sqrt{2}}\left(\left\langle 0\right|+\left\langle 1\right|\right)\\
 & =\hf\left(\left|0\right\rangle \left\langle 0\right|+\left|0\right\rangle \left\langle 1\right|+\left|1\right\rangle \left\langle 0\right|+\left|1\right\rangle \left\langle 1\right|\right)\\
 & =\hf\left(\left(\begin{array}{cc}
1 & 0\\
0 & 0
\end{array}\right)+\left(\begin{array}{cc}
0 & 1\\
0 & 0
\end{array}\right)+\left(\begin{array}{cc}
0 & 0\\
1 & 0
\end{array}\right)+\left(\begin{array}{cc}
0 & 0\\
0 & 1
\end{array}\right)\right)\\
 & =\hf\left(\begin{array}{cc}
1 & 1\\
1 & 1
\end{array}\right),
\end{align*}
and similarly
\[
\left|-\right\rangle \left\langle -\right|=\hf\left(\begin{array}{cc}
1 & -1\\
-1 & 1
\end{array}\right).
\]
If a qubit is just in the pure state $\left|0\right\rangle $ with probability 1, then the density matrix is trivial:
\[
\rho=\left|0\right\rangle \left\langle 0\right|=\left(\begin{array}{cc}
1 & 0\\
0 & 0
\end{array}\right).
\]
What about superposition? The pure state $\left|+\right\rangle $ yields 0 or 1 with probability 1/2 each when measured in the computational basis. If the qubit is in this superposition state with probability 1, then the corresponding density matrix is
\[
\rho=\left|+\right\rangle \left\langle +\right|=\hf\left(\begin{array}{cc}
1 & 1\\
1 & 1
\end{array}\right).
\]
On the other hand, if the qubit is in an ensemble where it is prepared either in the state $\left|0\right\rangle $ or $\left|1\right\rangle $ with probability 1/2 each, then its density matrix will be
\begin{align*}
\rho & =\hf\left|0\right\rangle \left\langle 0\right|+\hf\left|1\right\rangle \left\langle 1\right|\\
 & =\hf\left(\begin{array}{cc}
1 & 0\\
0 & 0
\end{array}\right)+\hf\left(\begin{array}{cc}
0 & 0\\
0 & 1
\end{array}\right)\\
 & =\hf\left(\begin{array}{cc}
1 & 0\\
0 & 1
\end{array}\right).
\end{align*}
This is \textbf{not }a pure state, and it is \textbf{not }a superposition! As you can see, a \textbf{superposition }of $\left|0\right\rangle $ and $\left|1\right\rangle $ with probability 1/2 each and an \textbf{ensemble }of $\left|0\right\rangle $ and $\left|1\right\rangle $ with probability 1/2 each are represented by different density matrices.

Finally, consider a qubit prepared in the state $\left|+\right\rangle $ with probability 1/2 and the state $\left|-\right\rangle $ with probability 1/2. Then the density matrix is
\begin{align*}
\rho & =\hf\left|+\right\rangle \left\langle +\right|+\hf\left|-\right\rangle \left\langle -\right|\\
 & =\hf\left(\hf\left(\begin{array}{cc}
1 & 1\\
1 & 1
\end{array}\right)\right)+\hf\left(\hf\left(\begin{array}{cc}
1 & -1\\
-1 & 1
\end{array}\right)\right)\\
 & =\fr\left(\left(\begin{array}{cc}
1 & 1\\
1 & 1
\end{array}\right)+\left(\begin{array}{cc}
1 & -1\\
-1 & 1
\end{array}\right)\right)\\
 & =\fr\left(\begin{array}{cc}
2 & 0\\
0 & 2
\end{array}\right)\\
 & =\hf\left(\begin{array}{cc}
1 & 0\\
0 & 1
\end{array}\right),
\end{align*}
which is actually the same as the previous case, an ensemble prepared either in the state $\left|0\right\rangle $ or $\left|1\right\rangle $ with probability 1/2 each! We will try to understand what this means in \zcref[S]{subsec:Equivalent-Ensembles} below.

\subsubsection{Density Operators as States}

Pure states cannot be any vector in a Hilbert space; they must be normalized to 1, and they are only defined up to a phase factor. More precisely, as we mentioned in \zcref[S]{fn:Equiv}, a pure state is an equivalence class of vectors such that $\left|\Psi\right\rangle $ is equivalent to $\lambda\left|\Psi\right\rangle $ for any scalar $\lambda\in\BBC\setminus\left\{ 0\right\} $, and the choice of normalization to 1 (which is required in order for probabilities to sum to 1) selects the unit vectors from that class, which still differ by an overall phase factor. So pretty much any vector in a Hilbert space, except the zero vector, can be considered a quantum state.

The same is not true for density operators; not every operator on a Hilbert space can be a density operator, that is, represent a state. First of all, consider a density operator constructed explicitly as an ensemble of pure states, as in \zcref[S]{subsec:Density-Operators-and}:
\begin{equation}
\rho=\sum_{i}p_{i}\left|\Psi_{i}\right\rangle \left\langle \Psi_{i}\right|.\label{eq:ensemble}
\end{equation}
Taking the trace, we find that
\[
\tr\rho=\sum_{i}p_{i}\tr\left(\left|\Psi_{i}\right\rangle \left\langle \Psi_{i}\right|\right)=\sum_{i}p_{i}\tr\left(\langle\Psi_{i}|\Psi_{i}\rangle\right)=\sum_{i}p_{i}\langle\Psi_{i}|\Psi_{i}\rangle=\sum_{i}p_{i}=1.
\]
In the first step we used the fact that the trace is linear. In the next step we used \zcref[S]{eq:trace-outer}. Next we used the fact that the trace of a $1\times1$ matrix is equal to the single element of that matrix (in other words, the trace of a scalar is the same scalar). Then we used that the states $\left|\Psi_{i}\right\rangle $ are normalized to 1, and finally, that the probabilities sum to 1.

Furthermore, let $\left|\Phi\right\rangle $ be an arbitrary vector. Then
\begin{align*}
\langle\Phi|\rho|\Phi\rangle & =\langle\Phi|\left(\sum_{i}p_{i}\left|\Psi_{i}\right\rangle \left\langle \Psi_{i}\right|\right)|\Phi\rangle\\
 & =\sum_{i}p_{i}\langle\Phi|\Psi_{i}\rangle\langle\Psi_{i}|\Phi\rangle\\
 & =\sum_{i}p_{i}\left|\langle\Phi|\Psi_{i}\rangle\right|^{2}\\
 & \ge0,
\end{align*}
since the probabilities $p_{i}$ are all non-negative and the magnitudes-squared $\left|\langle\Phi|\Psi_{i}\rangle\right|^{2}$ are also non-negative. Therefore, we see that $\rho$ constructed as an ensemble of pure states is a positive semi-definite operator (as defined in \zcref[S]{subsec:Positive-Definite-and-Positive}) with trace 1.

Note that the ensemble states $\left|\Psi_{i}\right\rangle $ may be non-orthogonal, so \zcref[S]{eq:ensemble} is not necessarily a spectral decomposition! However, now, suppose we start with an \textbf{arbitrary }positive semi-definite operator $\rho$ with trace 1. Since it is positive semi-definite (and thus also Hermitian according to \zcref[S]{prob:pos-hermitian}), it does have (according to \zcref[S]{subsec:Spectral-Decomposition}) a spectral decomposition
\[
\rho=\sum_{i}q_{i}\left|\Theta_{i}\right\rangle \left\langle \Theta_{i}\right|,
\]
where $\left|\Theta_{i}\right\rangle $ is an orthonormal eigenbasis and $q_{i}$ are the non-negative eigenvalues of the eigenvectors $\left|\Theta_{i}\right\rangle $. Since $\tr\rho=1$, the eigenvalues $q_{i}$ sum to 1, so they can be treated as probabilities. We thus see that any positive semi-definite operator with trace 1 represents an ensemble of pure states, given by its eigenvectors, with the probability of each pure state in the ensemble given by its eigenvalue. Note that this is very different from an observable, where the eigenvalues of the states represent the measured values (and can also be negative).

In conclusion, an operator $\rho$ is a density operator\index{Density operator}, and represents a quantum state, if and only if it satisfies the following two conditions:
\begin{enumerate}
\item \emph{The trace condition}\index{Trace condition}: $\tr\rho=1$.
\item \emph{The positivity condition}\index{Positivity condition}: $\rho$ is a positive semi-definite operator.
\end{enumerate}

\subsubsection{Measuring Density Operators}\label{subsec:Measuring-Density-Operators}

Consider an observable $A$ on an $n$-dimensional Hilbert space, with $k\le n$ distinct eigenvalues $\left\{ \lambda_{1},\ldots,\lambda_{k}\right\} $, corresponding projectors $\left\{ P_{1},\ldots,P_{k}\right\} $, and spectral decomposition
\[
A=\sum^{k}_{i=1}\lambda_{i}P_{i}.
\]
The projective Measurement Axiom (see \zcref[S]{subsec:The-Measurement-Axiom-Projective}) says that if the system is in the (pure) state $\left|\Psi\right\rangle $, then the probability to measure the eigenvalue $\lambda_{i}$ is given by
\[
\langle\Psi|P_{i}|\Psi\rangle,
\]
and after the measurement, if the eigenvalue $\lambda_{i}$ was measured, then the system will collapse to the state
\begin{equation}
\left|\Psi\right\rangle \mt\frac{P_{i}\left|\Psi\right\rangle }{\sqrt{\langle\Psi|P_{i}|\Psi\rangle}}.\label{eq:collapse}
\end{equation}
In addition, the expectation value of $A$ when the system is in the state $\left|\Psi\right\rangle $ is given by
\[
\left\langle A\right\rangle _{\Psi}\equiv\langle\Psi|A|\Psi\rangle.
\]

To reformulate this in terms of density operators, let the system be prepared in the ensemble
\[
\rho=\sum_{j}p_{j}\left|\Psi_{j}\right\rangle \left\langle \Psi_{j}\right|.
\]
To remind you, this means that the system is prepared in the state $\left|\Psi_{j}\right\rangle $ with probability $p_{j}$ for each $j$, and the states $\left|\Psi_{j}\right\rangle $ are not necessarily orthogonal. Let's assume first that the system was, in fact, prepared in the specific pure state $\left|\Psi_{j}\right\rangle $ for some $j$. Then by the Measurement Axiom for pure states, the probability $P\mr{\lambda_{i}}{\Psi_{j}}$ to measure the eigenvalue $\lambda_{i}$ is given by
\[
P\mr{\lambda_{i}}{\Psi_{j}}=\langle\Psi_{j}|P_{i}|\Psi_{j}\rangle,
\]
which we can write in terms of a trace, using \zcref[S]{eq:trace-outer}, as follows:
\[
P\mr{\lambda_{i}}{\Psi_{j}}=\tr\left(P_{i}|\Psi_{j}\rangle\langle\Psi_{j}|\right).
\]
This is the \textbf{conditional }probability (recall \zcref[S]{subsec:Conditional-Probability}) to measure $\lambda_{i}$ given that the system was prepared in the state $\left|\Psi_{j}\right\rangle $, but we know that the system can also be prepared in other states. According to the law of total probability (see \zcref[S]{subsec:Conditional-Probability}), the total probability $P\left(\lambda_{i}\right)$ to measure $\lambda_{i}$, taking into account every possible state $\left|\Psi_{j}\right\rangle $ prepared with probability $p_{j}$, will be the sum of the conditional probabilities weighted by the probability for each condition:
\begin{align*}
P\left(\lambda_{i}\right) & =\sum_{j}P\mr{\lambda_{i}}{\Psi_{j}}P\left(\Psi_{j}\right)\\
 & =\sum_{j}\tr\left(P_{i}|\Psi_{j}\rangle\langle\Psi_{j}|\right)p_{j}\\
 & =\tr\left(P_{i}\sum_{j}p_{j}|\Psi_{j}\rangle\langle\Psi_{j}|\right)\\
 & =\tr\left(P_{i}\rho\right).
\end{align*}
Next, we must find out what state the system collapses to. Assume again that the system was prepared in the specific pure state $\left|\Psi_{j}\right\rangle $ for some $j$. If the eigenvalue $\lambda_{i}$ was measured, the state of the system after the measurement is then given, according to \zcref[S]{eq:collapse}, by
\begin{equation}
|\Psi^{i}_{j}\rangle\equiv\frac{P_{i}\left|\Psi_{j}\right\rangle }{\sqrt{\langle\Psi_{j}|P_{i}|\Psi_{j}\rangle}}.\label{eq:Psi_i_j}
\end{equation}
The density operator after this measurement will be
\[
\rho_{i}\equiv\sum_{j}P\mr{\Psi_{j}}{\lambda_{i}}|\Psi^{i}_{j}\rangle\langle\Psi^{i}_{j}|,
\]
where $P\mr{\Psi_{j}}{\lambda_{i}}$ is the conditional probability that the system was prepared in the state $\left|\Psi_{j}\right\rangle $ given that the eigenvalue $\lambda_{i}$ was measured. This simply comes from the definition of the density operator as the sum of states weighted by their probabilities. Plugging in \zcref[S]{eq:Psi_i_j}, we get
\begin{align}
\rho_{i} & =\sum_{j}P\mr{\Psi_{j}}{\lambda_{i}}\frac{P_{i}\left|\Psi_{j}\right\rangle }{\sqrt{\langle\Psi_{j}|P_{i}|\Psi_{j}\rangle}}\frac{\left\langle \Psi_{j}\right|P^{\dagger}_{i}}{\sqrt{\langle\Psi_{j}|P_{i}|\Psi_{j}\rangle}}\nonumber \\
 & =\sum_{j}P\mr{\Psi_{j}}{\lambda_{i}}\frac{P_{i}\left|\Psi_{j}\right\rangle \left\langle \Psi_{j}\right|P^{\dagger}_{i}}{\langle\Psi_{j}|P_{i}|\Psi_{j}\rangle}.\label{eq:rho_i}
\end{align}

Recall that the relationship between the conditional probability and the joint probability is given by \zcref[S]{eq:conditional-joint}:
\[
P\mr XY=\frac{P\left(X\cap Y\right)}{P\left(Y\right)}.
\]
Therefore
\begin{equation}
P\mr{\Psi_{j}}{\lambda_{i}}=\frac{P\left(\Psi_{j}\cap\lambda_{i}\right)}{P\left(\lambda_{i}\right)}.\label{eq:P-j-i}
\end{equation}
Similarly, we have
\[
P\mr{\lambda_{i}}{\Psi_{j}}=\frac{P\left(\Psi_{j}\cap\lambda_{i}\right)}{P\left(\Psi_{j}\right)}\soosp P\left(\Psi_{j}\cap\lambda_{i}\right)=P\mr{\lambda_{i}}{\Psi_{j}}P\left(\Psi_{j}\right),
\]
and plugging that into \zcref[S]{eq:P-j-i} we get
\[
P\mr{\Psi_{j}}{\lambda_{i}}=\frac{P\mr{\lambda_{i}}{\Psi_{j}}P\left(\Psi_{j}\right)}{P\left(\lambda_{i}\right)}=\frac{\langle\Psi_{j}|P_{i}|\Psi_{j}\rangle p_{j}}{\tr\left(P_{i}\rho\right)}.
\]
We can use this to simplify \zcref[S]{eq:rho_i}:
\begin{align*}
\rho_{i} & =\sum_{j}\frac{\langle\Psi_{j}|P_{i}|\Psi_{j}\rangle p_{j}}{\tr\left(P_{i}\rho\right)}\frac{P_{i}\left|\Psi_{j}\right\rangle \left\langle \Psi_{j}\right|P^{\dagger}_{i}}{\langle\Psi_{j}|P_{i}|\Psi_{j}\rangle}\\
 & =\frac{P_{i}\left(\sum_{j}p_{j}\left|\Psi_{j}\right\rangle \left\langle \Psi_{j}\right|\right)P^{\dagger}_{i}}{\tr\left(P_{i}\rho\right)}\\
 & =\frac{P_{i}\rho P^{\dagger}_{i}}{\tr\left(P_{i}\rho\right)}.
\end{align*}
This is the final expression for the state the system collapses to if the eigenvalue $\lambda_{i}$ was measured. In analogy with the case of the state vector, \zcref[S]{eq:collapse}, we see that the density operator after collapse is given by projecting the original density operator onto the eigenspace of the eigenvalue that was measured, and then dividing by the probability of that eigenvalue, which normalizes the state so that it has trace 1, as required.

Finally, the expectation value of the observable $A$, given that the system is in the state $\rho$, is the sum of the expectation values of each pure state in the ensemble, weighted by their preparation probabilities (not to be confused with the measurement probabilities for each eigenvalue of $A$, which are not actually used here):
\begin{align*}
\left\langle A\right\rangle _{\rho} & =\sum_{j}p_{j}\langle\Psi_{j}|A|\Psi_{j}\rangle\\
 & =\sum_{j}p_{j}\tr\left(A|\Psi_{j}\rangle\langle\Psi_{j}|\right)\\
 & =\tr\left(A\sum_{j}p_{j}|\Psi_{j}\rangle\langle\Psi_{j}|\right)\\
 & =\tr\left(A\rho\right).
\end{align*}

\subsubsection{The Axioms of Quantum Theory for Density Operators}

\index{Axioms of quantum mechanics!With density operators}The axioms of quantum theory for discrete systems with finite-dimensional Hilbert spaces can be reformulated in terms of density operators. The System Axiom remains the same:
\begin{quote}
\textbf{The System Axiom}\index{System Axiom!Density operators}\textbf{: }Physical systems are represented by complex $n$-dimensional Hilbert spaces $\BBC^{n}$, where $n$ depends on the specific system.
\end{quote}
However, the State Axiom changes to:
\begin{quote}
\textbf{The State Axiom (Density Operators)}\index{State Axiom!Density operators}\textbf{: }The states of the system are represented by positive semi-definite operators with trace 1 on the system's Hilbert space. These are known as density operators, and can be represented in a particular basis as $n\times n$ matrices called density matrices.
\end{quote}
The Operator Axiom must also be modified to clarify how the operators act on the states:
\begin{quote}
\textbf{The Operator Axiom (Density Operators)}\index{Operator Axiom!Density operators}\textbf{: }The operators on the system, which act on states, are represented by $n\xx n$ matrices in the system's Hilbert space. The action of an operator $A$ on a state $\rho$ is given by $A\rho A^{\dagger}$.
\end{quote}
Indeed, this follows from the Operator Axiom on pure states by noticing that if the action of an operator $A$ on a pure state is given by $\left|\Psi\right\rangle \mt A\left|\Psi\right\rangle $, then the action of an operator $A$ on a density operator $\rho$ is given by
\[
\rho=\sum_{i}p_{i}\left|\Psi_{i}\right\rangle \left\langle \Psi_{i}\right|\mt\sum_{i}p_{i}A\left|\Psi_{i}\right\rangle \left\langle \Psi_{i}\right|A^{\dagger}=A\rho A^{\dagger}.
\]
If $U$ is unitary, then $U\rho U^{\dagger}$ is also a state.

The Observable Axiom stays the same, but we remove the corollary regarding superposition, as that only applies to pure states:
\begin{quote}
\textbf{The Observable Axiom}\index{Observable Axiom!Density operators}\textbf{: }Physical observables in the system are represented by Hermitian operators on the system's Hilbert space. The (real) eigenvalues of the observable represent its possible measured values. The eigenstates of the observable can be used to form an orthonormal eigenbasis of the Hilbert space.
\end{quote}
The Composite System Axiom stays the same, but we now also specify that independently prepared subsystems have a composite state given by a tensor product of density operators. The corollary regarding entanglement is also rewritten in terms of density operators.
\begin{quote}
\textbf{The Composite System Axiom (Density Operators)}\index{Composite System Axiom!Density operators}\textbf{:} The Hilbert space of a composite system is represented by the tensor product of the Hilbert spaces of the individual systems. If each individual system $\HH_{i}$, where $i\in\left\{ 1,\ldots,n\right\} $, is independently prepared in the state $\rho_{i}$, then the composite state of the system is given by $\rho_{1}\otimes\cdots\otimes\rho_{n}$. Such a state is called a \emph{product state}\index{Product state}.
\begin{itemize}
\item \textbf{Entanglement}\index{Quantum entanglement!Density operators}\textbf{:} A state of a composite system $\HH_{1}\otimes\cdots\otimes\HH_{n}$ is called \emph{separable}\index{Separable state!Density operators} if it can be written as a \emph{convex sum}\index{Convex sum} of any number of product states:
\[
\rho=\sum_{j}\chi_{j}\rho^{\left(j\right)}_{1}\otimes\cdots\otimes\rho^{\left(j\right)}_{n},
\]
where $\rho^{\left(j\right)}_{i}$ is a density operator on $\HH_{i}$ and by “convex sum” we mean that the coefficients $\chi_{j}$ are all non-negative and sum to 1:
\[
\sum_{j}\chi_{j}=1\sp\chi_{j}\ge0.
\]
A state that is not separable is called \emph{entangled}\index{Entangled state!Density operators}.
\end{itemize}
\end{quote}
The Evolution Axiom is modified in a straightforward way:
\begin{quote}
\textbf{The Evolution Axiom (Density Operators)}\index{Evolution Axiom!Density operators}\textbf{:} Isolated quantum systems evolve in time via the action of a unitary operator. More precisely, given two points in time $t_{1}$ and $t_{2}$, there is a unitary operator $U$ such that any state $\rho$ at $t_{1}$ evolves to the state $U\rho U^{\dagger}$ at time $t_{2}$.
\end{quote}
The Measurement Axiom is the most complicated one, but we already derived all the results in \zcref[S]{subsec:Measuring-Density-Operators}:
\begin{quote}
\textbf{The Measurement Axiom (Density Operators)}\index{Measurement Axiom!Density operators}\textbf{:} Consider an observable $A$ on an $n$-dimensional Hilbert space, with $k\le n$ distinct eigenvalues $\left\{ \lambda_{1},\ldots,\lambda_{k}\right\} $, corresponding projectors $\left\{ P_{1},\ldots,P_{k}\right\} $, and spectral decomposition
\[
A=\sum^{k}_{i=1}\lambda_{i}P_{i}.
\]
If the system is in the state $\rho$, then the probability to measure the eigenvalue $\lambda_{i}$ is given by
\[
P\left(\lambda_{i}\right)=\tr\left(P_{i}\rho\right).
\]
After the measurement, if the eigenvalue $\lambda_{i}$ was measured, then the system will collapse to the state
\[
\rho\mt\frac{P_{i}\rho P^{\dagger}_{i}}{\tr\left(P_{i}\rho\right)}.
\]
\begin{itemize}
\item \textbf{Expectation Value}\index{Expected (or expectation) value!With density operators}\textbf{:} If the system is in the state $\rho$, the expectation value for the measurement of the observable $A$ is given by
\[
\left\langle A\right\rangle _{\rho}=\tr\left(A\rho\right).
\]
\end{itemize}
\end{quote}

\subsection{Applications of Density Operators}

\subsubsection{Mixed States}\label{subsec:Mixed-States}

If the density operator is composed of just one pure state $\left|\Psi\right\rangle $ with probability 1, that is,
\[
\rho=\left|\Psi\right\rangle \left\langle \Psi\right|,
\]
then $\rho$ is also called a \emph{pure state}\index{Pure state}\index{Quantum state!Pure} – or more precisely, a quantum system described by either $\left|\Psi\right\rangle $ or $\rho=\left|\Psi\right\rangle \left\langle \Psi\right|$ is said to be in a pure state. If the system is not in a pure state, then it is said to be in a \emph{mixed state}\index{Mixed state}\index{Quantum state!Mixed}. Note that by definition, mixed states can \textbf{only} be described using density operators; if the system is described by a vector, then it is necessarily in a pure state. However, if the system is described by a density operator, it could be either pure or mixed. A density operator representing a mixed state is said to be a \emph{mixture}\index{Mixture} of two or more pure states.

Given an arbitrary density operator $\rho$, we can determine whether it is pure or mixed by calculating the trace of its square, $\tr\left(\rho^{2}\right)$. This is also called the \emph{purity}\index{Purity of a state} of the state. Since $\rho$ is Hermitian, it has the spectral decomposition
\[
\rho=\sum_{i}q_{i}\left|\Theta_{i}\right\rangle \left\langle \Theta_{i}\right|,
\]
where $q_{i}\ge0$, $\sum_{i}q_{i}=1$, and the states $\left|\Theta_{i}\right\rangle $ form an orthonormal eigenbasis. Then we have:
\begin{align*}
\rho^{2} & =\left(\sum_{i}q_{i}\left|\Theta_{i}\right\rangle \left\langle \Theta_{i}\right|\right)\left(\sum_{j}q_{j}\left|\Theta_{j}\right\rangle \left\langle \Theta_{j}\right|\right)\\
 & =\sum_{i,j}q_{i}q_{j}\left|\Theta_{i}\right\rangle \langle\Theta_{i}|\Theta_{j}\rangle\left\langle \Theta_{j}\right|\\
 & =\sum_{i,j}q_{i}q_{j}\left|\Theta_{i}\right\rangle \delta_{ij}\left\langle \Theta_{j}\right|\\
 & =\sum_{i}q^{2}_{i}\left|\Theta_{i}\right\rangle \left\langle \Theta_{i}\right|.
\end{align*}
Therefore
\[
\tr\left(\rho^{2}\right)=\sum_{i}q^{2}_{i}.
\]
If the state is pure, then we only have one non-zero eigenvalue in the sum and it must be equal to 1, thus $\tr\left(\rho^{2}\right)=1$. If the state is mixed, then we have two or more eigenvalues $q_{i}$ such that $q_{i}\in\left(0,1\right)$, and therefore $q^{2}_{i}<q_{i}$. Since $\sum_{i}q_{i}=1$, we see that for a mixed state, $\tr\left(\rho^{2}\right)<1$. So in conclusion:
\begin{itemize}
\item A state $\rho$ is pure if and only if $\tr\left(\rho^{2}\right)=1$.
\item A state $\rho$ is mixed if and only if $\tr\left(\rho^{2}\right)<1$.
\end{itemize}
Note also that if the state is pure, then it has a state vector $\left|\Psi\right\rangle $, and we have
\[
\rho=\left|\Psi\right\rangle \left\langle \Psi\right|\soosp\rho^{2}=|\Psi\rangle\langle\Psi|\Psi\rangle\langle\Psi|=\left|\Psi\right\rangle \left\langle \Psi\right|=\rho.
\]
This both provides a trivial proof that $\tr\left(\rho^{2}\right)=1$ for pure states (since $\tr\rho=1$ for any density operator by definition) and provides an additional characterization of pure states: a state $\rho$ is pure if and only if $\rho=\rho^{2}$ (that is, if $\rho$ is idempotent, as defined in \zcref[S]{subsec:Spectral-Decomposition}).

For an $n$-dimensional Hilbert space, you will prove in \zcref[S]{prob:purity} that the lowest possible purity is $\tr\left(\rho^{2}\right)=1/n$. This is achieved for a \emph{maximally mixed state}\index{Maximally mixed state}, which has the form
\[
\rho=\frac{1}{n}\sum_{i}\left|\Theta_{i}\right\rangle \left\langle \Theta_{i}\right|,
\]
where $\left|\Theta_{i}\right\rangle $ is an orthonormal basis. Indeed, it is easy to see that in that basis, the density matrix will be the identity matrix 1 divided by $n$:
\[
\rho=\frac{1}{n},
\]
so $\rho^{2}=1/n^{2}$ and thus
\[
\tr\left(\rho^{2}\right)=n\cdot\frac{1}{n^{2}}=\frac{1}{n}.
\]
Note that a maximally mixed state corresponds to an ensemble where the system is prepared in each one of the eigenstates $\left|\Theta_{i}\right\rangle $ with equal probability.
\begin{problem}\leavevmode
\label{prob:purity}Prove that $\tr\left(\rho^{2}\right)\ge1/n$, that is, a maximally mixed state has the lowest possible purity.
\end{problem}

\subsubsection{Ensembles of Density Operators}

Mixed states can be ensembles of pure states represented by vectors, but they can also be written as ensembles of density operators, whether these operators represent pure or mixed states. Indeed, suppose we have several density operators $\rho_{i}$ defined as ensembles of one or more pure states:
\[
\rho_{i}\equiv\sum_{j}p^{i}_{j}\left|\Psi^{i}_{j}\right\rangle \left\langle \Psi^{i}_{j}\right|.
\]
Note that for each $\rho_{i}$, the corresponding ensemble states $\left|\Psi^{i}_{j}\right\rangle $ do not necessarily need to form an orthonormal basis.

If we prepare a quantum system in the state $\rho_{i}$ with probability $p_{i}$, then the probability that the system is in the pure state $\left|\Psi^{i}_{j}\right\rangle $ is given by the product of the probability $p_{i}$ to prepare the density operator containing that state and the probability $p^{i}_{j}$ to prepare the pure state $\left|\Psi^{i}_{j}\right\rangle $ within that density operator:
\[
P\left(\Psi^{i}_{j}\right)=p_{i}p^{i}_{j}.
\]
Thus we can describe the state of the system as
\begin{align*}
\rho & =\sum_{i,j}P\left(\Psi^{i}_{j}\right)\left|\Psi^{i}_{j}\right\rangle \left\langle \Psi^{i}_{j}\right|\\
 & =\sum_{i,j}p_{i}p^{i}_{j}\left|\Psi^{i}_{j}\right\rangle \left\langle \Psi^{i}_{j}\right|\\
 & =\sum_{i}p_{i}\sum_{j}p^{i}_{j}\left|\Psi^{i}_{j}\right\rangle \left\langle \Psi^{i}_{j}\right|\\
 & =\sum_{i}p_{i}\rho_{i}.
\end{align*}

\subsubsection{Reduced Density Operators}

When two systems are entangled with each other, neither can be described individually by a pure state. Consider, for example, the entangled Bell state (recall \zcref[S]{subsec:The-Bell-States})
\[
\left|\beta_{00}\right\rangle \equiv\frac{1}{\sqrt{2}}\left(\left|0\right\rangle \otimes\left|0\right\rangle +\left|1\right\rangle \otimes\left|1\right\rangle \right).
\]
What is the state of the first qubit? It is clearly not $\left|0\right\rangle $, $\left|1\right\rangle $, or any superposition thereof, since its state also depends on the state of the second qubit. However, we can describe the state of the first qubit, or in general of any subsystem of a composite system, in terms of density operators.

Consider two physical systems $\HH_{A}$ and $\HH_{B}$. Let $\left|A_{1}\right\rangle ,\left|A_{2}\right\rangle \in\HH_{A}$ and $\left|B_{1}\right\rangle ,\left|B_{2}\right\rangle \in\HH_{B}$. We define the \emph{partial trace}\index{Partial trace} with respect to the system $B$ using the relation
\begin{align*}
\tr_{B}\left(\left|A_{1}\right\rangle \left\langle A_{2}\right|\otimes\left|B_{1}\right\rangle \left\langle B_{2}\right|\right) & \equiv\left|A_{1}\right\rangle \left\langle A_{2}\right|\tr\left(\left|B_{1}\right\rangle \left\langle B_{2}\right|\right)\\
 & =\left|A_{1}\right\rangle \left\langle A_{2}\right|\langle B_{2}|B_{1}\rangle.
\end{align*}
In other words, the partial trace operation only takes the trace of the term from system $B$, leaving the term from system $A$ intact. Requiring the partial trace operation to be linear then allows us to define it on arbitrary density operators on the composite system, which can always be written as a sum over operators of the above form.

Now, let the state of the composite system $\HH_{A}\otimes\HH_{B}$ be described by the density operator $\rho^{AB}$. The \emph{reduced density operator}\index{Reduced density operator} for system $A$ is defined as
\[
\rho^{A}\equiv\tr_{B}\left(\rho^{AB}\right).
\]
We have traced out system $B$, so this should be the state of system $A$, even if they were entangled. However, we should make sure this specific way of obtaining the state of system $A$ is consistent.

First, for a separable state of the form $\rho^{AB}=\rho^{A}\otimes\rho^{B}$, where
\[
\rho^{A}=\sum_{i}p^{A}_{i}\left|\Psi^{A}_{i}\right\rangle \left\langle \Psi^{A}_{i}\right|\sp\rho^{B}=\sum_{j}p^{B}_{j}\left|\Psi^{B}_{j}\right\rangle \left\langle \Psi^{B}_{j}\right|,
\]
we get from the linearity of the partial trace:
\begin{align*}
\tr_{B}\left(\rho^{AB}\right) & =\tr_{B}\left(\rho^{A}\otimes\rho^{B}\right)\\
 & =\tr_{B}\left(\left(\sum_{i}p^{A}_{i}\left|\Psi^{A}_{i}\right\rangle \left\langle \Psi^{A}_{i}\right|\right)\otimes\left(\sum_{j}p^{B}_{j}\left|\Psi^{B}_{j}\right\rangle \left\langle \Psi^{B}_{j}\right|\right)\right)\\
 & =\sum_{i,j}p^{A}_{i}p^{B}_{j}\tr_{B}\left(\left|\Psi^{A}_{i}\right\rangle \left\langle \Psi^{A}_{i}\right|\otimes\left|\Psi^{B}_{j}\right\rangle \left\langle \Psi^{B}_{j}\right|\right)\\
 & =\sum_{i,j}p^{A}_{i}p^{B}_{j}\left|\Psi^{A}_{i}\right\rangle \left\langle \Psi^{A}_{i}\right|\langle\Psi^{B}_{j}|\Psi^{B}_{j}\rangle\\
 & =\left(\sum_{i}p^{A}_{i}\left|\Psi^{A}_{i}\right\rangle \left\langle \Psi^{A}_{i}\right|\right)\left(\sum_{j}p^{B}_{j}\right)\\
 & =\rho^{A},
\end{align*}
where we used the fact that $\langle\Psi^{B}_{j}|\Psi^{B}_{j}\rangle=1$ since the states are normalized and $\sum_{j}p^{B}_{j}=1$ since the probabilities sum to 1. Actually, what we did here was just a very explicit way of verifying that
\[
\tr_{B}\left(\rho^{A}\otimes\rho^{B}\right)=\rho^{A}\tr\rho^{B}=\rho^{A},
\]
since $\tr\rho^{B}=1$ by definition. Of course, this is trivial, as we already knew that system $A$ is described by $\rho^{A}$ even without the partial trace. The more interesting case is when the composite state of the system is entangled. Consider, for example, the pure state
\[
\left|\Psi\right\rangle =\alpha\left|0\right\rangle \otimes\left|0\right\rangle +\beta\left|1\right\rangle \otimes\left|1\right\rangle ,
\]
where $\left|\alpha\right|^{2}+\left|\beta\right|^{2}=1$. It is entangled as long as $\alpha,\beta\ne0$; measuring 0 or 1 for qubit $A$ means that the same value will also be measured for qubit $B$, and vice versa. The density operator for this pure state is
\begin{align*}
\rho^{AB} & =\left(\alpha\left|0\right\rangle \otimes\left|0\right\rangle +\beta\left|1\right\rangle \otimes\left|1\right\rangle \right)\left(\alpha^{*}\left\langle 0\right|\otimes\left\langle 0\right|+\beta^{*}\left\langle 1\right|\otimes\left\langle 1\right|\right)\\
 & =\left|\alpha\right|^{2}\left|0\right\rangle \left\langle 0\right|\otimes\left|0\right\rangle \left\langle 0\right|+\alpha\beta^{*}\left|0\right\rangle \left\langle 1\right|\otimes\left|0\right\rangle \left\langle 1\right|+\alpha^{*}\beta\left|1\right\rangle \left\langle 0\right|\otimes\left|1\right\rangle \left\langle 0\right|+\left|\beta\right|^{2}\left|1\right\rangle \left\langle 1\right|\otimes\left|1\right\rangle \left\langle 1\right|.
\end{align*}
Tracing out qubit $B$, we get the state of qubit $A$:
\[
\rho^{A}=\tr_{B}\left(\rho^{AB}\right)=\left|\alpha\right|^{2}\left|0\right\rangle \left\langle 0\right|+\left|\beta\right|^{2}\left|1\right\rangle \left\langle 1\right|,
\]
or in matrix form
\begin{equation}
\rho^{A}=\left(\begin{array}{cc}
|\alpha|^{2} & 0\\
0 & |\beta|^{2}
\end{array}\right).\label{eq:rhoA-alpha-beta}
\end{equation}
The trace of the square is
\[
\tr\left(\left(\rho^{A}\right)^{2}\right)=|\alpha|^{4}+|\beta|^{4}.
\]
If the state is not entangled, then either $\alpha=0$ or $\beta=0$, and the other coefficient is 1 up to a phase factor, so the purity will be 1 and the reduced state will remain pure. However, if the state is entangled, we have
\[
\left|\alpha\right|^{2}+\left|\beta\right|^{2}=1\soosp\left|\beta\right|^{2}=1-\left|\alpha\right|^{2}\soosp\left|\beta\right|^{4}=\left(1-\left|\alpha\right|^{2}\right)^{2}=1-2\left|\alpha\right|^{2}+\left|\alpha\right|^{4},
\]
so
\[
\tr\left(\left(\rho^{A}\right)^{2}\right)=1+2\left(\left|\alpha\right|^{4}-\left|\alpha\right|^{2}\right).
\]
Taking the derivative with respect to $\left|\alpha\right|^{2}$, we get:
\[
\frac{\d}{\d\left|\alpha\right|^{2}}\tr\left(\left(\rho^{A}\right)^{2}\right)=2\left(2\left|\alpha\right|^{2}-1\right),
\]
which has a zero at
\[
\left|\alpha\right|^{2}=\hf\soosp\left|\alpha\right|=\frac{1}{\sqrt{2}}.
\]
Furthermore, since the second derivative is positive, this is a local minimum. By plotting this function, you can convince yourself that the purity is equal to 1 for $\left|\alpha\right|=0$ or $\left|\alpha\right|=1$, otherwise it is always less than 1, with the minimum obtained at $\left|\alpha\right|=1/\sqrt{2}$ and equal to 1/2. In other words, the (pure) entangled state with equal probabilities for 0 and 1 ($\left|\alpha\right|^{2}=\left|\beta\right|^{2}=1/2$) produces the most mixed reduced state for each individual qubit. Indeed, in this case the state of each qubit will be the maximally mixed state $\rho=1/2$, which has the minimum purity possible (the dimension of the Hilbert space is 2, so the minimum purity is $1/2$).

What does this mean? We can think of a pure state as one where we have complete knowledge of the state of the quantum system. A mixed state is one where we \textbf{do not} have complete knowledge; we don't know which pure state the system is in, we only know the \textbf{probability} for the system to be in each pure state in the ensemble. In the case of an entangled system, we have the seemingly paradoxical situation where we know more about the state of both particles together than we do about each individual particle on its own!

If the entangled state is $\alpha\left|00\right\rangle +\beta\left|11\right\rangle $, then the reduced state for each qubit will be an ensemble where $\left|0\right\rangle $ is prepared with probability $\left|\alpha\right|^{2}$ and $\left|1\right\rangle $ with probability $\left|\beta\right|^{2}$. In this mixed state, the probabilities match the ones obtained from the pure state, which makes sense. However, note that we have lost information about the phases of $\alpha$ and $\beta$ in the process; we only have classical probabilities.

Let $\HH_{A}$ and $\HH_{B}$ be two Hilbert spaces of the same dimension $n$. A state $\left|\Psi\right\rangle \in\HH_{A}\otimes\HH_{B}$ of a composite system is called \emph{maximally entangled}\index{Maximally entangled state} if the reduced states with respect to both Hilbert spaces are maximally mixed:
\[
\rho^{A}=\rho^{B}=\frac{1}{n}.
\]
In the example discussed here, we see that the state is maximally entangled if and only if $\left|\alpha\right|^{2}=\left|\beta\right|^{2}=1/2$.

\subsubsection{Equivalent Ensembles}\label{subsec:Equivalent-Ensembles}

In \zcref[S]{subsec:Density-Operators-and} we saw that the density matrix
\[
\rho=\hf\left(\begin{array}{cc}
1 & 0\\
0 & 1
\end{array}\right)
\]
represents two different ensembles:
\begin{enumerate}
\item A qubit prepared either in the state $\left|0\right\rangle $ or $\left|1\right\rangle $ with probability 1/2 each: $\rho=\hf\left|0\right\rangle \left\langle 0\right|+\hf\left|1\right\rangle \left\langle 1\right|$.
\item A qubit prepared either in the state $\left|+\right\rangle $ or $\left|-\right\rangle $ with probability 1/2 each: $\rho=\hf\left|+\right\rangle \left\langle +\right|+\hf\left|-\right\rangle \left\langle -\right|$.
\end{enumerate}
This seems a bit weird, but we must ask: is there any way to distinguish between these two ensembles by performing a measurement on the system? For example, let us measure in the $\left\{ \left|0\right\rangle ,\left|1\right\rangle \right\} $ basis. According to the Measurement Axiom, the probability to measure the eigenvalue $\lambda_{i}$ is given by
\[
P\left(\lambda_{i}\right)=\tr\left(P_{i}\rho\right),
\]
where
\[
\lambda_{0}=+1\sp P_{0}=\left|0\right\rangle \left\langle 0\right|,
\]
\[
\lambda_{1}=-1\sp P_{1}=\left|1\right\rangle \left\langle 1\right|.
\]
Therefore, the probability to measure 0 is
\begin{align*}
P\left(\lambda_{0}\right) & =\tr\left(\left|0\right\rangle \left\langle 0\right|\left(\hf\left|0\right\rangle \left\langle 0\right|+\hf\left|1\right\rangle \left\langle 1\right|\right)\right)\\
 & =\tr\left(\hf\left|0\right\rangle \left\langle 0\right|\right)\\
 & =\hf,
\end{align*}
and similarly
\[
P\left(\lambda_{1}\right)=\hf.
\]
Since we only used the density operator itself in these calculations, the probabilities apply to both ensembles. This result makes sense in the case of ensemble 1, that is, when the qubit was prepared either in the state $\left|0\right\rangle $ or $\left|1\right\rangle $ with classical probability 1/2 each. For ensemble 2, where the qubit was prepared either in the state $\left|+\right\rangle $ or $\left|-\right\rangle $ with classical probability 1/2 each, let's recall that the states $\left|+\right\rangle $ and $\left|-\right\rangle $ are both equal superpositions of $\left|0\right\rangle $ and $\left|1\right\rangle $, so the probability to measure either 0 or 1 will be 1/2 in either case. The total probability to measure 0 can be calculated by multiplying the probability for each state in the ensemble with the probability of measuring 0 for that specific state and adding up the results:
\begin{align*}
P\left(\lambda_{0}\right) & =P\left(\textrm{prepared }\left|+\right\rangle \right)P\left(\textrm{measured }0\textrm{ in }\left|+\right\rangle \textrm{ state}\right)+\\
 & \qquad+P\left(\textrm{prepared }\left|-\right\rangle \right)P\left(\textrm{measured }0\textrm{ in }\left|-\right\rangle \textrm{ state}\right)\\
 & =\hf\cdot\hf+\hf\cdot\hf\\
 & =\hf,
\end{align*}
and similarly for $P\left(\lambda_{1}\right)$. We conclude that even though the ensembles are prepared differently, they predict the exact same measurement probabilities; this is true for any measurement, since the same density operator predicts all the same probabilities for any observable. Therefore, no measurement can distinguish between the two ensembles.

In fact, it turns out that there is an \textbf{infinite }number of different ensembles that correspond to the same density operator, and are therefore indistinguishable. Let $\left\{ p_{i},\left|\Psi_{i}\right\rangle \right\} $ be an ensemble, then any ensemble $\left\{ p^{\prime}_{i},\left|\Psi^{\prime}_{i}\right\rangle \right\} $ such that
\[
\sqrt{p^{\prime}_{i}}\left|\Psi^{\prime}_{i}\right\rangle =\sum_{j}U_{ij}\sqrt{p_{j}}\left|\Psi_{j}\right\rangle ,
\]
where $U_{ij}$ are the components of some unitary matrix, generates the same density operator. In the example considered in this section, we have
\[
\left|\Psi_{0}\right\rangle =\left|0\right\rangle =\left(\begin{array}{c}
1\\
0
\end{array}\right)\sp\left|\Psi_{1}\right\rangle =\left|1\right\rangle =\left(\begin{array}{c}
0\\
1
\end{array}\right),
\]
\[
\left|\Psi^{\prime}_{0}\right\rangle =\left|+\right\rangle =\frac{1}{\sqrt{2}}\left(\begin{array}{c}
1\\
1
\end{array}\right)\sp\left|\Psi^{\prime}_{1}\right\rangle =\left|-\right\rangle =\frac{1}{\sqrt{2}}\left(\begin{array}{c}
1\\
-1
\end{array}\right),
\]
\[
p_{0}=p_{1}=p^{\prime}_{0}=p^{\prime}_{1}=\hf.
\]
What is the unitary matrix relating them? We have for $i=0$:
\[
\frac{1}{\sqrt{2}}\left|+\right\rangle =\frac{1}{2}\left(\begin{array}{c}
1\\
1
\end{array}\right)=\frac{1}{\sqrt{2}}\left(U_{00}\left|0\right\rangle +U_{01}\left|1\right\rangle \right)=\frac{1}{\sqrt{2}}\left(\begin{array}{c}
U_{00}\\
U_{01}
\end{array}\right),
\]
and for $i=1$:
\[
\frac{1}{\sqrt{2}}\left|-\right\rangle =\frac{1}{2}\left(\begin{array}{c}
1\\
-1
\end{array}\right)=\frac{1}{\sqrt{2}}\left(U_{10}\left|0\right\rangle +U_{11}\left|1\right\rangle \right)=\frac{1}{\sqrt{2}}\left(\begin{array}{c}
U_{10}\\
U_{11}
\end{array}\right).
\]
We thus conclude that
\[
U=\left(\begin{array}{cc}
U_{00} & U_{01}\\
U_{10} & U_{11}
\end{array}\right)=\frac{1}{\sqrt{2}}\left(\begin{array}{cc}
1 & 1\\
1 & -1
\end{array}\right),
\]
which is indeed a unitary matrix.

In the examples above, we used orthonormal basis states, $\left\{ \left|0\right\rangle ,\left|1\right\rangle \right\} $ or $\left\{ \left|+\right\rangle ,\left|-\right\rangle \right\} $, to define the ensemble. However, in general, the states need not be orthogonal. For example, if we want, we can prepare a qubit with probability 1/2 in each of the \textbf{non-orthogonal} states
\[
\left|\Psi_{\pm}\right\rangle \equiv\frac{\sqrt{3}}{2}\left|0\right\rangle \pm\hf\left|1\right\rangle .
\]
Indeed, you can easily check that $\langle\Psi_{+}|\Psi_{-}\rangle\ne0$, so they are not orthogonal. The density operator will be
\[
\rho=\hf\left|\Psi_{+}\right\rangle \left\langle \Psi_{+}\right|+\hf\left|\Psi_{-}\right\rangle \left\langle \Psi_{-}\right|,
\]
and you will show in \zcref[S]{exer:Calculate-the-matrix} that the corresponding density matrix is
\begin{equation}
\rho=\fr\left(\begin{array}{cc}
3 & 0\\
0 & 1
\end{array}\right).\label{eq:non-orthogonal-rho}
\end{equation}
It is obvious from the matrix representation that the equivalent orthogonal ensemble is one where we prepare $\left|0\right\rangle $ with probability $3/4$ or $\left|1\right\rangle $ with probability 1/4:
\[
\rho=\frac{3}{4}\left|0\right\rangle \left\langle 0\right|+\fr\left|1\right\rangle \left\langle 1\right|.
\]
We therefore see that while a statistical ensemble can be defined using arbitrary states, there is always an equivalent ensemble corresponding to an orthonormal basis, obtained via the spectral decomposition, with the same predictions for all measurement outcomes.
\begin{xca}\leavevmode
\label{exer:Calculate-the-matrix}Calculate \zcref[S]{eq:non-orthogonal-rho} from the outer product representation in terms of $\left|\Psi_{\pm}\right\rangle $.
\end{xca}

\subsubsection{Von Neumann Entropy}

Shannon entropy, which we defined in \zcref[S]{subsec:Information-and-Entropy}, only applies to classical random variables. However, it is straightforward to generalize it to quantum states, provided we use density operators. We define the \emph{von Neumann entropy}\index{von Neumann entropy} of a quantum state $\rho$, denoted $S\left(\rho\right)$, as follows:
\[
S\left(\rho\right)\equiv-\tr\left(\rho\log\rho\right).
\]
However, we generally cannot use the matrix logarithm defined in \zcref[S]{subsec:Matrix-and-Operator} to calculate $\log\rho$, since the matrix logarithm only exists for invertible matrices, but $\rho$ is often not invertible (as it may have the eigenvalue 0). Therefore, we will instead use the spectral decomposition
\[
\rho=\sum_{i}q_{i}\left|\Theta_{i}\right\rangle \left\langle \Theta_{i}\right|,
\]
where as usual $q_{i}$ are the eigenvalues, with $\sum_{i}q_{i}=1$, and $\left|\Theta_{i}\right\rangle $ is an orthonormal eigenbasis. Now we can simply define
\[
\rho\log\rho\equiv\sum_{i}q_{i}\log q_{i}\left|\Theta_{i}\right\rangle \left\langle \Theta_{i}\right|.
\]
As in \zcref[S]{subsec:Information-and-Entropy}, we are using the base-2 logarithm and $0\log0\equiv0$. Taking the trace, we can express the von Neumann entropy in terms of the eigenvalues of $\rho$:
\[
S\left(\rho\right)=-\sum_{i}q_{i}\log q_{i}.
\]
Since the eigenvalues $q_{i}$ are the (classical) probabilities to prepare each of the orthonormal eigenstates $\left|\Theta_{i}\right\rangle $ in the ensemble given by the spectral decomposition, we see that $S\left(\rho\right)$ is identical to the Shannon entropy of these probabilities. However, as we saw in \zcref[S]{subsec:Equivalent-Ensembles}, the same density operator can also be prepared using other ensembles, whose probabilities may not be the eigenvalues of $\rho$; hence, their Shannon entropy can differ from $S\left(\rho\right)$. The von Neumann entropy is only the Shannon entropy of the ensemble obtained from the eigenstates.

A density operator representing a pure state only has one non-zero eigenvalue, and it must be equal to 1. Thus, the von Neumann entropy of a state is 0 if and only if the state is pure. This means that the von Neumann entropy can be interpreted as measuring “how mixed” the state is. For a maximally mixed state (as defined in \zcref[S]{subsec:Mixed-States}) of the form $\rho=1/n$ where $n$ is the dimension of the Hilbert space and 1 is the identity matrix, the von Neumann entropy is the same as that of a fair $n$-sided die:
\[
S\left(\rho\right)=n\left(-\frac{1}{n}\log\frac{1}{n}\right)=\log n.
\]
One can show that this is in fact the maximum possible entropy of any state in an $n$-dimensional Hilbert space. Thus, a maximally mixed state has \textbf{maximal entropy}, just as it has \textbf{minimal purity}.

\subsection{The Bloch Sphere}\label{subsec:The-Bloch-Sphere}

\subsubsection{Pure States}\label{subsec:Pure-States}

Consider a qubit in the general superposition
\[
\left|\Psi\right\rangle =a\left|0\right\rangle +b\left|1\right\rangle ,
\]
where the complex coefficients $a$ and $b$ satisfy the usual normalization condition
\[
\left|a\right|^{2}+\left|b\right|^{2}=1.
\]
Recall that this condition means $\left\Vert \Psi\right\Vert =1$, so $\left|\Psi\right\rangle $, seen as a vector in the 2-dimensional space $\BBC^{2}$, has “length” 1. This intuitively creates a 1-dimensional “circle” in $\BBC^{2}$, but that's not exactly what happens, because $\BBC^{2}$ is a complex space and thus actually has 4 \textbf{real }dimensions. The requirement that $\left|a\right|^{2}+\left|b\right|^{2}=1$ eliminates 1 out of the 4 degrees of freedom, and therefore chooses a 3-dimensional surface in this 4-dimensional space.

We should also remember that quantum states are defined only up to an overall phase factor, meaning that $\left|\Psi\right\rangle $ and $\e^{\i\gamma}\left|\Psi\right\rangle $ represent the same physical state and the overall phase factor $\e^{\i\gamma}$ has no observable effect. This eliminates 1 additional degree of freedom, so in fact, the qubit is represented by a 2-dimensional surface. This surface is called the \emph{Bloch sphere}\index{Bloch sphere}.

To make this more precise, let us decompose the two complex numbers $a,b\in\BBC$ in a polar decomposition:
\[
a=\alpha\e^{\i\gamma}\sp b=\beta\e^{\i\delta}\sp\alpha,\beta\geq0\sp\gamma,\delta\in\left[0,2\pi\right).
\]
Since a quantum state is only defined up to an overall phase factor, we can divide the entire state by the phase factor $\e^{\i\gamma}$ without changing its meaning, so we get:
\[
\left|\Psi\right\rangle =\alpha\left|0\right\rangle +\beta\e^{\i\phi}\left|1\right\rangle \sp\alpha,\beta\geq0\sp\phi\in\left[0,2\pi\right),
\]
where we defined
\[
\phi\equiv\left(\delta-\gamma\right)\mod 2\pi,
\]
that is, the difference between the angles, modulo $2\pi$ so it will be within the right range. The condition $\left|a\right|^{2}+\left|b\right|^{2}=1$ now becomes simply
\[
\alpha^{2}+\beta^{2}=1,
\]
which means the point $\left(\alpha,\beta\right)$ is on the unit circle in $\BBR^{2}$, so it can be parametrized by some angle $\theta\in\left[0,2\pi\right)$. But remember that we also required that $\alpha,\beta\geq0$, so in fact $\left(\alpha,\beta\right)$ is only in the first quadrant of the unit circle. Therefore, we can parametrize $\alpha$ and $\beta$ as follows\footnote{We could also have written $\alpha=\cos\theta$ and $\beta=\sin\theta$ for $\theta\in\left[0,\pi/2\right]$, but we parametrize using $\theta/2$ instead so that $\theta$ can be in the range $\left[0,\pi\right]$, because then the coordinates $\theta$ and $\phi$ have the right ranges for the familiar spherical coordinates.}:
\[
\alpha=\cos\frac{\theta}{2}\sp\beta=\sin\frac{\theta}{2}\sp\theta\in\left[0,\pi\right].
\]

In conclusion, the pure state of an arbitrary qubit can be parametrized as
\[
\left|\Psi\right\rangle =\cos\frac{\theta}{2}\left|0\right\rangle +\e^{\i\phi}\sin\frac{\theta}{2}\left|1\right\rangle \sp\theta\in\left[0,\pi\right]\sp\phi\in\left[0,2\pi\right),
\]
where the coordinates $\theta,\phi$ represent points on the unit sphere in $\BBR^{3}$, which in this context is referred to as the \emph{Bloch sphere}\index{Bloch sphere}.

Let us consider some examples. The state $\left|0\right\rangle $ clearly corresponds to $\theta=0$, which is the north pole of the sphere, and $\left|1\right\rangle $ clearly corresponds to $\theta=\pi$, which is the south pole. In both cases, $\phi$ can take any value, since it becomes a degenerate coordinate. The state
\[
\left|+\right\rangle =\frac{\left|0\right\rangle +\left|1\right\rangle }{\sqrt{2}}
\]
corresponds to $\theta=\pi/2$ and $\phi=0$, so it is on the equator. The state
\[
\left|-\right\rangle =\frac{\left|0\right\rangle -\left|1\right\rangle }{\sqrt{2}}
\]
also corresponds to $\theta=\pi/2$, but the phase of $\left|1\right\rangle $ is inverted by taking $\phi=\pi$, so this state is also on the equator, but on the opposite side.

\subsubsection{Mixed States and the Bloch Ball}

The density operator for a qubit in a mixed state is represented by a 2-dimensional Hermitian matrix. From \zcref[S]{prob:Consider-the-real} we know that $\left\{ 1,\sigma_{x},\sigma_{y},\sigma_{z}\right\} $ is a basis for the space of such matrices over the reals, so any 2-dimensional Hermitian matrix (with non-zero trace) can be written in the form
\[
\rho=a\left(1+\r\cdot\si\right),
\]
where $a\in\BBR$ is an overall normalization factor, $\r\equiv\left(r_{x},r_{y},r_{z}\right)\in\BBR^{3}$, and $\si\equiv\left(\sigma_{x},\sigma_{y},\sigma_{z}\right)$ is a slight abuse of notation so we can write succinctly
\[
\r\cdot\si=r_{x}\sigma_{x}+r_{y}\sigma_{y}+r_{z}\sigma_{z}.
\]
However, remember also that density matrices must have trace 1. The Pauli matrices are traceless, and the identity matrix has trace 2, so
\[
\tr\rho=\tr\left(a\left(1+\r\cdot\si\right)\right)=2a=1,
\]
and we conclude that we must take $a=1/2$. Therefore, the density operator of an arbitrary qubit can be written in the form
\[
\rho=\hf\left(1+\r\cdot\si\right).
\]
The real 3-vector $\r$ is called the \emph{Bloch vector}\index{Bloch vector}. Explicitly, the density matrix is
\begin{align*}
\rho & =\hf\left(\left(\begin{array}{cc}
1 & 0\\
0 & 1
\end{array}\right)+r_{x}\left(\begin{array}{cc}
0 & 1\\
1 & 0
\end{array}\right)+r_{y}\left(\begin{array}{cc}
0 & -\i\\
\i & 0
\end{array}\right)+r_{z}\left(\begin{array}{cc}
1 & 0\\
0 & -1
\end{array}\right)\right)\\
 & =\hf\left(\begin{array}{cc}
1+r_{z} & r_{x}-\i r_{y}\\
r_{x}+\i r_{y} & 1-r_{z}
\end{array}\right).
\end{align*}
Calculating its square, we get
\[
\rho^{2}=\fr\left(\begin{array}{cc}
r^{2}_{x}+r^{2}_{y}+\left(1+r_{z}\right)^{2} & 2\left(r_{x}-\i r_{y}\right)\\
2\left(r_{x}+\i r_{y}\right) & r^{2}_{x}+r^{2}_{y}+\left(1-r_{z}\right)^{2}
\end{array}\right).
\]
Therefore, the purity of the state is
\begin{align*}
\tr\left(\rho^{2}\right) & =\fr\left(r^{2}_{x}+r^{2}_{y}+\left(1+r_{z}\right)^{2}+r^{2}_{x}+r^{2}_{y}+\left(1-r_{z}\right)^{2}\right)\\
 & =\hf\left(1+r^{2}_{x}+r^{2}_{y}+r^{2}_{z}\right)\\
 & =\hf\left(1+\left|\r\right|^{2}\right).
\end{align*}
If the state is pure, then we must have $\tr\left(\rho^{2}\right)=1$. Therefore, the state is pure if and only if $\left|\r\right|=1$. We also know that the purity cannot be larger than 1, so we must have $\left|\r\right|\le1$. It follows that $\left|\r\right|<1$ corresponds to mixed states. Conversely, the eigenvalues of $\rho$ are $\left(1\pm\left|\r\right|\right)/2$ (as you can check), so every vector with $\left|\r\right|\le1$ gives a positive semi-definite matrix with trace 1, and therefore a valid density operator.

In conclusion, density operators for arbitrary qubits can be parametrized in terms of the Bloch vector $\r$ with $\left|\r\right|\le1$. In $\BBR^{3}$, this corresponds to the unit ball, which in this context is called the \emph{Bloch ball}\index{Bloch ball}. The surface of this ball, at $\left|\r\right|=1$, is the Bloch sphere, and corresponds to pure states. The interior of the ball, with $\left|\r\right|<1$, corresponds to mixed states. This is a neat geometrical interpretation of qubits, which also allows you to visualize the difference between pure and mixed states. The Bloch ball is visualized in \zcref[S]{fig:cat}.

\subsection{Quantum Decoherence}

\subsubsection{Coherent Superpositions}\label{subsec:Coherence}

Consider again the Bloch sphere parametrization of a qubit, defined in \zcref[S]{subsec:Pure-States}:
\[
\left|\Psi\right\rangle =\cos\frac{\theta}{2}\left|0\right\rangle +\e^{\i\phi}\sin\frac{\theta}{2}\left|1\right\rangle \sp\theta\in\left[0,\pi\right]\sp\phi\in\left[0,2\pi\right).
\]
This is the most general superposition state of a qubit, and it emphasizes that there is a relative phase of $\phi$ between the two possible states in the superposition. A superposition with such a well-defined phase is called a \emph{coherent superposition}\index{Coherent superposition}. The most important property of a coherent superposition is that it can exhibit \emph{interference}\index{Interference} between the terms in the superposition.

Clearly, a measurement of $\left|\Psi\right\rangle $ in the computational basis $\left\{ \left|0\right\rangle ,\left|1\right\rangle \right\} $ will yield 0 with probability
\[
P\left(0\right)=\cos^{2}\frac{\theta}{2}=\hf\left(1+\cos\theta\right),
\]
and 1 with probability
\[
P\left(1\right)=\sin^{2}\frac{\theta}{2}=\hf\left(1-\cos\theta\right),
\]
where we used known trigonometric identities.

However, let us now apply a Hadamard gate $H$, given by \zcref[S]{eq:Hadamard}, before we perform a measurement. Recall that
\[
H\left|0\right\rangle =\left|+\right\rangle \sp H\left|1\right\rangle =\left|-\right\rangle ,
\]
so we get
\begin{align*}
H\left|\Psi\right\rangle  & =\cos\frac{\theta}{2}\left|+\right\rangle +\e^{\i\phi}\sin\frac{\theta}{2}\left|-\right\rangle \\
 & =\cos\frac{\theta}{2}\left(\frac{\left|0\right\rangle +\left|1\right\rangle }{\sqrt{2}}\right)+\e^{\i\phi}\sin\frac{\theta}{2}\left(\frac{\left|0\right\rangle -\left|1\right\rangle }{\sqrt{2}}\right)\\
 & =\frac{1}{\sqrt{2}}\left(\left(\cos\frac{\theta}{2}+\e^{\i\phi}\sin\frac{\theta}{2}\right)\left|0\right\rangle +\left(\cos\frac{\theta}{2}-\e^{\i\phi}\sin\frac{\theta}{2}\right)\left|1\right\rangle \right).
\end{align*}
The probability to measure 0 is now
\begin{align*}
P\left(0\right) & =\hf\left|\cos\frac{\theta}{2}+\e^{\i\phi}\sin\frac{\theta}{2}\right|^{2}\\
 & =\hf\left(\cos\frac{\theta}{2}+\e^{\i\phi}\sin\frac{\theta}{2}\right)\left(\cos\frac{\theta}{2}+\e^{-\i\phi}\sin\frac{\theta}{2}\right)\\
 & =\hf\left(\cos^{2}\frac{\theta}{2}+\sin^{2}\frac{\theta}{2}+\left(\e^{\i\phi}+\e^{-\i\phi}\right)\cos\frac{\theta}{2}\sin\frac{\theta}{2}\right)\\
 & =\hf\left(1+2\cos\phi\cos\frac{\theta}{2}\sin\frac{\theta}{2}\right),
\end{align*}
where we used Euler's formula (see also \zcref[S]{prob:e-1})
\[
\e^{\i\phi}=\cos\phi+\i\sin\phi.
\]
Using the identity $\sin\theta=2\sin\frac{\theta}{2}\cos\frac{\theta}{2}$, we can simplify this to
\[
P\left(0\right)=\hf\left(1+\cos\phi\sin\theta\right).
\]
Obviously, the probability to measure 1 will be $P\left(1\right)=1-P\left(0\right)$, so
\[
P\left(1\right)=\hf\left(1-\cos\phi\sin\theta\right).
\]
Before acting with the Hadamard gate, the probabilities depended only on $\theta$. However, now they also depend on the relative phase $\phi$. When this happens we say that there is \emph{interference}\index{Interference} between the terms. As in the familiar \emph{double-slit experiment}\index{Double-slit experiment}, this can be either \emph{constructive interference}\index{Interference!Constructive} or \emph{destructive interference}\index{Interference!Destructive}. For example, consider the case $\theta=\pi/2$, so that
\[
P\left(0\right)=\hf\left(1+\cos\phi\right)\sp P\left(1\right)=\hf\left(1-\cos\phi\right).
\]
If $\phi=0$ then $\cos\phi=1$ and therefore we get fully constructive interference for 0, yielding probability 1, and fully destructive interference for 1, yielding probability 0. On the other hand, if $\phi=\pi$ then $\cos\phi=-1$ and we get the opposite interference. If $\phi=\pi/2$ then $\cos\phi=0$ and we get no interference at all, with both measurements having probability 1/2. For intermediate values of $\phi$, we get partially constructive or destructive interference.

Let us emphasize that although the relative phase $\phi$ had no effect on the computational-basis measurement probabilities before we acted with the Hadamard gate, it does have an observable effect after acting with this gate, and in fact completely determines the measurement probabilities if $\theta$ is kept fixed. Of course, $\phi$ still has no effect if $\theta=0$ or $\theta=\pi$, but those are trivial cases where we did not have a superposition in the computational basis in the first place. As long as $\theta\in\left(0,\pi\right)$, the phase $\phi$ will have some observable effect, which becomes more and more significant as $\theta$ approaches $\pi/2$.

\subsubsection{Decoherence Due to Noise}

If a qubit is part of an actual quantum computer, then it is subject to noise from the environment. Indeed, this is one of the main difficulties in constructing quantum computers. Let us imagine, for example, starting again with a qubit of the form
\[
\left|\Psi\right\rangle =\cos\frac{\theta}{2}\left|0\right\rangle +\e^{\i\phi}\sin\frac{\theta}{2}\left|1\right\rangle \sp\theta\in\left[0,\pi\right]\sp\phi\in\left[0,2\pi\right).
\]
Noise from the environment (heat, particles, radiation, etc.) could introduce a random relative phase\footnote{Technically the noise will introduce a random phase for each term, but since an overall phase factor doesn't have any effect, we can cancel it out and assume only a relative phase without loss of generality.} $\gamma$, so that after some time the qubit will be in the state
\[
\left|\Psi\right\rangle =\cos\frac{\theta}{2}\left|0\right\rangle +\e^{\i\left(\phi+\gamma\right)}\sin\frac{\theta}{2}\left|1\right\rangle \sp\gamma\in\left[0,2\pi\right).
\]
As before, the relative phase has no effect if we measure the qubit in the computational basis without modifying it, but if we first act with a Hadamard gate, we will get
\[
P\left(0\right)=\hf\left(1+\cos\left(\phi+\gamma\right)\sin\theta\right)\sp P\left(1\right)=\hf\left(1-\cos\left(\phi+\gamma\right)\sin\theta\right).
\]
Let us assume that the phase $\gamma$ is uniformly distributed, so no specific value of $\gamma$ is more likely than any other. This means that if we wish to know the actual probabilities to expect, taking into account the random noise, we must average over all possible values of $\gamma$. Since $\gamma$ is a continuous variable, this averaging can be performed by integrating over it. But
\[
\int^{2\pi}_{0}\cos\left(\phi+\gamma\right)\d\gamma=\int^{\phi+2\pi}_{\phi}\cos\gamma\thinspace\d\gamma=0,
\]
since we're integrating on a full period of the cosine, so the integral over the term which includes $\gamma$ vanishes. Therefore, we get:
\[
P\left(0\right)=P\left(1\right)=\hf.
\]
There is no interference, since the original relative phase $\phi$ has been “drowned out” by the noise. In other words, the interference terms vanish after taking into account the noise, so there is no interference. We say that the state has undergone \emph{decoherence}\index{Decoherence}.

\subsubsection{Decoherence Due to Entanglement}

Consider now a composite state of two qubits, parametrized similarly:
\[
\left|\Psi\right\rangle =\cos\frac{\theta}{2}\left|0\right\rangle \otimes\left|0\right\rangle +\e^{\i\phi}\sin\frac{\theta}{2}\left|1\right\rangle \otimes\left|1\right\rangle \sp\theta\in\left[0,\pi\right]\sp\phi\in\left[0,2\pi\right).
\]
This state is entangled as long as $\theta\in\left(0,\pi\right)$. The reduced density operator for the first qubit is
\[
\rho^{A}=\left(\begin{array}{cc}
\cos^{2}\frac{\theta}{2} & 0\\
0 & \sin^{2}\frac{\theta}{2}
\end{array}\right)=\hf\left(\begin{array}{cc}
1+\cos\theta & 0\\
0 & 1-\cos\theta
\end{array}\right),
\]
which follows from \zcref[S]{eq:rhoA-alpha-beta} with $\alpha\equiv\cos\frac{\theta}{2}$ and $\beta\equiv\e^{\i\phi}\sin\frac{\theta}{2}$. Notice that, in the process of taking the partial trace, we lost information about the phase $\phi$. But that's fine, since in \zcref[S]{subsec:Coherence} we saw that the phase has no observable effect before acting with the Hadamard gate.

If we act with the Hadamard gate on the first qubit, the composite state becomes (using $\alpha,\beta$ instead of $\theta,\phi$ for brevity):
\begin{align*}
H_{A}\left|\Psi\right\rangle  & =\alpha\left|+\right\rangle \otimes\left|0\right\rangle +\beta\left|-\right\rangle \otimes\left|1\right\rangle \\
 & =\frac{1}{\sqrt{2}}\left(\alpha\left(\left|0\right\rangle +\left|1\right\rangle \right)\otimes\left|0\right\rangle +\beta\left(\left|0\right\rangle -\left|1\right\rangle \right)\otimes\left|1\right\rangle \right).
\end{align*}
Thus the density operator is
\begin{align*}
\rho^{AB} & =\hf\left(\alpha\left(\left|0\right\rangle +\left|1\right\rangle \right)\otimes\left|0\right\rangle +\beta\left(\left|0\right\rangle -\left|1\right\rangle \right)\otimes\left|1\right\rangle \right)\left(\alpha^{*}\left(\left\langle 0\right|+\left\langle 1\right|\right)\otimes\left\langle 0\right|+\beta^{*}\left(\left\langle 0\right|-\left\langle 1\right|\right)\otimes\left\langle 1\right|\right)\\
 & =\hf\left(\left|\alpha\right|^{2}\left(\left|0\right\rangle +\left|1\right\rangle \right)\left(\left\langle 0\right|+\left\langle 1\right|\right)\otimes\left|0\right\rangle \left\langle 0\right|+\beta^{*}\alpha\left(\left|0\right\rangle +\left|1\right\rangle \right)\left(\left\langle 0\right|-\left\langle 1\right|\right)\otimes\left|0\right\rangle \left\langle 1\right|\right)+\\
 & \qquad+\hf\left(\alpha^{*}\beta\left(\left|0\right\rangle -\left|1\right\rangle \right)\left(\left\langle 0\right|+\left\langle 1\right|\right)\otimes\left|1\right\rangle \left\langle 0\right|+\left|\beta\right|^{2}\left(\left|0\right\rangle -\left|1\right\rangle \right)\left(\left\langle 0\right|-\left\langle 1\right|\right)\otimes\left|1\right\rangle \left\langle 1\right|\right).
\end{align*}
If we take the partial trace with respect to $B$, we get
\begin{align*}
\rho^{A} & =\hf\left(\left|\alpha\right|^{2}\left(\left|0\right\rangle +\left|1\right\rangle \right)\left(\left\langle 0\right|+\left\langle 1\right|\right)+\left|\beta\right|^{2}\left(\left|0\right\rangle -\left|1\right\rangle \right)\left(\left\langle 0\right|-\left\langle 1\right|\right)\right)\\
 & =\hf\left(\left|\alpha\right|^{2}\left(\left|0\right\rangle \left\langle 0\right|+\left|0\right\rangle \left\langle 1\right|+\left|1\right\rangle \left\langle 0\right|+\left|1\right\rangle \left\langle 1\right|\right)+\left|\beta\right|^{2}\left(\left|0\right\rangle \left\langle 0\right|-\left|0\right\rangle \left\langle 1\right|-\left|1\right\rangle \left\langle 0\right|+\left|1\right\rangle \left\langle 1\right|\right)\right),
\end{align*}
which corresponds to the density matrix
\[
\rho^{A}=\hf\left(\begin{array}{cc}
\left|\alpha\right|^{2}+\left|\beta\right|^{2} & \left|\alpha\right|^{2}-\left|\beta\right|^{2}\\
\left|\alpha\right|^{2}-\left|\beta\right|^{2} & \left|\alpha\right|^{2}+\left|\beta\right|^{2}
\end{array}\right).
\]
Restoring $\alpha=\cos\frac{\theta}{2}$ and $\beta=\e^{\i\phi}\sin\frac{\theta}{2}$, we get:
\[
\rho^{A}=\hf\left(\begin{array}{cc}
1 & \cos\theta\\
\cos\theta & 1
\end{array}\right),
\]
where we used the identity
\[
\cos\theta=\cos^{2}\frac{\theta}{2}-\sin^{2}\frac{\theta}{2}.
\]
If we now perform a measurement in the computational basis, the probability to get 0 will be
\begin{align*}
P\left(0\right) & =\tr\left(\left|0\right\rangle \left\langle 0\right|\rho^{A}\right)\\
 & =\hf\tr\left(\left(\begin{array}{cc}
1 & 0\\
0 & 0
\end{array}\right)\left(\begin{array}{cc}
1 & \cos\theta\\
\cos\theta & 1
\end{array}\right)\right)\\
 & =\hf\tr\left(\begin{array}{cc}
1 & \cos\theta\\
0 & 0
\end{array}\right)\\
 & =\hf,
\end{align*}
and the probability to get 1 will thus also be 1/2. Now there is \textbf{no interference} from the phase $\phi$, and indeed, there cannot be such interference since $\rho^{A}$ doesn't even depend on $\phi$ in the first place.

We can summarize the three cases discussed so far in this section as follows. In the coherent superposition, to calculate probabilities after applying the Hadamard gate, we first added or subtracted the complex amplitudes and then took the magnitude-squared:
\[
P\left(0\right)=\hf\left|\alpha+\beta\right|^{2}\sp P\left(1\right)=\hf\left|\alpha-\beta\right|^{2},
\]
where $\alpha=\cos\frac{\theta}{2}$ and $\beta=\e^{\i\phi}\sin\frac{\theta}{2}$. This allowed us to get constructive or destructive interference, depending on the relative phase $\phi$ between $\alpha$ and $\beta$.

However, in the other two cases, either due to noise or due to entanglement, we essentially calculated probabilities in the opposite order – first took the magnitude-squared, and only then added the probabilities:
\[
P\left(0\right)=P\left(1\right)=\hf\left(\left|\alpha\right|^{2}+\left|\beta\right|^{2}\right).
\]
In this case we could have no interference, since we are just adding two non-negative numbers. This is just a classical, or “incoherent”, addition of probabilities, which cannot lead to interference.

\section{Continuous Quantum Systems}

Quantum mechanics is a confusing and unintuitive theory, and requires the introduction of many new concepts. One of the main goals of these notes is to introduce quantum mechanics to students in a way that is mathematically as simple as possible, so that they won't have to struggle with complicated math on top of trying to understand new physical concepts.

It is quite remarkable that we have managed to describe all of the axioms of quantum theory, and almost all of its important aspects such as superposition, entanglement, and the uncertainty principle, using only linear algebra – without any calculus. Moreover, by focusing on discrete two-state systems, or qubits, we actually managed to do everything almost exclusively in $\BBC^{2}$, the simplest non-trivial complex vector space.

Unfortunately, in real life not all systems are this simple, and the time has finally come to start introducing some calculus and talking about continuous quantum systems, which are described by infinite-dimensional Hilbert spaces. However, the student may take comfort in the fact that this is going to be merely a straightforward generalization of what we've already learned. The main difference is that now states are going to be represented as functions instead of $n$-vectors, and operators such as position and momentum will act by multiplication or differentiation instead of multiplication by $n\xx n$ matrices.

\subsection{Continuous Time Evolution, Hamiltonians, and the Schrödinger Equation}\label{subsec:Continuous-Time-Evolution}

When we described the Evolution Axiom, we only talked about evolution from one discrete point in time to another. As a first step toward quantum mechanics of continuous systems, let us discuss time evolution with a continuous time variable.

\subsubsection{The Schrödinger Equation and Hamiltonians: Preface}

Usually, in introductory quantum mechanics textbooks, the \emph{Schrödinger equation}\index{Schrödinger equation} is introduced at the very beginning – as a fundamental postulate, without any explanation or motivation. The student is simply told that this is the equation they are going to be working with, and typically, much of the rest of the textbook consists of solving the Schrödinger equation for different systems.

In these notes, I chose to do the exact opposite, and introduce the Schrödinger equation only near the end of the notes. The reason is that the Schrödinger equation is actually \textbf{not }a fundamental component of the modern 21st-century formulation of quantum theory\footnote{In fact, entire books have been written about fields such as quantum computation and quantum gravity without mentioning the Schrödinger equation even once!}. What is truly fundamental about quantum theory is what we studied throughout most of these notes: the abstract formulation of the theory in terms of Hilbert spaces, states, and operators, using the axioms summarized in \zcref[S]{subsec:The-Foundations-of}. The Schrödinger equation turns out to be merely a \textbf{special case }of the Evolution Axiom – which, as you recall, simply says that quantum states evolve by the action of unitary operators.

The Evolution Axiom applies to \textbf{any }kind of evolution, whether in time or due to some transformation performed on the system; and with regard to evolution in time, the time variable can be either discrete or continuous. Quantum gates, which are the main type of evolution we have seen so far, correspond to discrete time evolution – the qubit is in some state now, and will be in another state after it passes through the gate, but these are two \textbf{discrete} points in time, and nothing of interest is happening in the gap between them\footnote{At least, this is the case in the idealized discrete quantum gate model we considered in these notes. Real quantum gates need to be realized via some kind of physical apparatus; for example, an oscillating magnetic field can flip the direction of electron spins, thus realizing an $X$ gate. In this case, the underlying system will have continuous evolution, even if the gate itself represents a discrete operation.}.

In the specific case when the evolution is the system's natural evolution in time (so not a result of some explicit transformation, like a rotation) and with respect to a \textbf{continuous }time variable, it is useful in practice to replace the Evolution Axiom, which is very abstract, with the Schrödinger equation, which is a concrete differential equation that can be solved, either exactly or approximately, for a variety of different systems. The focus then shifts from the unitary evolution operator of the Evolution Axiom to a Hermitian operator called the \emph{Hamiltonian}\index{Hamiltonian!Quantum}. We will see in \zcref[S]{subsec:Hamiltonians-and-the} precisely how these two operators are related to each other.

Finally, consider the fact that the Evolution Axiom is an almost inevitable result of the mathematical framework of quantum theory. Probabilities must always sum to 1, or in other words, states must have norm 1. The only type of linear operator that preserves the norm of every possible initial state is a unitary operator\footnote{In infinite-dimensional Hilbert spaces, there are non-unitary operators that preserve probabilities. One example is the operator $S\left|n\right\rangle =\left|n+1\right\rangle $ in the space spanned by an infinite orthonormal basis $\left|n\right\rangle $ where $n\in\left\{ 0,1,2,\ldots\right\} $. This operator preserves probabilities because it keeps all the amplitudes the same, but it is not unitary because $S^{\dagger}\left|0\right\rangle =0$, so $SS^{\dagger}\left|0\right\rangle =0$, therefore $SS^{\dagger}\ne1$. However, if we require the operator to be invertible in addition to preserving probabilities, the operator is guaranteed to be unitary.}. While the Schrödinger equation also preserves probabilities (as it must), this fact is not immediately obvious from the form of the equation, which is why the Evolution Axiom may be considered more fundamental.

\subsubsection{Derivation of the Schrödinger Equation}\label{subsec:Hamiltonians-and-the}

Let us recall the Evolution Axiom from \zcref[S]{subsec:Unitary-Transformations-and}, with slightly different notation. In an isolated system, given two points in time $t_{1}$ and $t_{2}$, there is a unitary operator $U\left(t_{2}\ot t_{1}\right)$ such that any state $\left|\Psi\left(t_{1}\right)\right\rangle $ at $t_{1}$ evolves to a state $\left|\Psi\left(t_{2}\right)\right\rangle $ at time $t_{2}$ by acting on it with the unitary operator:
\begin{equation}
\left|\Psi\left(t_{2}\right)\right\rangle =U\left(t_{2}\ot t_{1}\right)\left|\Psi\left(t_{1}\right)\right\rangle .\label{eq:unitary-t1}
\end{equation}
The main difference between this formulation and the one we had for discrete systems is that now we are letting $U$ be a \textbf{differentiable function}{\bfseries\footnote{In an infinite-dimensional Hilbert space, it may be that $U\left(t_{2}\ot t_{1}\right)\left|\Psi\left(t_{1}\right)\right\rangle $ is differentiable with respect to $t_{2}$ only if $\left|\Psi\left(t_{1}\right)\right\rangle $ belongs to a certain subspace of the Hilbert space. This happens, for example, whenever the Hamiltonian, which we will define in \zcref[S]{eq:Hamiltonian-t0}, has arbitrarily large eigenvalues, as in the case of the harmonic oscillator. \zcref[S]{eq:Schr}, the Schrödinger equation, then holds only for states in that subspace, even though all states still evolve according to \zcref[S]{eq:unitary-t1}.}} of $t_{1}$ and $t_{2}$, so that we can encode the unitary evolution of the system from any point in time to any other point in time. This is very different than what we discussed in the discrete case, where for example, a quantum gate is not a function of time – it is the same quantum gate at all times.

However, this is still just a special case of the Evolution Axiom; the axiom simply states that evolution between any two points in time must be encoded in some unitary operator, but it will in general be a different operator for different start and end times, so here we are explicitly encoding the different operators as one universal function $U\left(t_{2}\ot t_{1}\right)$.

In \zcref[S]{eq:unitary-t1}, if we assume that $t_{2}=t_{1}$ (that is, no time has passed) then we get
\[
\left|\Psi\left(t_{1}\right)\right\rangle =U\left(t_{1}\ot t_{1}\right)\left|\Psi\left(t_{1}\right)\right\rangle .
\]
Since this must be true for \textbf{every }state $\left|\Psi\left(t_{1}\right)\right\rangle $ and for \textbf{every }time $t_{1}$, we see\footnote{Recall \zcref[S]{prob:divide}!} that if no time has passed, $U\left(t_{1}\ot t_{1}\right)$ must be the identity operator:
\begin{equation}
U\left(t_{1}\ot t_{1}\right)=1\sp\fa t_{1}\in\BBR.\label{eq:Uid}
\end{equation}
Let us now assume that the system is in the state $\left|\Psi\left(t_{3}\right)\right\rangle $ at time $t_{3}$. Then from \zcref[S]{eq:unitary-t1} we must have on the one hand
\[
\left|\Psi\left(t_{3}\right)\right\rangle =U\left(t_{3}\ot t_{1}\right)\left|\Psi\left(t_{1}\right)\right\rangle ,
\]
but on the other hand
\[
\left|\Psi\left(t_{3}\right)\right\rangle =U\left(t_{3}\ot t_{2}\right)\left|\Psi\left(t_{2}\right)\right\rangle =U\left(t_{3}\ot t_{2}\right)U\left(t_{2}\ot t_{1}\right)\left|\Psi\left(t_{1}\right)\right\rangle .
\]
Therefore, $U$ must satisfy the \emph{composition property}\index{Composition property}\footnote{This property is the reason we used the notation $U\left(t_{2}\ot t_{1}\right)$: we wanted the $t_{2}$ in $U\left(t_{2}\ot t_{1}\right)$ and the $t_{2}$ in $U\left(t_{3}\ot t_{2}\right)$ to be adjacent. If the times were arranged from left to right, we would have had $U\left(t_{2}\to t_{3}\right)U\left(t_{1}\to t_{2}\right)$, which does not make it clear that the operator on the left starts when the operator on the right ends. Note that when applying operators to a ket, the operators always act from right to left. So in $U\left(t_{3}\ot t_{2}\right)U\left(t_{2}\ot t_{1}\right)\left|\Psi\left(t_{1}\right)\right\rangle $ the operator $U\left(t_{2}\ot t_{1}\right)$ acts on the state first, to take it to $t_{2}$, and then $U\left(t_{3}\ot t_{2}\right)$ acts on the result, to take it to $t_{3}$.}:
\begin{equation}
U\left(t_{3}\ot t_{1}\right)=U\left(t_{3}\ot t_{2}\right)U\left(t_{2}\ot t_{1}\right)\sp\fa t_{1},t_{2},t_{3}\in\BBR.\label{eq:composition}
\end{equation}
In particular, if $t_{3}=t_{1}$ we get
\[
1=U\left(t_{1}\ot t_{1}\right)=U\left(t_{1}\ot t_{2}\right)U\left(t_{2}\ot t_{1}\right).
\]
Therefore we must have
\[
U\left(t_{1}\ot t_{2}\right)=U^{-1}\left(t_{2}\ot t_{1}\right)=U^{\dagger}\left(t_{2}\ot t_{1}\right)\sp\fa t_{1},t_{2}\in\BBR,
\]
or in other words, evolution to the past is given by the adjoint (or inverse) of the evolution to the future, as we discussed in \zcref[S]{subsec:Unitary-Transformations-and}.

We now change notation slightly by taking $t_{1}\mt t_{0}$ and $t_{2}\mt t$ in \zcref[S]{eq:unitary-t1}:
\begin{equation}
\left|\Psi\left(t\right)\right\rangle =U\left(t\ot t_{0}\right)\left|\Psi\left(t_{0}\right)\right\rangle .\label{eq:Utt0}
\end{equation}
For any \textbf{arbitrary }time $t$, the evolution of the system from a \textbf{fixed }time $t_{0}$ is given by this equation. Let us take the time derivative of the equation:
\begin{equation}
\frac{\d}{\d t}\left|\Psi\left(t\right)\right\rangle =\frac{\d U\left(t\ot t_{0}\right)}{\d t}\left|\Psi\left(t_{0}\right)\right\rangle ,\label{eq:ddtU}
\end{equation}
where we consider $\left|\Psi\left(t_{0}\right)\right\rangle $ to be independent of $t$ since $t_{0}$ is a fixed time. From \zcref[S]{eq:Utt0} we find, by multiplying both sides by $U^{\dagger}\left(t\ot t_{0}\right)$ from the left, that
\[
\left|\Psi\left(t_{0}\right)\right\rangle =U^{\dagger}\left(t\ot t_{0}\right)\left|\Psi\left(t\right)\right\rangle .
\]
We plug that into \zcref[S]{eq:ddtU} and find
\begin{equation}
\frac{\d}{\d t}\left|\Psi\left(t\right)\right\rangle =\frac{\d U\left(t\ot t_{0}\right)}{\d t}U^{\dagger}\left(t\ot t_{0}\right)\left|\Psi\left(t\right)\right\rangle ,\label{eq:ddtU2}
\end{equation}
where the time derivative only acts on $U$ and not on $U^{\dagger}$. Now, let us define a new operator $H$ called the \emph{Hamiltonian}\index{Hamiltonian!Quantum} as follows:
\begin{equation}
H\left(t\right)\equiv\i\frac{\d U\left(t\ot t_{0}\right)}{\d t}U^{\dagger}\left(t\ot t_{0}\right).\label{eq:Hamiltonian-t0}
\end{equation}
Note that $H$ can in general be a function of $t$, but it is \textbf{independent }of $t_{0}$, which is why we called it $H\left(t\right)$ and not $H\left(t,t_{0}\right)$ or $H\left(t\ot t_{0}\right)$. Also, the Hamiltonian is Hermitian. You will prove both of these facts in \zcref[S]{prob:Hamiltonian-independent}.

In terms of the Hamiltonian, \zcref[S]{eq:ddtU2} becomes
\begin{equation}
\i\frac{\d}{\d t}\left|\Psi\left(t\right)\right\rangle =H\left(t\right)\left|\Psi\left(t\right)\right\rangle .\label{eq:Schr}
\end{equation}
This equation is called the \emph{Schrödinger equation}\index{Schrödinger equation}\footnote{In non-natural units, this equation features the reduced Planck constant $\hbar$:
\[
\i\hbar\frac{\d}{\d t}\left|\Psi\left(t\right)\right\rangle =H\left(t\right)\left|\Psi\left(t\right)\right\rangle .
\]
Of course, as we discussed in \zcref[S]{subsec:Dimensionless-and-Dimensionful}, $\hbar$ is dimensionful and therefore its numerical value doesn't matter, so we can just choose units such as the Planck units, where it simply has the value $\hbar=1$.}.

In this equation, a time derivative $\d/\d t$ is acting on the state $\left|\Psi\left(t\right)\right\rangle $. Therefore, one might wonder whether $\d/\d t$ is an operator on the Hilbert space. However, the answer is no. This is because in (non-relativistic) quantum mechanics, while $x$ is an operator (as we will see in \zcref[S]{subsec:The-Position-Operator}), $t$ is just a \textbf{label}. In this section we defined a function $|\Psi(t)\rangle$, which takes a real number $t$ as input, and returns a state in the Hilbert space as output. The derivative $\d/\d t$ doesn't act on the vectors in the Hilbert space, which is what operators do; instead, it acts on this function. Therefore, $\d/\d t$ is not an operator on the Hilbert space.
\begin{problem}\leavevmode
\label{prob:Hamiltonian-independent}
\begin{problemparts}
\item Prove that $H\left(t\right)$ as defined in \zcref[S]{eq:Hamiltonian-t0} is independent of $t_{0}$, thus justifying the notation $H\left(t\right)$, as well as its use in \zcref[S]{eq:Schr}, the Schrödinger equation, where $t$ is the only variable.
\item Prove that $H\left(t\right)$ is a Hermitian operator.
\end{problemparts}
\end{problem}

\subsubsection{Time-Independent Hamiltonians}

Let us now assume that the Hamiltonian is constant, that is, \emph{time-independent}\index{Time-independent Hamiltonian}\index{Hamiltonian!Time-independent}. Although in some quantum systems the Hamiltonian does depend on time, this is not very common; most quantum systems have time-independent Hamiltonians.

We can rewrite \zcref[S]{eq:Hamiltonian-t0} as follows:
\begin{equation}
\frac{\d U\left(t\ot t_{0}\right)}{\d t}=-\i HU\left(t\ot t_{0}\right).\label{eq:Udiff}
\end{equation}
Compare this with \zcref[S]{eq:exp-At}:
\[
\frac{\d}{\d t}\e^{At}=A\e^{At},
\]
which holds assuming that $A$ is constant. If the Hamiltonian $H$ is constant, then $A\equiv-\i H$ is also constant. In addition, we can replace $t$ with $t-t_{0}$ in the exponent, since that does not change the derivative (because $t_{0}$ is constant). Hence, we see that the solution\footnote{Even if the Hamiltonian is time-dependent, it is still possible to solve \zcref[S]{eq:Udiff}; however, the solution is then much more complicated and involves \emph{time-ordered exponentials\index{Time-ordered exponential}\index{Exponential!Time-ordered}}, which we will not cover in these notes.} to \zcref[S]{eq:Udiff} is
\begin{equation}
U\left(t\ot t_{0}\right)=\e^{-\i H\left(t-t_{0}\right)}.\label{eq:Uexp}
\end{equation}
Don't get confused by the notation: $H\left(t-t_{0}\right)$ in the exponent is the \textbf{constant }$H$ \textbf{times }the \textbf{number }$t-t_{0}$, not the function $H$ evaluated at $t-t_{0}$!

The reason we wrote $t-t_{0}$ instead of $t$ in the exponential is that from \zcref[S]{eq:Uid} we have the initial condition
\[
U\left(t_{0}\ot t_{0}\right)=1.
\]
This condition is indeed satisfied for $U$ as given in \zcref[S]{eq:Uexp}, since
\[
U\left(t_{0}\ot t_{0}\right)=\e^{-\i H\left(t_{0}-t_{0}\right)}=\e^{0}=1,
\]
by \zcref[S]{eq:exp0}. However, it would \textbf{not} be satisfied if we just wrote $\e^{-\i Ht}$, since then we would have $U\left(t_{0}\ot t_{0}\right)=\e^{-\i Ht_{0}}\ne1$. In general, when solving differential equations, the solution always depends on the initial (or boundary) conditions.

We can rewrite \zcref[S]{eq:Uexp} to match the notation in the beginning of \zcref[S]{subsec:Hamiltonians-and-the} as follows:
\begin{equation}
U\left(t_{2}\ot t_{1}\right)=\e^{-\i H\left(t_{2}-t_{1}\right)}.\label{eq:U21}
\end{equation}
The evolution operator between any two arbitrary points in time, $t_{1}$ and $t_{2}$, is given by \zcref[S]{eq:U21}.

It is interesting that, since $H$ is constant, the unitary evolution operator is not a function of both $t_{1}$ and $t_{2}$, but only of the \textbf{difference} between them, $t_{2}-t_{1}$. So for example, the evolution from time $t_{1}=3$ to time $t_{2}=4$ and from time $t_{1}=4$ to time $t_{2}=5$ will be given by the \textbf{same }unitary operator, $\e^{-\i H}$, since in both cases the time difference is $t_{2}-t_{1}=1$.

In \zcref[S]{prob:Hamiltonian-independent}, you proved that the Hamiltonian is a Hermitian operator. Therefore, it might correspond to an \textbf{observable}. Indeed, it does; this observable is the \emph{energy}\index{Energy} of the system. Its (real) eigenvalues correspond to \emph{energy eigenstates\index{Energy eigenstate}}:
\begin{equation}
H\left|E\right\rangle =E\left|E\right\rangle ,\label{eq:ham-en}
\end{equation}
where we denote by $\left|E\right\rangle $ the eigenstate with energy $E$. This is often referred to as the \emph{time-independent Schrödinger equation}\index{Time-independent Schrödinger equation}, but it's really just an eigenvalue equation!

As we have seen, the Hamiltonian is used to evolve continuous systems in time. What does energy have to do with time, you ask? Well, from relativity we know that momentum in spacetime is described by a 4-vector called the \emph{4-momentum}\index{4-momentum}, which is defined as follows:
\[
\p\left(t,x,y,z\right)\equiv\left(\begin{array}{c}
E\\
p_{x}\\
p_{y}\\
p_{z}
\end{array}\right).
\]
Here, $p_{x}$, $p_{y}$, and $p_{z}$ are the momenta in the $x$, $y$, and $z$ directions respectively. In the first component, which is the one in the time direction, we have the energy $E$. Thus energy is actually “momentum in the time direction”! Indeed, in relativity we will often write $p_{0}$ for the energy. Just like momentum moves you in space, so does energy move you in time. This is exactly why the Hamiltonian is responsible for evolution in time. It is also why Hamiltonians are usually time-independent – if they are not, then energy is not conserved!
\begin{problem}\leavevmode
For the unitary operator given in \zcref[S]{eq:U21}:
\begin{problemparts}
\item Verify that it satisfies the composition property in \zcref[S]{eq:composition}.
\item Verify that it is invariant under the \emph{time shift}\index{Time shift} transformation
\[
t_{1}\mt t_{1}+t\sp t_{2}\mt t_{2}+t,
\]
where $t\in\BBR$.
\item Verify that under the transformation
\[
t_{1}\mt-t_{1}\sp t_{2}\mt-t_{2},
\]
the evolution operator is replaced with its adjoint (or inverse). This is an explicit example of the \emph{reversibility}\index{Unitary evolution!Reversibility} of unitary evolution, which we discussed in \zcref[S]{subsec:Unitary-Transformations-and}.
\end{problemparts}
\end{problem}

\begin{problem}\leavevmode
In \zcref[S]{subsec:Quantum-Logic-Gates} we discussed several unitary operators which act on qubits. For example, the quantum $Z$ gate is given by the Pauli matrix $\sigma_{z}$:
\[
Z\equiv\sigma_{z}=\left(\begin{array}{cc}
1 & 0\\
0 & -1
\end{array}\right),
\]
and its action is to leave $\left|0\right\rangle $ unchanged but flip the phase of $\left|1\right\rangle $. Find a Hamiltonian corresponding to this unitary evolution operator. Since this gate is a single discrete operation, we can simply take the time interval to be 1. In other words, you need to find the $H$ in the equation $Z=\e^{-\i H}$.
\end{problem}

\subsection{Hamiltonian Mechanics and Canonical Quantization}

We have seen that in order to create a model for a specific physical system in quantum theory, we must choose a specific Hilbert space with specific states and specific operators. But how do we know which Hilbert space, states, and operators to use for a given physical system? This is often a hard question to answer. For example, we currently do not have a consistent and experimentally verified quantum model for general relativity; the problem of finding such a model is known as \emph{quantum gravity}\index{Quantum gravity}, and it is one of the hardest problems in physics.

Luckily, it turns out that there is a certain prescription that allows us to take a classical theory and turn it into a quantum theory in a straightforward way. The properties of the classical theory will dictate the type of Hilbert space, states, and operators we should use in the corresponding quantum theory. This process is known as \emph{quantization}\index{Quantization}. It doesn't work for every classical theory; for example, when applied to general relativity, it doesn't produce a complete and consistent quantum theory, which is why quantizing gravity is so hard. However, it does work, in an experimentally verifiable way, for most classical theories of interest.

\subsubsection{A Quick Review of Classical Hamiltonian Mechanics}

Classical mechanics can be reformulated using a quantity called the \emph{(classical) Hamiltonian}\index{Hamiltonian!Classical}. This is basically the total energy of the system, usually written as \emph{kinetic energy\index{Kinetic energy}\index{Energy!Kinetic}} plus \emph{potential energy\index{Potential energy}\index{Energy!Potential} }and in terms of the \emph{canonical coordinates}\index{Canonical coordinates}\index{Coordinates!Canonical} $q$ and $p$. Here we will consider the case where $q$ and $p$ represent position and momentum respectively, and therefore we will label them $x$ and $p$ instead.

The \emph{phase space}\index{Phase space} of the system consists of all the possible values of the canonical coordinates; for a particle, the phase space includes both “actual” space (all the values of $x$) and \emph{momentum space}\index{Momentum space} (all the values of $p$).

Since we are interested in quantum mechanics, not classical mechanics, we will not go over the Hamiltonian formulation in detail. Instead, we will just review certain important definitions and results. We will consider only Hamiltonians of the form
\[
H=K\left(p\right)+V\left(x\right),
\]
where $K$ is the kinetic energy, which depends only on the momentum $p$, and $V$ is the potential energy, which depends only on the position $x$.

Let us consider the specific case of a single particle of mass $m$. In Newtonian mechanics, the particle's momentum is defined as
\[
p\equiv mv\sp\textrm{where }v\equiv\xd\equiv\frac{\d x}{\d t}.
\]
The kinetic energy is defined as $\hf mv^{2}$, and we can write it in terms of the momentum as follows:
\[
K=\hf mv^{2}=\frac{1}{2m}\left(mv\right)^{2}=\frac{p^{2}}{2m}.
\]
We conclude that for a particle of mass $m$, the Hamiltonian\index{Hamiltonian!Of a point particle} will be of the form
\begin{equation}
H=\frac{p^{2}}{2m}+V\left(x\right).\label{eq:H_particle}
\end{equation}
The kinetic energy of such a particle will always be $p^{2}/2m$, but the potential energy $V\left(x\right)$ depends on the forces acting on the particle, such as gravitational or electrostatic forces.

Now, let us define the \emph{Poisson brackets}\index{Poisson brackets} of two functions $f,g$ of position $x$ and momentum $p$ as follows:
\[
\left\{ f,g\right\} \equiv\frac{\partial f}{\partial x}\frac{\partial g}{\partial p}-\frac{\partial g}{\partial x}\frac{\partial f}{\partial p}.
\]
In \zcref[S]{prob:Poisson} you will prove some properties of these brackets; in particular they are anti-symmetric, $\left\{ g,f\right\} =-\left\{ f,g\right\} $, which means that $\left\{ f,f\right\} =0$ for any $f$. For $x$ and $p$ themselves we have
\[
\left\{ x,x\right\} =\left\{ p,p\right\} =0,
\]
and
\begin{equation}
\left\{ x,p\right\} =\frac{\partial x}{\partial x}\frac{\partial p}{\partial p}-\frac{\partial p}{\partial x}\frac{\partial x}{\partial p}=1,\label{eq:xp1}
\end{equation}
since $x$ and $p$ are assumed to be \textbf{independent }variables, so their derivatives with respect to each other vanish. Even though in Newtonian mechanics we \textbf{define }the momentum to be $p=m\xd$, in Hamiltonian mechanics we “forget” about this relation and just assume that $x$ and $p$ are two completely independent variables of the system, thus generalizing the concept of momentum to any kind of system.

The dynamics of the system in Hamiltonian mechanics are determined as follows. If $A$ is any function of $x$ and $p$, then its time derivative is given by\footnote{Here we are assuming that $A$ does not depend on $t$ \textbf{explicitly}, but only \textbf{implicitly }via its dependence on $x$ and $p$. If $A$ does have explicit dependence on $t$, then this equation becomes
\[
\frac{\d A}{\d t}=\left\{ A,H\right\} +\frac{\partial A}{\partial t}.
\]
}
\begin{equation}
\Adt\equiv\frac{\d A}{\d t}=\left\{ A,H\right\} .\label{eq:Poisson-dt}
\end{equation}
For $x$ and $p$ themselves, we get
\begin{equation}
\xd=\frac{\d x}{\d t}=\left\{ x,H\right\} =\frac{\partial x}{\partial x}\frac{\partial H}{\partial p}-\frac{\partial H}{\partial x}\frac{\partial x}{\partial p}=\frac{\partial H}{\partial p},\label{eq:Ham-1}
\end{equation}
\begin{equation}
\pd=\frac{\d p}{\d t}=\left\{ p,H\right\} =\frac{\partial p}{\partial x}\frac{\partial H}{\partial p}-\frac{\partial H}{\partial x}\frac{\partial p}{\partial p}=-\frac{\partial H}{\partial x}.\label{eq:Ham-2}
\end{equation}
In other words, the evolution of each parameter depends on the derivative of the Hamiltonian with respect to the other parameter. \zcref[S]{eq:Ham-1,eq:Ham-2} are called \emph{Hamilton's equations}\index{Hamilton's equations}.

For a point particle with the Hamiltonian in \zcref[S]{eq:H_particle}, we get
\begin{equation}
\xd=\frac{\d x}{\d t}=\frac{\partial}{\partial p}\left(\frac{p^{2}}{2m}+V\left(x\right)\right)=\frac{p}{m},\label{eq:point-x}
\end{equation}
\begin{equation}
\pd=\frac{\d p}{\d t}=-\frac{\partial}{\partial x}\left(\frac{p^{2}}{2m}+V\left(x\right)\right)=-V'\left(x\right).\label{eq:point-p}
\end{equation}
The first equation relates the two independent variables $x$ and $p$ to each other: $p=m\xd$. Of course, this is just the definition of the momentum of a particle in Newtonian mechanics, but Hamiltonian mechanics allows us to consider more general systems and define a \emph{generalized momentum}\index{Generalized momentum} for any kind of system. For example, in a rotating system $p$ will be the angular momentum, and so on.

The second equation is \emph{Newton's second law}\index{Newton's second law}: the time derivative of momentum is the force, and the force is given by minus the derivative of the potential\footnote{Here we are working in one spatial dimension, for simplicity. In the $3$-dimensional case, the force is minus the gradient of the potential:
\[
\F=-\na V=-\left(\frac{\partial V}{\partial x},\frac{\partial V}{\partial y},\frac{\partial V}{\partial z}\right).
\]
}. We can take the derivative of \zcref[S]{eq:point-x} and plug \zcref[S]{eq:point-p} into it to get
\[
\xdd\equiv\frac{\d^{2}x}{\d t^{2}}=\frac{\d\xd}{\d t}=\frac{1}{m}\frac{\d p}{\d t}=-\frac{1}{m}V'\left(x\right).
\]
Multiplying by $m$, we get the familiar form of Newton's law:
\begin{equation}
F=ma=m\xdd=m\frac{\d^{2}x}{\d t^{2}}=-V'\left(x\right),\label{eq:2nd-law}
\end{equation}
where $a$ is the acceleration.
\begin{problem}\leavevmode
\label{prob:Poisson}Prove the following properties of the Poisson brackets:
\begin{itemize}
\item \textbf{Anti-symmetry:} For all functions $f,g$
\[
\left\{ f,g\right\} =-\left\{ g,f\right\} .
\]
\item \textbf{Linearity:} For all functions $f,g,h$ and numbers $\alpha,\beta$
\[
\left\{ \alpha f+\beta g,h\right\} =\alpha\left\{ f,h\right\} +\beta\left\{ g,h\right\} .
\]
\item \textbf{Leibniz rule:} For all functions $f,g,h$
\[
\left\{ fg,h\right\} =f\left\{ g,h\right\} +\left\{ f,h\right\} g.
\]
\item \textbf{Jacobi identity}\index{Jacobi identity}\textbf{:}
\[
\left\{ f,\left\{ g,h\right\} \right\} +\left\{ g,\left\{ h,f\right\} \right\} +\left\{ h,\left\{ f,g\right\} \right\} =0.
\]
\end{itemize}
\end{problem}

\subsubsection{Canonical Quantization}

Recall the definition of the expectation value for the measurement of an observable $A$ when the system is in the state $\left|\Psi\right\rangle $:
\[
\left\langle A\right\rangle \equiv\langle\Psi|A|\Psi\rangle.
\]
Let us take the time derivative of this, assuming that the state $\left|\Psi\right\rangle $ depends on time but the observable $A$ doesn't:
\begin{equation}
\frac{\d\left\langle A\right\rangle }{\d t}=\left(\frac{\d}{\d t}\langle\Psi|\right)A|\Psi\rangle+\langle\Psi|A\left(\frac{\d}{\d t}|\Psi\rangle\right).\label{eq:ddtA}
\end{equation}
By \zcref[S]{eq:Schr}, the Schrödinger equation, we have
\[
\frac{\d}{\d t}\left|\Psi\right\rangle =-\i H\left|\Psi\right\rangle .
\]
We can take the adjoint of this equation to get (remember that $H$ is Hermitian so $H=H^{\dagger}$)
\[
\frac{\d}{\d t}\left\langle \Psi\right|=\i\left\langle \Psi\right|H.
\]
Plugging into \zcref[S]{eq:ddtA}, we get
\begin{align*}
\frac{\d\left\langle A\right\rangle }{\d t} & =\i\langle\Psi|HA|\Psi\rangle-\i\langle\Psi|AH|\Psi\rangle\\
 & =-\i\langle\Psi|\left(AH-HA\right)|\Psi\rangle\\
 & =-\i\langle\Psi|\left[A,H\right]|\Psi\rangle\\
 & =-\i\left\langle \left[A,H\right]\right\rangle ,
\end{align*}
so in conclusion,
\begin{equation}
\frac{\d\left\langle A\right\rangle }{\d t}=-\i\left\langle \left[A,H\right]\right\rangle .\label{eq:dA-AH}
\end{equation}
Comparing this with \zcref[S]{eq:Poisson-dt},
\[
\frac{\d A}{\d t}=\left\{ A,H\right\} ,
\]
we find a very interesting result: the \textbf{quantum }expectation value of the observable $A$ evolves in time in a way similar to what \textbf{classical} Hamiltonian mechanics predicts. This motivates us to propose a relation between the Poisson brackets of classical functions and the commutator of the corresponding quantum operators:
\[
\left[A,H\right]\tot\i\left\{ A,H\right\} ,
\]
or more generally, for two classical observables $A$ and $B$,
\begin{equation}
\left[A,B\right]\tot\i\left\{ A,B\right\} .\label{eq:AB-P}
\end{equation}
This is consistent, because in \zcref[S]{prob:Poisson} you proved some properties of Poisson brackets, and these properties also happen to be satisfied by the commutator, as you proved in \zcref[S]{subsec:The-Matrix-Commutator}.

In particular, for $x$ and $p$ themselves, according to \zcref[S]{eq:xp1} we have $\left\{ x,p\right\} =1$, so in the quantum theory we will have\footnote{With $\hbar$, this equation will take the form $\left[x,p\right]=\i\hbar$.}
\begin{equation}
\left[x,p\right]=\i.\label{eq:canonical-xp}
\end{equation}
This is called the \emph{canonical commutation relation}\index{Canonical commutation relation}\index{Commutation relation!Canonical}.

What we have derived (or at least, motivated) here is called \emph{canonical quantization\index{Canonical quantization}\index{Quantization!Canonical}}. Given a classical system described by a Hamiltonian, we can turn it into a quantum system – \emph{quantize\index{Quantization}} it – by \emph{“promoting”\index{Promotion to operators}} classical functions on the phase space, including the variables $x$ and $p$ themselves, to Hermitian operators. We are not provided with any specific information about these operators, except that they are Hermitian (which they must be, since in classical physics all variables are real!) and that the quantum commutators should be related to the classical Poisson brackets according to the prescription in \zcref[S]{eq:AB-P}.

These Hermitian operators now represent observables in the quantum theory; they have eigenstates and eigenvalues which represent possible measurement outcomes. This means that the values of $x$ and $p$ are no longer uniquely determined from some initial conditions, as in the classical theory; they become \textbf{probabilistic}. In addition, the time evolution of the system is no longer described by Hamilton's equations, but rather, by the Schrödinger equation.

Note that what we did here does not constitute a proof\textbf{ }that \textbf{all }classical theories are related to quantum theories in this way. Canonical quantization merely ensures that expectation values of the observables in the quantum theory evolve in time similarly to the observables in the classical theory, which is something that we expect to be true, but it is not by itself a sufficient condition for creating a sensible quantum theory. Indeed, there are known cases where canonical quantization doesn't quite work, or is at least ambiguous, because two Poisson brackets which in the classical theory are equal to each other will have different values in the quantum theory, generating an inconsistency.

Nevertheless, canonical quantization works incredibly well in the vast majority of cases – and indeed, most classical theories, from a single point particle to very complicated systems with many different particles and forces, can be quantized in this way, and the results have been verified experimentally to high precision!

Just as in the case of the Schrödinger equation, in introductory quantum mechanics textbooks canonical quantization is usually just presented as an arbitrary axiom. I hope I managed to motivate it and give you some intuition as to why classical and quantum theories are related in this way.

\subsection{The Harmonic Oscillator}

\subsubsection{The Classical Harmonic Oscillator}\label{subsec:The-Classical-Harmonic-Oscillator}

The \emph{quantum harmonic oscillator\index{Quantum harmonic oscillator}\index{Harmonic oscillator!Quantum}} is the quantization of the \emph{classical harmonic oscillator}\index{Classical harmonic oscillator}\index{Harmonic oscillator!Classical}, and just like the qubit, it is a reasonably simple quantum system which turns out to describe many different realistic physical systems, either exactly or approximately. This makes sense, because the classical harmonic oscillator itself describes or approximates many different classical systems!

In particular, quantum harmonic oscillators form the basis of \emph{quantum field theory}\index{Quantum field theory}, which is the theory describing all of the known elementary particles, including matter particles (such as electrons and \emph{quarks}\index{Quark}), particles which mediate fundamental interactions (such as photons, which mediate the electromagnetic force, and \emph{gluons}\index{Gluon}, which mediate the \emph{strong nuclear force}\index{Strong nuclear force}), and others (such as the \emph{Higgs boson}\index{Higgs boson}).

The (simple) classical harmonic oscillator has the Hamiltonian
\begin{equation}
H=\frac{p^{2}}{2m}+\hf m\omega^{2}x^{2}.\label{eq:har-ham}
\end{equation}
We have the standard kinetic energy term $K\left(p\right)=p^{2}/2m$, where $m$ is the mass of the particle, and the potential energy
\[
V\left(x\right)\equiv\hf m\omega^{2}x^{2},
\]
where $\omega$ is a positive constant called the \emph{frequency\index{Frequency}} or \emph{angular frequency}\index{Angular frequency}, because it represents the frequency at which the oscillator oscillates.

It is easy to find the \emph{equations of motion}\index{Equations of motion!From Hamiltonian} using \zcref[S]{eq:Ham-1,eq:Ham-2}, Hamilton's equations. Alternatively, since this is a particle with a Hamiltonian of the standard form given in \zcref[S]{eq:H_particle}, we can just use Newton's second law in \zcref[S]{eq:2nd-law} directly:
\[
\frac{\d^{2}x}{\d t^{2}}=-\frac{1}{m}\frac{\partial}{\partial x}V\left(x\right)=-\frac{1}{m}\frac{\partial}{\partial x}\left(\hf m\omega^{2}x^{2}\right)=-\omega^{2}x.
\]
To solve this differential equation, we can use the fact that
\[
\frac{\d}{\d t}\cos t=-\sin t\sp\frac{\d}{\d t}\sin t=\cos t,
\]
which means that
\[
\frac{\d^{2}}{\d t^{2}}\cos t=-\frac{\d}{\d t}\sin t=-\cos t.
\]
If we replace $t$ by $\omega t+\phi$, where both $\omega$ and $\phi$ are constant (independent of $t$), then since
\[
\frac{\d}{\d t}\left(\omega t+\phi\right)=\omega,
\]
we get, by the chain rule, that each derivative generates a factor of $\omega$, so
\[
\frac{\d^{2}}{\d t^{2}}\cos\left(\omega t+\phi\right)=-\omega\frac{\d}{\d t}\left(\vphantom{\bll}\sin\left(\omega t+\phi\right)\right)=-\omega^{2}\cos\left(\omega t+\phi\right).
\]
Therefore, this differential equation has the solution:
\begin{equation}
x\left(t\right)=A\cos\left(\omega t+\phi\right),\label{eq:osc-sol}
\end{equation}
where the \emph{integration constants}\index{Integration constants} $A$ and $\phi$ are real numbers determined by the initial conditions. Now we see why this is called a harmonic oscillator: the position of the particle oscillates repeatedly between $+A$ and $-A$ over time.
\begin{problem}\leavevmode
Prove that the most general solution for the classical harmonic oscillator can also be written as
\[
x\left(t\right)=B\cos\left(\omega t\right)+C\sin\left(\omega t\right),
\]
where $B$ and $C$ are integration constants, or as
\[
x\left(t\right)=D\e^{\i\omega t}+E\e^{-\i\omega t},
\]
where $D$ and $E$ are integration constants. All of these solutions are equivalent; find the relationships between the integration constants $\left\{ A,\phi\right\} $, $\left\{ B,C\right\} $, and $\left\{ D,E\right\} $ – that is, write each pair in terms of another pair.
\end{problem}

\begin{problem}\leavevmode
As an example of solving the equation of motion for specific initial conditions, if the particle starts at time $t=0$ at position $x\left(0\right)=1$ with velocity $\xd\left(0\right)=0$, then we have from the velocity
\[
\xd\left(0\right)=-\omega A\sin\phi=0,
\]
therefore $\phi=\pi k$ for some integer $k$; we choose without loss of generality $\phi=0$. Then, from the position,
\[
x\left(0\right)=A=1,
\]
and thus the solution is
\[
x\left(t\right)=\cos\left(\omega t\right).
\]
Similarly, find a solution for the classical harmonic oscillator with the initial conditions $x\left(0\right)=0$ and $\xd\left(0\right)=\omega$.
\end{problem}

\begin{problem}\leavevmode
By plugging the general solution in \zcref[S]{eq:osc-sol} into the Hamiltonian in \zcref[S]{eq:har-ham}, show that the total energy of the system is
\[
H=\hf m\omega^{2}A^{2}.
\]
Thus the Hamiltonian is time-independent, and energy is conserved.
\end{problem}

\subsubsection{Quantizing the Harmonic Oscillator}

Let us now quantize the simple harmonic oscillator by promoting $x$ and $p$ to operators. We are interested in finding the energy eigenstates of this quantum system. Instead of finding them by solving a differential equation, we will use an easier and more intuitive method. We define the \emph{ladder operators}\index{Ladder operators}:
\[
a=\sqrt{\frac{m\omega}{2}}\left(x+\frac{\i}{m\omega}p\right)\sp a^{\dagger}=\sqrt{\frac{m\omega}{2}}\left(x-\frac{\i}{m\omega}p\right),
\]
where $a^{\dagger}$ is called the \emph{creation operator\index{Creation operator} }and $a$ is called the \emph{annihilation operator}\index{Annihilation operator}. Notice that $a^{\dagger}$ is indeed the adjoint of $a$, since the numbers $m,\omega$ are real and the operators $x,p$ are Hermitian. These definitions may be inverted to get the position and momentum operators in terms of the ladder operators:
\[
x=\sqrt{\frac{1}{2m\omega}}\left(a^{\dagger}+a\right)\sp p=\i\sqrt{\frac{m\omega}{2}}\left(a^{\dagger}-a\right).
\]
Now, notice that
\begin{align*}
\omega a^{\dagger}a & =\omega\sqrt{\frac{m\omega}{2}}\left(x-\frac{\i}{m\omega}p\right)\cdot\sqrt{\frac{m\omega}{2}}\left(x+\frac{\i}{m\omega}p\right)\\
 & =\hf m\omega^{2}\left(x-\frac{\i}{m\omega}p\right)\left(x+\frac{\i}{m\omega}p\right)\\
 & =\hf m\omega^{2}\left(x^{2}+\frac{\i}{m\omega}xp-\frac{\i}{m\omega}px-\left(\frac{\i}{m\omega}p\right)^{2}\right)\\
 & =\hf m\omega^{2}\left(\frac{p^{2}}{m^{2}\omega^{2}}+x^{2}+\frac{\i}{m\omega}\left[x,p\right]\right)\\
 & =\frac{p^{2}}{2m}+\hf m\omega^{2}x^{2}+\hf\i\omega\left[x,p\right].
\end{align*}
Recall that in the classical theory we have $\left\{ x,p\right\} =1$, so in the quantum theory we have $\left[x,p\right]=\i$. Therefore:
\[
\omega a^{\dagger}a=\frac{p^{2}}{2m}+\hf m\omega^{2}x^{2}-\hf\omega.
\]
Comparing this to the Hamiltonian operator in \zcref[S]{eq:har-ham}:
\[
H=\frac{p^{2}}{2m}+\hf m\omega^{2}x^{2},
\]
we see that we can write
\[
H=\omega\left(a^{\dagger}a+\hf\right).
\]
Finally, we define a new operator called the \emph{number operator}\index{Number operator}:
\[
N\equiv a^{\dagger}a.
\]
Now the Hamiltonian may be written as
\[
H=\omega\left(N+\hf\right).
\]
The Hamiltonian has been simplified considerably! Since both $\omega$ and $1/2$ are just numbers, the problem of finding the eigenvalues and eigenstates of $H$ now reduces to finding the eigenvalues and eigenstates of $N$.
\begin{problem}\leavevmode
\label{prob:number}
\begin{problemparts}
\item Show that $N$ is Hermitian.
\item Show that if $\left|n\right\rangle $ is an eigenstate of $N$ with the eigenvalue $n$, that is,
\[
N\left|n\right\rangle =n\left|n\right\rangle ,
\]
then $\left|n\right\rangle $ is also an eigenstate of $H$ with the eigenvalue $\omega\left(n+\hf\right)$.
\item Calculate, using the canonical commutation relation $\left[x,p\right]=\i$, the following commutators:
\[
[a,a^{\dagger}]=1\sp[N,a^{\dagger}]=a^{\dagger}\sp[N,a]=-a.
\]
\end{problemparts}
\end{problem}

\subsubsection{The Energy Eigenstates of the Harmonic Oscillator}

Let $\left|n\right\rangle $ be an eigenstate of $N$ with eigenvalue $n$:
\begin{equation}
N\left|n\right\rangle =n\left|n\right\rangle .\label{eq:Nn}
\end{equation}
Since $N$ is Hermitian, we know that $n$ must be a real number. Let us calculate the expectation value of the observable $N$ with respect to the eigenstate $\left|n\right\rangle $:
\[
\left\langle N\right\rangle _{n}=\langle n|N|n\rangle=\langle n|a^{\dagger}a|n\rangle=\left\Vert an\right\Vert ^{2},
\]
where we used the fact that $\left\langle n\right|a^{\dagger}$ is the bra of $a\left|n\right\rangle $. On the other hand, we have
\[
\left\langle N\right\rangle _{n}=\langle n|N|n\rangle=n\langle n|n\rangle=n,
\]
where we used \zcref[S]{eq:Nn} and the fact that the state $\left|n\right\rangle $ is normalized to 1. By comparing the two equations, we see that
\begin{equation}
n=\left\Vert an\right\Vert ^{2}\ge0,\label{eq:an-norm}
\end{equation}
that is, $n$ is not only real but also non-negative. Therefore, $N$ is positive semi-definite (as defined in \zcref[S]{subsec:Positive-Definite-and-Positive}).

Next, we act with $Na$ and $Na^{\dagger}$ on $\left|n\right\rangle $. In \zcref[S]{prob:number} you showed that
\[
Na-aN=[N,a]=-a,
\]
\[
Na^{\dagger}-a^{\dagger}N=[N,a^{\dagger}]=a^{\dagger},
\]
so we have
\[
Na=aN-a=a\left(N-1\right)\sp Na^{\dagger}=a^{\dagger}N+a^{\dagger}=a^{\dagger}\left(N+1\right),
\]
and thus
\[
Na\left|n\right\rangle =a\left(N-1\right)\left|n\right\rangle =\left(n-1\right)a\left|n\right\rangle ,
\]
\[
Na^{\dagger}\left|n\right\rangle =a^{\dagger}\left(N+1\right)\left|n\right\rangle =\left(n+1\right)a^{\dagger}\left|n\right\rangle ,
\]
where we used \zcref[S]{eq:Nn} and the fact that since $n\pm1$ is a number, it commutes with operators and can be moved to the left. Writing this result in a different way, we see that
\[
N\left(\vphantom{\bll}a\left|n\right\rangle \right)=\left(n-1\right)\left(\vphantom{\bll}a\left|n\right\rangle \right),
\]
\[
N\left(\vphantom{\bll}a^{\dagger}\left|n\right\rangle \right)=\left(n+1\right)\left(\vphantom{\bll}a^{\dagger}\left|n\right\rangle \right),
\]
or in other words, $a\left|n\right\rangle $ for $n>0$ is an eigenstate of $N$ with eigenvalue $n-1$, and $a^{\dagger}\left|n\right\rangle $ is an eigenstate of $N$ with eigenvalue $n+1$. ($a\left|n\right\rangle $ for $n=0$ is just the zero vector, which can't be an eigenstate.) However, since the eigenvalues of $N$ are non-degenerate, the normalized eigenstates of $N$ with eigenvalues $n-1$ and $n+1$ are $\left|n-1\right\rangle $ and $\left|n+1\right\rangle $ respectively. Thus, we conclude that $a\left|n\right\rangle $ is proportional to $\left|n-1\right\rangle $ and $a^{\dagger}\left|n\right\rangle $ is proportional to $\left|n+1\right\rangle $. The proportionality factors must be chosen so that the states are normalized, and we fix the phases of the eigenstates so that these factors are positive. Let us therefore calculate the norms. The norm-squared $\left\Vert an\right\Vert ^{2}$ was already calculated in \zcref[S]{eq:an-norm}:
\[
\left\Vert an\right\Vert ^{2}=\langle n|a^{\dagger}a|n\rangle=\langle n|N|n\rangle=n.
\]
To calculate $\Vert a^{\dagger}n\Vert^{2}$, we recall from \zcref[S]{prob:number} that
\[
aa^{\dagger}-a^{\dagger}a=[a,a^{\dagger}]=1,
\]
and thus
\[
aa^{\dagger}=a^{\dagger}a+1=N+1.
\]
We therefore get
\[
\Vert a^{\dagger}n\Vert^{2}=\langle n|aa^{\dagger}|n\rangle=\langle n|\left(N+1\right)|n\rangle=\langle n|N|n\rangle+\langle n|n\rangle=n+1.
\]
To summarize, the norms are
\[
\left\Vert an\right\Vert =\sqrt{n}\sp\Vert a^{\dagger}n\Vert=\sqrt{n+1}.
\]
The normalized eigenstates are now obtained, as usual, by dividing by the norm:
\[
\left|n-1\right\rangle =\frac{1}{\sqrt{n}}a\left|n\right\rangle \sp\left|n+1\right\rangle =\frac{1}{\sqrt{n+1}}a^{\dagger}\left|n\right\rangle .
\]
(Note that $\left|n-1\right\rangle $ is only defined if $n>0$.)

Another way to write this, from a different point of view, is as the action of the operators $a$ and $a^{\dagger}$ on the state $\left|n\right\rangle $:
\[
a\left|n\right\rangle =\sqrt{n}\left|n-1\right\rangle \sp a^{\dagger}\left|n\right\rangle =\sqrt{n+1}\left|n+1\right\rangle .
\]
We see that $a$ \textbf{reduces} the number eigenvalue by $1$, while $a^{\dagger}$ \textbf{increases} the number eigenvalue by $1$. In other words, $a^{\dagger}$ gets us to the state of next higher energy (it “creates one quantum of energy”) while $a$ gets us to the state of next lower energy (it “annihilates one quantum of energy”). This is the reason we called $a^{\dagger}$ the \emph{creation operator\index{Creation operator}} and $a$ the \emph{annihilation operator}\index{Annihilation operator}. We call them the \emph{ladder operators\index{Ladder operators}} because they let us “climb the ladder” of energy eigenstates.

The eigenvalues of $N$ must be non-negative integers. To see this, consider that a non-integer eigenvalue $n$ could be decreased by integer steps using $a$ until it becomes negative, but this contradicts the fact that $N$ is positive semi-definite. However, if we try to do the same with an integer eigenvalue $n$, we'll find that once we get down to $n=0$, further applications of $a$ will result in the zero vector, which is not a valid state (as it has norm $0$ and thus cannot be normalized). Therefore, $n$ must be a non-negative integer, and the valid eigenstates are
\[
\left|0\right\rangle ,\left|1\right\rangle ,\left|2\right\rangle ,\left|3\right\rangle ,\ldots.
\]

Going back to the expression for the Hamiltonian in terms of the number operator, we see that
\[
H\left|n\right\rangle =\omega\left(n+\hf\right)\left|n\right\rangle ,
\]
and thus, as you proved in \zcref[S]{prob:number}, $\left|n\right\rangle $ is an energy eigenstate with eigenvalue
\[
E_{n}\equiv\omega\left(n+\hf\right).
\]

We found that the energy of the quantum harmonic oscillator is discrete, or \emph{quantized}\index{Quantized energy}\index{Energy!Quantized}, and the system can only have energy which differs from $\omega/2$ by equal steps of $\omega$. The state of lowest energy, also called the \emph{ground state}\index{Ground state}, is $\left|0\right\rangle $. It has the energy eigenvalue
\[
E_{0}=\hf\omega.
\]
If we act on the ground state with the annihilation operator, we get
\[
a\left|0\right\rangle =0,
\]
which, as noted above, is \textbf{not }a state. This means that we cannot generate states with energy lower than that of the ground state. If we act on $\left|0\right\rangle $ with the creation operator, we get
\[
a^{\dagger}\left|0\right\rangle =\left|1\right\rangle .
\]
We say that $a^{\dagger}$, which takes us from $\left|0\right\rangle $ to $\left|1\right\rangle $, \emph{excites}\index{Excitation} the harmonic oscillator from the ground state to the \emph{first excited state}\index{First excited state}\index{Excited state}, which has exactly one \emph{“quantum”}\index{Quantum (of energy)} of energy. The state $\left|n\right\rangle $ has exactly $n$ quanta, while the ground state $\left|0\right\rangle $ has no quanta.

As we mentioned in \zcref[S]{subsec:The-Classical-Harmonic-Oscillator}, the quantum harmonic oscillator may be used to describe many different physical systems. In quantum field theory\index{Quantum field theory}, we can think of the operator $N$ as corresponding to the number of particles with certain properties excited from the field. So $\left|0\right\rangle $ is the \emph{vacuum state}\index{Vacuum state}, or a state with no particles\footnote{Notice that the vacuum state, despite having no particles, still has non-zero energy $\omega/2$! This is called \emph{zero-point energy}\index{Zero-point energy}, and it is simply part of the energy of the field itself.}; $\left|1\right\rangle $ is a state with one particle (e.g. one photon of a certain type has been excited from the electromagnetic field); $\left|2\right\rangle $ is a state with two particles\footnote{Note that quantum fields with half-integer spins, like electron or quark fields, cannot have more than one particle with given properties, due to the \emph{Pauli exclusion principle}\index{Pauli exclusion principle}, so they can only have the states $\left|0\right\rangle $ or $\left|1\right\rangle $. Quantum fields with integer spins, like the electromagnetic or Higgs fields, can have higher excitations such as $\left|2\right\rangle $. We won't go into more details here, since this is beyond the scope of our notes.}; and so on.
\begin{problem}\leavevmode
Prove that
\[
\left|n\right\rangle =\frac{\left(a^{\dagger}\right)^{n}}{\sqrt{n!}}\left|0\right\rangle .
\]
This means that, once we know the ground state, we can create any energy eigenstate by simply applying $n$ times the operator $a^{\dagger}$ and normalizing.
\end{problem}

\begin{problem}\leavevmode
\begin{problemparts}
\item Find $\left\langle V\right\rangle $ for the harmonic oscillator given that the system is in the energy eigenstate $\left|n\right\rangle $.
\item How is the expectation value of the potential energy related to the total energy?
\item What is the expectation value of the kinetic energy?
\end{problemparts}
\end{problem}

\subsection{Wavefunctions, Position, and Momentum}\label{subsec:Wavefunctions}

\subsubsection{The Position Operator}\label{subsec:The-Position-Operator}

When canonically quantizing a particle, the position function $x$ is promoted to a Hermitian \emph{position operator}\index{Position operator}. We usually denote this operator with a \emph{hat}\index{Hat notation for operators}, $\hat{x}$, to distinguish it from its eigenvalues, which are confusingly also written as $x$. Even more confusingly, we denote the \emph{position eigenstate}\index{Position eigenstate} corresponding to a measurement of position $x$ as $\left|x\right\rangle $:
\[
\hat{x}\left|x\right\rangle =x\left|x\right\rangle \sp x\in\BBR.
\]
As usual, since $\hat{x}$ is a Hermitian operator, its eigenstates $\left|x\right\rangle $ form an orthonormal basis\footnote{Our Hilbert space is now infinite-dimensional, and a rigorous discussion of such a space requires dealing with many mathematical subtleties, but we will mostly ignore them in these notes, since infinite-dimensional Hilbert spaces are not our main focus.}. Recall that for an orthonormal basis $\left|B_{i}\right\rangle $, $i\in\left\{ 1,\ldots,n\right\} $ in a finite-dimensional Hilbert space, the orthonormality condition is given by \zcref[S]{eq:Kron}:
\[
\langle B_{i}|B_{j}\rangle=\delta_{ij}=\begin{cases}
0 & \textrm{if }i\ne j,\\
1 & \textrm{if }i=j.
\end{cases}
\]
The \emph{Kronecker delta}\index{Kronecker delta} $\delta_{ij}$ has the property that, when evaluated inside a sum over an index $i$, it “chooses” the term in the sum with index $j$:
\begin{equation}
\sum^{n}_{i=1}f_{i}\delta_{ij}=f_{j},\label{eq:Kron-f}
\end{equation}
where $f_{i}$ represents the terms to be summed over. You don't actually need to evaluate the sum, since all of the terms with $i\ne j$ vanish, and you are left with just one term, the one with $i=j$.

The continuous version of this is that for two basis states $\left|x\right\rangle $ and $\left|x'\right\rangle $, where $x,x'\in\BBR$, we have
\begin{equation}
\langle x|x'\rangle=\delta\left(x-x'\right),\label{eq:x-xprime}
\end{equation}
where $\delta\left(x-x'\right)$ is the \emph{Dirac delta function}\index{Dirac delta function}. Intuitively, you can think of $\delta\left(x-x'\right)$ as a function that is zero everywhere except when $x=x'$, in which case it is divergent. More precisely, the Dirac delta isn't actually a function; it is a \emph{distribution}\index{Distribution (generalized function)}, which basically means it is only well-defined when used inside an integral. For any function $f$ that is continuous at $x'$, the Dirac delta satisfies the condition
\[
\int^{+\infty}_{-\infty}f\left(x\right)\delta\left(x-x'\right)\d x=f\left(x'\right).
\]
In other words, when evaluated inside an integral over a variable $x$, the delta function $\delta\left(x-x'\right)$ simply “chooses” the value of the integrand for which $x=x'$. This is simply a generalization of the property of the Kronecker delta in \zcref[S]{eq:Kron-f}. You don't need to evaluate the integral, since all of the terms with $x\ne x'$ vanish, and you are left with just one term, the one with $x=x'$.

Note that the position eigenstates $\left|x\right\rangle $ cannot be normalized, since the norm-squared
\[
\left\Vert x\right\Vert ^{2}\equiv\langle x|x\rangle=\delta\left(0\right)
\]
is not a finite number; as mentioned above, intuitively “$\delta\left(0\right)=\infty$”. However, every vector in a Hilbert space must have a finite norm, since the inner product of any two vectors is a complex number (recall \zcref[S]{subsec:Dual-Vectors}), and $\infty$ is not a number. Therefore, the position eigenstates are not actually elements of the Hilbert space and do not represent physical states; they form an orthonormal basis only in the generalized sense of \zcref[S]{eq:x-xprime}. As we will see in \zcref[S]{subsec:Wavefunctions-in-the-Position-Basis}, the physical states of a particle are normalizable superpositions (integrals) of position eigenstates.
\begin{problem}\leavevmode
Prove the following properties of the Dirac delta function:
\begin{problemparts}
\item 
\[
\int^{+\infty}_{-\infty}f\left(x\right)\delta\left(x\right)\d x=f\left(0\right).
\]
\item 
\[
\int^{+\infty}_{-\infty}\delta\left(x\right)\d x=1.
\]
\item 
\[
\delta\left(x\right)=\delta\left(-x\right).
\]
\item 
\[
\delta\left(\lambda x\right)=\frac{1}{\left|\lambda\right|}\delta\left(x\right)\sp\lambda\in\BBR\setminus\left\{ 0\right\} .
\]
\end{problemparts}
\end{problem}

\begin{problem}\leavevmode
Let us define the \emph{Heaviside step function}\index{Heaviside step function}:
\[
\Theta\left(x\right)\equiv\begin{cases}
0 & x<0,\\
\hf & x=0,\\
1 & x>0.
\end{cases}
\]
Prove that
\[
\frac{\d}{\d x}\Theta\left(x\right)=\delta\left(x\right),
\]
where $\delta\left(x\right)$ is the Dirac delta function. Note that since $\Theta$ is not differentiable at $x=0$, its derivative is understood as a distribution, and it is therefore – like the Dirac delta – defined only inside an integral, using integration by parts:
\[
\int^{+\infty}_{-\infty}f\left(x\right)\frac{\d}{\d x}\Theta\left(x\right)\d x\equiv-\int^{+\infty}_{-\infty}\frac{\d f}{\d x}\Theta\left(x\right)\d x,
\]
for any smooth function $f$ that vanishes at $\pm\infty$.
\end{problem}

\subsubsection{Wavefunctions in the Position Basis}\label{subsec:Wavefunctions-in-the-Position-Basis}

Since $\left|x\right\rangle $ is an orthonormal eigenbasis, we should be able to write down any state $\left|\Psi\right\rangle $ as a linear combination – or superposition – of the basis eigenstates. Let us recall that in the finite-dimensional case, with a finite orthonormal basis $\left|B_{i}\right\rangle $, we have
\[
\left|\Psi\right\rangle =\sum^{n}_{i=1}|B_{i}\rangle\langle B_{i}|\Psi\rangle.
\]
In \zcref[S]{subsec:Representing-Vectors-in} we said that $\langle B_{i}|\Psi\rangle$ – the probability amplitudes – are the coordinates of the representation of the vector $\left|\Psi\right\rangle $ with respect to the basis $\left|B_{i}\right\rangle $, and they can be collected into an $n$-dimensional vector:
\[
\left|\Psi\right\rangle \bllll_{B}\equiv\left(\begin{array}{c}
\langle B_{1}|\Psi\rangle\\
\vdots\\
\langle B_{n}|\Psi\rangle
\end{array}\right).
\]
In the continuous case, we simply replace the sum with an integral (and optionally add time dependence, since we now have a continuous time coordinate):
\begin{equation}
\left|\Psi\left(t\right)\right\rangle =\int^{+\infty}_{-\infty}\left|x\right\rangle \ma x{\Psi\left(t\right)}\d x.\label{eq:wf}
\end{equation}
In this case, $\langle x|\Psi\left(t\right)\rangle$ are the \emph{coordinates}\index{Coordinates of a vector in a basis!Continuous case} of the \emph{representation}\index{Representing a vector in a basis!Continuous case} of the vector $\left|\Psi\left(t\right)\right\rangle $ with respect to the basis $\left|x\right\rangle $. Since there is one coordinate for each real number $x$, we cannot collect them into a vector; instead, we define a function:
\begin{equation}
\psi\left(t,x\right)\equiv\ma x{\Psi\left(t\right)}.\label{eq:wavefunction}
\end{equation}
The complex-valued function $\psi\left(t,x\right)$, which returns the probability amplitude to measure the particle at position $x$ at time $t$, is called the \emph{wavefunction}\index{Wavefunction}.

Given a wavefunction $\psi\left(t,x\right)$, the \emph{probability density}\index{Probability density} to find the particle at position $x$ at time $t$ is given by the magnitude-squared of the probability amplitude:
\[
\left|\psi\left(t,x\right)\right|^{2}=\left|\ma x{\Psi\left(t\right)}\right|^{2}.
\]
The reason this is a probability \textbf{density}, and not a probability, is that continuous probability distributions behave a bit differently than discrete ones. The probability to find the particle somewhere in the \emph{real interval}\index{Real interval} $\left[a,b\right]\subset\BBR$ at time $t$ is given by the integral
\[
\int^{b}_{a}\left|\psi\left(t,x\right)\right|^{2}\d x.
\]
If $a=b$, then the integral evaluates to zero. This means that the probability to find a particle at any one specific point $x$ is actually \textbf{zero}! A set containing just one point, or even a countable number of discrete points, is a set of \emph{Lebesgue measure zero}\index{Set of measure zero}\index{Lebesgue measure}, which means it has no length. It only makes sense to talk about finding a particle inside an interval such as $\left[a,b\right]$ with $a\ne b$, which has non-zero Lebesgue measure and thus non-zero length.

Also, instead of the probabilities summing to 1, we must demand that the integral of the probability densities over the entire real line evaluates to 1:
\begin{equation}
\int^{+\infty}_{-\infty}\left|\psi\left(t,x\right)\right|^{2}\d x=1.\label{eq:inf-norm}
\end{equation}
This makes sense, because there is 100\% probability to find the particle \textbf{somewhere }on the real line, that is, inside the interval $\left(-\infty,+\infty\right)$.

Using the wavefunction $\psi\left(t,x\right)=\ma x{\Psi\left(t\right)}$, we can rewrite \zcref[S]{eq:wf} as follows:
\begin{equation}
\left|\Psi\left(t\right)\right\rangle =\int^{+\infty}_{-\infty}\psi\left(t,x\right)\left|x\right\rangle \d x.\label{eq:state-wf}
\end{equation}
If we are given a state $\left|\Psi\left(t\right)\right\rangle $, we can use \zcref[S]{eq:wavefunction} to convert it to a wavefunction, and conversely, if we are given a wavefunction $\psi\left(t,x\right)$, we can use \zcref[S]{eq:state-wf} to convert it to a state. This is, of course, a consequence of the wavefunction being a representation of the state in a specific basis. For this reason, you will sometimes hear the term “wavefunction” used as a synonym for “state”; for systems where a wavefunction description exists, such as a quantized particle, these two descriptions are equivalent.

However, it should be noted that wavefunctions are \textbf{not} fundamental entities in modern quantum theory. The fundamental entities are the states, since any quantum system has states, but only some systems have wavefunctions. For example, there is no wavefunction for a qubit, since there are no continuous variables with respect to which the wavefunction can be defined\footnote{This is why, in the discussion of the Measurement Axiom, I used the term “collapse” rather than the more popular “wavefunction collapse”. Qubits also collapse, but they do not have wavefunctions!}. Even for systems that do have wavefunctions, the description using states is more general, since a state is independent of a basis, while a wavefunction is only defined in a particular basis.

Next, recall the completeness relation in \zcref[S]{eq:completeness}:
\[
\sum^{n}_{i=1}|B_{i}\rangle\langle B_{i}|=1.
\]
We can use \zcref[S]{eq:wf} to derive a continuous analogue. We simply note that $\left|\Psi\left(t\right)\right\rangle $ does not explicitly depend on the variable $x$, so it can actually be taken out of the integral, and we get:
\[
\left|\Psi\left(t\right)\right\rangle =\left(\int^{+\infty}_{-\infty}\left|x\right\rangle \left\langle x\right|\d x\right)\left|\Psi\left(t\right)\right\rangle ,
\]
from which we get the \emph{continuous completeness relation}\index{Completeness relation!Continuous case}
\begin{equation}
\int^{+\infty}_{-\infty}\left|x\right\rangle \left\langle x\right|\d x=1.\label{eq:comp-inf}
\end{equation}
This relation allows us to define an explicit \emph{inner product}\index{Inner product!Continuous case} between states on our infinite-dimensional Hilbert space as follows:
\begin{align*}
\langle\Psi\left(t\right)|\Phi\left(t'\right)\rangle & =\langle\Psi\left(t\right)|\left(\int^{+\infty}_{-\infty}\left|x\right\rangle \left\langle x\right|\d x\right)|\Phi\left(t'\right)\rangle\\
 & =\int^{+\infty}_{-\infty}\langle\Psi\left(t\right)|x\rangle\langle x|\Phi\left(t'\right)\rangle\d x\\
 & =\int^{+\infty}_{-\infty}\psi^{*}\left(t,x\right)\phi\left(t',x\right)\d x,
\end{align*}
where $\psi^{*}\left(t,x\right)=\langle\Psi\left(t\right)|x\rangle$ is the complex conjugate of the wavefunction for $\left|\Psi\left(t\right)\right\rangle $ defined in \zcref[S]{eq:wavefunction} (since as usual, switching the order of states in the inner product turns it into its complex conjugate), and $\phi\left(t',x\right)\equiv\langle x|\Phi\left(t'\right)\rangle$ is the wavefunction for the state $|\Phi\left(t'\right)\rangle$.

This is really nothing more than the familiar inner product we defined all the way back in \zcref[S]{subsec:Dual-Vectors}, except instead of summing on the components of a vector, we are integrating on the values of a function! The vector in the discrete and finite-dimensional case was the representation of the state in a particular basis (such as the standard basis), while the function in the continuous and infinite-dimensional case is also the representation of the state in a particular basis, in this case the position basis.

Now we can see that the normalization condition in \zcref[S]{eq:inf-norm} simply says that the \emph{norm}\index{Norm!Continuous case} of a state has to be 1, as usual:
\[
\left\Vert \Psi\left(t\right)\right\Vert =\sqrt{\langle\Psi\left(t\right)|\Psi\left(t\right)\rangle}=\sqrt{\int^{+\infty}_{-\infty}\left|\psi\left(t,x\right)\right|^{2}\d x}=1.
\]

\begin{problem}\leavevmode
\label{prob:x-exp}The expectation value of the position, given that the state of the system is $\left|\Psi\left(t\right)\right\rangle $, is defined as usual by
\[
\left\langle x\right\rangle =\langle\Psi\left(t\right)|\hx|\Psi\left(t\right)\rangle.
\]
By inserting the completeness relation in \zcref[S]{eq:comp-inf}, show that, in terms of the wavefunction $\psi\left(t,x\right)$, the expectation value of $x$ is
\[
\left\langle x\right\rangle =\int^{+\infty}_{-\infty}x\left|\psi\left(t,x\right)\right|^{2}\d x.
\]
\end{problem}

\begin{problem}\leavevmode
\label{prob:Vx}Let $V\left(x\right)$ be an arbitrary smooth function of $x$ whose Taylor series converges to $V\left(x\right)$ for any $x\in\BBR$. When we promote $x$ to an operator, $V\left(x\right)$ becomes the operator $V\left(\hx\right)$. (For example, if $V\left(x\right)=x^{2}$, then $V\left(\hx\right)$ is the operator $\hx^{2}$.) By expanding $V\left(\hx\right)$ in a Taylor series, show that $\left|x\right\rangle $ is an eigenstate of $V\left(\hx\right)$ with eigenvalue $V\left(x\right)$:
\[
V\left(\hx\right)\left|x\right\rangle =V\left(x\right)\left|x\right\rangle .
\]
As a corollary, show that
\[
\left\langle x\right|V\left(\hx\right)\left|\Psi\left(t\right)\right\rangle =V\left(x\right)\psi\left(t,x\right).
\]
\end{problem}

\begin{xca}\leavevmode
A wavefunction is given by
\[
\psi\left(t,x\right)=A\e^{-x^{2}}\sp A\in\BBC.
\]
Find a value of $A$ for which the wavefunction is properly normalized, that is, \zcref[S]{eq:inf-norm} is satisfied. Then, calculate the expectation value $\left\langle x\right\rangle $ for this wavefunction.
\end{xca}

\subsubsection{The Momentum Operator}

When we canonically quantize a particle, in addition to the position operator, we also promote the momentum function to a Hermitian \emph{momentum operator}\index{Momentum operator} $\hat{p}$. This operator has \emph{momentum eigenstates}\index{Momentum eigenstate} $\left|p\right\rangle $, which correspond to measurements of momentum $p$:
\[
\hat{p}\left|p\right\rangle =p\left|p\right\rangle .
\]
Everything that we discussed in \zcref[S]{subsec:The-Position-Operator,subsec:Wavefunctions-in-the-Position-Basis} also applies to the momentum operator and its eigenstates – simply replace $x$ with $p$. This also includes the wavefunction: representing the state in the momentum basis gives the \emph{momentum-space wavefunction}\index{Momentum-space wavefunction}\index{Wavefunction!Momentum-space}
\[
\psi\left(t,p\right)\equiv\ma p{\Psi\left(t\right)}.
\]
Now, let us recall that in \zcref[S]{subsec:Continuous-Time-Evolution} we found out that, for a time-independent Hamiltonian, the unitary operator responsible for shifts in time can be written as the exponential of the Hamiltonian. This can be written in slightly different notation as
\[
\e^{-\i Ht_{0}}\left|\Psi\left(t\right)\right\rangle =\left|\Psi\left(t+t_{0}\right)\right\rangle .
\]
This relation is equivalent to \zcref[S]{eq:Schr}, the Schrödinger equation, which tells us that the Hamiltonian – the Hermitian operator corresponding to energy – acts on states as a time derivative:
\[
H\left|\Psi\left(t\right)\right\rangle =\i\frac{\d}{\d t}\left|\Psi\left(t\right)\right\rangle .
\]
Since the energy is just the momentum in the time direction, we expect, in analogy, that the momentum operator will act on states as a derivative with respect to \textbf{position}, and that its exponential will translate states in \textbf{space}. However, here we encounter a complication: in (non-relativistic) quantum mechanics, time is considered to be just a \textbf{label }on the states $\left|\Psi\left(t\right)\right\rangle $, while position is the \textbf{eigenvalue }of the position operator. Due to this complication, we won't give the derivation here, but simply state the result:
\begin{equation}
\langle x|\hp|\Psi\left(t\right)\rangle=-\i\frac{\partial}{\partial x}\ma x{\Psi\left(t\right)}=-\i\frac{\partial}{\partial x}\psi\left(t,x\right).\label{eq:mom-rep}
\end{equation}
This means that the \textbf{representation }of the momentum operator in the position basis is given by the derivative with respect to position (times a scalar $-\i$, which is fixed by the canonical commutation relation in \zcref[S]{eq:canonical-xp}). This result will be very useful in \zcref[S]{subsec:Solutions-of-the}, when we discuss solutions to the Schrödinger equation. \zcref[S]{eq:mom-rep} is often written simply as
\[
\hp=-\i\frac{\partial}{\partial x},
\]
but this is actually incorrect (or at the very least, serious abuse of notation), since the momentum operator is an abstract operator, and only becomes a derivative when represented in the position basis!

By exponentiating the momentum operator, we get the \emph{translation operator}\index{Translation operator} $\e^{-\i\hp a}$, a unitary operator (as it has to be, since it must preserve norms) which translates position eigenstates a distance $a$ in space:
\[
\e^{-\i\hp a}\left|x\right\rangle =\left|x+a\right\rangle .
\]
By taking the adjoint of this expression and acting on a state $\left|\Psi\left(t\right)\right\rangle $, we get
\[
\left\langle x\right|\e^{\i\hp a}\left|\Psi\left(t\right)\right\rangle =\langle x+a|\Psi\left(t\right)\rangle=\psi\left(t,x+a\right).
\]
Therefore, the translation operator translates not only position eigenstates but also wavefunctions.
\begin{problem}\leavevmode
\label{prob:p-exp}Calculate the expectation value of the momentum, $\left\langle p\right\rangle $, given that the state of the system is $\left|\Psi\left(t\right)\right\rangle $, in terms of the wavefunction $\psi\left(t,x\right)$.
\end{problem}

\subsubsection{Quantum Interference}

Let us consider the familiar \emph{double-slit experiment}\index{Double-slit experiment}. Schematically, the particle's state can be described as a superposition of passing through slit $A$ and passing through slit $B$:
\[
\left|\Psi\right\rangle =a\left|\Psi_{A}\right\rangle +b\left|\Psi_{B}\right\rangle \sp\left|a\right|^{2}+\left|b\right|^{2}=1.
\]
We suppress the time dependence here, for brevity, and we assume $\left|\Psi_{A}\right\rangle $ and $\left|\Psi_{B}\right\rangle $ are orthonormal. The probability amplitude to measure the particle at the position $x$ is given by
\[
\psi\left(x\right)=\langle x|\Psi\rangle=a\langle x|\Psi_{A}\rangle+b\langle x|\Psi_{B}\rangle\equiv a\psi_{A}\left(x\right)+b\psi_{B}\left(x\right).
\]
The probability density is then, as usual, the magnitude-squared of the amplitude:
\begin{align*}
\left|\psi\left(x\right)\right|^{2} & =\left|a\psi_{A}\left(x\right)+b\psi_{B}\left(x\right)\right|^{2}\\
 & =\left(a^{*}\psi^{*}_{A}\left(x\right)+b^{*}\psi^{*}_{B}\left(x\right)\right)\left(a\psi_{A}\left(x\right)+b\psi_{B}\left(x\right)\right)\\
 & =a^{*}a\psi^{*}_{A}\left(x\right)\psi_{A}\left(x\right)+b^{*}b\psi^{*}_{B}\left(x\right)\psi_{B}\left(x\right)+a^{*}b\psi^{*}_{A}\left(x\right)\psi_{B}\left(x\right)+b^{*}a\psi^{*}_{B}\left(x\right)\psi_{A}\left(x\right)\\
 & =\left|a\right|^{2}\left|\psi_{A}\left(x\right)\right|^{2}+\left|b\right|^{2}\left|\psi_{B}\left(x\right)\right|^{2}+2\re\left(a^{*}b\psi^{*}_{A}\left(x\right)\psi_{B}\left(x\right)\right).
\end{align*}
The terms $\left|a\right|^{2}\left|\psi_{A}\left(x\right)\right|^{2}$ and $\left|b\right|^{2}\left|\psi_{B}\left(x\right)\right|^{2}$ are always non-negative, for any $x$. However, the third term $2\re\left(a^{*}b\psi^{*}_{A}\left(x\right)\psi_{B}\left(x\right)\right)$, called the \emph{interference term}\index{Interference} or sometimes the \emph{cross term}\index{Cross term} (because it “crosses” $\psi_{A}$ and $\psi_{B}$), is a real number which can be either positive or negative, depending on the specific values of $a$ and $b$, as well as the specific position $x$ at which $\psi^{*}_{A}\left(x\right)$ and $\psi_{B}\left(x\right)$ are calculated.

The interference term will either increase or decrease the probability density to find the particle at $x$. If it increases the probability density, this is \emph{constructive interference}\index{Interference!Constructive}, and if it decreases the probability density, this is \emph{destructive interference}\index{Interference!Destructive}. This is precisely what is responsible for the interference pattern in the double-slit experiment; for different values of $x$, there will be different amounts of constructive and destructive interference. This is very similar to the qubit interference we discussed in \zcref[S]{subsec:Coherence}, but here the effect is more complicated, as we also have dependence on a continuous variable, which generates a continuous interference pattern.

\subsection{Solutions of the Schrödinger Equation}\label{subsec:Solutions-of-the}

\subsubsection{The Schrödinger Equation for a Particle}

Recall \zcref[S]{eq:Schr}, the Schrödinger equation:
\[
\i\frac{\d}{\d t}\left|\Psi\left(t\right)\right\rangle =H\left|\Psi\left(t\right)\right\rangle .
\]
For a particle, we have the Hamiltonian in \zcref[S]{eq:H_particle}:
\[
H=\frac{p^{2}}{2m}+V\left(x\right).
\]
Therefore, the Schrödinger equation becomes
\[
\i\frac{\d}{\d t}\left|\Psi\left(t\right)\right\rangle =\left(\frac{\hp^{2}}{2m}+V\left(\hx\right)\right)\left|\Psi\left(t\right)\right\rangle ,
\]
where we promoted the position and momentum to operators. To find the representation of this equation in the position basis, we multiply by $\left\langle x\right|$ from the left:
\[
\left\langle x\right|\i\frac{\d}{\d t}\left|\Psi\left(t\right)\right\rangle =\left\langle x\right|\left(\frac{\hp^{2}}{2m}+V\left(\hx\right)\right)\left|\Psi\left(t\right)\right\rangle .
\]
On the left-hand side, since the position eigenstate $\left|x\right\rangle $ is independent of time, we can move the time derivative out of the inner product:
\[
\left\langle x\right|\i\frac{\d}{\d t}\left|\Psi\left(t\right)\right\rangle =\i\frac{\partial}{\partial t}\ma x{\Psi\left(t\right)}=\i\frac{\partial}{\partial t}\psi\left(t,x\right).
\]
On the right-hand side, since in the position representation we have
\[
\hp=-\i\frac{\partial}{\partial x},
\]
the first term will be
\begin{align*}
\left\langle x\right|\frac{\hp^{2}}{2m}\left|\Psi\left(t\right)\right\rangle  & =\frac{1}{2m}\left(-\i\frac{\partial}{\partial x}\right)^{2}\psi\left(t,x\right)\\
 & =\frac{1}{2m}\left(-\i\frac{\partial}{\partial x}\right)\left(-\i\frac{\partial}{\partial x}\right)\psi\left(t,x\right)\\
 & =-\frac{1}{2m}\frac{\partial^{2}}{\partial x^{2}}\psi\left(t,x\right).
\end{align*}
As for the second term, in \zcref[S]{prob:Vx} you showed that a potential whose Taylor series converges to it everywhere satisfies
\begin{equation}
V\left(\hx\right)\left|x\right\rangle =V\left(x\right)\left|x\right\rangle ,\label{eq:Vx-Vx}
\end{equation}
and used that to prove that
\begin{equation}
\left\langle x\right|V\left(\hx\right)\left|\Psi\left(t\right)\right\rangle =V\left(x\right)\psi\left(t,x\right).\label{eq:Vpsi-Vpsi}
\end{equation}
For other types of potentials, such as discontinuous ones, we simply take \zcref[S]{eq:Vx-Vx} to be the \textbf{definition} of $V\left(\hx\right)$: it's the operator whose action on a position eigenstate $\left|x\right\rangle $ gives $V\left(x\right)\left|x\right\rangle $ for all $x\in\BBR$. \zcref[S]{eq:Vpsi-Vpsi} then follows similarly.

In total, we get:
\begin{equation}
\i\frac{\partial}{\partial t}\psi\left(t,x\right)=\left(-\frac{1}{2m}\frac{\partial^{2}}{\partial x^{2}}+V\left(x\right)\right)\psi\left(t,x\right).\label{eq:Schr-pos}
\end{equation}
This is the Schrödinger equation\index{Schrödinger equation} in the position basis. It is a concrete differential equation that one can solve for a variety of different potentials $V\left(x\right)$.
\begin{problem}\leavevmode
In this problem you will prove \emph{Ehrenfest's theorem}\index{Ehrenfest's theorem}, which states that:
\begin{equation}
\left\langle p\right\rangle =m\frac{\d\left\langle x\right\rangle }{\d t},\label{eq:pmv}
\end{equation}
\begin{equation}
\frac{\d\left\langle p\right\rangle }{\d t}=-\left\langle V'\left(x\right)\right\rangle .\label{eq:pvp}
\end{equation}
\zcref[S]{eq:pmv} shows that the expectation values of the position and momentum in the quantum theory satisfy the same relation as the position and momentum in the classical theory. \zcref[S]{eq:pvp} is Newton's second law in \zcref[S]{eq:point-p} in terms of the expectation values of the momentum and the force $F\equiv-V'\left(x\right)$.
\begin{problemparts}
\item Using the identities you proved in \zcref[S]{subsec:The-Matrix-Commutator}, calculate the commutator
\[
\left[x^{n},p\right],
\]
where $n$ is a positive integer.

\textbf{Hint:} Make an educated guess and prove it by induction.
\item Using the result of (A), calculate the commutator
\[
\left[f\left(x\right),p\right],
\]
where $f\left(x\right)$ is a function whose Taylor series converges to it everywhere.
\item Recall \zcref[S]{eq:dA-AH}:
\[
\frac{\d\left\langle A\right\rangle }{\d t}=-\i\left\langle \left[A,H\right]\right\rangle .
\]
Use this equation, and the result of (B), to prove \zcref[S]{eq:pmv,eq:pvp}.
\item In \zcref[S]{prob:x-exp}, you showed that
\[
\left\langle x\right\rangle =\int^{+\infty}_{-\infty}x\left|\psi\left(t,x\right)\right|^{2}\d x,
\]
and in \zcref[S]{prob:p-exp}, you calculated $\left\langle p\right\rangle $. Prove \zcref[S]{eq:pmv,eq:pvp} using these results and the Schrödinger equation in the position basis, \zcref[S]{eq:Schr-pos}.

\textbf{Hint:} You will have to use integration by parts, and assume\footnote{This is pretty much always assumed to be true about wavefunctions in quantum mechanics. It can be justified in two ways. First, according to \zcref[S]{eq:inf-norm}, $\left|\psi\left(t,x\right)\right|^{2}$ has to integrate to 1 so that the state is normalized. Therefore, it makes sense that $\psi\left(t,x\right)$ should vanish at infinity – although, if you look hard enough (you are encouraged to try!), you can find normalized wavefunctions which nonetheless do not vanish at infinity. Second, if we create a particle in the lab, we would expect the probability to find this particle a trillion light years away to be very close to zero...} that $\psi\to0$ and $\partial\psi/\partial x\to0$ as $x\to\pm\infty$ (and that they tend to $0$ fast enough to make any boundary terms vanish).
\item Does \zcref[S]{eq:pvp} imply that the expectation values of $x$ and $p$ obey Newton's laws? If so, prove this for a general $V\left(x\right)$. If not, find some $V\left(x\right)$ which provides a counterexample.
\end{problemparts}
\end{problem}

\begin{problem}\leavevmode
Recall \zcref[S]{eq:ham-en}, the time-independent Schrödinger equation, which is just the eigenvalue equation for the Hamiltonian:
\[
H\left|E_{i}\right\rangle =E_{i}\left|E_{i}\right\rangle ,
\]
where we assume that the energy eigenvalues are discrete. Let us denote the wavefunctions corresponding to the energy eigenstates as follows:
\[
\psi_{i}\left(x\right)\equiv\langle x|E_{i}\rangle.
\]
They don't depend on $t$, since we are assuming the Hamiltonian doesn't depend on $t$ either, and energy is constant. Show that (for a point particle with mass $m$) these wavefunctions satisfy the equation
\begin{equation}
\left(-\frac{1}{2m}\frac{\partial^{2}}{\partial x^{2}}+V\left(x\right)\right)\psi_{i}\left(x\right)=E_{i}\psi_{i}\left(x\right).\label{eq:time-ind}
\end{equation}
\end{problem}

\subsubsection{Separation of Variables}

Let us assume that the wavefunction can be separated into a part which depends only on $x$ and a part which depends only on $t$:
\[
\psi\left(t,x\right)=\psi_{i}\left(x\right)\psi_{t}\left(t\right).
\]
By plugging this into \zcref[S]{eq:Schr-pos}, the Schrödinger equation in the position basis, and dividing by $\psi$, we obtain the equation
\[
\frac{\i}{\psi_{t}}\frac{\d\psi_{t}}{\d t}=-\frac{1}{2m}\frac{1}{\psi_{i}}\frac{\partial^{2}\psi_{i}}{\partial x^{2}}+V\left(x\right).
\]
Since the left-hand side only depends on $t$ and the right-hand side only depends on $x$, we conclude that they must in fact both be \textbf{constant}, that is, independent of \textbf{both }$t$ and $x$ – otherwise, if for example the left-hand side was a function of $t$, then the right-hand side would have to be a function of $t$ also, in contradiction with our assumption that it only depends on $x$. This is called \emph{separation of variables}\index{Separation of variables}.

Let $E_{i}$ be the constant that both sides are equal to. Then we get two equations. The first equation will just be \zcref[S]{eq:time-ind}, the eigenvalue equation, which therefore implies that $E_{i}$ is the energy (and thus must be real). The other equation will be
\begin{equation}
\frac{\d\psi_{t}}{\d t}=-\i E_{i}\psi_{t}.\label{eq:time-dep}
\end{equation}
Recalling \zcref[S]{eq:exp-lambda-t}, we see that the solution to \zcref[S]{eq:time-dep} is simply
\[
\psi_{t}=\e^{-\i E_{i}t},
\]
up to a multiplicative constant, which can be absorbed into $\psi_{i}\left(x\right)$. Therefore, any \textbf{separable }solution to the Schrödinger equation is given by a wavefunction of the form
\[
\psi\left(t,x\right)=\psi_{i}\left(x\right)\e^{-\i E_{i}t}.
\]
These are called \emph{stationary states}\index{Stationary state}. Since these states are energy eigenstates, they have a well-defined energy $E_{i}$.

As it turns out, since the Schrödinger equation is linear and the energy eigenstates form a basis, the \textbf{most general solution} to the equation is a linear combination of stationary states:
\[
\psi\left(t,x\right)=\sum_{i}\alpha_{i}\psi_{i}\left(x\right)\e^{-\i E_{i}t},
\]
where $\alpha_{i}\in\BBC$ are constant coefficients and $E_{i}$ are all the possible energy eigenvalues, of which there can be infinitely many. If the energy eigenvalues are continuous rather than discrete, then the sum is replaced with an integral. Of course, this is nothing other than a \textbf{superposition }of energy eigenstates, represented in the position basis, and therefore the coefficients $\alpha_{i}$ are none other than the \textbf{probability amplitudes }to measure each energy $E_{i}$ given the state $\left|\Psi\left(0\right)\right\rangle $.

In other words, the general solution to the Schrödinger equation simply amounts to writing the state of the system as a superposition with respect to the eigenbasis of a particular observable – the Hamiltonian. With the time dependence out of the way, all that remains is to solve \zcref[S]{eq:time-ind}, the time-independent Schrödinger equation in the position basis, for $\psi_{i}$, and find the coefficients $\alpha_{i}$. The solution will depend on the explicit form of the potential $V\left(x\right)$. However, this is the hard part! Thousands upon thousands of pages have been written in the last 100 years or so about solutions (or even just approximations of solutions) to the Schrödinger equation for all kinds of different potentials. We will not work out any specific solutions in these notes, as that is beyond our scope.
\begin{problem}\leavevmode
Show that the probability density of a stationary state, as well as the expectation value of any time-independent observable $A$ with respect to that state, is independent of $t$.
\end{problem}

\begin{xca}\leavevmode
A wavefunction is given at time $t=0$ by
\[
\psi\left(0,x\right)=\alpha_{1}\psi_{1}\left(x\right)+\alpha_{2}\psi_{2}\left(x\right).
\]
What is the wavefunction $\psi\left(t,x\right)$ at some other time $t$, and what is the corresponding probability density?
\end{xca}

\subsection{Epilogue}

Unfortunately, our notes have now come to an end. I hope you had fun, and that the notes helped you develop a deep intuition for quantum theory, understand its most important concepts and consequences, and demystify common misconceptions.

\subsection{Further Reading}

The readers who successfully made it through these lecture notes, and wish to learn more about quantum theory, are invited to check out volume III of the lectures by Feynman \cite{feynman_online} for more intuition, the undergraduate-level quantum mechanics textbook by Griffiths and Schroeter \cite{griffiths_schroeter_2018} for more on continuous systems, and the quantum computation textbook by Nielsen and Chuang \cite{nielsen_chuang_2010} for more on quantum information and computing.

\printbibliography

@misc{feynman_online,
    author       = {Feynman, Richard P. and Leighton, Robert B. and Sands, Matthew},
    howpublished = {California Institute of Technology},
    note         = {Online edition},
    title        = {The {Feynman} Lectures on Physics},
    url          = {https://www.feynmanlectures.caltech.edu/},
}

@book{griffiths_schroeter_2018,
    author    = {Griffiths, David J. and Schroeter, Darrell F.},
    doi       = {10.1017/9781316995433},
    edition   = {3},
    place     = {Cambridge},
    publisher = {Cambridge University Press},
    title     = {Introduction to Quantum Mechanics},
    year      = {2018},
}

@book{nielsen_chuang_2010,
    author    = {Nielsen, Michael A. and Chuang, Isaac L.},
    doi       = {10.1017/CBO9780511976667},
    place     = {Cambridge},
    publisher = {Cambridge University Press},
    title     = {Quantum Computation and Quantum Information: 10th Anniversary Edition},
    year      = {2010},
}

\printindex
\end{document}